\documentclass[aps,pra,twocolumn,nofootinbib,superscriptaddress,longbibliography,10pt]{revtex4-2} 
\usepackage[T1]{fontenc}
\usepackage[utf8]{inputenc}
\usepackage[english]{babel}
\usepackage[dvipsnames]{xcolor}
\usepackage[ruled,vlined]{algorithm2e}
\usepackage{bm,graphicx,mathrsfs,amsmath,amssymb,mathtools,makecell,bbm, amsthm,dsfont,times,txfonts,nicefrac,framed,enumitem,tikz,physics,wrapfig,amsfonts,lipsum,nicematrix,wrapfig,halloweenmath,tabularray}
\usepackage{booktabs}
\usepackage{tabularx}
\usepackage[colorlinks=true,linkcolor=NavyBlue,citecolor=NavyBlue]{hyperref}
\usepackage[most]{tcolorbox}
\tcbuselibrary{breakable,skins}
\usepackage{pifont}
\usepackage{multirow}
\usepackage{etoolbox}
\usepackage{xargs}    

\makeatletter
\patchcmd{\@bibdataout@aps}
  {author="08"}
  {author="48"}
  {}{}
\makeatother
 
\makeatletter
\def\blfootnote{\gdef\@thefnmark{}\@footnotetext}
\makeatother

\renewcommand{\v}[1]{\ensuremath{\boldsymbol #1}}
\newcommand{\bracket}[3]{\langle #1 \lvert #2 \rvert #3 \rangle}

\newcommand{\drm}{\textrm{d}}

\RequirePackage{ifthen}
\RequirePackage{xkeyval}
\usetikzlibrary{positioning}
\RequirePackage{calc}
\newcommand{\iden}{\mathds{1}}

\hypersetup{urlcolor=NavyBlue}
\definecolor{equationcolor}{RGB}{222,94,100}
\definecolor{alecolor}{RGB}{198,113,190}

\hypersetup{urlcolor=NavyBlue}
\definecolor{equationcolor}{RGB}{222,94,100}
\definecolor{alecolor}{RGB}{198,113,190}
\definecolor{changescolor}{rgb}{0, 0, 0.7}
\newcommand{\alex}[1]{{\color{alecolor}#1}}

\definecolor{def_color_frame}{RGB}{199,215,231}
\colorlet{def_color_back}{def_color_frame!30}
\definecolor{def_color_text}{RGB}{81,106,124}

\definecolor{def_color_frame2}{RGB}{204,234,222}
\colorlet{def_color_back2}{def_color_frame2!30}
\definecolor{def_color_text2}{RGB}{109,124,114}

\definecolor{def_color_frame3}{RGB}{255,197,197}
\colorlet{def_color_back3}{def_color_frame3!30}
\definecolor{def_color_text3}{RGB}{165,96,99}

\newcommand{\NoEquationNumbersInBox}{%
  \let\originalendequation\endequation
  \def\endequation{\nonumber\originalendequation}%
  \let\originalendalign\endalign
  \def\endalign{\nonumber\originalendalign}%
}

\newtcolorbox[auto counter]{backgroundbox}[2][]{%
  breakable,
  sharp corners,
  toprule at break=0pt,
  bottomrule at break=0pt,
  colback=def_color_back3,
  colframe=def_color_frame3,
  fonttitle=\bfseries,
  coltitle=def_color_text3,
  title={Concept~\thetcbcounter: #2},
  before upper={\NoEquationNumbersInBox},
  #1
}

\newtcolorbox[auto counter]{examplebox}[2][]{%
  breakable,
  sharp corners,
  toprule at break=0pt,
  bottomrule at break=0pt,
  colback=def_color_back,
  colframe=def_color_frame,
  fonttitle=\bfseries,
  coltitle=def_color_text,
  title={Example~\thetcbcounter: #2},
  before upper={\NoEquationNumbersInBox},
  #1
}

\usepackage{hyperref}

\begin{document}

\title{Quantum prepare-and-measure scenarios: concepts and methods}

\author{A. de Oliveira Junior$^*$}
\affiliation{Center for Macroscopic Quantum States bigQ, Department of Physics,
Technical University of Denmark, Fysikvej 307, 2800 Kgs. Lyngby, Denmark}

\author{Santiago Zamora$^*$}
\affiliation{International Institute of Physics, Federal University of Rio Grande do Norte, 59078-970, Natal, RN, Brazil}
\affiliation{Department of Mathematics and Statistics, University of Ottawa, Ottawa, Ontario K1N 6N5, Canada}

\author{Jonatan Bohr Brask}
\affiliation{Center for Macroscopic Quantum States bigQ, Department of Physics,
Technical University of Denmark, Fysikvej 307, 2800 Kgs. Lyngby, Denmark}

\author{Rafael Chaves}
\affiliation{International Institute of Physics, Federal University of Rio Grande do Norte, 59078-970, Natal, RN, Brazil}

\author{Armin Tavakoli}
\affiliation{Physics Department and NanoLund, Lund University, Box 118, 22100 Lund, Sweden.}

\date{\today}

\begin{abstract}
Prepare-and-measure scenarios are among the simplest and most widely used settings for quantum communication. A sender prepares a physical system in one of several possible states, which a receiver then measures. Despite this simple structure, these scenarios offer a rich framework for exploring the differences between classical and quantum communication, revealing quantum advantages, and developing protocols for quantum cryptography and device certification. In this tutorial, we give a ground-up introduction to prepare-and-measure scenarios and the methods used to study them. We introduce different frameworks for constraining and characterising communication, with an emphasis on semi-device-independent approaches to quantum information. We also cover modern convex-optimisation techniques for characterising the correlations achievable in these scenarios and illustrate how these tools can be applied in practice. Our aim is to give newcomers the conceptual and technical foundations needed to approach the current literature and apply prepare-and-measure methods in their own research. We conclude with a discussion of open problems and directions for future work.
\end{abstract}

\maketitle
\tableofcontents

\newpage

\section{Introduction}\label{Sec:introduction}
\begin{NoHyper}
\blfootnote{$^*$\hspace{1pt}These authors contributed equally to this work.}
\end{NoHyper}

Quantum theory predicts correlation phenomena with no explanation within the confines of classical physics. The most famous and rigorous example of this is Bell's theorem~\cite{bell1964einstein}, which stipulates that the outcome statistics obtained by two observers who measure parts of an entangled state can elude any conceivable local hidden variable model~\cite{BrunnerReview2014,Scarani2019}. Bell nonlocality rules out local hidden-variable models under remarkably weak assumptions, namely measurement independence and space-like separation. This motivates the question of whether quantum correlations can defy classical models in more general physical settings. Such experiments could possibly be simpler than Bell inequality tests, and not even require entanglement.

The prepare-and-measure (PM) scenario is the most common setting for studying this question. In a PM scenario, the observers do not operate in isolation but instead actively influence one another. It features two observers: a sender named Alice and a receiver named Bob. Alice selects some data, $x$, which she encodes into a message that she sends to Bob. Bob's job is to read the message and provide an answer, labelled $b$, to a question, labelled $y$, about Alice's data. After the experiment has been repeated many times, its results can be described by a conditional probability distribution $p(b|x,y)$, i.e.~the probability that Bob answers $b$ given that he asked question $y$ and Alice held the data $x$. 

It is important to highlight early on that any form of genuinely quantum phenomenon that might arise in the PM scenario would require a stronger assumption than what is encountered in Bell nonlocality. This is a direct consequence of removing space-like separation from the scenario. If Alice were able to simply send $x$ as her message to Bob, then Bob would know both $x$ and $y$ and hence be able to generate any distribution $p(b|x,y)$. Clearly this leaves no room for observing quantum phenomena. Consequently, in order to meaningfully study PM scenarios, we must introduce an assumption on the communication between Alice and Bob. 

The question of how to choose the communication assumption in the PM scenario is an important one, and we will return to it later, as the answer depends on what one precisely aims to achieve. It is worth appreciating that since the PM scenario can be studied under a variety of communication assumptions, each with their own motivation, it becomes a versatile platform that can be adapted to many different physics purposes. Combined with its clear operational setting, this makes PM scenarios suitable for addressing several different aspects of quantum theory and quantum information science. The leading directions in which the PM scenarios are commonly studied are the following.
\begin{enumerate}
	\item \textbf{Fundamental physics:} why, how, and to what extent quantum theory eludes classical models. 
	
	\item \textbf{Quantum advantages:} boosting performance in communication tasks by exploiting quantum resources.
	
	\item \textbf{Certification of devices:} verifying, under mild assumptions, that quantum devices operate as intended.
	
	\item \textbf{Semi-device-independent protocols:} quantum cryptography that combines the practicality of conventional protocols with the principles of device-independence. 
\end{enumerate}

Let us outline, in a nutshell,   why these four directions all connect to the PM scenario. Quantum theory turns out to predict correlations in PM scenarios that have no classical model. This basic physics phenomenon is revealed in tests that, just like  Bell inequalities, most often have no truly appealing task-interpretation of their own. From the fundamental physics point of view, this is no issue. However, sometimes the direction of the reasoning can be reversed: we can start with an interesting communication problem involving Alice and Bob, and then show that quantum theory would allow it to be performed with a distinct advantage over classical implementations. Naturally, this is no longer motivated by understanding the physics itself, but by performing the task as well as possible. Moreover, as we will later see, an important feature of (sufficiently strong) quantum correlations is that they are only compatible with a rather limited set of physical realisations. This means that if we go to the lab and observe these correlations, we can deduce significant information about the inner workings of the quantum devices  used to generate them. This paves the way for using PM scenarios as tools for certifying the functionality of quantum devices. Finally, since the quantum correlations in PM scenarios arise under assumptions that are intended to be as mild as possible for communicating observers, it is natural to use them as workhorses for quantum cryptography. Protocols of this kind, most prominently quantum key distribution and quantum random number generation, are said to be semi-device-independent, because they export the idea of device-independence from Bell experiments to communication experiments. This comes with the price of having to accept the associated communication assumption (hence the word \emph{``semi''}) but can in return offer simpler setups and less demanding performance requirements.

While PM scenarios are found in the quantum physics literature dating back to the early days of quantum information science, the underlying prepare-and-measure idea is even more familiar. Textbook experiments such as Stern--Gerlach~\cite{Gerlach1989,grossi2023one}, the double slit and delayed-choice interferometry~\cite{Young1804,Wheeler1978,rossi2017restrictions,chaves2018causal} already follow the pattern of preparing a physical system and later probing it by a measurement. In quantum information science, the same pattern appears in several important protocols, such as quantum state discrimination~\cite{bae2015quantum} and quantum cryptography~\cite{Pirandola2020}. Since the introduction of dimension witnesses around 2010, these scenarios have been investigated systematically from a modern quantum information perspective across fundamental, applied and experimental physics. For an overview of the state of the art, we refer the reader to the recent review article~\cite{Brask2026}. 

The goal of this tutorial is to provide a ground-up and hands-on introduction to the study of quantum correlations in the PM scenario and its quantum information applications. It is therefore not meant to serve as a comprehensive review of the research topic, but as an accessible and useful entry point for researchers who want to understand PM scenarios or themselves become active in the topic. Therefore, the focus is placed on the topic's major themes. Central ideas, results, methods, and applications are discussed with sufficient detail to allow the reader to use them independently. The tutorial is structured as follows. 

We begin by introducing the prepare-and-measure scenario and its motivations (Sec.~\ref{Sec:introduction}), which set the stage for the rest of the tutorial. We then develop the core framework of dimension-restricted PM scenarios (Sec.~\ref{S:PM-setting}), forming the backbone of our discussion. From there, we turn to entanglement-assisted scenarios~(Sec.~\ref{sec:EAPM}), where shared quantum resources come into play. We then broaden the perspective by surveying alternative assumptions of PM scenarios (Sec.~\ref{sec:non-dimension_constraints}). Finally, we wrap up the main ideas discussed in the tutorial and outline open problems and future directions (Sec.~\ref{S:conclusions}). 

\begin{figure}[t]
    \centering
    \includegraphics{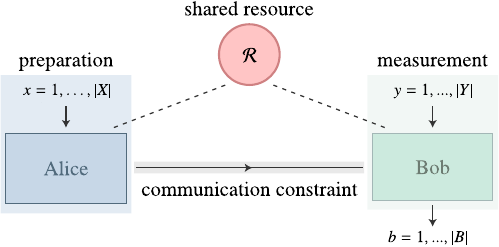}
    \caption{\emph{Schematic representation of a PM scenario.}
    Alice receives an input $x \in \{1, ..., |X|\}$ and prepares a message for Bob, possibly assisted by a shared resource $\mathcal{R}$. The communication is subject to a physical constraint, and may be either classical $\mathcal{C}$ or quantum $\mathcal{Q}$. Bob then chooses an input $y \in \{1, ..., |Y|\}$, performs his measurement, and produces an output $b\in \{1, ..., |B|\}$.}
    \label{F:PAM-setting}
\end{figure}

\subsection{Prepare-and-measure scenarios}\label{PMstructure}

We begin with a general description of the structure of PM scenarios, which we illustrate in Fig.~\ref{F:PAM-setting}. For clarity, we split it into three parts.

\textbf{Alphabets.---} Alice selects her input $x$ from a finite set $X$, which we identify with $\{1,2,\ldots,|X|\}$. Similarly, Bob selects his input $y$ from a set $Y = \{1,\ldots,|Y|\}$ and produces $b$ in $B=\{1,\ldots, |B|\}$. The cardinalities $|X|, |Y|$ and $|B|$ are the respective alphabet sizes. The tuple $(|X|, |Y|, |B|)$ specifies the number of preparation inputs, measurement inputs and outcomes. We will simply refer to this  as \emph{``the $(|X|,|Y|,|B|)$ scenario''}.

\textbf{Shared resource.---} Before the experiment begins, a shared resource can be distributed to Alice and Bob, which they may use to assist their communication.  A common example of such a resource is that Alice and Bob share a classical random variable, which they can use to stochastically coordinate their communication strategies. If we instead are interested in allowing Alice and Bob to share a quantum resource, then a natural choice is to let them each hold a share of an entangled state. Going further, one can even consider that Alice and Bob share resources that are not allowed in quantum theory. A prime example of this is a Bell nonlocal probability distribution that is required only to comply with the nonsignalling principle~\cite{popescu1994quantum}~(see Box~\ref{Box:PR}). Naturally, it is also an option to allow no shared resource at all between Alice and Bob. Over the course of this tutorial, we will encounter each of these resources and describe them in detail. For now, we simply denote the choice of shared resource by $\mathcal{R}$.

\begin{table}[t]
\centering
\renewcommand{\arraystretch}{1.15}
\setlength{\tabcolsep}{6pt}

\begin{tabular}{@{}lc@{}}
\toprule
\textbf{Assumption $\mathcal{A}$} 
& \textbf{Discussed in Sec.} \\
\midrule

Dimension bound 
& \ref{S:PM-setting} \\

Small high-dimensional components
& \hyperref[Sub:almost-dimension]{IV-B} \\

Energy restrictions 
& \hyperref[Sub:energy-restrictions]{IV-C} \\

Fidelity constraints 
& \hyperref[Sub:fidelity-constraints]{IV-D} \\

Overlap constraints 
& \hyperref[Sub:overlaps]{IV-E} \\

\makecell[l]{Spacetime/dynamical constraints} 
& \hyperref[Sub:spacetime-symmetries-speed-limits]{IV-F} \\

Informational restrictions 
& \hyperref[Sub:information-restriction]{IV-G} \\
\bottomrule
\end{tabular}
\caption{\emph{Examples of communication assumptions in PM scenarios.}
The symbol $\mathcal{A}$ denotes the physical restriction imposed on Alice's message.}
\label{Tab:communication-assumptions}
\end{table}
\textbf{Communication constraint.---} Lastly, we must specify the type of communication between Alice and Bob. This communication can be either classical or quantum. In the classical case, it means that Alice's messages have no superposition properties, i.e.~they are all jointly diagonalisable and can therefore, without loss of generality, be associated with distinguishable symbols $\{1,2,\ldots\}$. In the quantum case, Alice's messages are quantum states, i.e.~they are density matrices in an appropriate Hilbert space. We denote classical messages by $\mathcal{C}$ and quantum messages by $\mathcal{Q}$. Furthermore, as emphasised before, we must introduce a physical restriction on the communication, regardless of whether it is classical or quantum. The most commonly considered restriction is that the dimension of the messages is bounded~\cite{Gallego2010}. For classical systems, this means that the messages are selected from a set $\{1,\ldots,d\}$ and for quantum systems, it means that the messages are selected from the set of density matrices on a common $d$-dimensional Hilbert space. However, as we will discuss later, many other communication assumptions are possible (see Table~\ref{Tab:communication-assumptions}), and we will label the choice of assumption by $\mathcal{A}$.

In summary, to define a PM scenario we must specify its input/output alphabet, $(|X|,|Y|,|B|)$, the shared resource $\mathcal{R}$ between Alice and Bob, the physical limitation, $\mathcal{A}$, imposed on the communication, and lastly whether the communicated system is classical ($\mathcal{C}$) or quantum ($\mathcal{Q}$). 

\subsection{How to use this tutorial}
This tutorial can be read in different ways. Readers who are new to prepare-and-measure scenarios may benefit from following the text in order, since the early sections introduce the basic framework, examples, and methods that reappear throughout the tutorial. Readers who are already familiar with these basics may instead use the tutorial more selectively, moving directly to sections that best match their interests. 

With that in mind, we have tried to keep the main discussion reasonably self-contained while placing complementary material in separate boxes, each with its own colour code. Background boxes, shown in red, introduce concepts, tools, or definitions that clarify the discussion without interrupting the main flow. 
\begin{backgroundbox}{Concept A}
Background material is separated from the main discussion.
\end{backgroundbox}
We will also discuss examples, often through explicit calculations and simple derivations. These are highlighted in worked example boxes, shown in blue.
\begin{examplebox}{Calculation B}
Worked examples include explicit calculations.
\end{examplebox}

\section{Dimension-restricted scenarios}\label{S:PM-setting}

We begin our journey into PM scenarios by focusing on their most commonly studied form, namely when the communication is either a classical or quantum system restricted only by its dimension. Specifically, we consider PM scenarios in which Alice selects her input $x$, encodes it into a message $m\in\{1,\ldots,d\}$ that is sent to Bob, who decodes it with respect to his input $y$ to produce his output $b$. Here, the integer $d\geq 2$ is a limitation imposed on the message's dimension, which corresponds to the assumption $\mathcal{A}$ defined in Sec.~\hyperref[PMstructure]{I-A}. Since the dimension is now part of the specification of the scenario, we will write it as $(|X|, d, |Y|, |B|)$. In order to make sure the scenario is not trivial, we must  select Alice's input alphabet to be larger than the message's dimension, namely $|X|>d$. This is because otherwise she could send the input as the message ($m=x$) and Bob could simply read the value of $x$ and then sample his output from any distribution $p(b|x,y)$ that he desires\footnote{This applies to the standard PM scenario, where the message is unobserved and the data are $p(b|x,y)$. If the message is itself observed, as in outcome-communication \cite{brask2017bell} or instrumental-type scenarios \cite{chaves2018quantum,van2019quantum}, one instead studies distributions such as $p(m,b|x,y)$, and nonclassicality may arise even when $|X|=|M|$ (see the related outcome communication model in Box~\ref{Box:LHVcomm}).}.

Moreover, following the most common approach in the literature, we will select the shared resource between Alice and Bob to be a classical random variable, $\lambda$. We refer to it as shared randomness. This corresponds to the choice of $\mathcal{R}$ in Sec.~\hyperref[PMstructure]{I-A}. While this will be our main focus, we note that in Sec.~\hyperref[sec:NoSR]{II-E} we will discuss the PM scenario without any resource shared between Alice and Bob, and later in Sec.~\ref{sec:EAPM} we will also consider quantum resources shared between the parties.

\subsection{Classical models}\label{sec:ClassicalModels}
We start by discussing classical  models for the dimension-restricted PM scenario with shared randomness. In general, Alice can employ some arbitrary rule for encoding $x$ into her message $m$. Since, in principle, this rule could be probabilistic, we label it by the conditional probability distribution $p_A(m|x,\lambda)$. Note that since $\lambda$ is shared between the parties, Alice can use it when choosing her encoding. Similarly, when Bob receives $m$, he can use it together with his input, $y$, and his knowledge of $\lambda$ to (perhaps stochastically) decide his output $b$. This is represented by a decoding rule $p_B(b|m,y,\lambda)$. Since $\lambda$ could in principle be continuous and have any probability density, $q(\lambda)$, the conditional probability distribution observed between Alice and Bob becomes
\begin{equation}\label{PMdimclassical2}
p(b|x,y)=\sum_{m=1}^d \int \drm \lambda\ q(\lambda) p_A(m|x,\lambda) p_B(b|m,y,\lambda).
\end{equation}
We will refer to $p(b|x,y)$ as the correlations. The form of Eq.~\ref{PMdimclassical2} can also be understood as the Markov decomposition associated with the causal structure of a classical PM experiment (see Box~\ref{Box:DAG}).

\begin{backgroundbox}[label={Box:DAG}]{Classical causal structure of a PM scenario}
A useful way of seeing Eq.~\eqref{PMdimclassical2} is as the Markov decomposition of the causal graph underlying a classical PM experiment:
\begin{center}
    \includegraphics{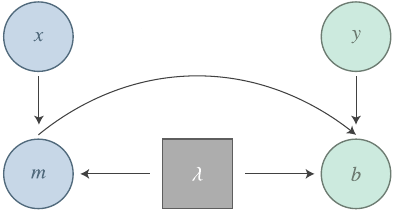}
\end{center}

\vspace{0.2cm}

Here $x$ and $y$ are freely chosen inputs, independent of each other and of the shared random variable $\lambda$. The variable $\lambda$ is distributed before the round begins and can influence Alice's encoding and Bob's output, while Alice's input reaches Bob only via the communicated message $m$, whose alphabet is restricted to $d$ possible values. Bob's output depends on $m$, $y$ and $\lambda$. The Markov decomposition of the joint distribution is therefore
\begin{equation*}
    p(x,y,\lambda,m,b) = p(x) p(y) q(\lambda)p_A(m|x,\lambda)p_B(b|m,y,\lambda).
\end{equation*}
Since the PM correlations are described by the conditional distribution $p(b|x,y)$, we condition on the chosen values of $x$ and $y$, sum over the unobserved message $m$, and integrate over the shared randomness $\lambda$. This gives Eq.~\eqref{PMdimclassical2}.
\end{backgroundbox}

Our task is to understand the set of classical correlations defined via Eq.~\eqref{PMdimclassical2}. To this end, we can make the useful observation that since $x$ can only take $|X|$ different values, and $m$ can only take $d$ different values, there are exactly $d^{|X|}$ different functions that deterministically map $x$ into $m$. Let us list them as $f_\mu$, with $\mu=1,\ldots,d^{|X|}$. Similarly, since $y$ takes $|Y|$ possible values and $b$ takes $|B|$ possible values, there exist precisely $|B|^{d|Y|}$ functions that map the pair $(m,y)$ into $b$. We list them as $g_\nu$ for $\nu=1,\ldots,|B|^{d|Y|}$. Let us now associate deterministic probability distributions to each of Alice and Bob's response functions. These are defined as
\begin{align}\nonumber\label{deterministicdistributions}
D_\mu(m|x)&= \delta(m=f_\mu(x)),\\
D_\nu(b|m,y)&=\delta(b=g_\nu(m,y)),
\end{align}
where $\delta(u)=1$ if the statement $u$ is true and $\delta(u)=0$ otherwise. Since for any given $\lambda$ the encoding and decoding rules of Alice and Bob can be viewed as convex combinations of deterministic encoding and decoding rules, we can write
\begin{align}\nonumber
p_A(m|x,\lambda)&= \sum_{\mu=1}^{d^{|X|}} q(\mu|\lambda) D_\mu(m|x),\\
p_B(b|m,y,\lambda)&=\sum_{\nu=1}^{|B|^{d|Y|}} q(\nu|\lambda) D_\nu(b|m,y).
\end{align}
Inserting this in Eq.~\eqref{PMdimclassical2}, we obtain
\begin{equation}
p(b|x,y)\!=\!\sum_{m,\mu,\nu}\int \drm \lambda\ q(\lambda) q(\mu|\lambda) q(\nu|\lambda)D_\mu(m|x) D_\nu(b|m,y).
\end{equation}
We may choose the auxiliary variables $\mu$ and $\nu$ independently conditional on $\lambda$. With this choice, we have
\begin{equation}
 q(\lambda)q(\mu|\lambda) q(\nu|\lambda)=q(\lambda)q(\mu,\nu|\lambda)=q(\mu,\nu,\lambda)   
\end{equation}
and hence the integral simply marginalises this distribution to $q(\mu,\nu)$. This leaves us with
\begin{align}\nonumber\label{PMdimclassical}
p(b|x,y)&=\sum_{m=1}^d\sum_{\mu=1}^{d^{|X|}} \sum_{\nu=1}^{|B|^{d|Y|}} q(\mu,\nu)D_\mu(m|x) D_\nu(b|m,y)\\\nonumber
& =\sum_{\lambda=1}^{d^{|X|}|B|^{d|Y|}} q(\lambda) \sum_{m=1}^{d} D_\lambda(m|x) D_\lambda(b|m,y)\\
& = \sum_{\lambda=1}^{d^{|X|}|B|^{d|Y|}}q(\lambda) p_\lambda(b|x,y),
\end{align}
where in the second line we have for simplicity switched labels by writing  $\lambda=(\mu,\nu)$ for the shared randomness. In the last line we have defined the deterministic probability distributions, $p_\lambda(b|x,y)=\sum_{m=1}^dD_\lambda(m|x) D_\lambda(b|m,y)$, associated with each choice of $\lambda$. This is our final and simplest general expression for the correlations attainable in a classical model for the PM scenario.

\subsubsection{The classical polytope}\label{sec:ClassicalPolytope}

Let us now better understand the result in Eq.~\eqref{PMdimclassical}. Firstly, it teaches us that we can without loss of generality restrict the shared randomness to a finite alphabet of size at most $d^{|X|}|B|^{d|Y|}$. 

\begin{figure}[t]
    \centering
    \includegraphics{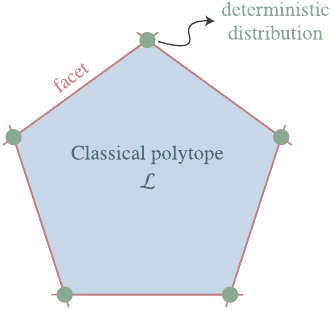}
    \caption{\emph{Classical polytope.} Toy illustration of the set of classical correlations in dimension-restricted PM scenarios with shared randomness. The set is a polytope characterised by the convex hull of the probability distributions associated with all deterministic encoding/decoding strategies for Alice and Bob.}
    \label{fig_classical_polytope}
\end{figure}

Secondly, it provides us with a geometric understanding of the space of classical correlations. Specifically, we can think about any correlation $p(b|x,y)$ as a list of $N=|X||Y||B|$ numbers\footnote{If we use that the probabilities are normalised, we can reduce it to $N=|X||Y|(|B|-1)$ numbers.}, i.e.~$p(b|x,y)$ is a point in the space $\mathbb{R}^N$. In particular, each deterministic distribution, $p_\lambda(b|x,y)$, is  a point in this space.  The last line in Eq.~\eqref{PMdimclassical} tells us that by varying $q(\lambda)$, we can realise any convex combination of these distributions. In other words, the space of classical correlations is the subset of $\mathbb{R}^N$ corresponding to the convex hull of the points associated with $\{p_\lambda(b|x,y)\}_{\lambda}$. Such an object is known as a polytope --- it can be thought of as a high-dimensional generalisation of a polyhedron --- and we will label it by $\mathcal{L}$. We therefore refer to $\mathcal{L}$ as the classical polytope and we provide a qualitative illustration of it in Fig.~\ref{fig_classical_polytope}.

Thirdly, if given an arbitrary correlation $p(b|x,y)$ and asked to determine if it admits a classical model (i.e.~whether $p\in\mathcal{L}$), we find the answer by solving the following linear program\footnote{Notice that the normalisation of the distribution $q(\lambda)$ is implicitly enforced by the last condition.},
\begin{align}\label{dimLP}
\text{find } & \quad \{q(\lambda)\}_{\lambda} \nonumber \\[1ex]
\text{s.t.} &\quad q(\lambda) \ge 0, \quad \forall\, \lambda,\nonumber\\
& \quad p(b|x,y) = \sum_{\lambda} q(\lambda)\, p_{\lambda}(b|x,y), \quad \forall\, b,x,y.
\end{align}
Linear programs (LPs) are an important class of problems in convex optimisation that can also be solved exactly with standard algorithms. We briefly introduce them in Box~\ref{Box:LP}. For our specific LP, given above in Eq.~\eqref{dimLP}, evaluating it will quickly become computationally expensive due to how rapidly the cardinality of $\lambda$ increases with the size of the scenario's alphabet. However, this cost can be significantly reduced. In Appendix~\ref{AppFastLP} we discuss a method introduced in Ref.~\cite{Renner2023}, which  allows us to obtain the solution in a more efficient manner. Furthermore, a general method to systematically characterise whether a set of prepared states has a classical model or not irrespective of the number of measurements has been introduced in Ref.~\cite{Goeisletter}.

\begin{backgroundbox}[label={Box:LP}]{Linear programs}
A linear program (LP) is the  maximisation (or minimisation) of a linear function subject to linear equalities and inequalities. It is in general written in the form
\begin{equation}
    \max_{\v x} \v c^\top \v x \:\:\text{subject to} \:\: A \v x \leq  \v b \:\:\text{and}\:\: \v x \geq 0.
\end{equation}
That is, our goal is to find a vector $\v x$, whose components are the real-valued variables, while $\v c$ and $\v b$ are given vectors and $A$ is a given matrix. The function being maximised, $\v c^\top \v x$, is called the objective function. The constraints $A\v x \leq \v b$ and $\v x \geq 0$ define a convex polyhedron over which the objective function is to be optimised. If bounded, this polyhedron is a polytope.

\medskip

The dual of the maximisation above is the following minimisation problem~(see Appendix~\ref{App:weak-and-strong-duality} for details)
\begin{equation}\label{LPdualdefinition}
    \min_{\v y} \v b^\top \v y \:\:\text{subject to} \:\: A^\top \v y \geq  \v c \:\:\text{and}\:\: \v y \geq 0.
\end{equation}
When both programs have finite optima, their optimal values agree by the theorem known as strong duality of LPs. For a proof, we refer the reader to the literature on convex optimisation~\cite{boyd2004convex}.

\medskip

LPs can be solved exactly. Most practically, they can be evaluated on computers (up to solver precision) with publicly available libraries such as CVX~\cite{CVX}, YALMIP~\cite{YALMIP}, JuMP~\cite{JuMP} and CVXPY~\cite{CVXPY}.
\end{backgroundbox}

There are two equivalent ways to describe a polytope: by its vertices, as in Eq.~\eqref{dimLP}, or by its facet inequalities together with its affine-hull equations. The facets of a polytope can loosely be thought of as its ``boundaries''. More specifically, each facet is an $(N-1)$-dimensional hyperplane in $\mathbbm{R}^N$ with the property that the intersection of all the half-spaces defined by these hyperplanes is the polytope itself (see Fig.~\ref{fig_classical_polytope}). In general, we can write a linear functional defining such a hyperplane in the form
\begin{equation}\label{linearWitness}
W\equiv \sum_{b,x,y} c_{bxy}p(b|x,y),
\end{equation}
for some real coefficients $c_{bxy}$. The hyperplane is given by $W=\beta$ for a real constant $\beta$. Converting vertex representation to facet representation is a well-known problem in computational geometry and several algorithms are available to  address it. However, for the classical correlation polytope there is an important additional consideration to keep in mind. Imagine that $W$ and $W'$ are two facets of $\mathcal{L}$, but they only differ by Alice and/or Bob individually relabelling their inputs and outputs. While these indeed are two distinct facets, their distinction is uninteresting to us because the choice of how to label the inputs and outputs will not affect the physics of the correlations. As a result, we are interested not in identifying all facets of $\mathcal{L}$ but only those that are inequivalent under such relabellings. The most naive way of dealing with this is to use a standard vertex-to-facet conversion method to  generate all the facets,  then group them into equivalence classes and select one element from each to obtain the complete list of non-equivalent  facets. However, this is computationally inefficient. A  better method is to use a library (e.g.~PANDA~\cite{PANDA2015}) that directly takes the equivalence classes into account, thereby circumventing the exhaustive facet list for $\mathcal{L}$ and returning directly the list of inequivalent facets.

\begin{examplebox}[label={Box:3213}]{Scenario $(3,2,1,3)$}
To determine the classical polytope in the $(3,2,1,3)$ scenario, we first enumerate all deterministic behaviours and then compute the facets of their convex hull.

\medskip

We start by defining $u_x:=p(1|x,1), v_x:=p(2|x,1)$ and consequently $p(3|x,1) = 1-u_x - v_x$, for $x=1,2, 3$. Thus $\mathcal{L} \subset \mathbbm{R}^6$ with coordinates $(u_1, v_1,u_2, v_2, u_3, v_3)$. Because Bob has only one input, a deterministic strategy simply assigns one output label to each of Alice's three inputs. Hence, a deterministic point can be represented as a triple $(o_1, o_2, o_3)\in\{1,2,3\}^3$, where $o_x$ is Bob's output when Alice uses input $x$. However, Alice sends only one bit. This means that she can create at most two distinct messages, which implies that Bob can produce at most two distinct output labels across the three inputs. Therefore, the allowed deterministic triples are exactly those that use at most two distinct symbols. The total number of such distinct deterministic vertices is
\begin{equation}
    3+\binom{3}{2}(2^3-2)=21.
\end{equation}
Indeed, there are 3 constant triples. For each unordered pair of outputs $\{o,o'\}\subset\{1,2,3\}$, there are $2^3-2 = 6$ non-constant assignments using only $o$ and $o'$. Thus, although there are $2^3 \cdot 3^2 = 72$ deterministic pairs $(f,g)$, they collapse to only $21$ distinct vertices in correlation space. 

\medskip

A vertex-to-facet conversion of these 21 vertices gives 15 facets. They can be grouped as follows:
\begin{itemize}
    \item \textbf{9 trivial facets}, namely the three simple constraints for each input $x$: $$u_x \geq 0, v_x \geq 0, u_x+v_x \leq 1$$
    \item \textbf{6 non-trivial facets}, one for each permutation $\pi \in S_3$:
\begin{equation}
    p(\pi(1)|1,1) + p(\pi(2)|2,1) + p(\pi(3)|3,1)\leq 2.
\end{equation}
\end{itemize}
Up to relabelling of Alice's inputs and Bob's outputs, these six inequalities are equivalent. Choosing the identity permutation recovers Eq.~\eqref{facet3213}.
\end{examplebox}

\subsubsection{Facets in simple scenarios}
Let us discuss some examples of the classical polytope for small alphabets. Consider that Alice sends one bit to Bob, i.e.~that $d=2$. Then we need at least $|X|=3$ inputs for Alice. Let us select $|Y|=1$ and $|B|=3$. The complete list of facets of $\mathcal{L}$ turns out to have 15 elements (see Box~\ref{Box:3213}), but once the equivalence classes associated with the relabellings of Alice and Bob are taken into account, we are left with just a single non-trivial\footnote{The polytope $\mathcal{L}$ has some facets that are trivial from a physical point of view. An example is $p(b|x,y)\geq 0$. Our presentation will ignore these.} facet. It is~\cite{Doolittle2021}
\begin{equation}\label{facet3213}
p(1|1,1)+p(2|2,1)+p(3|3,1)\leq 2.
\end{equation}
We see that this facet can be interpreted as a state discrimination task: Bob tries to guess Alice's input ($b=x$) and one third of the left-hand side is the average success probability for uniformly distributed inputs. By analogy, one can extend this state discrimination task to  scenarios with larger alphabets and obtain corresponding facet inequalities. This is shown in Box~\ref{Box:statediscriminationfacets}.

\begin{examplebox}[label={Box:statediscriminationfacets}]{Facets from state discrimination}
Consider that Alice has $n$ inputs that she encodes into a classical $d$-valued message, with $n>d$. From the message, Bob must guess Alice's input. We show that the sum of the success probabilities satisfies
\begin{equation}\label{mesd}
    \sum_{x=1}^n p(b=x|x) \leq d,
\end{equation}
and that this is a facet of the classical polytope in the scenario $(n,d,1,n)$ for $2\leq d < n$. A deterministic $d$-message strategy can guess at most $d$ distinct inputs correctly, which proves the inequality. Note that convex mixtures preserve it.

\medskip

Let us think of the probability distribution $p(b|x)$ as the $n\times n$ matrix whose entries are $P_{xb}=p(b|x)$. Accounting for normalisation, namely $\sum_b P_{xb}=1$, this matrix has $n(n-1)$ free variables. To show that the inequality is a facet, we therefore need to identify $n(n-1)$ classical probability distributions $P^{(\lambda)}$, for $\lambda=1,\ldots,n(n-1)$, such that i) every $P^{(\lambda)}$ saturates the inequality  $\sum_{x=1}^n p(b=x|x) = d$, and ii) all $P^{(\lambda)}$ are linearly independent as full matrices. Because every row is normalised, this is equivalent to affine independence here. The classical polytope is full-dimensional in the $n(n-1)$-dimensional normalised space for $d\geq2$. Appendix~\ref{App:ElnasConstruction} proves that the equality face has the required dimension and gives an explicit construction for $n=3$, $d=2$.
\end{examplebox}

However, a facet possessing such a natural task-interpretation is the exception to the rule. We can begin to see this by considering the scenario $(3,2,2,2)$. Up to relabellings, its only non-trivial facet is~\cite{Gallego2010}\footnote{Note that while we use 1-based indexing for these examples, we will frequently index alphabets starting from zero in subsequent sections.} 
\begin{equation}\label{Gallego3222}
p(1|1,1) +p(1|1,2)+p(1|2,1)+p(2|2,2)+p(2|3,1) \leq 4.
\end{equation}
If we were to force an interpretation here, we could say that Alice and Bob play a game in which they are awarded one point if Bob correctly computes the function $f:(x,y)\rightarrow \{1,2\}$ for $f(1,1)=f(1,2)=f(2,1)=1$ and $f(2,2)=f(3,1)=2$, but this is  rather artificial. It underlines the fact that optimal tests of classicality (i.e.~facets of $\mathcal{L}$) do not have to be useful from a communication-tasks point of view.

As a final example,  let us consider the scenario $(4,2,2,2)$. In this scenario, we can still view Eq.~\eqref{Gallego3222} as a facet --- we need only to ignore the fourth input of Alice. This is an example of a so-called lifting~\cite{pironio2005lifting}, i.e.~the possibility to embed the classical polytope for a given scenario into the classical polytope for  another scenario that has a larger alphabet. While lifted facets are entirely valid and necessary for a complete description of $\mathcal{L}$, they rarely offer new insights as compared to the scenario from which they were lifted. Fortunately, there is also another, genuinely new, facet appearing in the $(4,2,2,2)$ scenario. It is  
\begin{align}\nonumber\label{RACprime}
p(1|1,1) + p(1|1,2)+p(1|2,1)+p(2|2,2)\\
&\hspace{-5.47cm}+p(2|3,1)+p(1|3,2)+p(2|4,1)+p(2|4,2)\leq 6.
\end{align}
At first sight, this might look quite arbitrary. However, a clear-cut interpretation emerges from relabelling Alice's four inputs in terms of a pair of bits, namely $\{1,2,3,4\}\rightarrow \{00,01,10,11\}$, and relabelling Bob's outputs by the shift $\{1,2\}\rightarrow\{0,1\}$. With this relabelling, and a convenient normalisation, the inequality \eqref{RACprime} takes the form
\begin{equation}\label{RAC4222}
W_{\text{RAC}}\equiv\frac{1}{8}\sum_{x_1,x_2=0,1}\sum_{y=1,2} p(b=x_y|x,y)\leq \frac{3}{4}.
\end{equation}
We can interpret this as follows. Alice selects uniformly at random a pair of bits $x=(x_1,x_2)$, which she may encode into just a single bit that she sends to Bob. Bob selects uniformly at random whether he wants to learn the first bit or the second bit. This choice is labelled by $y$. His goal is then to output the value of the $y$th bit, namely $b=x_y$. The left-hand-side of Eq.~\eqref{RAC4222} represents the average success probability of succeeding with this guess. This task is known in the literature as a random access code (RAC) and the inequality \eqref{RAC4222} tells us that in a classical model it can never be performed with an average success probability that exceeds $\tfrac34$. Saturating this bound is simple. We can for example let Alice send the message $m=x_1$. Then, if $y=1$,  Bob correctly outputs $b=m$ whereas if $y=2$ he can make a random guess for the value of $x_2$, which succeeds with $50\%$. On average, the success probability becomes $75\%$.

\subsection{Quantum correlations}
Having familiarised ourselves with the correlations predicted by classical models in the PM scenario, we now ask how we can understand the correlations predicted by quantum theory. Specifically, we want to compare the classical models based on sending $d$-valued messages assisted by shared randomness to quantum models based on sending $d$-dimensional quantum messages assisted by shared randomness. In other words, we are upgrading the message from classical to quantum, but with the same dimensionality. Following Born's rule, such quantum correlations are written in the form
\begin{equation}\label{quantumModel}
    p(b|x,y)=\int \drm\lambda\  q(\lambda) \tr\left(\rho_x^{(\lambda)} M_{b|y}^{(\lambda)}\right),
\end{equation}
where $\rho^{(\lambda)}_x$ are  $d$-dimensional quantum states and $M_{b|y}^{(\lambda)}$ are quantum measurement operators. Formally, a quantum state $\rho$ is characterised by the conditions $\tr(\rho)=1$ and $\rho\succeq 0$, and a measurement $\{M_b\}_b$, known as a positive operator-valued measure (POVM), is characterised by  $\sum_b M_b=\mathds{1}$ and $M_b\succeq 0$. In Eq.~\eqref{quantumModel}, Alice uses $\lambda$ and her input $x$ to select a $d$-dimensional quantum  state $\rho_x^{(\lambda)}$ which she sends to Bob, who applies to it his $y$th measurement choice for the given value of $\lambda$, which is the POVM $\{M_{b|y}^{(\lambda)}\}_b$, and returns the result $b$ as his output. The fact that shared randomness is an allowed resource means that the set of quantum correlations, i.e.~all $p(b|x,y)$ that admit the model~\eqref{quantumModel}, is a convex set.

\begin{figure}[t]
    \centering
    \includegraphics{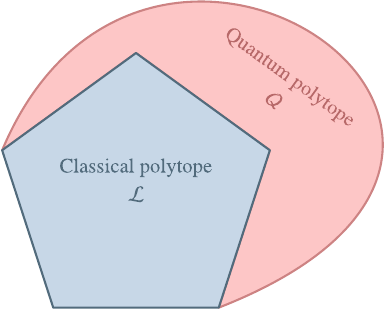}
    \caption{\emph{Classical and quantum correlation sets.}
    Schematic illustration of the classical set $\mathcal{L}$ and the quantum set $\mathcal{Q}$ in a PM scenario with shared randomness. The inclusion $\mathcal{L}\subseteq\mathcal{Q}$ follows because classical messages can be represented by quantum states diagonal in a common basis. Whether the inclusion is strict, as illustrated above, ultimately depends on the specific input-output scenario.}
    \label{F-classical-quantum-polytope}
\end{figure}

It is always possible to recover the classical correlation model from the quantum one. This amounts merely to selecting all states as being diagonal in the same basis. To see this, let us write $\rho_x^{(\lambda)}=\sum_{m} p_A(m|x,\lambda) \ketbra{e_m}{e_m}$, where $\{\ket{e_m}\}_{m=1}^d$ is some orthonormal basis. Inserting this in Eq.~\eqref{quantumModel}, we obtain
\begin{equation}
    p(b|x,y)=\sum_{m}\int \drm \lambda\ q(\lambda) p_A(m|x,\lambda) \bracket{e_m}{M_{b|y}^{(\lambda)}}{e_m}.
\end{equation}
This is precisely the classical model in Eq.~\eqref{PMdimclassical2}, once we realise that any conditional distribution $p_B(b|m,y,\lambda)$ may be written in the form  $p_B(b|m,y,\lambda)=\bracket{e_m}{M_{b|y}^{(\lambda)}}{e_m}$. Hence, the classical set $\mathcal{L}$ is contained in the quantum set, which we denote by $\mathcal{Q}$ (see Fig.~\ref{F-classical-quantum-polytope}). What is, however,  less clear is whether there exist correlations that admit a quantum model but have no classical model. We now set out to explore the answer to this question.

\subsubsection{The Frenkel--Weiner theorem}\label{sec:FWtheorem}

We found in Eq.~\eqref{facet3213} that in the simplest non-trivial scenario, corresponding to $(3,2,1,3)$, the classical polytope only has one interesting facet and that it corresponds to a state discrimination task. Moreover, in Box~\ref{Box:statediscriminationfacets} we showed that for any scenario of the form $(n,d,1,n)$ with $n>d$ we can obtain the inequality $ \sum_{x=1}^n p(b=x|x) \leq d$, which is a state discrimination task for $n$ states, as a facet of the classical polytope. It therefore appears natural to ask whether any of these facet inequalities can be violated in quantum theory. Perhaps surprisingly, the answer is negative, i.e.~all these optimal tests of classical correlations are also satisfied in quantum theory. Proving this is unexpectedly straightforward. Consider the following chain of inequalities
\begin{align}\nonumber\label{mesdquantum}
W&=\sum_{x=1}^{n} p(b=x|x)=\sum_{x=1}^{n} \tr\left(\rho_x M_x\right) \\
& \leq \sum_{x=1}^{n} \lambda_{\max}(M_x)\leq \sum_{x=1}^{n} \tr(M_x)=d, 
\end{align}
where for the first inequality we have used that an optimal $\rho_x$ can be chosen as an eigenstate of $M_x$ that corresponds to its largest eigenvalue, $\lambda_{\max}$. In the second inequality we have used that $\lambda_{\max}(A)\leq \tr(A)$ for any $A\succeq0$, and in the final step we have used the definition of POVMs. The argument applies at each fixed value of $\lambda$. Averaging over shared randomness preserves the bound.

This lack of nonclassicality is not an accident. It is, in fact, a consequence of a cornerstone result known as the Frenkel-Weiner theorem~\cite{Frenkel2015}. The content of this theorem is the following.
\begin{quote}
\textit{In any dimension-restricted PM scenario where shared randomness is allowed and $|Y|=1$, all quantum correlations admit a classical model.}
\end{quote}
While the theorem holds for every $d$, we  give a construction for qubits, namely $d=2$, following the construction found by M.~J.~Renner~\cite{Renner2026private}. It suffices to treat pure preparations and rank-one POVMs: mixed states can be obtained by local randomisation, while a general POVM can be refined into rank-one outcomes and then coarse-grained.

Let us write a pure qubit prepared by Alice as \mbox{$\rho=\frac{1}{2}\left(\mathds{1}+\vec{x}\cdot\vec{\sigma}\right)$} and Bob's rank-one measurement operators as $M_b=y_b\left(\mathds{1}+\vec{y}_b\cdot\vec{\sigma}\right)$. To form a valid POVM of this form, we require that $0\leq y_b\leq \frac{1}{2}$, $\sum_b y_b=1$ and $\sum_b y_b \vec{y}_b=0$, with $\|\vec x\|=\|\vec y_b\|=1$ for every nonzero effect. From Born's rule, we find that the correlations become
\begin{equation}\label{qcorrFW}
p(b|\vec{x},\{y_b,\vec{y}_b\})=\tr\left(\rho M_b\right)=y_b \left(1+\vec{x}\cdot \vec{y}_b\right).
\end{equation}
Our job is to reproduce these correlations with a classical model. To this end, let us define the Heaviside function $H(z)=1$ for $z>0$ and $H(z)=0$ otherwise, and also $\Theta(z)=zH(z)$. Alice and Bob now proceed as follows. 
\begin{enumerate}
	\item They share a random variable $\vec{\lambda}\in\mathbb{R}^3$ that is distributed over the unit-radius sphere with respect to its surface measure of total area $4\pi$ according to
	\begin{equation}
	q(\vec{\lambda})=\frac{1}{N}\sum_b y_b \Theta(\vec{y}_b\cdot \vec{\lambda}),
	\end{equation}
    where the normalisation is $N=\int \drm\vec{\lambda}\ \Theta(\vec{y}_b\cdot \vec{\lambda}) = \pi$, which is independent of $b$ for unit $\vec{y}_b$. This ensures that $q(\vec{\lambda})\geq0 $ and $\int \drm\vec{\lambda}\ q(\vec{\lambda})=1$, so we conclude that $q(\vec{\lambda})$ is a valid probability density.
	
	\item Alice encodes her state into the bit $m\!=\!H(\vec{x}\cdot \vec{\lambda})$ and sends it to Bob. This means choosing 
    \begin{equation}
        p_A(m|\vec{x},\vec\lambda)=\delta(m=H(\vec{x}\cdot \vec{\lambda})).
    \end{equation}
	
	\item Bob reads the message: if $m=1$ he keeps $\vec{\lambda}$ whereas if $m=0$ he flips its direction. Thus, he can define $\vec{\lambda}_m=(-1)^{m+1}\vec{\lambda}$.  Then, he determines his output by sampling from  the  distribution
	\begin{equation}
	p_B(b|m,\{y_b,\vec{y}_b\},\vec\lambda)=\frac{y_b\Theta(\vec{y}_b\cdot\vec{\lambda}_m)}{\sum_j y_j\Theta(\vec{y}_j\cdot \vec{\lambda}_m)}.
	\end{equation}
\end{enumerate}
	Using that $q(\vec{\lambda})=q(-\vec{\lambda})$, which follows from $\Theta(z)-\Theta(-z)=z$ and $\sum_b y_b \vec y_b=0,$ this protocol gives us classical correlations of the form
\begin{equation}
p(b|\vec{x},\{y_b,\vec{y}_b\})=\frac{2y_b}{N}\int \drm\vec{\lambda}\ H(\vec{x}\cdot\vec{\lambda}) \Theta(\vec{y}_b\cdot \vec{\lambda}).
\end{equation}
Calculating this integral reveals that these are precisely the quantum correlations in Eq.~\eqref{qcorrFW}.

The restriction to a fixed measurement is important. A related but distinct question is whether a single classical protocol, using shared randomness independent of the inputs, can reproduce the statistics of all preparations and measurements in a given quantum dimension. Recent work by de Gois \textit{et al.}~\cite{deGois2026infinite} shows that the answer depends on the dimension. For qutrit communication, an exact simulation is possible using at most $357$ classical bits, whereas for $d\geq4$ no finite classical message alphabet suffices. The latter result holds even for pure ququart preparations and binary projective measurements with a rank-one effect. These results concern a uniform simulation of the entire family of preparations and measurements. Conversely, for any fixed finite preparation alphabet, Alice can communicate the preparation label, so a finite classical simulation always exists.

\subsubsection{Quantum violations}\label{sec:Qviolations}
In order to find quantum correlations without a classical model, we must look to scenarios where Bob has an input, i.e.~$|Y|\geq 2$. The simplest such scenario corresponds to $(3,2,2,2)$ and in Eq.~\eqref{Gallego3222} we identified the only relevant facet inequality. Thus, we ask ourselves whether quantum theory can violate that inequality. The answer turns out to be affirmative \cite{Gallego2010}. 

To this end, consider that Alice's qubit states \mbox{$\rho_x=\frac{1}{2}\left(\mathds{1}+\vec{n}_x\cdot\vec{\sigma}\right)$} are selected so that their Bloch vectors form the following isosceles triangle,
\begin{align}\label{eq:optimal_states_gallego}
\vec{n}_0\!=\!\frac{1}{\sqrt{2}}\left(1,0,1\right), \: \vec{n}_1\!=\!\frac{1}{\sqrt{2}}\left(1,0,-1\right),\: \vec{n}_2\!=\!-\left(1,0,0\right).
\end{align}
Then, we select Bob's first measurement as the observable $\sigma_X$ and his second measurement as the observable $\sigma_Z$, associating the outcome $b=0$ with the eigenvalue $+1$. With these choices, the left-hand-side of Eq.~\eqref{Gallego3222} becomes $3+\sqrt{2}$, which evidently violates the inequality. The first experimental demonstration of the violation of this inequality, using both the polarisation and orbital angular momentum of entangled photons, was reported in \cite{Hendrych2012,Ahrens2012}.

In a similar spirit, we can also find quantum violations of the facet inequality in Eq.~\eqref{RAC4222}. An elegant violation is obtained as follows~\cite{Ambainis1999, Ambainis2002}. Select the Bloch vectors of Alice's four different qubit states so that they form a rotated square in the $XZ$-disc of the Bloch sphere, i.e.~
\begin{align}\nonumber\label{RACvectors}
	&\vec{n}_{00}=\frac{1}{\sqrt{2}}\left(1,0,1\right), && \vec{n}_{01}=\frac{1}{\sqrt{2}}\left(1,0,-1\right), \\
	& \vec{n}_{10}=\frac{1}{\sqrt{2}}\left(-1,0,1\right), && \vec{n}_{11}=\frac{1}{\sqrt{2}}\left(-1,0,-1\right).
\end{align}
Then, we choose the two measurements of Bob as the observables $\sigma_X$ and $\sigma_Z$, i.e.~their Bloch vectors are $\vec{m}_1=(1,0,0)$ and $\vec{m}_2=(0,0,1)$. A direct calculation gives that
\begin{eqnarray}
    p(b=x_y|x,y)=\frac{1}{2}(1+(-1)^{x_y}\vec{n}_x\cdot \vec{m}_y)=\frac{1}{2}\left(1+\frac{1}{\sqrt{2}}\right).
\end{eqnarray}
Hence, we have found that the left-hand-side of Eq.~\eqref{RAC4222} is  $\frac{1}{2}\left(1+\frac{1}{\sqrt{2}}\right)\approx 0.8536$, which violates the inequality. The first experimental demonstration of this quantum random access code was reported in Ref.~\cite{Spekkens2009} and used polarisation qubits to achieve the value $W_\text{RAC}\approx 0.852$. Later, other demonstrations have followed (see e.g.~\cite{Ahrens2014, Ribezzo2024, Karthik2025}). 

\subsubsection{Quantum random access codes}\label{sec:QRAC}
Given the task-interpretation of the random access code, it is natural to extend it to larger alphabets. In Eq.~\eqref{RAC4222} we considered that Alice has two bits and Bob wants to learn one of them, while Alice may only communicate one bit to him. An appealing generalisation is to let  Alice have $n$ different $d$-valued symbols, i.e.~$x=(x_1,\ldots,x_n)\in\{0,\ldots,d-1\}^n$, which she must encode into a $d$-valued message, from which Bob wants to recover the $y$th element, namely $b=x_y$. Such $(n,d)$ random access codes were proposed in~\cite{Tavakoli2015}. Their average success probability is 
\begin{equation}\label{RACwitness}
W_{n,d}\equiv \frac{1}{nd^n} \sum_{x=x_1\ldots x_n}\sum_{y=1}^n p(b=x_y|x,y).
\end{equation}
Notice that we recover the left-hand-side of Eq.~\eqref{RAC4222} by selecting $(n,d)=(2,2)$. However, for general $(n,d)$, it is less straightforward to determine the largest value of $W_{n,d}$ achievable in classical models. It turns out that the optimal strategy is remarkably simple: Alice selects the most frequently occurring  symbol in the string $x=(x_1,\ldots,x_n)$ and sends it to Bob, who simply outputs this message as his value of $b$ independently of $y$~\cite{Tavakoli2015}. The proof of this optimality and the resulting value of $W_{n,d}$ can be found in Refs.~\cite{Ambainis2024, Saha2023}. For instance, when $n=2$ one finds
\begin{equation}\label{RACdits}
    W_{2,d}\leq \frac{1}{2}\left(1+\frac{1}{d}\right),
\end{equation}
and when $d=2$ one finds
\begin{equation}\label{RACbits}
    W_{n,2}\leq \frac{1}{2}+\frac{1}{2^n}\binom{n-1}{\lfloor \frac{n-1}{2}\rfloor}.
\end{equation}

Our focus is on the quantum advantages in the random access codes. As a first step beyond the $(2,2)$ code, we can consider the $(3,2)$ code, where Eq.~\eqref{RACbits} gives  $W_{3,2}\leq \frac{3}{4}$. Due to the geometry of the Bloch sphere, it is straightforward to generalise the quantum violation for the $(2,2)$ case:  select Alice's eight Bloch vectors so that they form the vertices of a cube, namely $\vec{n}_x=\frac{1}{\sqrt{3}}(\pm1,\pm1,\pm1)$, and select Bob's measurements as $\sigma_X$, $\sigma_Y$ and $\sigma_Z$.  We then find that $p(b=x_y|x,y)\!=\!\frac{1}{2}\left(1+\frac{1}{\sqrt{3}}\right)\!\approx\! 0.7887$ for all $x,y$ and the same value for $W_{3,2}$. This exceeds the classical limit. 

However, a moment of thought reveals that there is no straightforward way to further extend this to the natural next case of $(n,d)=(4,2)$. The reason is that we simply run out of dimensions --- we cannot fit a four-dimensional cube into the Bloch sphere. In fact, we face the same problem for any $n>3$ when $d=2$. The consequence is that one must find the quantum violation on a case-by-case basis, with several of these cases discussed in detail in Ref.~\cite{Ambainis2008}. For example, for $n=4$, the optimal classical value is $W_{4,2}=\frac{11}{16}=0.6875$, while the quantum construction gives $W_{4,2}=\frac{1}{2}+\frac{1+\sqrt{3}}{8\sqrt{2}}\approx0.7415$.

Instead of diving into the zoo of constructing quantum random access codes, let us show that there exists a simple upper bound on $W_{n,2}$ which is respected by all quantum strategies. We write Alice's qubits as $\rho_x=\frac{1}{2}\left(\mathds{1}+\vec{n}_x\cdot \vec{\sigma}\right)$ and Bob's observables as $\vec{m}_y\cdot \vec{\sigma}$. The corresponding probabilities are then given by $p(b|x,y)=\frac{1}{2}\left(1+(-1)^b\vec{n}_x\cdot\vec{m}_y\right)$. Therefore, we have
\begin{align}\nonumber\label{RACcalculation}
    W_{n,2}&=\frac{1}{n 2^n}\sum_{x,y}\frac{1}{2}\left(1+(-1)^{x_y}\vec{n}_x\cdot\vec{m}_y\right)\\\nonumber
    &=\frac{1}{2}+\frac{1}{n2^{n+1}} \sum_{x} \vec{n}_x \cdot\sum_y (-1)^{x_y}\vec{m}_y\\ 
    &\leq \frac{1}{2}+\frac{1}{n2^{n+1}} \sum_{x} \big\lVert \sum_y (-1)^{x_y}\vec{m}_y\big\rVert,
\end{align}
where in the last step we have maximised the value of $W_{n,2}$ by selecting $\vec{n}_x$ as a unit vector aligned with $\sum_y (-1)^{x_y}\vec{m}_y$ whenever this vector is non-zero. When it vanishes, any state is optimal for that input. We now bound the right-hand-side by  expanding the norm and using Cauchy--Schwarz in the form $\sum_{x} \sqrt{a_x}\leq 2^{\tfrac{n}{2}}\sqrt{\sum_{x} a_x}$. This gives
\begin{align}\nonumber\label{RACcalculation2}
\sum_{x} \big\lVert \sum_y (-1)^{x_y}\vec{m}_y\big\rVert&=\sum_x \sqrt{\sum_{y,y'}(-1)^{x_y+x_{y'}}\vec{m}_y\cdot \vec{m}_{y'}}\\\nonumber
&\leq \sqrt{2^n}\sqrt{\sum_{y,y'}\vec{m}_y\cdot \vec{m}_{y'}\sum_x(-1)^{x_y+x_{y'}}}\\
&=2^n \sqrt{\sum_y \lVert \vec{m}_y\rVert^2}\leq 2^n\sqrt{n},
\end{align}
where we have used that $\sum_x(-1)^{x_y+x_{y'}}=2^n\delta_{y,y'}$. Inserting this into Eq.~\eqref{RACcalculation}, we arrive at the bound
\begin{equation}\label{RACbound}
    W_{n,2}\leq \frac{1}{2}\left(1+\frac{1}{\sqrt{n}}\right).
\end{equation}
This holds for all quantum models and, for nontrivial codes with $n\geq 2$, is tight only for $n=2,3$. Equality requires unit-length, pairwise orthogonal measurement Bloch vectors. At most three such vectors exist in $\mathbbm{R}^3$. The trivial case $n=1$ also saturates the bound. This means that both classical and quantum strategies are, for large $n$ and $d=2$, ultimately limited by a scaling in $W_{n,2}$ which converges to $\tfrac12$ as $O(n^{-\nicefrac12})$.

\subsubsection{Limitations of qubit advantage}\label{sec:limitations-qubit-advantage}

The Frenkel-Weiner theorem~[Sec.~\hyperref[sec:FWtheorem]{II-B-1}] tells us that Bob must have several measurements for a quantum advantage to exist, but it does not tell us how large that violation can be and how robust it is to imperfections. For the qubit case with binary outcomes, a quantification of the size and robustness of the advantage emerges from a connection between the PM scenario and the Grothendieck constant~\cite{Grothendieck1952}.

Consider the scenario $(n, 2, m, 2)$, i.e.~$n$ preparations, $m$ binary measurements with a qubit message. Since outcomes are dichotomic, it is convenient to work with the expectation values $E_{x,y}=p(+1|x,y)-p(-1|x,y)$. For pure qubit states with Bloch vectors $\vec{a}_x$ and rank-one projective measurements with Bloch vectors $\vec{b}_y$, Born's rule gives $E_{x,y} = \vec{a}_x\cdot \vec{b}_y$. Any linear witness takes the form $W=\sum_{x,y} M_{x,y}E_{x,y}$ for a real $n\times m$ matrix $M$, and the largest value attainable with these rank-one qubit measurements  is
\begin{equation}\label{qMgroth}
    Q(M) = \max_{\vec{a}_x, \vec{b}_y} \sum_{x,y}M_{x,y} \vec{a}_x \cdot \vec{b}_y.
\end{equation}
For fixed preparations, an optimal $\vec{b}_y$ is a unit vector along $\sum_x M_{x,y}\vec{a}_x$ when this vector is nonzero; otherwise, any unit vector is optimal. The classical bit benchmark must allow two independent deterministic outputs, $u_y, v_y \in\{-1,1\}$, for each setting, one for each received message. Encoding $x$ as a sign $a_x$ gives $E_{x,y}=\tfrac{1}{2}[(1+a_x) u_y + (1-a_x)v_y]$. Maximising over $u_y$, $v_y$ gives
\begin{equation}
    C(M)=\max_{a_x=\pm 1}\bigg(\big\lVert \sum_{x:a_x=+1} M_x \big\rVert_1 + \big\lVert \sum_{x:a_x=-1} M_x \big\rVert_1\bigg),
\end{equation}
where $M_x$ is the $x$th row of $M$ and $\lVert \cdot \rVert_1$ is the vector $\ell_1$ norm. One can show that $C(M)$ is itself a matrix norm.

An appealing quantity is the ratio between the quantum and classical values, and in particular its largest value considered over all possible choices of $M$. That is,
\begin{equation}\label{KPM}
 K_\text{PM}=\sup_{M\neq0} \frac{Q(M)}{C(M)}.   
\end{equation}
One can show that this constant equals the Grothendieck constant of order three~\cite{Divianszky2023}. This constant appears in the analysis of local hidden variable models for two-qubit Werner states in Bell experiments \cite{Acin2006G}. Its value is not exactly known but it has been  confined to the interval $1.4367\leq K_\text{PM}\leq 1.4546$ \cite{Designolle2023}. The coincidence of the same constant in Bell and PM settings is not accidental. It captures a correspondence between one bit of classical communication with shared randomness and local hidden-variable models, made explicit in Box~\ref{Box:GisinGisin}. Physically,  $K_{\text{PM}}$ can be interpreted as determining the critical amount of noise that qubit states can tolerate before the experiment generates classical correlations. Specifically, Alice's states are passed through the depolarisation map $\rho\rightarrow v \rho+\frac{1-v}{2}\mathds{1}$ and the critical value is given by $v=\frac{1}{K_\text{PM}}$. This means that for  $ v\leq 0.687$ qubit correlations are always classical under projective measurements whereas for $v> 0.696$ quantum correlations are possible. 

A closely related constant plays the analogous role for detection inefficiency, quantifying robustness against losses instead of noise. Suppose Bob's detector registers only with probability $\eta$, and that on a no-click event he outputs a fixed result $+1$. The expectation value then becomes $E_{x,y}(\eta) = \eta E_{x,y}+(1-\eta)$, and the relevant ratio is
\begin{eqnarray}\label{KD}
 K_D = \sup_{M:C(M)>S(M)} \frac{Q(M)-S(M)}{C(M)-S(M)},
\end{eqnarray}
where $S(M) = \sum_{x,y} M_{x,y}$, with critical efficiency $\eta_{\text{crit}} = \tfrac{1}{K_D}$. Since the offset $S(M)$ can only help, $K_D \geq K_{\text{PM}}$. Its exact value remains open. A numerical search using a $70\times 70$ witness establishes  $1.5682\leq K_D \leq 2$. Consequently, $\eta_{\text{crit}} \leq 0.6377$ for the symmetric detection efficiency of qubit measurements in this scenario~\cite{Divianszky2023}. For a broader overview of detection efficiency thresholds and how these bounds are established across various (semi-)device-independent scenarios, we refer the reader to the review in Ref.~\cite{Sarubi2026}. 

\begin{backgroundbox}[label={Box:GisinGisin}]{One bit of communication as a  local model}

The identification of $K_\text{PM}$ and the Grothendieck constant of order three rests on a correspondence between the relevant classical-bit and Bell-local witness bounds. A related one-bit construction is via the Gisin-Gisin protocol~\cite{Gisin1999}.

\medskip

Alice and Bob share a unit vector $\vec{\lambda}$, drawn uniformly. Given a preparation Bloch vector $\vec{a}$, Alice sends the one bit $c=\text{sgn}(\vec{a}\cdot\vec{\lambda})$. Bob, holding a measurement vector $\vec{b}$, outputs $b=\text{sgn}(c\,\vec{b}\cdot\vec{\lambda})$ with probability $|\vec{b}\cdot\vec{\lambda}|$, and otherwise outputs a fixed symbol that is coarse-grained into one of the two outcomes (here $+1$). Integrating over the sphere gives the expectation value $E(\vec{a},\vec{b})=\tfrac{\vec{a}\cdot\vec{b}+1}{2}.$
    
\medskip
Thus a single transmitted bit reproduces the qubit correlations $\vec{a}\cdot\vec{b}$ up to a shift and rescaling. It is this shift, captured by the term $S(M)$, that separates the noise constant $K_{\text{PM}}$ from the efficiency constant $K_D$.
\end{backgroundbox}

\subsubsection{Seesaw method}\label{sec:seesaw}
Our  examples of quantum violations have so far been based on explicit choices of states and measurements, but how does one know how to choose them in the first place? Sometimes this can be answered analytically, but more often we need numerical tools. Here we present a simple heuristic method to find quantum violations~\cite{Werner2001}. 

Consider that we are given some arbitrary linear correlation quantity $W$, as defined in Eq.~\eqref{linearWitness}. How can we systematically search for quantum strategies to maximise its value? An important observation is that since $W$ is linear, its optimal value can be achieved without shared randomness. This means that we can safely ignore $\lambda$ and consider only correlations of the form $p(b|x,y)=\tr(\rho_xM_{b|y})$. Still, it is not clear how to best choose $\{\rho_x\}_x$ and $\{M_{b|y}\}_{b,y}$. 

On the one hand, consider that we are given some arbitrary measurements. The best choice of $\rho_x$ can then be determined as follows
\begin{align}\label{maxeig}
    \max_{\{\rho_x\}} W &= \max_{\{\rho_x\}} \sum_x \tr\left(\rho_x \sum_{b,y} c_{bxy}M_{b|y}\right) \nonumber\\&=\sum_x \lambda_{\max}\left(\sum_{b,y} c_{bxy}M_{b|y}\right).
\end{align}
In other words, the optimal choice of $\rho_x$ is the rank-one projector onto an eigenvector of the operator $\sum_{b,y}c_{bxy}M_{b|y}$ associated with its largest eigenvalue. Thus, this optimisation is just a largest-eigenvalue calculation. 

On the other hand, consider that we are given some arbitrary states. The best choice of measurements means solving the optimisation problem
\begin{align}\label{measurementSDP} \nonumber
\max_{\{M_{b|y}\}}\quad &\sum_{b,x,y} c_{bxy}\tr(\rho_x M_{b|y})\\ \nonumber
\text{s.t.}\quad &\sum_b M_{b|y}=\mathds{1} \quad \forall y\\
&M_{b|y}\succeq0 \hspace{1cm}\forall b,y.
\end{align}
This is an instance of a so-called semidefinite program (SDP). SDPs are a class of convex optimisation problems which are particularly important for quantum information science. In Box~\ref{Box:SDP} we briefly introduce them. For a thorough introduction to SDPs in quantum information science, we refer the reader to the textbook~\cite{SkrzypczykCavalcanti2023}.

Since the two optimisation problems are individually solvable, we can combine them iteratively to search for the maximal value of $W$. This is known as a seesaw heuristic, or more formally as an alternating convex search. 
\begin{algorithm}
	\caption{Seesaw search}
	\label{algorithm_seesaw}
    \begin{enumerate}[label=\textbf{Input:}]
        \item Coefficients $\{c_{bxy}\}$ and random choice of $\{M_{b|y}\}$
    \end{enumerate}
	\begin{enumerate}[label=\emph{Step \arabic*.}]
	    \item Solve largest-eigenvalue problem in Eq.~\eqref{maxeig} to obtain optimal states and a value $W$.
        \item Using the states in \textit{step} 1, solve the SDP in Eq.~\eqref{measurementSDP} to obtain optimal measurements and an updated value of $W$.
        \item Iterate \textit{steps} 1 and 2 until desired convergence in the value of $W$.
	\end{enumerate}
\end{algorithm}
There is no guarantee that the seesaw heuristic actually returns the largest value of $W$ possible with quantum states and measurements. Its result can often depend strongly on the initial random choice of $\{M_{b|y}\}$ and in practice delivers 
a lower bound on the optimal quantum value. Nevertheless, it is often a powerful tool that, perhaps surprisingly often, ends up returning an optimal or near-optimal value of $W$. Since it is constructive, it also returns the explicit quantum states and measurements used to achieve the final value of $W$. 

\begin{backgroundbox}[label={Box:SDP}]{Semidefinite programs}
A semidefinite program (SDP) is an optimisation of a linear function over the cone of positive semidefinite matrices intersected with affine spaces. This can in general be written as
\begin{align}\nonumber
\max_X & \quad  \langle C, X \rangle \\ \nonumber
\text{s.t.}&  \quad \langle A_i, X \rangle = b_i, \quad i = 1, \dots, m \\
& \quad X \succeq 0,
\end{align}
where $X$ is the square-matrix optimisation variable, $C$ and $A_i$ are matrices and $b_i$ are numbers. SDPs can also include inequality constraints, taking the form $\langle F_j,X \rangle \leq f_j$, but these can without loss of generality be included in the above equality constraints. For that, we need only to introduce auxiliary variables $s_j\geq 0$ and impose that $\langle F_j,X \rangle +s_j = f_j$. Notice that LPs are the special case of SDPs in which the matrix $X$ is diagonal.

\vspace{0.2cm}

The importance of SDPs in quantum information stems from the fact that the domain of positive semidefinite matrices is essential to the structure of quantum operations. This frequently makes SDPs  the appropriate tool for quantum information problems. There are many available SDP solvers --- often they are built into the same  libraries as those mentioned in Box~\ref{Box:LP} for LPs.
\end{backgroundbox}

\subsubsection{Quantum communication complexity}\label{sec:quantumCCP}
An important aspect of quantum correlations in the PM scenario is that they are closely related to advantages in communication complexity. Communication complexity problems are tasks in which $n$ different parties each hold some data $x_i$ and want to compute a function $f(x_1,\ldots,x_n)$ that depends on all their data. To compute $f$, they must communicate, and the smallest amount of communication necessary is called the communication complexity. The use of quantum resources to reduce communication complexity is often called quantum communication complexity~\cite{Yao1993}.

In the PM scenario, we can focus on  one-way communication complexity problems: Bob wants to compute a function $f(x,y)$ but $x$ is held by Alice. She must  send to him as few bits as possible to enable the computation. Here, we are interested in comparing the number of communicated bits needed to solve the problem with the number of communicated qubits needed to solve the problem. Already for small alphabets, one can find tasks in which the number of necessary qubits (or equivalently, the dimension of the message) is smaller than in the classical case (see e.g.~\cite{Emeriau2022, Cruzeiro2026}). However, the most interesting comparison in communication complexity is the scaling behaviour, that is, how the classical versus quantum communication complexity scales when the alphabets become large.

Remarkably, there exist problems that have large gaps between the classical and quantum communication complexity. The earliest example of this  was a distributed version of the well-known Deutsch-Jozsa algorithm: $x=(x_1,\ldots,x_n)$ and $y=(y_1,\ldots,y_n)$ are $n$-bit strings with the promise that either $x=y$ or $x$ and $y$ are different in exactly $\tfrac{n}{2}$ positions ($n$ is a power of two). The task is to determine which is the case. It was shown that the number of bits needed to solve the problem is exponentially larger than the number of necessary qubits~\cite{Buhrman1998}. Specifically, only $\log n$ qubits are needed whereas a classical protocol needs $O(n)$ bits. However, this gap disappears if we are willing to permit a small error probability in the task. In that case, Alice needs only to randomly send some pairs $(i,x_i)$ to Bob. If $x$ and $y$ are different in $\tfrac{n}{2}$ positions, then the probability of Bob detecting this is high already if Alice sends $O(\log n)$ bits. This shows that the exponential quantum advantage is sensitive to the requirement of an exact solution.

Fortunately, this obstacle can be overcome. As first shown in Ref.~\cite{Raz1999}, there exist communication complexity problems with an exponential quantum-to-classical gap that persists even under bounded errors. A particularly elegant example of this is the so-called  hidden-matching problem~\cite{Yossef2004}. In this problem, Alice has an $n$-bit string $x=(x_1,\ldots,x_n)\in\{0,1\}^n$ and Bob is randomly given a set $H=\{(i_1,j_1),\ldots,(i_{n/2},j_{n/2})\}$ of disjoint pairs selected from $\{1,\ldots,n\}$. For example, if $n=4$ there are three possible such sets, $\{(1,2),(3,4)\}$, $\{(1,3),(2,4)\}$, $\{(1,4),(2,3)\}$. Bob's job is to output a triple $(i,j,x_i\oplus x_j)$ for some $(i,j)\in H$. Thus, the task has  $\tfrac n2$ possible winning answers. It was proven in Ref.~\cite{Yossef2004} that a classical protocol cannot solve the problem with bounded error unless Alice sends $O(\sqrt{n})$ bits. A way to saturate this scaling is to let Alice and Bob use shared randomness to select indices $i$, and let Alice send the corresponding bits $x_i$ to Bob. Due to the so-called birthday paradox~\cite{Abramson01101970}, it suffices that Alice selects roughly $\sqrt{n}$ such indices in order for Bob to have a high probability of receiving both $x_i$ and $x_j$ for one of the pairs in $H$. In contrast, consider the following quantum protocol. Alice sends the $n$-dimensional state $\ket{\psi_x}=\frac{1}{\sqrt{n}}\sum_{i=1}^n (-1)^{x_i} \ket{i}$ to Bob. Bob partitions the $n$-dimensional Hilbert space into $\frac{n}{2}$ two-dimensional subspaces according to his set $H$, that is, each outcome is associated with a  projector $\ketbra{i}+\ketbra{j}$ whose indices are taken from $H$. After this measurement, the post-measurement state is $\frac{1}{\sqrt{2}}\left[\ket{i}+(-1)^{x_i+x_j}\ket{j}\right]$. He can read out the phase $x_i\oplus x_j$ by measuring $\sigma_X$ in this subspace. Thus, the quantum protocol consumed $\log n$ qubits and we have found an exponential separation.

\subsubsection{Network scenarios}
\label{subsubsec:networkquantummessage}

\begin{figure}[t]
    \centering
    \includegraphics{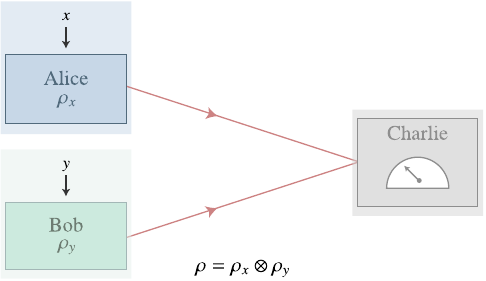}
    \caption{\emph{Prepare-and-measure networks with quantum messages.} (a) Two preparation devices send quantum messages to a single measurement device, enabling the certification of whether the receiver performs an entangled measurement. (b) The quantum state prepared by the sender is transmitted through a broadcasting channel to two spatially separated measurement devices, enabling the activation of the nonclassicality of the prepared states.}
    \label{fig_networks1}
\end{figure}

As in the case of Bell's theorem, prepare-and-measure scenarios with quantum messages have also inspired a variety of network generalisations. An important example is the Pusey-Barrett-Rudolph (PBR) theorem~\cite{pusey2012reality}, which considers two independent preparation devices followed by a joint entangled measurement. Under the assumption of preparation independence, namely that independently prepared systems are associated with independent ontic states, the resulting measurement statistics are incompatible with interpreting the quantum state as merely representing incomplete knowledge of an underlying physical state (see Fig.~\ref{fig_networks1}).

Another application of the PM scenario with multiple preparation devices is the certification of entangled measurements~\cite{Bennet2014,Abbott2018}. The central idea is that a single measurement device receives quantum states from different senders and the violation of a Bell-like inequality beyond a certain threshold is only possible if the measurement is entangled. The scenario proposed in Ref.~\cite{Bennet2014} consists of two preparation devices, Alice and Bob, each producing a qubit state $\rho_x^A$ and $\rho_y^B$, respectively. These states are sent to a third party, Charlie, who performs a joint measurement described by the POVM $\{M_{c|z}\}$. The observed statistics are therefore
\begin{equation}
p(c|x,y,z)=\tr\left[(\rho_x^A\otimes\rho_y^B)M_{c|z}\right],
\end{equation}
where the preparations are independent and the communicated systems are qubits. The goal is to infer, only from the observed statistics, whether Charlie's measurement is entangled, that is, whether at least one POVM element $M_{c|z}$ is non-separable.

To certify entangled measurements, Ref.~\cite{Bennet2014} considers a single three-outcome measurement, so that the measurement input $z$ can be omitted. The witness defined in Ref.~\cite{Bennet2014} is given by $w=\sum_{c=1}^{2}\sum_{x,y=0}^{2}W_{c|x,y}\,p(c|x,y)$, where, with rows labelled by $x$ and columns by $y$,
\begin{equation}
W_{1|x,y}= \begin{pmatrix}
1&-1&-1\\-1&1&-1\\-1&-1&1
\end{pmatrix} \:\: \text{and} \:\:
W_{2|x,y}=
\begin{pmatrix}
1&-1&-1\\-1&-1&1\\-1&1&-1
\end{pmatrix}.
\end{equation}
For every unentangled (separable) measurement it follows that $w\leq 1$. Hence, any observed value $w>1$ certifies that Charlie's measurement is entangled.

The optimal quantum strategy consists of Alice and Bob preparing three pure qubit states lying in the $x$--$z$ plane of the Bloch sphere, namely $\rho^A_x=\tfrac{1}{2}(\mathds{1}+\vec{r}_x\cdot\vec{\sigma})$, $\rho^B_y=\tfrac{1}{2}(\mathds{1}+\vec{q}_y\cdot\vec{\sigma})$ with $\vec{r}_x=(\cos\alpha_x,0,\sin\alpha_x)$
and $\vec{q}_y=(\cos\beta_y,0,\sin\beta_y)$, where $\alpha_j=\beta_j=\frac{2\pi j}{3}$ for $j=0,1,2$. Charlie performs the three-outcome POVM
\begin{align}
M_1&=|\Phi^+\rangle\!\langle\Phi^+|, \nonumber \\
M_2&=|\Psi^+\rangle\!\langle\Psi^+|, \nonumber\\
M_3&=\mathds{1}-M_1-M_2,
\end{align}
where $\ket{\Phi^+}=2^{-\tfrac{1}{2}}(\ket{00}+\ket{11})$ and $\ket{\Psi^+}=2^{-\tfrac{1}{2}}(\ket{01}+\ket{10})$. This leads to the conditional probabilities
\begin{equation}
p(c|x,y)=\frac14\left[1+\cos\!\left(\alpha_x+(-1)^c\beta_y\right)\right],
\end{equation}
with $c=1,2$, from which one obtains $w_{\rm Q}=\frac32$, therefore establishing the presence of an entangled measurement without requiring any characterisation of the preparation devices beyond the dimension assumptions.

Another direction of generalisation concerns multiple receivers, as in the sequential PM scenario discussed in Sec.~\hyperref[subsubsec:instruments]{II-D-5}, or in spatially distributed measurement networks. An example is the prepare-and-broadcast scenario~\cite{wang2019characterising,ioannou2022receiver,jia2025characterizing,sarubi2026prepare,sarubi2026randomness}, which extends the standard prepare-and-measure framework by replacing the single measurement device with a broadcasting transformation followed by multiple local measurement devices. More precisely, Alice prepares a bounded-dimensional physical system according to an input $x$, which is processed by a broadcasting channel and distributed to two spatially separated parties, Bob and Charlie. They independently choose measurement settings $y$ and $z$, producing outcomes $b$ and $c$, respectively. Similarly to the standard PM scenario, the observed behaviour is therefore described by the conditional distribution $p(b,c|x,y,z)$.

In the classical model, Alice sends a $d$-valued classical message $m$, while the broadcaster distributes the information through a classical channel. Allowing for shared randomness $\lambda$, the resulting correlations admit the decomposition
\begin{equation}
p(b,c|x,y,z)=\sum_{m,t,\lambda}q(\lambda)p(m|x,\lambda)p(t|m,\lambda)p(b|t,y,\lambda)p(c|t,z,\lambda).
\end{equation}
 Since the broadcast variable $t$ is unrestricted, it can be absorbed into the local response functions, yielding the equivalent form
\begin{equation}
p(b,c|x,y,z)=\sum_{m,\lambda}q(\lambda)D^A_\lambda(m|x)D^B_\lambda(b|m,y)D^C_\lambda(c|m,z).
\end{equation}

The quantum model replaces Alice's classical message by a $d$-dimensional quantum state $\rho_x$, while the broadcasting device is described by a quantum channel
$\mathcal T:\mathcal L(\mathcal H_M)\rightarrow
\mathcal L(\mathcal H_B\otimes\mathcal H_C)$.
The correlations are then given by
\begin{equation}
p(b,c|x,y,z)= \tr\left[ \mathcal T(\rho_x) \left( M^B_{b|y}\otimes M^C_{c|z} \right) \right].
\end{equation}

Unlike the standard PM scenario, the prepare-and-broadcast (PB) scenario naturally separates the nonclassicality associated with the communicated system from that of the broadcasting transformation. This motivates a hierarchy of models according to whether Alice's message and the broadcaster are classical ($\mathcal{CC}$), quantum ($\mathcal{QQ}$), or hybrid resources, where the first (second) letter refers to Alice's message (the broadcaster). In particular, one finds $\mathcal{CC}=\mathcal{QC}\subsetneq\mathcal{CQ}\subsetneq\mathcal{QQ}$, where the equality $\mathcal{QC}=\mathcal{CC}$ follows from the Frenkel-Weiner theorem~\cite{Frenkel2015}~(see Sec.~\hyperref[sec:FWtheorem]{II-B-1}).

A feature of the PB scenario is the activation of quantum communication. If Bob and Charlie have no measurement choices, i.e.~$y=z=1$, the broadcasting stage is operationally equivalent to a single POVM acting on Alice's message. Consequently, quantum communication offers no advantage over a classical message of the same dimension, implying that all four models coincide, yet another consequence of the Frenkel-Weiner theorem~\cite{Frenkel2015}. However, this equivalence is broken if the receivers are allowed measurement choices.

Within this context, the PB scenario allows for the activation of nonclassicality in the communicated system. That is, there exist prepared states that appear classically simulable in the standard PM tests considered, but whose correlations become nonclassical after the broadcasting channel. To certify this effect, it is not sufficient to compare with the fully classical model, since a violation could still be explained by a classical preparation message followed by a quantum broadcaster. Instead, the relevant benchmark is the classical-quantum ($\mathcal{CQ}$) model, in which Alice communicates only a classical message while Bob and Charlie are allowed to share arbitrary quantum resources.

To observe the activation effect, one can consider the witness of Ref.~\cite{sarubi2026prepare}, valid for the $\mathcal{CC}$ model and given by
\begin{align}
W_{\mathcal{CC}} = 3(E_{00}^{[0]}&-E_{00}^{[2]})+(E_{11}^{[2]}-E_{11}^{[0]}) \nonumber\\ &+\sum_{y=0}^{1}\left( E_{y,1-y}^{[0]} -2E_{y,1-y}^{[1]} +E_{y,1-y}^{[2]} \right)\le 8.
\label{eq:CC}
\end{align}
where $E_{yz}^{[x]}\equiv\langle B_y C_z\rangle^{[x]} =\sum_{b,c\in\{0,1\}}(-1)^{b+c}p(b,c\vert x,y,z)$. The optimal quantum strategy employs three qubit preparations followed by an isometry $1\rightarrow2$ mapping an input qubit to two qubits, achieving $W_{\mathcal{CC}}^{QQ}\simeq12.944$, while every classical preparation assisted by an arbitrary quantum broadcasting device satisfies the bound $W_{\mathcal{CC}} \leq W_{\mathcal{CC}}^{CQ} \simeq9.238$. Therefore, the observed violation certifies that the communicated system itself must be quantum. Moreover, the witness remains robust against white noise in the preparations down to a critical visibility of $v_{\mathrm{PB}}^{*}\simeq0.713$, whereas the corresponding prepare-and-measure scenario certifies the same preparations only for $v_{\mathrm{PM}}^{*}\simeq0.769$. Hence, the interval $0.713<v<0.769$ provides numerical evidence of activation relative to the tested PM measurements: these measurements do not certify the preparations as nonclassical, while the prepare-and-broadcast witness does.

\subsection{Tests of physical dimension}
One application of quantum correlations in PM scenarios is that they can be used to test the dimension of an uncharacterised quantum system. Consider that Alice prepares some unknown quantum states $\rho_x$ and that she sends them to Bob who performs some unknown measurements on them. By studying the correlations $p(b|x,y)$ from this PM experiment, we can deduce a lower bound on the Hilbert space dimension of $\rho_x$. Specifically, imagine that we could prove that the correlations admit a quantum model where Alice's states have dimension $d$, but that there exists no possible quantum strategy that could do the same if the dimension had been $d-1$. We would thus have certified that the system has dimension at least $d$. This process is known as device-independent dimension witnessing~\cite{Gallego2010}.

To realise a device-independent dimension witness, it is not sufficient to simply violate a test of classicality. Instead, we must prove the unattainability of correlations in a $d$-dimensional quantum realisation. In general, characterising the quantum set of correlations in dimension $d$ is far too complicated to be achieved analytically. Therefore, the methods we discuss here are either partial criteria that can be established analytically, or methods based on convex programming. Let us begin with the analytical approach.

\subsubsection{Analytical criteria}
From the outset, it is important to underline that trivial examples of dimension witnesses can easily be constructed. In fact, we have already seen one: the facet related to the state discrimination game discussed in Box~\ref{Box:statediscriminationfacets}. Since we know that all quantum correlations satisfy the same inequality [see Eq.~\eqref{mesdquantum}], any violation thereof implies a lower bound on the dimension.  However, this is of secondary interest because we cannot distinguish quantum from classical correlations in this test, i.e.~the dimension certification could be entirely due to a classical system having a larger dimension. Therefore, let us consider genuinely quantum examples.

Since we are already familiar with the random access code, let us consider whether it can be further used as a tool for device-independent dimension witnessing. In other words, our task is to consider the quantity $W_{n,d}$ in Eq.~\eqref{RACwitness} and determine its maximal value possible for any quantum strategy of dimension $d$. By exceeding that limit, we certify a dimension no smaller than $d+1$. Many different bounds on quantum random access codes are found in the literature (see for instance Ref.~\cite{Farkas2025}). For our purposes here, we focus on the family $(n,d)=(2,d)$. For simplicity of the derivation, we follow the proof of Ref.~\cite{Aguilar2018}, which makes the extra assumption  that Bob performs rank-one projective measurements\footnote{This assumption can be lifted (without changing the result) by  following the approach of Ref.~\cite{Farkas2025}.}, namely $M_{b|y}=\ketbra*{e_{b|y}}{e_{b|y}}$ where $\{\ket*{e_{b|y}}\}_b$ forms a basis. Using Eq.~\eqref{maxeig} for the random access code gives
\begin{align}
W_{2,d}&\leq \frac{1}{2d^2}\sum_{x_1,x_2=0}^{d-1} \lambda_{\max} (M_{x_1|1}+M_{x_2|2}).
\end{align}
Due to the measurements being rank-one projective, we now have to deal with expressions of the form $\lambda_{\max}(\ketbra{\psi}+\ketbra{\phi})$. Since this is invariant under unitaries, we can equally well choose $\ket{\psi}=\ket{0}$ and $\ket{\phi}=\cos\theta \ket{0}+\sin\theta\ket{1}$. Calculating the eigenvalues then gives $\lambda_{\max}(\ketbra{\psi}+\ketbra{\phi})=1+|\braket{\psi}{\phi}|$, which we write for our case as $ \lambda_{\max} (M_{x_1|1}+M_{x_2|2})=1+\sqrt{\tr\left(M_{x_1|1}M_{x_2|2}\right)}$. Hence, 
\begin{align}\nonumber\label{QRACbound}
    W_{2,d}&\leq \frac{1}{2} + \frac{1}{2d^2}\sum_{x_1,x_2} \sqrt{\tr\left(M_{x_1|1}M_{x_2|2}\right)}\\ 
    & \leq \frac{1}{2} + \frac{1}{2d} \sqrt{\sum_{x_1,x_2}\tr\left(M_{x_1|1}M_{x_2|2}\right)}
    =\frac{1}{2}\left(1+\frac{1}{\sqrt{d}}\right).
\end{align}
This bound is in fact tight for any choice of $d$. Explicit quantum strategies for this are given in Ref.~\cite{Tavakoli2015}. Notice also the comparison to the classical limit in Eq.~\eqref{RACdits}: in the classical case the bound tends to $\frac{1}{2}$ as $O(\frac{1}{d})$ but in the quantum case it goes as $O(\frac{1}{\sqrt{d}})$. By observing $W_{2,d}$ larger than the above bound, we conclude that the quantum dimension is at least $d+1$. 

The first demonstration of a high-dimensional random access code focused on dimension $d=4$~\cite{Tavakoli2015} but demonstrations motivated by dimension witnessing have been reported in much higher dimension~\cite{Aguilar2018b}. In contrast, the first experiments that tested physical dimension were mainly focused on the few lowest dimensions~\cite{Ahrens2012, Hendrych2012}.

Let us give a second example of a family of dimension witnesses. This time we consider a different task, which in spirit is a randomised version of two-state discrimination~\cite{Brunner2013}. Consider that Alice selects $x\in\{1,\ldots, N\}$. Bob selects a pair of inputs $y,y'\in\{1,\ldots,N\}$ such that $y<y'$ and outputs $b\in\{0,1\}$. Their task is as follows. We count only rounds where either $x=y$ or $x=y'$, and Bob must decide whether the former equality holds ($b=0$) or the latter ($b=1$). This can be written as
\begin{align}\nonumber\label{DIDWBrunner}
W&=\sum_{x<x'} [p(0|x,(x,x'))+p(1|x',(x,x'))]\\
&= \binom{N}{2}+\sum_{x<x'}\tr\left[(\rho_x-\rho_{x'})M_{x,x'}\right] \nonumber\\ &\leq \binom{N}{2}+\sum_{x<x'}\lambda_+\left(\rho_x-\rho_{x'}\right)
\end{align}
where we have defined the measurement operator for outcome $b=0$ as $M_{x,x'}$. In the last step we have optimally selected $M_{x,x'}$ as the projector onto the positive eigenspace of $\rho_x-\rho_{x'}$ and labelled the sum of all positive eigenvalues by $\lambda_+$. Furthermore, due to linearity of $W$ we can without loss of generality restrict to pure states, which means that $\rho_x-\rho_{x'}$ has rank no larger than two and at most one positive eigenvalue. Hence, using an argument analogous to the one  above Eq.~\eqref{QRACbound}, we obtain $\lambda_+\left(\rho_x-\rho_{x'}\right)=\lambda_{\max}\left(\rho_x-\rho_{x'}\right)=\sqrt{1-|\braket{\psi_x}{\psi_{x'}}|^2}$, where $\rho_x=\ketbra{\psi_x}$. Using the concavity of the square-root leads us to
\begin{equation}
    \sum_{x<x'}\lambda_+\left(\rho_x-\rho_{x'}\right) \leq \sqrt{\binom{N}{2}^2-\binom{N}{2}\sum_{x<x'} |\braket{\psi_x}{\psi_{x'}}|^2}.
\end{equation}
We can write $\sum_{x<x'} |\braket{\psi_x}{\psi_{x'}}|^2=\frac{1}{2}\sum_{x,x'} |\braket{\psi_x}{\psi_{x'}}|^2-\frac{N}{2}$. Our task is now to minimise the value of $t=\sum_{x,x'} |\braket{\psi_x}{\psi_{x'}}|^2$. To this end, let us write $t=\tr(S^2)$ where $S=\sum_x \ketbra{\psi_x}$. We denote by $\lambda_k$ the non-zero eigenvalues of $S$, of which there can be at most $d$. Our job is to minimise $\tr(S^2)=\sum_k \lambda_k^2$ when $\tr(S)=\sum_k \lambda_k=N$. By Cauchy-Schwarz, $t\geq\tfrac{N^2}{d}$, with equality when $S=\tfrac{N}{d}\iden$. Putting it all together gives us the dimension witness
\begin{equation}
    W\leq \sqrt{\frac{N^3(N-1)(d-1)}{4d}}+\binom{N}{2}.
\end{equation}

Linear correlation quantities that serve as device-independent dimension witnesses for any choice of $d$ are rare in the literature. In the few cases where analytical solutions are possible, they are (as in our examples above) tailored for specific states and measurements that have the potential to violate the resulting inequalities. In contrast, general and useful criteria that can be applied directly to the full correlation $p(b|x,y)$ are difficult to find. Typically, generality comes at the expense of the tightness of the bound. Let us discuss such a criterion by following Ref.~\cite{Vicente2019}. For this, we define two matrices
\begin{align}\label{dataTableMatrix}
    P&=\sum_{b,x,y} p(b|x,y)\ketbra{x}{y,b},\\
    Z_{AB}&=\left(\sum_{x} \ket{x}\otimes\rho_x\right)\left(\sum_{b,y} \bra{y,b}\otimes M_{b|y}\right),
\end{align}
where $p(b|x,y)=\tr(\rho_xM_{b|y})$. Observe that $\tr_B(Z_{AB})=P$. Since $P$ is obtained from $Z$ by a partial trace, it follows that
\begin{align}
\lVert P \rVert_1 \leq \lVert Z\rVert_1 \leq \big\lVert \sum_{x} \ket{x}\otimes\rho_x \big\rVert_2\ \big\lVert \sum_{b,y} \bra{y,b}\otimes M_{b|y} \big\rVert_2,
\end{align}
where we used the H\"older inequality for Schatten norms, $\lVert AB\rVert_1 \leq  \lVert A\rVert_2 \lVert B\rVert_2$. Using the orthogonality of the basis vectors, the first factor is bounded by
\begin{equation}
\big\lVert \sum_{x} \ket{x}\otimes\rho_x \big\rVert_2^2  = \sum_{x} \big\lVert  \ket{x}\otimes\rho_x \big\rVert_2^2= \sum_x \tr(\rho_x^2)\leq |X|.
\end{equation}
The analogous calculation for the second factor gives $\lVert \sum_{b,y} \bra{y,b}\otimes M_{b|y} \rVert_2^2\leq d|Y|$. Thus, we conclude that if $p(b|x,y)$ admits a quantum model, the Hilbert space dimension must satisfy $d\geq \frac{\lVert P\rVert_1^2}{|X||Y|}$. While this is fully general, it is  not a tight criterion. For instance, applying this to the random access code will return a weaker bound larger than the one derived in Eq.~\eqref{QRACbound}.

\subsubsection{Bounding quantum correlations}\label{sec:boundingNV}

Since analytical criteria have limited capability to describe the set of quantum correlations, it is natural to ask if a more general method is possible. The leading approach to this is based on semidefinite programming (SDP) relaxations (see Box~\ref{Box:SDP} for an introduction to SDPs). The core idea consists in identifying outer relaxations to otherwise intractable optimisation problems, which can be computed as semidefinite programs (see the review article~\cite{SDPreview}). Several such SDP relaxation tools, each with their own pros and cons, have been proposed to analyse dimension-restricted quantum correlations in PM scenarios. Here, we will follow the heuristic approach of Refs.~\cite{Navascues2015, Navascues2015b}.

Consider that we are given some arbitrary linear correlation quantity as defined in Eq.~\eqref{linearWitness} and asked to determine an upper bound on its largest possible quantum value. This is a nonlinear optimisation, because it involves products of states and measurements. While we can without loss of generality restrict ourselves to pure states $\rho_x$, we will also choose to restrict ourselves only to projective measurements. Our goal is to bound $W$ by means of evaluating an SDP. To this end, consider  that we randomly sample a set of pure states and projective measurements and put them in the list $L^{(1)}=\{\mathds{1},\{\rho_x\}_x,\{M_{b|y}\}_{b,y}\}$. Next, we take this list of operators and compute a list of monomials, $J^{(1)}$, defined  over $L^{(1)}$. This means that $J^{(1)}$ is a list of products of the elements appearing in $L^{(1)}$, and the number of such products that we include in $J^{(1)}$ is a degree of freedom in our construction. We then use the monomial list $J^{(1)}$ to compute a matrix, which we  define as $ \Gamma^{(1)}_{u,v}=\tr\left(u^\dagger v\right)$, for every $u,v\in J^{(1)}$. Thus, the rows and columns are indexed by the monomials. Importantly, this matrix is by construction positive semidefinite because 
\begin{align}
 \sum_{u,v} c_u^*\Gamma^{(1)}_{u,v}c_v=\tr(A^\dagger A)\geq0,
\end{align}
where $A=\sum_u c_u u$ for arbitrary complex coefficients $c_u$.

Next, we repeat the above procedure many times independently, i.e.~we generate the $i$th set of random states and measurements, $L^{(i)}$, compute the associated monomial list, $J^{(i)}$, and then the corresponding matrix $\Gamma^{(i)}$. This continues until we find that the next sample is linearly dependent on all the previous ones. This can easily be checked by converting each $\Gamma^{(i)}$ to a vector, arranging all these vectors in a matrix and checking its rank. When we find linear dependence, we terminate the iteration  and use the set $\{\Gamma^{(i)}\}_i$ as a basis. Over this basis, we can now define our optimisation variables as  the affine combination,  $ \Gamma = \sum_i s_i \ \Gamma^{(i)}$, for $\sum_i s_i=1$. The reason we use an affine combination is to ensure normalisation in $\Gamma$, namely that its top-left element is $\Gamma_{\mathds{1},\mathds{1}}=d$. The reason that we wanted to construct $\Gamma$ is that we can express our objective function, $W$, as a linear combination over its elements, namely $W(\Gamma)=\sum_{b,x,y} c_{bxy} \Gamma_{\rho_x,M_{b|y}}$. An upper bound on $W$ can now be obtained from the following SDP
\begin{align}\nonumber\label{SDPMiguelTamas}
\max_{\{s_i\}}\quad &\sum_{b,x,y} c_{bxy} \Gamma_{\rho_x,M_{b|y}} \\ \nonumber
\text{s.t.}\quad &\Gamma = \sum_i s_i \ \Gamma^{(i)}\quad \\ \nonumber
&\sum_i s_i = 1\\
& \Gamma \succeq 0.
\end{align}

The main idea behind this construction is to indirectly capture the constraints associated with $d$-dimensional Hilbert space by sampling states and measurements over it. The method works well for PM scenarios with small alphabets. It can for example most easily prove that the quantum violations reported in Sec.~\hyperref[sec:Qviolations]{II-B-2} are optimal and a number of other case studies based on this method can be found in for instance Ref.~\cite{Navascues2015b}. The above construction assumed projective measurements, but this assumption can be lifted by employing a Naimark dilation. Concretely, this means embedding the states into a larger Hilbert space and sampling the projective measurements therein. 

However, the method also has a few noteworthy drawbacks. Firstly, since it is based on sampling, it is only a heuristic method. Secondly, it can often be that the number of samples required to make the basis is large and therefore just setting up the SDP is itself time- and memory-consuming. Thirdly, in order to obtain good bounds in practice, one typically has to break up the set of projective measurements into its distinct rank combinations. Specifically, each projective measurement can be selected so that the measurement operator associated with outcome $b$ has rank $r_b\in\{0,\ldots,d\}$, such that $\sum_b r_b=d$. Typically, we want to compute the SDP \eqref{SDPMiguelTamas} for  one specific choice of ranks (for all the measurements of Bob), obtain a bound on $W$ for that specific rank combination, and then repeat the process for every rank combination. In the end, we select the largest bound returned by all the SDPs. Clearly, since the number of rank combinations explodes with the size of the scenario, this becomes a costly procedure. 

\begin{figure}[t]
    \centering
    \includegraphics{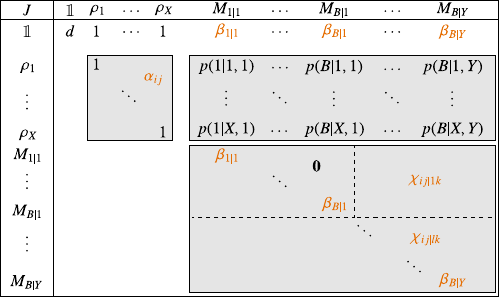}
    \caption{\emph{Lowest-level moment matrix}. Illustration of the SDP matrix $\Gamma$ for $J=L=\{\iden,\{\rho_x\}_x,\{M_{b|y}\}_{b,y}\}$. The entries encode normalisation, purity, projectivity and the probabilities $p(b|x,y) = \Gamma_{\rho_x, M_{b|y}}$. Shaded blocks highlight the different structural blocks of the matrix and coloured symbols indicate free parameters subject to these linear constraints.}
    \label{fig_moment_matrix}
\end{figure}

One way to overcome these drawbacks is to consider a different SDP relaxation method. For instance, Ref.~\cite{Pauwels2022} suggests doing away with the sampling and simply defining an abstract operator list $L=\{\mathds{1},\{\rho_x\}_x,\{M_{b|y}\}_{b,y}\}$ to which we associate the monomial list $J$ and the matrix $\Gamma_{u,v}=\tr\left(u^\dagger v\right)$. Then, we constrain the entries of this matrix using the basic rules of quantum theory. For instance, since we can without loss of generality assume the state to be pure, we have $\rho_x^2=\rho_x$. This means for instance $\Gamma_{\rho_x,\rho_x}=\Gamma_{\mathds{1},\rho_x}$. Furthermore, since $\tr(\rho_x)=1$ we have $\Gamma_{\mathds{1},\rho_x}=1$. Projectivity of measurements means that $M_{b|y}M_{b'|y}=M_{b|y}\delta_{b,b'}$, which implies for example $\Gamma_{M_{b|y},M_{b'|y}}=0$ for $b\neq b'$. Also, we can associate the probabilities with $p(b|x,y)=\Gamma_{\rho_x,M_{b|y}}$. However, none of these constraints capture the fact that Alice's states are limited to dimension $d$. In order to do this, we can consider the condition $\Gamma_{\mathds{1},\mathds{1}}=d$. At the lowest relaxation level, corresponding to $L=J$, the SDP matrix $\Gamma$ is illustrated in Fig.~\ref{fig_moment_matrix}. This approach is a crude relaxation of dimensionality, which sometimes is less accurate than that obtained from the SDP~\eqref{SDPMiguelTamas}, but as shown in Ref.~\cite{Pauwels2022} this can often return good and even optimal bounds when applied at the appropriate relaxation level.

Even though this method requires no sampling or separate consideration of the ranks of the projective measurements, it can still be expensive to compute due to the rapid increase in the size of the SDP matrix with the size of the scenario and/or the relaxation level. Instead, there is sometimes a way to make the evaluation of the SDP~\eqref{SDPMiguelTamas} dramatically faster, provided that $W$ features significant symmetry properties~\cite{Rosset2019}. In this context, a symmetry means that Alice and Bob both can relabel their inputs and/or outputs without changing either the objective function $W$ or the constraints associated with the quantum states and measurements. For example, the random access codes defined in Eq.~\eqref{RACwitness}, when evaluated with standard basis measurements, have these apparent symmetries. We can let $\omega$ be some permutation of $d$ elements and then let $\pi$ be a permutation of $n$ elements. If Alice were to make the relabelling $x_1\rightarrow \omega(x_1)$ and Bob to make the relabelling $b\rightarrow \omega(b)$ conditioned on $y=1$, then $W$ remains invariant. Similarly, Alice could make the relabelling of her input string $(x_1,\ldots,x_n)\rightarrow (x_{\pi(1)},\ldots,x_{\pi(n)})$ and Bob the relabelling $y\rightarrow \pi(y)$, to again keep $W$ invariant. Therefore, we can average each sample $\Gamma^{(i)}$ over this symmetry group and obtain a new, symmetrised, sample $\tilde{\Gamma}^{(i)}$. The space of symmetrised samples will be  smaller than that of the original samples, and therefore we expect the final basis $\{\tilde{\Gamma}^{(i)}\}_i$ to be smaller than the original, thereby reducing the number of variables in the SDP. Moreover, thanks to these symmetries, the final SDP matrix $\tilde{\Gamma} = \sum_i s_i \ \tilde{\Gamma}^{(i)}$ can be block-diagonalised. This means that the positivity condition $\tilde{\Gamma}\succeq 0$ splits into a number of positivity conditions, each for a smaller matrix. See Ref.~\cite{Rosset2019} for details of this procedure and the library RepLAB for numerical decompositions of representations of finite groups~\cite{Rosset2021}. All this brings significant computational advantage. For example, working with monomials of the type $J=\{\mathds{1}, \{\rho_x\}_x, \{M_{b|y}\}_{by}, \{\rho_xM_{b|y}\}_{bxy}\}$, we can for example consider  the random access code for $(n,d)=(3,3)$. Ref.~\cite{Rosset2019} shows that without using symmetries one requires over $11000$ samples and the SDP did not evaluate after 24 hours of  computing. In contrast, exploiting the  symmetries of the random access code, one needs only $82$ samples and the computation lasted four seconds. 

\subsection{Quantum certification}
An important application of PM scenarios is to view them as platforms for certifying quantum devices, i.e.~to verify that a quantum device works as intended. In Bell scenarios, it is possible to certify quantum devices under essentially minimal assumptions~\cite{Scarani2019}, but such tests are well-known to be experimentally demanding. The PM scenario offers a so-called semi-device-independent method for quantum certification. This means that certification is achieved up to only a mild assumption which in our present discussion is a limitation on the Hilbert space dimension. In return, the experiments are simpler and the benchmarks to meet are more modest. It can therefore be viewed as a compromise between the rigour of  device-independence and the practicality of having fully trusted devices. In this section, we explore different levels of this certification, starting with the strongest form: self-testing.

\subsubsection{Self-testing}\label{sec:self_test}
Self-testing is the highest standard of quantum certification. It is a process in which we can fully deduce (up to a modest equivalence class) the inner workings of quantum devices by using them to generate Bell nonlocal correlations~\cite{Supic2020}.  In the same spirit, it is possible to characterise initially unknown quantum devices in the PM scenario by observing the quantum correlations generated from them. Formally, consider that we have a set of quantum correlations $p(b|x,y)=\tr(\psi_x M_{b|y})$, where $\psi_x$ are $d$-dimensional pure states. Consider now that we apply to both the states and the measurements an operator $U$ that is either unitary or antiunitary. This gives states $U\psi_xU^\dagger$ and measurements  $UM_{b|y}U^\dagger$, but  the correlations $p(b|x,y)$ remain unchanged by $U$. Hence, given $p(b|x,y)$, we cannot deduce the states and measurements beyond such a unitary or antiunitary degree of freedom. 

Consider that we use $p(b|x,y)$ to compute a linear correlation quantity $W$ [see Eq.~\eqref{linearWitness}]. Suppose that using some unknown states $\{\bar{\psi}_x\}_x$ and some unknown measurements $\{\bar{M}_{b|y}\}_{b,y}$, we find that $W$ reaches its largest possible value for $d$-dimensional quantum systems, namely $W=W^Q$\footnote{Since $W$ is linear, its maximal value can always be attained with pure states, but purity of the unknown preparations must follow from the equality conditions.}. We say that this observation self-tests the states $\{\psi_x\}_x$ and  measurements $\{M_{b|y}\}_{b,y}$ if there exists a unitary or anti-unitary operator $U$ such that
\begin{align}\nonumber\label{selftest}
U\bar{\psi}_x U^\dagger &= \psi_x , \quad\:\: \forall x,\\
U\bar{M}_{b|y} U^\dagger &= M_{b|y}, \quad \forall b,y.
\end{align}
In other words, up to a possible $U$, we can deduce both the states and the measurements. If shared randomness is present this statement applies to each value of the shared variable separately, and $U$ may depend on that value.

Let us return to the simplest random access code \eqref{RAC4222} and show that it leads to self-testing~\cite{Tavakoli2018}. Recall that around Eq.~\eqref{RACvectors} we showed that it has a quantum violation of \mbox{$W_\text{RAC}=\frac{2+\sqrt{2}}{4}$}, and that via  Eq.~\eqref{RACbound} we proved that no larger value is possible using qubit states and measurements. In other words, $W_\text{RAC}^Q=\frac{2+\sqrt{2}}{4}$. This value of $W$ is indeed associated with a self-test of both the states and the measurements. To see that, return to Eq.~\eqref{RACcalculation} and observe that in order to saturate the inequality step, we require the states' Bloch vectors to satisfy $\vec{n}_x\propto (-1)^{x_1}\vec{m}_1+(-1)^{x_2}\vec{m}_2$. Furthermore, returning to Eq.~\eqref{RACcalculation2}, the inequality is saturated if and only if (i) the measurement Bloch vectors satisfy $\lVert \vec{m}_1\rVert=\lVert \vec{m}_2\rVert=1$, and (ii) all terms under the square-root on the first line are equal. The former means that Bob's measurements are projective and the latter means that  $\sum_{y,y'=1,2} (-1)^{x_y+x_{y'}}\vec{m}_y\cdot \vec{m}_{y'}$ is independent of $x$. The latter simplifies to $2+2(-1)^{x_1+x_2}\vec{m}_1\cdot \vec{m}_2$, which is independent of $x$ if and only if $\vec{m}_1\cdot \vec{m}_2=0$. This is effectively a pair of Pauli measurements, and we can choose  $U$ as the unitary that rotates $\{\vec{m}_1,\vec{m}_2\}$ to the Pauli vectors associated with $\sigma_X$ and $\sigma_Z$. For the states, we therefore obtain that $\vec{n}_x=\frac{1}{\sqrt{2}} \left((-1)^{x_1},0,(-1)^{x_2}\right)$. Hence, we have self-tested the quantum strategy proposed in and below Eq.~\eqref{RACvectors}. It is important to realise that this is a stronger result than what we discussed back then: we have shown that up to a global rotation, the quantum strategy compatible with an optimal quantum random access code is unique.

This type of reasoning has been extended beyond qubit systems. For instance, recall the family of random access codes defined in Eq.~\eqref{RACwitness}. For the family $(n,d)=(2,d)$, which was based on sending $d$-dimensional systems from Alice to Bob, we proved in Eq.~\eqref{QRACbound} that $W^Q=\frac{1}{2}\left(1+\frac{1}{\sqrt{d}}\right)$. If we revisit this argument, we see that the inequality is saturated if and only if $\tr(M_{x_1|1}M_{x_2|2})$ is a constant. For basis measurements, we can equally well write this as $\left|\braket{e_{x_1}}{f_{x_2}}\right|^2=c$, where $\{\ket{e_i}\}_i$ and $\{\ket{f_i}\}_i$ are orthonormal bases. By normalisation, the constant must be $c=\frac{1}{d}$. This leaves us with the definition of two mutually unbiased bases. At this point, it is important to remember that the derivation in Eq.~\eqref{QRACbound} assumed that Bob's measurements were basis measurements. Under this assumption, we have proven that an optimal quantum random access code certifies a pair of mutually unbiased bases. As shown in Ref.~\cite{Farkas2019}, it is possible to remove this assumption without changing the conclusion about the certification. It therefore gives us a  proper certification statement for the measurements. However, an important nuance here is that this is not a rigorous self-test for every $d$, because it is known that in some dimensions (e.g.~$d=4$) not all pairs of mutually unbiased bases are equivalent under unitaries~\cite{Brierley2009}.

Going further, it is possible to achieve self-testing in considerable generality. A recipe for this is given in Ref.~\cite{Navascues2023}. For self-testing states, the main idea is the following. Imagine we want to self-test the set of pure states $\Psi=\{\ket{\psi_x}\}_{x=1}^{m}$. To do this, we first extend the set with some other pure states, $\{\ket{\psi_x}\}_{x=m+1}^{n}$,  and find positive numbers $\{\alpha_x\}_{x=1}^{n}$ such that $\sum_{i=1}^n \alpha_i \ketbra{\psi_i} = \frac{\mathds{1}}{d}$. This can always be found because we can select $\mu$ as a sufficiently small positive number so that $\frac{\mathds{1}}{d}\succ \mu \sum_{i=1}^{m}\ketbra{\psi_i}$. Then, we can introduce a positive semidefinite operator defined as $S=\frac{\mathds{1}}{d}- \mu \sum_{i=1}^{m}\ketbra{\psi_i}$. We label its spectral decomposition by $S=\sum_{i=m+1}^{m+d} \beta_i \ketbra{\psi_i}$. Hence, we have obtained a construction by choosing $\alpha_i=\mu$ for $i=1,\ldots,m$ and $\alpha_i=\beta_i$ for $i=m+1,\ldots,m+d$ (so that $n=m+d$). Equipped with this property, we now consider a PM scenario with alphabet $\left(n,d,\frac{n(n-1)}{2},2\right)$ and a linear correlation quantity of the form 
\begin{equation}
    W_\Psi \!=\! \sum_{x< x'} \alpha_{x} \alpha_{x'} \lVert \ket{\psi_x}-\ket{\psi_{x'}}\rVert_1 \Big(p(0|x,(x,x'))\!-\!p(0|x',(x,x'))\Big).
\end{equation}
Notice the similarity between this and the dimension witness considered around Eq.~\eqref{DIDWBrunner}. The alphabets of the scenarios are the same and our witness here is also based on state discrimination between the input pair $(x,x')$\footnote{The state discrimination differs only by a normalisation.}, but we have changed the coefficients of the state discrimination from being uniform in Eq.~\eqref{DIDWBrunner} to being $\alpha_{x} \alpha_{x'} \lVert \ket{\psi_x}-\ket{\psi_{x'}}\rVert_1$, which is  determined by the above extension of the target set. One can then prove that for any quantum model it holds that 
\begin{equation}\label{miguelselftest}
    W_\Psi \leq 1-\frac{1}{d},
\end{equation}
which is saturated if and only if Alice's states  have the same projection overlaps as her target states $\{\ket{\psi_x}\}_{x=1}^n$, provided Bob's measurements are chosen appropriately. Thus, we have ended up certifying the projector overlaps of not only the states we were interested in, but also some auxiliary states needed for this approach. For $d\geq3$, equal projector overlaps do not in general imply unitary or antiunitary equivalence. The self-test in Ref.~\cite{Navascues2023} requires a further extension by auxiliary states with the Wigner property.

\subsubsection{Robust self-testing}\label{sec:robust_Self_test}
Self-testing, in the sense of Eq.~\eqref{selftest}, is only possible when quantum correlations belong to  the boundary of the set of possible correlations. Otherwise, their realisation would not be unique (modulo the freedom associated with $U$). However, no real-life experiment can achieve such perfect quantum correlations. It is thus relevant to ask whether we can make self-testing robust to small imperfections. The main idea of this enterprise is to identify well-motivated distance metrics between the quantum operations that we would ideally have self-tested --- if we had perfect quantum correlations --- and those that we could have had in the lab given the correlations we ended up observing. 

The first example of this in the PM scenario was focused on giving a robust self-testing statement for the states in the quantum random access code~\cite{Tavakoli2018}. Let $\psi_x$ represent the  states in Eq.~\eqref{RACvectors}, which we would have self-tested if we had observed the maximal value $W_\text{RAC}=\frac{2+\sqrt{2}}{4}$. Then, let $\rho_x$ be the states actually generated in the lab. Up to an arbitrary extraction channel $\Lambda$, the average fidelity between them is given by $f(\{\rho_x\})=\max_\Lambda \frac{1}{4}\sum_x\bracket{\psi_x}{\Lambda[\rho_x]}{\psi_x}$. Now, let $S(W_\text{RAC})$ denote the set of all qubit states $\{\rho_x\}$ for which there exist some measurements such that it is possible to achieve the random access code score $W_\text{RAC}$. To obtain a robust self-testing statement, we must find the smallest average fidelity compatible with $S(W_\text{RAC})$, i.e.~
\begin{equation}
    F(W_\text{RAC})=\min_{\{\rho_x\}\in S(W_\text{RAC}) }  f(\{\rho_x\}).
\end{equation}
To address this optimisation problem, one can borrow techniques from self-testing in Bell nonlocality~\cite{Kaniewski2016}. The idea is to find a dual channel $\Lambda^\dagger(M_1,M_2)$, which depends on Bob's two measurements, so that one can find inequalities of the form
\begin{equation}\label{ineqtools}
\Lambda^\dagger(M_1,M_2)[\psi_x] \geq sT_x + t_x(M_1,M_2)\mathds{1},
\end{equation}
where $t_x(M_1,M_2)$ are real-valued scalars that depend on the observables. Here, we have defined $T_x=\frac{1}{16}[(-1)^{x_1}M_1+(-1)^{x_2}M_2]$ and observables $M_y=M_{0|y}-M_{1|y}$ to simplify notation. Notice that this definition is motivated by the relation $W_\text{RAC}=\frac{1}{2}+\sum_x \tr\left(\rho_x T_x\right)$. The reason that we want to find these inequalities is that, if we had them, we could conclude that
\begin{align}
f(\{\rho_x\})&=\max_\Lambda \frac{1}{4}\sum_x \tr\left(\rho_x \Lambda^\dagger[\psi_x]\right) \nonumber\\  &\geq \frac{s}{4}\sum_x \tr\left(\rho_x T_x\right) +\sum_x \frac{t_x}{4}.
\end{align}
If we now define $t=\min_{M_1,M_2} \frac{1}{4}\sum_x t_x(M_1,M_2)$, we obtain the robust self-testing statement
\begin{equation}\label{robustQRAC}
  F(W_\text{RAC})\geq \frac{s}{4}\bigg(W_\text{RAC}-\frac{1}{2}\bigg)+t.
\end{equation}
The construction of the inequalities \eqref{ineqtools} was shown in Ref.~\cite{Tavakoli2018}.  It ultimately leads to $s=4(1+\sqrt{2})$ and $t=\frac{2-\sqrt{2}}{4}$. Notice that at the maximal value, $W_\text{RAC}=\frac{2+\sqrt{2}}{4}$, the formula gives a perfect fidelity, $F(W_{\text{RAC}})=1$, as expected from noise-free self-testing.

Let us turn to the measurements associated with the quantum random access code. We know from the Frenkel-Weiner theorem (see Sec.~\hyperref[sec:FWtheorem]{II-B-1}) that quantum advantages are impossible if Bob has only one measurement. This means that in any scenario where Bob has measurement choices, but those choices correspond to a set of measurements that are jointly measurable (i.e.~they can effectively be reduced to a single measurement), then again no quantum advantage is possible. See Box~\ref{Box:JM} for a brief summary of joint measurability. We can therefore ask whether it is possible to quantify the incompatibility of $M_1$ and $M_2$ solely from the observed value of $W_\text{RAC}$. As a maximal incompatibility is associated with a pair of Pauli observables, which we are guaranteed to self-test at the maximal value of $W_\text{RAC}$, we can view such a statement as a robust self-test. 

\begin{backgroundbox}[label={Box:JM}]{Joint measurability}
Let $\mathbf{M}\equiv \{M_{a|x}\}_{a,x}$ be a set of POVMs. It is said that $\mathbf{M}$ is jointly measurable if all the different measurements, associated with the index $x$, can be reconstructed by the following recipe. First, perform a single POVM $\{G_\mu\}_\mu$ with outcome $\mu$. Then, use $\mu$ and the choice of $x$ to classically determine the outcome via some stochastic rule $p(a|x,\mu)$. Formally, 
\begin{equation}
    M_{a|x} = \sum_\mu p(a|x,\mu)G_\mu.
\end{equation}
If this is not possible, it is said that the measurements in $\mathbf{M}$ are incompatible. Exact characterisations are known for (so-called unbiased) dichotomic qubit measurements, but it is only in the case of two observables that the formula is closed. In this simplest case, two qubit observables $\vec{m}_1\cdot \vec{\sigma}$ and $\vec{m}_2\cdot \vec{\sigma}$ are jointly measurable if and only if~\cite{Busch1986}
\begin{equation}\label{Buschquantifier}
    Q\equiv \lVert \vec{m}_1+\vec{m}_2\rVert +\lVert \vec{m}_1-\vec{m}_2\rVert \leq 2.
\end{equation}

\end{backgroundbox}

To this end, we follow Ref.~\cite{Carmeli2020} and consider the most general dichotomic qubit POVM, which is written $\frac{1}{2}\left((1\pm\alpha)\mathds{1}\pm\vec{m}\cdot\vec{\sigma}\right)$, where $\lVert\vec{m}\rVert \leq 1$ and $|\alpha|\leq 1-\lVert\vec{m}\rVert$. The measurement is called unbiased if $\alpha=0$, which in particular encompasses all basis measurements. In Box~\ref{Box:JMinQRAC} we do the small calculation that leads to the robust self-test
\begin{equation}\label{incompqrac}
   W_\text{RAC}\leq  \frac{1}{2}+\frac{Q}{8},
\end{equation}
where $Q$ is the quantifier of the incompatibility of a pair of unbiased qubit measurements defined in Box~\ref{Box:JM}. For jointly measurable observables ($Q\leq 2$), the bound becomes $W_\text{RAC}\leq \frac{3}{4}$, which is precisely the classical limit in the random access code of Eq.~\eqref{RAC4222}. This shows that every pair of incompatible unbiased observables makes possible a quantum advantage in the random access code, and that their degree of incompatibility, $Q$, determines the reachable values of $W_\text{RAC}$. In other words, by observing a nonclassical value of $W_\text{RAC}$, we can deduce a non-trivial lower bound on $Q$.

\begin{examplebox}[label={Box:JMinQRAC}]{Semi-device-independent quantification of measurement incompatibility}
Consider the quantum random access code and let $\rho_{x_1x_2}$ be Alice's qubits. We write Bob's qubit POVMs as $M_{b|y}=\frac{1}{2}\left[(1+(-1)^b\alpha_y)\mathds{1}+(-1)^b\vec{m}_y\cdot\vec{\sigma}\right]$ for $y\in\{1,2\}$ and $b\in\{0,1\}$. Inserting this in the definition of $W_{\text{RAC}}$ in Eq.~\eqref{RAC4222} gives
\begin{align}\nonumber
    W_\text{RAC} & = \frac{1}{8} \sum_{x_1,x_2,y} \tr(\rho_{x_1x_2}M_{x_y|y})\\ \nonumber
    & = \frac{1}{16}\sum_{x_1,x_2,y} \left( 1+(-1)^{x_y}\alpha_y\right) \\
    &+\frac{1}{16}\sum_{x_1,x_2,y} (-1)^{x_y} \tr\left( \rho_{x_1x_2} \vec{m}_y\cdot\vec{\sigma}\right).
\end{align}
Notice now that 
\begin{equation}
    \sum_{x_1,x_2,y} (-1)^{x_y}\alpha_y=\sum_{y}\sum_{x_{\bar{y}}} \sum_{x_y} (-1)^{x_y}\alpha_y,
\end{equation}
where $\bar{y}=3-y$, and that the innermost sum vanishes, i.e.~$\sum_{x_y} (-1)^{x_y}\alpha_y=0$. Therefore, the above simplifies to 
\begin{equation}
    W_\text{RAC} = \frac{1}{2}+\frac{1}{16}\sum_{x_1,x_2,y} (-1)^{x_y} \tr\left( \rho_{x_1x_2} \vec{m}_y\cdot\vec{\sigma}\right).
\end{equation}
The optimal choice of $\rho_{x_1x_2}$ is the rank-one projector onto an eigenvector of $\sum_y (-1)^{x_y} \vec{m}_y\cdot\vec{\sigma}$ corresponding to its largest eigenvalue. Since this is a $2\times 2$ matrix with zero trace, its largest eigenvalue is given by $\lVert \sum_y  (-1)^{x_y} \vec{m}_y \rVert$. Hence, a tight bound becomes
\begin{align}\nonumber
    W_\text{RAC} & \leq \frac{1}{2} + \frac{1}{16}\sum_{x_1,x_2} \bigg\lVert  (-1)^{x_1} \vec{m}_1+(-1)^{x_2} \vec{m}_2 \bigg\rVert \\ \nonumber
    & = \frac{1}{2} + \frac{1}{16}\Bigg(  \lVert \vec{m}_1+ \vec{m}_2 \rVert + \lVert \vec{m}_1- \vec{m}_2 \rVert \\
    & \quad + \lVert - \vec{m}_1+\vec{m}_2 \rVert +\lVert - \vec{m}_1-\vec{m}_2 \rVert\Bigg) = \frac{1}{2} + \frac{Q}{8},
\end{align}
which is the result appearing in Eq.~\eqref{incompqrac}.
\end{examplebox}

Going beyond the quantum random access code, robust self-testing can be established in a more general picture. For this, we revisit our discussion of Ref.~\cite{Navascues2023}. By saturating the bound in Eq.~\eqref{miguelselftest} we obtained matching projector overlaps as a step towards self-testing pure states. This can be upgraded into a robust self-testing statement~\cite{Navascues2023}. Specifically, for any value $W_\Psi$ such that  $W_\Psi^Q-W_\Psi\leq \epsilon$ it holds that the projector overlaps between any pair of different states deviate by at most $O(\sqrt{\epsilon})$ in absolute value from the projector overlaps associated with the target states. Their impurity is also no larger than $O(\epsilon)$. This approach can be further developed to self-test every extremal non-projective measurement, but in the next section we focus on simpler examples of this based on qubits, and on how the property of non-projectivity can be certified without self-testing.

\subsubsection{Non-projective measurements}\label{sec:POVMcertification}
Naimark’s theorem states that every non-projective measurement can be realised as a projective measurement on a suitably enlarged Hilbert space. Thus, non-projectivity is most naturally understood as a dimension-restricted property. This makes the use of dimension-restricted prepare-and-measure scenarios to certify non-projectivity a well-motivated approach.

\begin{figure}[!tbp]
    \centering
    \includegraphics{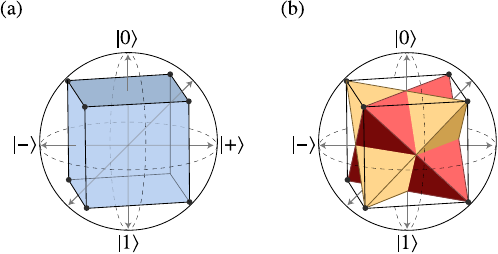}
    \caption{\emph{Geometry of the SIC certification}. (a) The optimal $(3,2)$ quantum random access code self-tests Alice's eight preparations as the vertices of a cube on the Bloch sphere, with Bob's three binary measurements aligned with the Pauli axes $\sigma_X$, $\sigma_Y$ and $\sigma_Z$ (grey arrows). (b) Adding the four-outcome input $\mathbf{inp}$ singles out four cube vertices, shown as the solid tetrahedron. The penalty term in $W_{\mathrm{SIC}}$ enforces zero probability for the corresponding outcomes, forcing the effects of $\mathbf{inp}$ to be antipodal to these states. These effects therefore form the dual, mirror-image tetrahedron, realising the qubit SIC-POVM geometry.}
    \label{F:SIC-geometry}
\end{figure}

The strongest way to approach this is via self-testing, i.e.~to deduce the non-projective measurement (modulo the unavoidable degrees of freedom) directly from the correlations. However, it is often relevant to pose a more relaxed question: is the measurement non-projective at all? Refs.~\cite{Mironowicz2019, Smania2020} showed that PM scenarios enable practical protocols for such certification. To showcase this, let us take aim at self-testing perhaps the most well-known non-projective qubit measurement, the so-called symmetric informationally complete measurement (SIC). It is given by $\{E_a\}_{a=1}^{4}$ where $E_a=\frac{1}{2}\ketbra{\psi_a}$ and the qubit states $\ket{\psi_a}$ have Bloch vectors
\begin{align}\nonumber\label{SICvectors}
&\vec{m}_1=\frac{1}{\sqrt{3}}\left(+1,+1,+1\right), && \vec{m}_2=\frac{1}{\sqrt{3}}\left(+1,-1,-1\right), \\
& \vec{m}_3=\frac{1}{\sqrt{3}}\left(-1,+1,-1\right), && \vec{m}_4=\frac{1}{\sqrt{3}}\left(-1,-1,+1\right).
\end{align}
One may note that these vectors form the vertices of a regular tetrahedron. To semi-device-independently certify such a measurement, we will consider  the quantum random access code associated with $(n,d)=(3,2)$ [as defined in Eq.~\eqref{RACwitness}]. Recall from below Eq.~\eqref{RACbits} that the maximal quantum value of this task is $W_{3,2}=\frac{1}{2}\left(1+\frac{1}{\sqrt{3}}\right)$ and that this is achieved by Alice preparing eight states whose Bloch vectors form a cube on the Bloch sphere, $\frac{1}{\sqrt{3}}\left(\pm1,\pm1,\pm1\right)$, and Bob performing measurements  $\sigma_X$, $\sigma_Y$ and $\sigma_Z$. In fact, this construction can be self-tested up to a global unitary or antiunitary $U$~\cite{Tavakoli2018} [see Fig.~\hyperref[F:SIC-geometry]{\ref{F:SIC-geometry}(a)}].

However, in this scenario we cannot certify non-projectivity because Bob's measurements have only binary outputs, and it is well-known that all extremal  binary POVMs are projective. Let us therefore supply Bob with one more input, which we name $\textbf{inp}$, for which he can have four possible outcomes. We now extend the quantum random access code to the following witness,
\begin{equation}
    W_\text{SIC} = \frac{1}{24}\sum_{x_1,x_2,x_3=0,1}\sum_{y=1}^3 p(b=x_y|x,y) - k \sum_{x'=1}^4 p(b=x'|x,\textbf{inp}),
\end{equation}
where we have defined $x'\in\{1,2,3,4\}$ as corresponding to $x\in\{(1,1,1),(1,0,0),(0,1,0),(0,0,1)\}$.  All we have done is to subtract a term from the original random access code. This term is constructed so that Bob's job is to avoid outputting $x'$, and the penalty for failing is some positive constant $k$. Clearly, the best quantum strategy for $W_{\text{SIC}}$ cannot achieve a value higher than that for $W_{3,2}$. The only way to have them both equal is if $p(b=x'|x,\textbf{inp})=0$ $\forall x'$. Hence, if we observe $W_{\text{SIC}}=\frac{1}{2}\left(1+\frac{1}{\sqrt{3}}\right)$, we conclude that this condition holds and also that $W_{3,2}=\frac{1}{2}\left(1+\frac{1}{\sqrt{3}}\right)$. The domain of $x'$ has been selected  so that the four corresponding vertices of the cube form a mirror-image tetrahedron as compared to that defined by the Bloch vectors in Eq.~\eqref{SICvectors} [see Fig.~\hyperref[F:SIC-geometry]{\ref{F:SIC-geometry}(b)}]. This means we have achieved our certification.  

We can then go further and ask how large the value of $W_\text{SIC}$ must be in order to certify that \textbf{inp} must have been a non-projective measurement. This means that we must compute the maximal value of $W_\text{SIC}$  under projective measurements. A projective qubit measurement can only have two outcomes, but \textbf{inp} has four. Therefore, we can select the two relevant outcomes in $\binom{4}{2}=6$ ways, but due to the symmetries of $W_\text{SIC}$ they are all ultimately equivalent. Tight analytical bounds can be placed on $W_\text{SIC}$ as a function of $k$, and one can then optimise the choice of $k$ to make the certification as noise-robust as possible~\cite{Smania2020}.

The key idea here is to first self-test a set of states that are (anti-) aligned with the non-projective measurement that we want to certify, then add an extra measurement setting for which we introduce a corresponding penalty (reward) term in the witness. This type of approach is broadly used: first appearing in the context of random number generation in Bell scenarios ~\cite{Acin2016}, then for the certification of SIC-POVMs as above~\cite{Mironowicz2019, Smania2020} which was extended to other elegant measurements~\cite{Drotos2024}, and finally extended further to certify non-projective measurements in the high-dimensional regime~\cite{Rosset2019, Martinez2023, Navascues2023}. One may object that these constructions all view the certification of non-projective measurements essentially as a perturbation that we add to a conventional self-test based on projective measurements. In Sec.~\hyperref[sec:simple_measurements]{\alex{III-C-3}}, we will encounter a PM scenario where non-projective measurements are not added post-hoc to the certification but instead are fundamental to a quantum advantage~\cite{Svegborn2026}. 

\subsubsection{Nonstabiliser states}
\label{subsubsec:nonstab}
In the context of quantum computation and quantum error correction, a fundamental resource is nonstabiliserness. It plays a central role in understanding the capabilities and advantages of quantum algorithms. The importance of nonstabiliserness is rooted in the Gottesman-Knill theorem~\cite{gottesman1998heisenberg, aaronson2004improved}, which establishes that stabiliser states~\cite{Veitch_2014}---those generated by Clifford circuits (composed of Hadamard, $\tfrac\pi2$-phase, and CNOT gates)---can be efficiently simulated on classical computers. As a consequence, to attain a genuine quantum computational advantage, it is necessary to use resources that go beyond the stabiliser framework. For the readers unfamiliar with the stabiliser formalism, we introduce the qubit stabiliser states in Box~\ref{Box:Qubit_Stab}.

\begin{backgroundbox}[label={Box:Qubit_Stab}]{Stabiliser polytope}
For a single qubit, the set of stabiliser states consists of the six eigenstates of the Pauli operators $\sigma_X, \sigma_Y, \sigma_Z$, namely $\{ \ket{0}, \ket{1}, \ket{+}, \ket{-}, \ket{+i}, \ket{-i} \}$. These correspond to the vertices of an octahedron inscribed in the Bloch sphere. The stabiliser polytope $\mathrm{STAB}_2$ is the convex hull of these six states. Geometrically:
\begin{center}
    \includegraphics{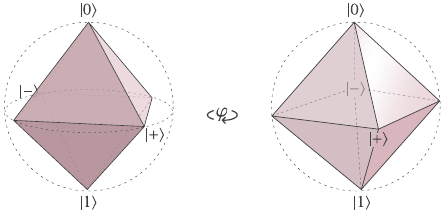}
\end{center}
Any state lying outside this octahedron--such as the $H$-state defined in Eq.~\eqref{eq:Hstate}--is a non-stabiliser state.
\end{backgroundbox}

What distinguishes a nonstabiliser state from a stabiliser one? A quantum state is nonstabiliser if it cannot be written as a convex mixture of pure stabiliser states~\cite{Veitch_2014}. For a single qubit, the pure stabiliser states are the six eigenstates of the Pauli operators $\vec\sigma=(\sigma_X, \sigma_Y, \sigma_Z)$, represented by the six points where the coordinate axes intersect the Bloch sphere. Their convex hull is an octahedron, known as the stabiliser polytope (see Box~\ref{Box:Qubit_Stab}). Hence, for a general qubit state $\rho=\tfrac12(\iden+\vec{a}\cdot \vec{\sigma})$, with Bloch vector $\vec a = (a_x, a_y, a_z)$, membership in the stabiliser polytope is completely characterised by 
\begin{eqnarray}
\|\vec{a}\|_1 =|a_x|+|a_y|+|a_z| \leq 1.
\end{eqnarray}
A qubit state is nonstabiliser when this inequality is violated. For $n$ qubits, the stabiliser polytope is the convex hull of all pure $n$-qubit stabiliser states and, in the same way, a state is nonstabiliser if it lies outside this polytope. However, the number of extremal points of this polytope grows super-exponentially with the number of qubits. Consequently, certifying the presence of nonstabiliserness becomes computationally costly, and the problem is NP-hard in general~\cite{leone2026,varela2026predicting}. 

Recent works have aimed to develop efficient methods to certify this resource within semi-device-independent frameworks~\cite{Zamora_2025, Macedo_2025,zamora2026logical}. While one might initially consider a fully device-independent approach using Bell inequalities to certify nonstabiliserness, the definition of stabiliser states is explicitly dimension-dependent. As shown in Ref.~\cite{Macedo_2025}, even in Bell scenarios, this dependency forces a semi-device-independent characterisation rather than a complete device-independent one, as the stabiliser set shifts with the assumed Hilbert space size. This fundamental limitation motivates the use of the dimension-restricted PM framework, where the dimension $d$ is an established assumption. Note that this is analogous to the semi-device-independent certification of non-projective measurements described previously: a dimension bound is strictly required to prevent the measurements from being trivially explained away as projective measurements in a larger Hilbert space via Naimark's dilation theorem.

Our goal is to find witnesses that separate the PM correlations arising only from stabiliser states from general quantum correlations, thereby certifying the presence of nonstabiliserness. As we shall see, a key takeaway from this section is that nonstabiliserness is not only essential for quantum computing, but it can also be necessary for quantum advantage in one-way communication protocols. This is because any restriction on the set of prepared states---such as confining them to the stabiliser polytope---naturally imposes a fundamental limit on the set of achievable correlations.

To showcase how nonstabiliserness can be certified in practice, we consider the simplest non-trivial qubit scenario $(|X|,d,|Y|,|B|) =(3,2,2,2)$.  As introduced in Sec.~\hyperref[sec:ClassicalModels]{II-A}, the only relevant facet in this scenario is Eq.~(\ref{Gallego3222}), with quantum maximum $W^Q = 3+\sqrt{2}$. 

\begin{examplebox}[label={Box:Stab_Bound}]{Stabiliser bound in the (3,2,2,2) scenario}
Define the set of stabiliser correlations $\mathcal{C}_{\mathrm{STAB},2}$ as 
\begin{equation}
    \mathcal{C}_{\mathrm{STAB},2} := \biggl\{p(b|x,y) = \tr(\rho_xM_{b|y})|\ \rho_x\in \mathrm{STAB}_2\biggl\},
\end{equation}
with $\{M_{b|y}\}$ arbitrary POVMs. Consider the general states $\rho_x =\frac{1}{2}(\iden + \vec{n}_x\cdot\vec\sigma)$. In Ref.~\cite{Poderini_2020} it was shown that Eq.~\eqref{Gallego3222} can be maximised analytically over all POVMs, resulting in 
\begin{equation}
   \max_{M_{b|y}} W=\frac{1}{2}\{ 5+\max\{1,\|\vec{n}_0+\vec{n}_1-\vec{n}_2 \|_2\} + \|\vec{n}_1-\vec{n}_0\|_2\},\label{eq:s3_vecto.rs}
\end{equation}
where the PM statistics resulting from a combination of states giving a value greater than $4$ in the above equation cannot be reproduced by a classical model. 
We now restrict the above expression to stabiliser states. Since Eq.~(\ref{Gallego3222}) is linear  on the convex set of states, its maximum is attained at extreme points of the set of states. In the stabiliser-restricted case, these are the vertices of the stabiliser octahedron, i.e.~the eigenstates of the Pauli matrices. Therefore, to maximise the above expression, in principle we need to consider all the possible $6^3 =216$ combinations $(\vec n_0, \vec n_1, \vec n_2)$ of the vectors $\{(\pm1,0,0),(0,\pm1,0),(0,0,\pm1) \}$. However, from a quick inspection one can conclude that the optimal combination is such that $\vec{n}_0\cdot\vec{n}_1 =0$ and $\vec{n}_0=-\vec{n}_2$. This gives the maximum value
\begin{equation}
    \frac{1}{2}\{ 5+\|\vec{n}_0+\vec{n}_1-\vec{n}_2 \|_2 + \|\vec{n}_1-\vec{n}_0\|_2\} = \frac{1}{2}\{ 5+\sqrt{5}+\sqrt{2}\}.
\end{equation}
Therefore,  it is proven that $\mathcal{C}_{\mathrm{STAB},2}$ is a proper subset of the set of quantum correlations in the scenario $(3,2,2,2)$ since $W^C = 4<W^{\mathrm{STAB_2}} = \frac{1}{2}\{ 5+\sqrt{5}+\sqrt{2}\}<W^Q = 3+\sqrt{2}$. Furthermore, it offers an explicit witness to detect non-stabiliser states. 
\end{examplebox}

As shown in Box~\ref{Box:Stab_Bound}, the maximum value of the left-hand side of Eq.~\eqref{Gallego3222} restricted to stabiliser preparations in the qubit case is~\cite{Zamora_2025},
\begin{equation}\label{eq:s3_vectors}
    W^{\mathrm{STAB_2}} =\frac{\sqrt{5} + \sqrt{2} + 5 }{2}.
\end{equation}
As a consequence, if an experimenter observes a value $W>W^{\mathrm{STAB_2}}$, she can be sure that at least one of the preparations is outside of the stabiliser polytope. A similar idea was later adapted for the detection of entanglement in the prepared states \cite{patra2026certification}.

A first relevant question is whether one can certify more accurately how many preparations are non-stabiliser states. For example, one can define the two-stabiliser value $ W^{\mathrm{2-STAB_2}} $ as the maximum of the LHS of Eq.~\eqref{Gallego3222} assuming $2$ of the $3$ preparations are stabiliser states. Unfortunately,  considering one nonstabiliser state is already enough to saturate the quantum bound. However, it is interesting to note that the only way to reach $W^Q$ is when the non-stabiliser preparation  corresponds to a Clifford rotation of the well-known $H$-state,
\begin{equation}
H =  \frac{1}{2}\qty(\iden +\frac{\sigma_X+\sigma_Z}{\sqrt{2}}).\label{eq:Hstate}
\end{equation}
This can be seen from the fact that, as exemplified with the isosceles triangle defined by Eq.~\eqref{eq:optimal_states_gallego}, the optimal value can be obtained (up to a unitary) by states with orthogonal unit Bloch vectors $\vec{n}_0, \vec{n}_1$, and $\vec{n}_2 = -\frac{\vec{n}_0+\vec{n}_1}{\sqrt{2}}$. Therefore, if $\vec{n}_0$ and $\vec{n}_1$ are restricted to be stabiliser states, $\vec{n}_2$ corresponds to the Bloch vector of one of the Clifford rotations of Eq.~\eqref{eq:Hstate}. Hence, under these assumptions, a maximum quantum violation of Eq.~\eqref{Gallego3222} self-tests the $H$-state.

Motivated by this lack of ``fine-grained detection'', a relevant generalisation is to consider more preparations. A natural candidate to consider is the average success probability for the $(n,2)$ random access code [see Eq.~\eqref{RACwitness}]. In this case we have $2^n$ preparations. The simplest RAC with $n=2$ is trivial, because the optimal states form a square in the XZ-plane, which can be rotated onto four vertices of the stabiliser octahedron. Hence, there is no gap between the stabiliser and the quantum scenarios. This is no longer true for $n=3$, since the cube defined by the optimal state vectors as described in Sec.~\hyperref[sec:QRAC]{II-B-3} cannot be rotated into the stabiliser polytope. Following similar reasoning one can see that there is a gap for $n\geq4$. In these cases a fine-grained certification can be found but more importantly, it shows that in order to obtain the maximum quantum value in a random access code, one requires nonstabiliserness. 
 
In fact, in Ref.~\cite{Chowdhury_2026} it was recently shown that any one-way communication protocol using a quantum system of prime dimension $d$ can be perfectly replicated by a classical channel of the same dimension $d$ (aided by shared randomness), provided the encoding is restricted to the stabiliser formalism and the decoding is limited to Clifford group operations.  In the same line of research, a recent article~\cite{stratton_2026} studies the relevance of nonstabiliserness in the measurements for quantum state discrimination. It shows that in order to obtain the Helstrom bound, non-stabiliser measurements are required in general. As a consequence, since an $(n,2)$ random access code  can be understood as a collection of quantum state discrimination  problems, these results imply bounds on the average success probabilities when restricted to stabiliser measurements. 

Finally, Ref.~\cite{Zamora_2025} offers the following optimisation problem to find the stabiliser value for systems of dimension $d$ using the $(2,d)$ RAC. Let $\{A^{\vec q} \}_{\vec q }$ be the facets in the H-representation of the stabiliser polytope $\mathrm{STAB}_d$ such that any stabiliser state $\rho_x$ obeys $\tr(\rho_{x} A^{\vec q}) \geq 0$. For a concrete description of these facets we refer to~\cite{Howard_2014, Zamora_2025}. Denoting $[d] =\{0,\dots,d-1\}$, the maximum stabiliser success probability is given by:
    \begin{align}
        W_{2,d}^\mathrm{STAB} = &\max_{\substack{\{\rho_{{x_1},{x_2}} \}, \{M_{b|y}\}}}  W_{2,d}(\{\rho_{x_1x_2} \},\{M_{b|y}\}) \nonumber\\
        \text{s.t.} \quad & \rho_{x_1x_2}  \succeq 0, \:\: \tr(\rho_{x_1x_2})=1, \:\: \forall x_1,x_2 \in [d] \nonumber\\
        & \tr(\rho_{x_1x_2}  A^{\vec q}) \geq 0, \:\: \forall \vec{q} \:\:\text{and}\:\: \forall x_1,x_2 \in [d]\nonumber\\
        & \sum_{b\in[d]} M_{b|y} = \iden, \quad \forall y \in \{1,2\} \nonumber\\
        & M_{b|y}  \succeq 0,  \:\: \forall b \in [d], \forall y \in \{1,2\}.
    \end{align}
The optimisation problem is bilinear, meaning it becomes a standard SDP if either the states or the measurements are fixed. To find the maximum over both, we employ the seesaw algorithm introduced in Sec.~\hyperref[sec:seesaw]{II-B-5}, alternating the optimisation between Alice and Bob until convergence. Although this numerical approach only guarantees a lower bound on the optimal value, it is highly effective for characterising these boundaries. The above optimisation gives a path to tackle the stabiliser value problem for higher-dimensional systems. However, given the heuristic nature of the seesaw algorithm, an outer approximation of the set of stabiliser correlations is required to guarantee the existence of a gap. This is an open problem. 
    
\subsubsection{Instruments}
\label{subsubsec:instruments}

Another type of quantum operation that is relevant for certification is the quantum instrument. These are devices that take as input a quantum state and provide as output both a classical outcome and a quantum state. Thus, they can be viewed as generalisations of measurements, in the sense that they not only produce the measurement outcome but also a post-measurement state. 

\begin{figure}[t]
    \centering
    \includegraphics{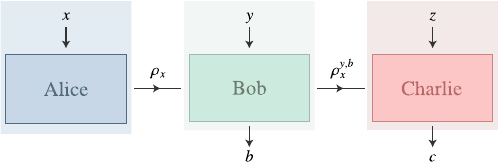}
    \caption{\emph{Prepare-transform-measure scenario.} Alice receives an input $x$ and prepares a message $m$, which is sent to Bob. Upon receiving the state, Bob chooses an input $y$ and applies an instrument, producing a classical outcome $b$ and a post-measurement state $\rho_x^{y,b}$. This state is then forwarded to Charlie, who selects an input $z$, performs a measurement and outputs $c$.
     }
    \label{fig_sequential_RAC}
\end{figure}

Quantum instruments cannot be certified in the PM scenario, because any hypothetical post-measurement state produced by Bob is irrelevant to the correlations $p(b|x,y)$ observed in the experiment. In order to make instruments relevant, we must extend the PM scenario into a three-party prepare-transform-measure scenario, as illustrated in Fig.~\ref{fig_sequential_RAC}. In this extension, Alice selects $x$ and prepares a dimension-restricted state $\rho_x$ and sends it to Bob. Bob selects $y$ and then applies a quantum instrument $\{I_{b|y}\}_b$ to $\rho_x$. The classical outcome of Bob is denoted $b$ and the quantum output is denoted $\rho_{x}^{y,b}$, which is still restricted to the same dimension as $\rho_x$. In Box~\ref{Box:instrument} we give a brief summary of quantum instruments. The post-measurement state is given to a new party, Charlie, who selects $z$ and measures it with a POVM $\{C_{c|z}\}_c$  to obtain an outcome $c$. Thus, the correlations in the scenario are of the form 
\begin{equation}
    p(b,c|x,y,z)=\tr(I_{b|y}(\rho_x) C_{c|z}).
\end{equation}

\begin{backgroundbox}[label={Box:instrument}]{Quantum instruments}
A quantum instrument is an operation that transforms a density matrix, $\rho$, into a classical outcome $a$ and a quantum state which is conditioned on the classical outcome. It can be represented as a collection of completely positive trace non-increasing maps $\{I_a\}_a$ with the property that $\sum_a I_a$ is trace preserving. The probability of the classical outcome can then be written as $p(a)=\tr\left(I_a(\rho)\right)$. Equivalently, we can write 
\begin{equation}
    p(a)=\tr[\mathds{1} I_a(\rho)]=\tr[I_a^\dagger(\mathds{1}) \rho],
\end{equation}
where $I_a^\dagger$ is the map dual to $I_a$. This allows us to interpret $\{I^\dagger_a(\mathds{1})\}_a$ as the POVM implemented by the instrument. It is important to understand that the same POVM can be implemented by many different instruments.

\vspace{0.2cm}

For SDP implementations, it is convenient to represent instruments in terms of their Choi operators. The Choi operator is given by
\begin{equation}
    \eta_a = (I_a \otimes \mathds{1})[\phi^+],
\end{equation}
where $\phi^+$ is the maximally entangled state in dimension $d$. From this, it follows that the Choi operators are characterised by $\eta_a\succeq 0$, $\tr_A \sum_a \eta_a=\frac{\mathds{1}_B}{d}$. One can then express the action of the map as
\begin{equation}
    I_a(\rho)= d \tr_B[\left(\mathds{1}_A\otimes \rho^T_B\right)\eta_a].
\end{equation}

\end{backgroundbox}
Prepare-transform-measure scenarios were explored in early sequential communication experiments~\cite{Trojek2005}, later studied for dimension and nonclassicality tests by Bowles et al.~\cite{Bowles2015testing} and further developed for sequential protocols in Refs.~\cite{Mohan2019, Miklin2020}. They are motivated by the fact that they feature an information-disturbance trade-off. Specifically, if Bob wants to extract more information about Alice's input by measuring her state, then he induces a larger disturbance in the post-measurement state, which then reduces Charlie's ability to extract information about Alice's input. Alternatively, Bob could extract very little information with his instrument, thereby enabling Charlie to extract more. These trade-offs can be used to certify quantum instruments.  

To this end, consider the following specific task. Alice takes as input $x=(x_1,x_2)\in\{0,1\}^2$ and wants to use her prepared states $\rho_{x_1x_2}$ for a random access code with both Bob and Charlie. Her average success probabilities with Bob and Charlie are, respectively,
\begin{align}
    & W_{AB} = \frac{1}{8}\sum_{x_1,x_2=0,1}\sum_{y=1,2} \tr\left( I_{x_y|y}(\rho_x )\right), \\
    & W_{AC} = \frac{1}{8}\sum_{x_1,x_2=0,1}\sum_{z=1,2} \tr\left( \tilde{\rho}_x C_{x_z|z}\right),
\end{align}
where we have defined the average post-measurement state of Bob as $\tilde{\rho}_x = \frac{1}{2}\sum_{b,y} I_{b|y}(\rho_x)$. 

\begin{figure}[t]
    \centering
    \includegraphics{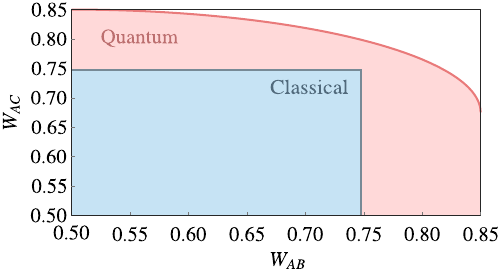}
    \caption{\emph{Trade-off for random access codes.} Achievable pairs of average success probabilities in the prepare-transform-measure scenario. Classical strategies are confined to the blue square, satisfying $W_{AB} \leq \tfrac34$ and $W_{AC}\leq \tfrac34$. Quantum strategies, shown in the red region, can exceed this region due to an information-disturbance trade-off, with the red curve representing the optimal information-disturbance trade-off between Bob's information gain and Charlie's remaining ability to recover Alice's input.
     }
    \label{fig_sequential_tradeoff}
\end{figure}

Because classical models have no information-disturbance relations, they are limited by the same constraints as the random access code in the ordinary PM scenario [Eq.~\eqref{RAC4222}], namely $W_{AB}\leq \frac{3}{4}$ and $W_{AC}\leq \frac{3}{4}$. This corresponds to the square in Fig.~\ref{fig_sequential_tradeoff}. Let us now consider the following quantum strategy. Alice prepares the same qubit states as for the ordinary quantum random access code and Charlie makes the same measurements as Bob used to do. This means the preparation Bloch vectors are given in  Eq.~\eqref{RACvectors} and that Charlie's measurements are $\sigma_X$ and $\sigma_Z$. The central question is how to choose Bob's instruments. We make the choice $I_{b|y}(\cdot)= K_{b|y} (\cdot) K_{b|y}^\dagger$, where the Kraus operators are
\begin{align}\nonumber
&  K_{b|1} = \sqrt{\frac{1+(-1)^b\gamma }{2}} \ketbra{+}{+} + \sqrt{\frac{1-(-1)^b\gamma }{2}} \ketbra{-}{-}, \\
&  K_{b|2} = \sqrt{\frac{1+(-1)^b\gamma }{2}} \ketbra{0}{0} + \sqrt{\frac{1-(-1)^b\gamma }{2}} \ketbra{1}{1},
\end{align}
where $\ket{\pm}$ are the eigenstates of $\sigma_X$ and $\gamma\in[0,1]$. A direct calculation shows that the average post-measurement states of Bob become identical to Alice's states, but with their Bloch vector shrunk  by a factor that depends monotonically on $\gamma$. In other words, by tuning $\gamma$, Bob can choose how the correlations are distributed between the two quantum random access codes. One finds that 
\begin{align}
    & W_{AB} = \frac{1}{4}\left(2+\sqrt{2}\gamma\right), \\
    & W_{AC} = \frac{1}{8}\left(4+\sqrt{2}\bigg(1+\sqrt{1-\gamma^2}\bigg)\right).
\end{align}
This turns out to be the optimal trade-off possible between the two random access code parameters (see Fig.~\ref{fig_sequential_tradeoff}). 

Furthermore, it turns out that correlations associated with the boundary of the quantum set (illustrated by a solid red line in Fig.~\ref{fig_sequential_tradeoff}) can be associated with a self-testing statement. From Sec.~\hyperref[sec:self_test]{II-D-1}, we already know this to be true when $\gamma=1$; this corresponds to Bob making projective measurements and the value of $W_{AB}$ becomes maximal, thereby achieving a self-test in the standard quantum random access code. This turns out to extend to  $0<\gamma\leq1$. A pair of values $(W_{AB},W_{AC})$ on the curved boundary of the quantum set with $0<\gamma\leq1$ self-tests that Alice's preparations are unitarily equivalent to those in Eq.~\eqref{RACvectors}, that Charlie's measurements are unitarily equivalent to $\sigma_X$ and $\sigma_Z$, and that Bob's instrument is unitarily equivalent to Kraus operators $K_{b|y}=\sqrt{M_{b|y}}$, where $M_{b|y}$ represent two POVMs corresponding to $\gamma \sigma_X$ and $\gamma \sigma_Z$. The proof of this result is somewhat lengthy and can be found in Ref.~\cite{Mohan2019}. There, it is also shown how to bound the sharpness of Bob's measurement in the presence of noise. If the value of the pair $(W_{AB},W_{AC})$ is not on the boundary of the quantum set, then its distance from the boundary allows one to determine upper and lower bounds on the unsharpness of the measurements implemented by Bob's instrument. Such certification via sequential quantum random access codes has also been reported in experiments that manipulate simultaneously the path and the polarisation of single photons~\cite{Anwer2020, Foletto2020}. Related demonstrations of quantum advantage in the sequential scenario, but without self-testing, have also been reported~\cite{Li2018, Anwer2021}.

\subsection{Independent devices}\label{sec:NoSR}
So far in this section, our discussion has concerned dimension-restricted scenarios where Alice and Bob share randomness. It is natural to ask how the discussion changes when  shared randomness is removed, i.e.~when Alice and Bob's devices are independent. As it turns out, it dramatically changes the capabilities of both classical and quantum models.  

For classical models, the deterministic distributions [see Eq.~\eqref{deterministicdistributions}] are still accessible, as these do not use shared randomness. However, their convex hull, which became the classical polytope (see Fig.~\ref{fig_classical_polytope}), is no longer accessible since this requires shared randomness. Nevertheless, Alice and Bob can still exploit local randomness, i.e.~they can generate random variables, $\mu$ and $\nu$, known only to each party individually, and use these to choose their encoding and decoding. This means that the correlations now take the form 
\begin{equation}\label{classicalNoSR}
    p(b|x,y) = \sum_{m=1}^d\sum_\mu \sum_\nu q_A(\mu) q_B(\nu) D_\mu(m|x) D_\nu(b|m,y),
\end{equation}
where $D_\mu(m|x)$ is the deterministic encoding of Alice and $D_\nu(b|m,y)$ is the deterministic decoding of Bob. We see that what was previously a shared random variable, with a joint distribution $q(\mu,\nu)$, is now factoring into $q_A(\mu)q_B(\nu)$ because of the independence of Alice and Bob's devices. The consequence of this is that the set of classical correlations is no longer a polytope but instead a non-convex body whose boundary is described by nonlinear constraints.

The same applies to the quantum set of correlations. Correlations of the form $p(b|x,y)=\tr(\rho_x M_{b|y})$, obtained without shared randomness, are still valid, but their convex hull [see Eq.~\eqref{quantumModel}] is no longer accessible. Instead, only local randomness can be used by Alice and Bob. Therefore, also the quantum set of correlations becomes a non-convex body. 

To see the impact of removing shared randomness, we need to go beyond linear correlation quantities, because these are insensitive to the non-convex nature of the correlations\footnote{Recall that the maximal value of any $W$, as defined in Eq.~\eqref{linearWitness}, is achieved without shared randomness.}. As a simple example, we follow Ref.~\cite{Ambainis1999} (see also Ref.~\cite{Nayak1999}) and consider again the random access code. In Eq.~\eqref{RAC4222} we defined it based on its average success probability, but to get rid of the linearity, let us now instead define it based on its worst-case success probability, i.e.~
\begin{equation}\label{QRACworstcase}
    W= \min_{x,y} p(b=x_y|x,y).
\end{equation}
We can immediately see that the quantum violation, discussed around Eq.~\eqref{RACvectors}, remains the same because that strategy achieved the same success probability for every $x,y$, namely $p(b=x_y|x,y)=\frac{2+\sqrt{2}}{4}$. However, the classical value collapses: one cannot in fact do better than $W=\frac{1}{2}$ which is the same as would have been achieved with random guessing. The reason for this is that the classical strategies that are good in the linear case, for instance achieving the bound in Eq.~\eqref{RAC4222}, do not do so uniformly but instead always have some instances of $(x,y)$ where the success probability is no larger than $\frac{1}{2}$. A consequence of this is that we can achieve a quantum violation when exposing our optimal states in Eq.~\eqref{RACvectors} to any rate of white noise that is smaller than $100\%$. This type of unbounded quantum advantage is standard when shared randomness is eliminated.

A similar collapse can also occur for quantum strategies without shared randomness, but this requires larger input/output alphabets. To this end, we follow Ref.~\cite{Hayashi2006} and consider the quantum random access code corresponding to $(n,d)=(4,2)$. Recall from Sec.~\hyperref[sec:QRAC]{II-B-3} that for $(n,d)=(2,2)$ we optimally chose state Bloch vectors forming a square to guarantee that $p(b=x_y|x,y)$ is constant. To do the same for $(n,d)=(3,2)$, we chose state Bloch vectors forming a cube. The key fact at play here is that we are dividing the Bloch sphere into four and eight non-empty regions by using two and three planes. To extend this to $(n,d)=(4,2)$, we would have wanted Bloch vectors forming a four-dimensional cube, or more generally that we could divide the Bloch sphere using four planes into 16 non-empty regions. This is geometrically impossible. The consequence becomes that this quantum random access code, by the worst-case metric in Eq.~\eqref{QRACworstcase}, cannot exceed a random guessing strategy, namely $W=\frac{1}{2}$. This means that if we were willing to assume the absence of  shared randomness, we could do dimension witnessing for quantum systems under unbounded noise conditions. 

At this point, we may ask how we can give criteria for whether a given correlation $p(b|x,y)$ belongs to the classical or quantum sets. In both cases, one can interpret the minimal dimension needed to generate the correlations as an appropriate notion of rank for the data table matrix defined in Eq.~\eqref{dataTableMatrix}.  We briefly explain these connections in Box~\ref{Box:rank} and there also refer to lower bounds known from the corresponding literature. Lower bounds on these ranks have also been obtained in the physics community by exploiting connections to Bell nonlocality~\cite{Sikora2016} or convenient nonlinear witnesses~\cite{Bowles2014, Tavakoli2020indep}. 

\begin{backgroundbox}[label={Box:rank}]{Dimension witness \& PSD rank}
Let $P$ be a matrix with nonnegative entries. The positive semidefinite rank~\cite{Fawzi2015} of $P$  is the smallest integer $k$ for which there exist positive semidefinite matrices $A_i$ and $B_j$ of size $k$ such that $P_{ij}=\tr(A_iB_j)$. Similarly, the nonnegative rank of $P$ is the smallest integer $k$ for which there exist nonnegative vectors $\vec{a}_i$ and $\vec{b}_j$ of size $k$ such that $P_{ij}=\vec{a}_i\cdot \vec{b}_j$.  For the type of matrices we encounter in dimension witnessing, namely those that contain the probabilities $p(b|x,y)$, these notions provide lower bounds on the minimal quantum and classical message dimensions, respectively, without shared randomness.

\vspace{0.2cm}
Several techniques have been developed outside the physics community for bounding the positive semidefinite rank of a matrix; see e.g.~\cite{Gouveia2013, Fawzi2015, Fiorini2015}. Some bounds are reasonably easy to compute but not very accurate, whereas others are more accurate but also more expensive to compute.  In general, determining these ranks is computationally hard.

\end{backgroundbox}

To gain some qualitative insight, we will  follow Ref.~\cite{Vicente2017} and discuss a bound that is far from optimal but very simple and generally applicable. To this end, let us start from the classical model in Eq.~\eqref{classicalNoSR} and write $s(m|x)=\sum_\mu q_A(\mu)D_\mu(m|x)$ and $t(b|m,y)=\sum_\nu q_B(\nu)D_\nu(b|m,y)$. Then we can write the data table matrix in Eq.~\eqref{dataTableMatrix} as 
\begin{align}\nonumber
    P&=\sum_{m=1}^d \sum_{b,x,y}  s(m|x)t(b|m,y) \ketbra{x}{y,b} \\
    & = \sum_{m=1}^d \left(\sum_x s(m|x)\ket{x}\right) \left(\sum_{b,y}  t(b|m,y)\bra{y,b}\right)\nonumber\\&=\sum_{m=1}^d \vec{u}_m\cdot \vec{v}^{\,T}_m,
\end{align}
where  $\vec{u}_m =\sum_x s(m|x)\ket{x}$ and $\vec{v}_m =\sum_{b,y} t(b|m,y)\ket{y,b}$. It is now clear that the rank of $P$ cannot exceed $d$. If we were to consider the quantum case, the states and measurements would act on a Hilbert space of dimension $d$. However, we know that Hermitian matrices can be represented in a real vector space of dimension $d^2$. In this real vector-space, the above bound holds once again. Therefore, we conclude that
\begin{align}
&    d \geq \text{rank}(P) && \text{for classical},\\
& d \geq \sqrt{\text{rank}(P)} && \text{for quantum}.
\end{align}
These rank constraints imply that both the classical and the quantum sets are limited to lower-dimensional spaces. Therefore, we can now understand the unbounded noise tolerance that we encountered before with the  random access code. In essence, only very contrived types of noise would influence the rank in such a way as to eliminate the extreme noise tolerance. In other words, arbitrarily small perturbations of $P$ are generically sufficient to make it full rank. This discussion, focused on the fact that lower-rank matrices are a measure-zero set that is nowhere dense, is formalised in Ref.~\cite{Vicente2017}.

From a physical point of view, the framework without shared randomness has some limitations. Firstly, it can only be applied when there is good reason to exclude the possibility of shared randomness. For instance, in adversarial experiments, such as those encountered in quantum cryptography, attacks that introduce shared randomness are comparatively cheap but greatly amplify the accessible set of correlations. One may argue that this becomes less relevant in non-malicious scenarios, where one wants to benchmark the functionality of quantum  devices or perform a task beyond classical limitations. For instance, it was proven in~\cite{Miklin2021} that all sets of pure states and projective measurements can be self-tested in dimension-restricted PM scenarios without shared randomness. However, for this, a second limitation also becomes relevant: the statistics generated from the devices may not have been obtained in independent and identically distributed rounds. This means that the preparation and measurement devices each can exploit internal memory effects to generate correlations that effectively converge to the convex hull of the non-convex set of correlations. This type of memory attack is a viable option for  adversarial devices that legitimately can be assumed to not have any shared randomness.

\subsection{Experiments}

Having discussed the dimension-restricted PM framework, it is useful to look at how it has been implemented in optical experiments. Photons are especially natural for this purpose. They can be prepared, transmitted and measured with low decoherence, and they provide several independently addressable degrees of freedom in which a finite-dimensional message can be encoded. This flexibility is also the main experimental caveat. The relevant dimension is not the number of detector labels or the number of modes drawn in a diagram, but the dimension of the controlled Hilbert space that actually carries Alice's message. In this sense, photonic experiments make the assumption $\mathcal A$ of Sec.~\hyperref[PMstructure]{I-A} both practically accessible and experimentally delicate. 

\begin{table*}[t]
\centering
\small

\begin{tabular}{lll}
\hline
\textbf{Experiment}
&
\textbf{Platform}
&
\textbf{Main significance}
\\
\hline

Quantum advantage~\cite{Spekkens2009}
&
Polarisation qubits
&
First demonstration of quantum random access codes.
\\

Delayed choice experiment~\cite{polino2019device,huang2019compatibility,yu2019realization}
&
Polarisation qubits
&
Certification of nonclassicality in the delayed-choice experiment.
\\

High-dimensional QRAC~\cite{Tavakoli2015}
&
Path/polarisation ququarts
&
First demonstration of high-dimensional \\ 
& & quantum random access codes.\\
Dimension witnessing~\cite{Ahrens2012,Hendrych2012}
&
Photonic qudits
&
First device-independent dimension witnesses.
\\

Extreme dimension certification~\cite{Aguilar2018b}
&
Transverse momentum
&
Certified dimension $1024$.
\\

Dimension witnessing without SR~\cite{Sun2020,Sun2026}
&
Orbital angular momentum
&
Nonclassical communication without SR.
\\

POVM self-testing~\cite{Smania2020}
&
Path-augmented photonic qubits
&
First PM self-test of a non-projective measurement.
\\

High-dimensional POVM certification~\cite{Martinez2023}
&
Multicore fibre
&
Certification of a seven-outcome POVM.
\\

Instrument self-testing~\cite{Anwer2020,Foletto2020}
&
Photonic qubits
&
Self-testing of quantum instruments.
\\
\hline
\end{tabular}
\caption{Selected photonic implementations of prepare-and-measure quantum information protocols, highlighting the physical platform used and the main achievement demonstrated in each experiment. SR denotes shared randomness.}
\label{tab:pm_landmarks}
\end{table*}

Before discussing platforms, let us separate the different conclusions that such experiments are used to support. A violation of a classical bound, as in the random access code of Sec.~\hyperref[sec:QRAC]{II-B-3}, demonstrates a quantum communication advantage. A dimension witness goes further in a different direction, namely it uses correlations $p(b|x,y)$ to rule out lower-dimensional models, as discussed in Sec.~\hyperref[sec:boundingNV]{II-C-2}. Measurement certification, as discussed in Sec.~\hyperref[sec:POVMcertification]{II-D-3}, instead asks whether Bob's measurement could have belonged to a simpler class, for instance projective measurements only. Certifying instruments, which leave a post-measurement state, requires the sequential extension of the PM scenario described in Sec.~\hyperref[subsubsec:instruments]{II-D-5}. These distinctions are important because the same photonic platform can be used for rather different certification claims. Table~\ref{tab:pm_landmarks} summarises representative photonic experiments. The entries are not meant to be exhaustive. They are selected to show how the same PM framework supports several conclusions.

\emph{Polarisation} is a common photonic qubit. It is easy to prepare, rotate and measure with high precision, and it gives a clean two-dimensional system for the basic PM tasks. This is why polarisation was used in early optical demonstrations of qubit communication advantages~\cite{Spekkens2009}. It is also well suited for qubit certification experiments, such as the causal-model version of Wheeler's delayed-choice experiment~\cite{polino2019device,huang2019compatibility,yu2019realization}, since the target states and measurements can be represented directly on the Bloch sphere. The limitation is equally clear. Polarisation alone cannot realise a system with $d>2$. Therefore, high-dimensional tests either use another degree of freedom or combine polarisation with path, spatial modes or fibre modes.

\emph{Path and spatial modes} provide another way to build a finite-dimensional photonic message. A single photon  coherently distributed over $d$ paths is a qudit in a basis that is physically transparent and experimentally addressable by interferometers, beam displacers, multiports and integrated circuits. This makes path modes particularly useful for few-dimensional dimension-witness experiments, as seen in Refs.~\cite{Ahrens2012,Hendrych2012}, as well as for high-dimensional random access codes where they can be combined with other degrees of freedom, such as the path-and-polarisation ququart implementation of Ref.~\cite{Tavakoli2015}. Their advantage is control, while their drawback is stability. As $d$ grows, phase drift, loss imbalance and cross-talk become part of the dimension-assumption problem, not merely technical imperfections.

\emph{Orbital angular momentum and transverse momentum}. In orbital angular momentum encodings, the basis states are modes with different azimuthal indices which can be prepared and analysed with spatial light modulators and mode-sorting techniques. They are natural for tests in which dimension itself is the resource, including dimension witnessing without shared randomness~(recall Sec.~\hyperref[sec:NoSR]{II-E}). Transverse momentum encoding uses a discretised set of spatial modes instead. It provides a large alphabet and has enabled the certification of irreducible photonic dimension 1024~\cite{Aguilar2018b}. The common message is that spatial degrees of freedom allow PM experiments to reach dimensions for which tomography would be impractical. The corresponding weakness is that one must carefully define the controlled mode subspace and exclude leakage into modes that are not part of the intended message.

\emph{Time-bin and frequency-bin modes} play a similar role to path modes. The message is distributed over a finite set of temporal or spectral modes. They are attractive because they are compatible with optical fibre and with phase-modulation techniques. In the present dimension-restricted context, however, their use requires a careful statement of which time or frequency modes constitute the message space. Detector timing, modulation bandwidth, dispersion and spectral filtering then become part of the physical implementation of the dimension assumption. For this reason, these encodings are often closer in spirit to the non-dimensional constraints of Sec.~\hyperref[Sub:energy-restrictions]{IV-C} than to the finite-dimensional experiments listed in Table~\ref{tab:pm_landmarks}.

\emph{Multicore fibre modes} provide a path-like encoding in a fibre-compatible architecture. Each core can be treated as one spatial mode, and light can be coupled into and out of the individual cores to prepare and analyse high-dimensional states. This makes the platform well suited to experiments where the goal is not only to use a high-dimensional system, but also to certify a structured measurement on it. In Ref.~\cite{Martinez2023}, multicore fibre modes were used to certify a seven-outcome non-projective measurement. Conceptually, this belongs to the measurement-certification part of the tutorial: the dimension of the system is assumed, and the conclusion concerns the type of measurement implemented by Bob.

\emph{Hybrid encodings} are very common in photonic experiments. The simplest reason is practical. One degree of freedom can carry the message, while another is used to implement the required optical transformation or measurement. For example, polarisation may encode the communicated qubit, while path modes are introduced only inside Bob's apparatus to realise a measurement with more than two outcomes. In that case, the path mode is not an additional part of Alice's message, but part of Bob's measurement device. This distinction is important for the dimension assumption. If the extra degree of freedom is already available at Alice's preparation stage, then it contributes to the communicated dimension. If it is introduced only at Bob's side, then it is an implementation tool for the measurement. This separation appears in the self-testing of non-projective measurements~\cite{Smania2020}. Hybrid encodings are also used in high-dimensional random access codes, where path and polarisation together define a four-dimensional system~\cite{Tavakoli2015}, and in sequential experiments, where the optical setup must control both the information extracted by a measurement and the disturbance left for the next observer~\cite{Anwer2020,Foletto2020}.

\emph{Binarised versus genuine multi-outcome measurements}. There is one further experimental point that is especially relevant for high-dimensional photonic implementations. Many PM protocols are formulated in terms of a measurement with several outcomes. However, in the lab, such a measurement is sometimes implemented by testing the outcomes one at a time: for each mode, Bob performs a binary measurement that asks whether the photon is found in that mode or not, and the multi-outcome statistics are reconstructed afterwards. This procedure is harmless if Bob's measurement device is trusted well enough to guarantee that the separate binary tests really form a single POVM. In a dimension-witness or semi-device-independent PM experiment, however, this is an additional assumption. Without it, the experiment should be analysed as a different PM scenario, with more measurement settings but binary outcomes. This is the binarisation loophole~\cite{TavakoliUolaPauwels2025}. It does not mean that binarised data are useless but that the corresponding classical or quantum bounds must be derived for the binarised input-output scenario itself. Genuine multi-outcome detection avoids this issue and is especially valuable in high-dimensional photonic certification experiments.

\section{Entanglement-assisted scenarios}\label{sec:EAPM}

In the previous section, we primarily studied PM scenarios where Alice and Bob were allowed to share randomness and the quantum resource was a $d$-dimensional state sent between them. In this section, we will introduce quantum resources also in the pre-shared correlations between Alice and Bob. Specifically, we will consider that Alice and Bob can share an entangled state, which they can use  to assist their $d$-dimensional communication. Hence, as compared to the previous section, we have changed the resource $\mathcal{R}$, from shared randomness to shared entanglement, but kept the assumption  $\mathcal{A}$, namely the dimension-restricted communication (as defined in Sec.~\hyperref[PMstructure]{I-A}).

There are two  natural frameworks in which to study entanglement in PM scenarios: either we allow classical messages over the channel, or we allow quantum messages. In the former case, we try to use entanglement as a resource to enhance the efficiency of classical communication, whereas in the latter case we want to achieve a synergy between the two distinct quantum resources, namely shared entanglement and quantum messages. These are inherently different settings, and we therefore refer to them as entanglement-assisted classical communication and entanglement-assisted quantum communication respectively. We begin by discussing the former.

\subsection{Classical communication}\label{sec:EACC}
The use of entanglement to assist classical communication between Alice and Bob  gives rise to correlations that defy classical models, i.e.~those associated with the classical polytope discussed in Sec.~\hyperref[sec:ClassicalPolytope]{II-A-1}. We follow Ref.~\cite{TavakoliPauwels2021} in formalising entanglement-assisted classical communication. To this end, let Alice and Bob share a state $\rho_{AB}$, where system $A$ is given to Alice and system $B$ to Bob. Alice uses her input $x$ to select a POVM, $\{A_{m|x}\}_m$, which she applies to system $A$. The outcome, $m\in\{1,\ldots,d\}$, is the message that she sends over the channel to Bob. Bob subsequently uses his input and the message to select  a POVM, $\{B_{b|y,m}\}_b$, which he applies to system $B$. The correlations in a quantum model are given by
\begin{equation}\label{EACCcorr}
    p(b|x,y)=\sum_{m=1}^d \tr\left(A_{m|x}\otimes B_{b|y,m}\ \rho_{AB}\right).
\end{equation}

\begin{backgroundbox}[label={Box:LHV}]{Local hidden variable models}
In a Bell nonlocality experiment, Alice and Bob are given independent inputs $x$ and $y$ respectively. They are tasked with producing outcomes $a$ and $b$ respectively, in space-like separated events. A classical model for the correlations in the experiment is called a local hidden variable model. This is based on the idea that a local hidden variable $\lambda$, subject to some probability density $q(\lambda)$, is shared between the parties. Alice and Bob's outcomes are then  determined by their respective inputs and the local hidden variable, using some arbitrary stochastic rule $p_A(a|x,\lambda)$ and $p_B(b|y,\lambda)$. The correlations therefore take the form
\begin{equation}\label{LHV}
    p(a,b|x,y)=\int \drm\lambda\ q(\lambda) p_A(a|x,\lambda) p_B(b|y,\lambda).
\end{equation}
Any set of correlations $p(a,b|x,y)$ arising in this scenario but not admitting such a model is called Bell nonlocal.
\end{backgroundbox}

A key observation is that these correlations are closely related to those encountered in Bell nonlocality tests~\cite{PauwelsAdaptive2022} as well as in Bell scenarios with communication~\cite{pironio2003violations,brask2017bell,chaves2015unifying,ringbauer2017probing} (see Box~\ref{Box:LHV} and Box~\ref{Box:LHVcomm} for a minimal overview of Bell nonlocality and its variant with communication). To see the connection, consider a Bell experiment in which Alice has an input $x$ and an output $m$, whereas Bob has a pair of inputs $(y,z)$, where $z\in\{1,\ldots,d\}$, and an output $b$. The correlations in the Bell test become $p(m,b|x,(y,z))=\tr\left(A_{m|x}\otimes B_{b|y,z}\ \rho_{AB}\right)$. Next, let us post-select the terms in which Bob's second input matches Alice's output, namely $z=m$, and marginalise over Alice. This selection gives $\sum_m p(m,b|x,(y,m))$ and returns the correlations in Eq.~\eqref{EACCcorr}. Indeed, in a local hidden variable model [as defined in  Eq.~\eqref{LHV}] we would have $p(m,b|x,(y,m))=\int \drm\lambda\ q(\lambda)p_A(m|x,\lambda)p_B(b|m,y,\lambda)$. Marginalising this gives $p(b|x,y)=\sum_{m=1}^d \int \drm\lambda\  q(\lambda)p_A(m|x,\lambda)p_B(b|m,y,\lambda)$ which  is precisely the definition of a classical model for the PM scenario given in Eq.~\eqref{PMdimclassical2}. We conclude that the Bell nonlocality of the full distribution $p(m,b|x,(y,z))$ is a necessary condition for quantum advantages in entanglement-assisted classical communication. This leads to the general conclusion that any such quantum advantage is fundamentally propelled by nonlocality, i.e.~ 
\begin{equation}
\begin{gathered}
\text{quantum advantage}\\[-3pt]
\text{in classical communication}
\end{gathered}
 \:\:\Rightarrow\:\: \text{quantum nonlocality}
\end{equation}
Here we focus on the entanglement-assisted PM scenario with classical communication, although analogous constructions can also be formulated for quantum messages \cite{Baroni2025}.

\begin{backgroundbox}[label={Box:LHVcomm}]{Bell scenarios with outcome-communication}
A Bell scenario with outcome-communication is described by the hidden-variable model
\begin{equation}
p_{\mathrm{Bell+Comm}}(a,b\vert x,y) = \sum_{\lambda} q(\lambda) p(a\vert x,\lambda) p(b\vert a,y,\lambda),
\label{eq:bellcomm}
\end{equation}
that is mathematically equivalent to a standard Bell scenario where Bob receives an additional input $\tilde y$:
\begin{equation}
p_{\mathrm{Bell}}(a,b\vert x,y,\tilde{y}) = \sum_{\lambda} q(\lambda) p(a\vert x,\lambda)p(b\vert y,\tilde{y},\lambda).
\end{equation}
The equivalence follows by selecting the probabilities with $\tilde y=a$, without renormalising them. In this way, the communicated outcome $a$ is effectively treated as an extra input for Bob. Geometrically, the polytope defined by the communication scenario can be understood as the projection of the standard Bell local polytope $p_{\mathrm{Bell}}(a,b\vert x,y,\tilde y)$ onto the subspace specified by $\tilde y=a$. An immediate consequence is that a violation of a Bell inequality in the extended scenario $p_{\mathrm{Bell}}(a,b\vert x,y,\tilde y)$ is a necessary condition for the violation of a Bell-like inequality in the communication-assisted scenario $p_{\mathrm{Bell+Comm}}(a,b\vert x,y)$. 

\medskip

The power of outcome communication depends on the measurements whose statistics must be reproduced. For nonsignalling correlations with binary outcomes for Alice, outcome communication offers no advantage over local hidden variable models if her measurement choices include a deterministic measurement~\cite{Vieira2025outcome}. This implies equivalence for all projective measurements on qubit--qudit states, including $\{\mathds{1},0\}$ on Alice's qubit, although restricted measurement families can admit an advantage.
\end{backgroundbox}

\subsubsection{Bell inequalities and PM scenarios}
Let us consider a first example of how the advantages of entanglement in classical communication can be related to violations of Bell inequalities. To this end, we return to our running example of the random access code. As in Sec.~\ref{S:PM-setting}, Alice and Bob want to implement a random access code. This means that Alice holds two bits $(x_0,x_1)\in\{0,1\}^2$ which she encodes into a binary message $m\in\{0,1\}$, from which Bob wants to extract the information $b=x_y$, where $y\in\{0,1\}$ marks his choice of information. Compared to Eq.~\eqref{RAC4222}, it is here convenient to count both the bits and Bob's inputs from zero. The average success probability in this task is given in Eq.~\eqref{RAC4222} together with the limit respected by all classical models. To violate this limit by means of sharing entanglement, let Alice and Bob share some quantum state $\rho_{AB}$. First, Alice computes the bit $x\equiv x_0\oplus x_1$. Then, the parties use $x$ and $y$ respectively as inputs for measuring their share of the state. Let these measurements produce binary outcomes $a'\in\{0,1\}$ and $b'\in\{0,1\}$ respectively. Next, Alice defines the message $m=a'\oplus x_0$ and sends it to  Bob, who outputs $b=b'\oplus m=a'\oplus b'\oplus x_0$. With this strategy, the average success probability in the random access code becomes
\begin{equation}
    W_\text{RAC} = \frac{1}{8}\sum_{x_0,x_1} \sum_y  p\left(x_y=a'\oplus b'\oplus x_0\big|x_0,x_1,y\right),
\end{equation}
where $p(x_y=a'\oplus b'\oplus x_0|x_0,x_1,y)$ is the sum of all conditional probabilities in which the condition $x_y=a'\oplus b'\oplus x_0$ is satisfied. Notice that if $y=0$ then the condition simplifies to $a'\oplus b'=0$ whereas if $y=1$ then the condition simplifies to $a'\oplus b'=x_0\oplus x_1=x$. Thus, we can compactly write it as $a'\oplus b'=xy$. Inserted above, this gives
\begin{align}\nonumber\label{chshrac}
   W_\text{RAC} &= \frac{1}{8}\sum_{x_0,x_1} \sum_y  p\left(a'\oplus b'=xy\big|x_0,x_1,y\right) \\
                &= \frac{1}{4}\sum_{x,y}  p\left(a'\oplus b'=xy\big|x,y\right),
\end{align}
where in the second step we have used that only the value of $x$ affects the probabilities. We can now recognise that the expression we have obtained is equivalent to that appearing in the well-known Clauser--Horne--Shimony--Holt (CHSH) Bell inequality. In fact, the CHSH inequality can equivalently be written as $W_\text{RAC}\leq \frac{3}{4}$ (see Box~\ref{Box:CHSH} for details). This means that if Alice and Bob can use $\rho_{AB}$ to generate a probability distribution $p(a',b'|x,y)$ which violates the CHSH inequality, then they can use these nonlocal correlations to assist their classical communication in a way that breaks classical constraints. Curiously, this was first shown experimentally in the context of improving strategies for winning competitive card games~\cite{Muhammad2014}.

It is worth appreciating that although entanglement on its own cannot be used to transmit information, this example shows that it nevertheless can be used to boost the power of classical communication. This basic insight was first realised in Ref.~\cite{Cleve1997} and the connection between the random access code and CHSH nonlocality was subsequently made in Ref.~\cite{Buhrman2001}. It is interesting to notice here that if Alice and Bob had access to a nonsignalling box~\cite{popescu1994quantum} --- a hypothetical post-quantum Bell nonlocal correlation which can achieve an algebraically maximal value of the CHSH parameter --- then they would achieve $W_\text{RAC}=1$. This would mean that Bob could deterministically recover either $x_0$ or $x_1$ even though Alice only sent him one bit. We will revisit this peculiarity in Sec.~\hyperref[sec:infocausality]{III-B-1}. 

\begin{backgroundbox}[label={Box:CHSH}]{CHSH Bell inequality}
The Clauser-Horne-Shimony-Holt (CHSH) inequality is one of the simplest Bell inequalities that can be violated in quantum theory. In the CHSH scenario, Alice and Bob receive independent and uniformly distributed binary inputs $x,y\in\{0,1\}$ and produce binary outputs $a,b\in\{0,1\}$, respectively. With the correlators defined as
\begin{equation}
    \langle A_x B_y \rangle = \sum_{a,b=0}^{1} (-1)^{a+b} p(a,b|x,y),
\end{equation}
the CHSH inequality reads
\begin{equation}\label{CHSHcorr}
   \mathrm{CHSH} = \langle A_0B_0\rangle +\langle A_0B_1\rangle +\langle A_1B_0\rangle -\langle A_1B_1\rangle \leq 2.
\end{equation}
This bound holds for all local hidden-variable models and, in particular, for quantum correlations obtained from separable states.

\vspace{0.2cm}

The CHSH inequality can equivalently be expressed in terms of the probability that the outputs satisfy $a\oplus b=xy$. Indeed, since
\begin{equation}
    p(a\oplus b=xy|x,y)=\frac{1}{2}\left(1+(-1)^{xy}\langle A_xB_y\rangle\right),
\end{equation}
averaging over the inputs gives
\begin{eqnarray}
    \frac{1}{4}\sum_{x,y}p(a\oplus b=xy|x,y)=\frac{1}{2}+\frac{\mathrm{CHSH}}{8}\leq \frac{3}{4}.
\end{eqnarray}
This is the probability form of the CHSH inequality, which directly connects to the random access code formulation.
\end{backgroundbox}

The above example is an instance of a more general mapping from Bell inequalities to PM scenarios. This is a connection that allows a broad class of Bell inequalities to be directly associated with classicality inequalities in the PM scenario, and it was  first made in Ref.~\cite{Brukner2004}. To understand it, consider a Bell scenario with $n$ parties who each receive independent inputs $x_1,\ldots,x_n$ and output respectively $o_1,\ldots,o_n\in\{0,1\}$. Consider that there exists some Bell inequality of the form 
\begin{equation}\label{Belltest}
    \mathcal{B}=\sum_{x} c_{x}\ p\left(o_1\oplus \ldots\oplus o_n=h(x)|x\right)\leq C,
\end{equation}
where $x=(x_1,\ldots,x_n)$, $h(x)$ is a Boolean function, $c_x$ are real coefficients and $C$ is the local bound. Bell inequalities of this form are often referred to as full-correlation inequalities, because by an argument analogous to that in Box~\ref{Box:CHSH}, they can be expressed exclusively in terms of joint expectation values of the $n$ parties. Note that more general Bell inequalities could involve expectation values of fewer than $n$ parties. 

To construct a corresponding PM task, let all but the final party have a pair of inputs $(x_i,z_i)$, where $z_i\in\{0,1\}$, and encode a message $m_i\in\{0,1\}$. The messages from the first $n-1$ parties are all sent to the final party. The final party has only an input $x_n$ and is then tasked with outputting a guess, $g$, for the value of the function $f(x,z)=h(x)\oplus z_1\oplus \ldots \oplus z_{n-1}$. Now, the parties use their inputs $x_1,\ldots,x_n$ as inputs to the Bell test  in Eq.~\eqref{Belltest}. Upon producing the outcome $o_i$, party number $i$ sends the message $m_i=o_i\oplus z_i$. The final party decides his guess as $g=m_1\oplus \ldots \oplus m_{n-1}\oplus o_n$. The condition of guessing correctly, namely $g=f$, becomes  $o_1\oplus \ldots\oplus o_n=h(x)$, which is that encountered in the Bell inequality \eqref{Belltest}. Therefore, the signed weighted score of computing $f$ is one-to-one with the value of $\mathcal{B}$. A normalised success probability is obtained by weighting inputs by $|c_x|$ and flipping the winning condition when $c_x<0$. For classical models, the score is limited by $C$, but quantum models may violate the Bell inequality and therefore achieve a higher value in the PM task. It is interesting to observe that by having a large number of parties ($n$) in the above discussion, the ratio of quantum to classical signed scores, or equivalently of the corresponding success biases, can become arbitrarily large. This follows from the long-known fact that some full-correlation Bell inequality violations can grow exponentially in the number of parties~\cite{Mermin1990}.

Notice that we have  assumed the specific encoding and decoding strategies used by the parties. One may wonder if the result still holds under general strategies. The question was answered in the affirmative in Ref.~\cite{Brukner2004}, thereby establishing the Bell inequality violation as a necessary and sufficient condition for an entanglement-assisted advantage for classical communication in the associated PM task.  In addition, if one continues with the above assumed encoding/decoding strategies, the result can also be extended to a class of $d$-outcome Bell inequalities~\cite{Brukner2002, Brukner2003, Zukowski2017}, but this is a significantly weaker result since it does not remain valid for more general strategies~\cite{TavakoliBrukner2020}. Moreover, taking a complementary perspective on mapping Bell scenarios onto PM scenarios, Ref.~\cite{Baroni2025} shows how self-tests achieved in the former scenario (for a broad class of Bell inequalities)  can be transformed into self-tests in an appropriately selected scenario of the latter type. 

Finally, back in Sec.~\hyperref[sec:quantumCCP]{II-B-6} we discussed how solving some communication problems could require exponentially more bits than qubits. Similar advantages exist in entanglement-assisted classical communication, i.e.~by sharing entanglement whose correlations violate an appropriate Bell inequality, Alice may need exponentially fewer bits to send to Bob in order to complete a communication task. We showcase this by returning to the hidden matching problem, but now solve it by means of shared entanglement and classical communication. Recall that in hidden matching, Alice receives an $n$-bit string $x=(x_1,\ldots,x_n)\in\{0,1\}^n$ and Bob a set $H=\{(i_1,j_1),\ldots,(i_{n/2},j_{n/2})\}$ of disjoint pairs selected from $\{1,\ldots,n\}$, with the task of outputting a triple $(i,j,x_i\oplus x_j)$ for some $(i,j)\in H$. To solve the problem, let Alice and Bob share the $n$-dimensional maximally entangled state, $\ket{\phi^+}=\frac{1}{\sqrt{n}}\sum_{k=1}^{n}\ket{k}_A\ket{k}_B$. Alice begins by applying the unitary $U_x=\sum_{j=1}^n (-1)^{x_j}\ketbra{j}$. Afterwards, the shared state becomes $\frac{1}{\sqrt{n}}\sum_{k=1}^{n}(-1)^{x_k} \ket{k}_A\ket{k}_B$. Let us for simplicity say that $n=2^K$ for some integer $K$, and identify each index $k\in\{1,\ldots,n\}$ with a $K$-bit string. Now, we let Alice apply the unitary $\mathsf{H}^{\otimes K}$, where $\mathsf{H}$ is the Hadamard matrix, to obtain the state $\frac{1}{n} \sum_{s} \ket{s}_A\sum_k(-1)^{x_k+s\cdot k}\ket{k}_B$, where $s$ and $k$ run over all $K$-bit strings and the inner product is evaluated modulo two. Alice now measures her system in the computational basis. Her outcome, denoted $m$, is a $K$-bit string whose distribution is uniform. Conditioned on this outcome, Bob's system is left in the state $\frac{1}{\sqrt{n}} \sum_k(-1)^{x_k+m\cdot k}\ket{k}$. Alice sends her outcome, which consists of $\log_2 n$ bits, to Bob. Bob then measures in the orthonormal basis $\left\{\frac{1}{\sqrt{2}}\left(\ket{i}+(-1)^b\ket{j}\right):(i,j)\in H,\ b\in\{0,1\}\right\}$. This produces some pair $(i,j)\in H$ together with a bit $b$ satisfying $b=x_i\oplus x_j\oplus m\cdot(i\oplus j)$. Since Bob knows $m$, $i$ and $j$, he can compute $x_i\oplus x_j=b\oplus m\cdot(i\oplus j)$, and therefore output the desired triple $(i,j,x_i\oplus x_j)$.

\subsubsection{Adaptive protocols}\label{sec:adaptive}
\begin{figure}[!tbp]
    \centering
    \includegraphics{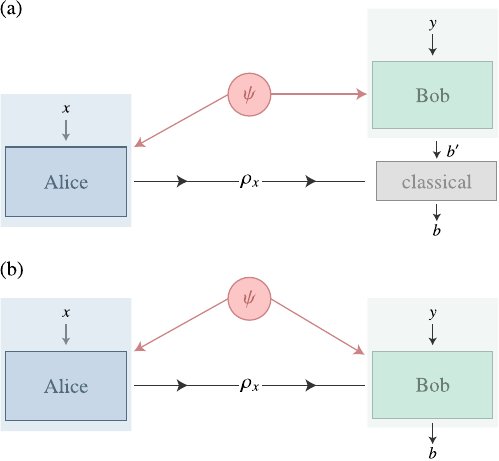}
    \caption{\emph{Non-adaptive and adaptive entanglement-assisted classical communication protocols}. In both cases, Alice receives an input $x$, performs a measurement on her share of a bipartite entangled state, and sends a classical message $m$ to Bob. (a) In a non-adaptive protocol, Bob’s quantum measurement is chosen only from his own input $y$, independently of Alice’s message. The message is used only in a final classical post-processing step that produces the output $b$. (b) In an adaptive protocol, Bob waits to measure his quantum system until after receiving $m$, and can therefore choose a message-dependent measurement.}
    \label{fig_adaptive}
\end{figure}

The examples of entanglement-assisted classical communication given in the above section all have one important property in common: they can be effectively separated into a first part where Alice and Bob perform a Bell inequality test and a second part where they use the nonlocal correlations to enhance their classical communication. In other words, Bob does not use Alice's message to choose his quantum measurement, but only to classically determine his final output. This is illustrated in Fig.~\hyperref[fig_adaptive]{\ref{fig_adaptive}(a)}. However, as highlighted in Ref.~\cite{PauwelsAdaptive2022}, more general protocols are possible in which Bob receives his share of the entangled state and withholds his measurement until Alice's message arrives. He can then use the message to adapt his choice of measurement. Protocols of this type are therefore called adaptive.

One may ask whether adaptive protocols are ever needed, in the sense that they outperform any non-adaptive protocol. The answer is positive. We can see the power of adaptivity by considering the relationship between using entanglement-assisted classical communication, as opposed to using quantum messages (as in Sec.~\ref{S:PM-setting}), to optimally perform a communication task. Say that Alice sends to Bob a single qubit, $\rho_x$, to which Bob applies some projective measurements $M_{b|y}$. It is possible to show the following resource inequality 
\begin{equation}\label{resourceineq}
    \text{qubit} \subset \text{bit} + \phi^+.
\end{equation}
In other words, any correlations in the PM scenario that can be achieved by sending a qubit can also be achieved by sending a bit assisted by a shared two-qubit maximally entangled state. Notice that this statement is different from quantum teleportation, as that would require two bits, and it would supply Bob with the qubit state. The above resource relation claims that with less communication, we cannot obtain the qubit itself but we can reproduce all its  statistical predictions. 

Let Bob perform rank-one projective measurements. Ref.~\cite{Pati2000} proposes a protocol in which Alice and Bob share $\phi^+$. Alice measures her part of the state with the binary POVM $\{\rho^T_x,\mathds{1}-\rho^T_x\}$  and Bob measures his part of the state with the measurement $\{M_{0|y},M_{1|y}\}$. Here $\rho^T_x$ means the matrix transpose of $\rho_x$. Depending on Alice's outcome, this leaves Bob's system either in the state $\rho_x$ or $\mathds{1}-\rho_x$ (we take $\rho_x$ pure without loss of generality), so his measurement leads to the statistics $\tr(\rho_x M_{b|y})$ and $1-\tr(\rho_x M_{b|y})$ respectively, where we used that for a rank-one projective measurement $\tr(M_{b|y})=1$. Alice now sends her binary outcome to Bob. If Bob receives  ``$0$'', he does nothing since he has already generated the statistics $\tr(\rho_x M_{b|y})$, which is indeed the same as obtained from sending a qubit. If Bob receives  ``$1$'', he flips this outcome $b$. Since the probability $1-\tr(\rho_x M_{b|y})$ is precisely that of observing ``not $b$'', the flipped outcome is sampled from the desired distribution $\tr(\rho_x M_{b|y})$. This protocol is non-adaptive, i.e.~Bob did not select his quantum measurement based on Alice's message. Nevertheless, it shows us that entanglement-assisted classical communication can in this scenario  be at least as powerful as sending qubits. In fact it is strictly more powerful, as can be shown for example by studying random access codes  and using larger-than-qubit entanglement~\cite{PawlowskiZukowski2010}\footnote{There exist scenarios with larger alphabets in which entanglement-assisted classical communication is strictly weaker than sending quantum messages. Examples of this are given in Refs.~\cite{Tavakoli2016, Martinez2018} based on specific protocols and a general proof appears in Ref.~\cite{TavakoliPauwels2021}.}.

The success of the simulation protocol relies on Bob making projective qubit measurements. However, the most general qubit measurements are POVMs. In Ref.~\cite{PauwelsAdaptive2022}, it was shown that in the input/output scenario $(3,2,2,3)$  there exists a facet of the classical polytope for which the violations achievable by sending qubits and making projective measurements are smaller than those achievable by sending qubits and making non-projective measurements. This means that restricting to projective measurements comes with a loss of generality. We must therefore ask if the resource inequality in Eq.~\eqref{resourceineq} can be shown also for general qubit POVMs. It has been proven in Ref.~\cite{PauwelsAdaptive2022} that the answer is negative, but only if we restrict ourselves to non-adaptive protocols.

We now show how to prove the resource inequality using an adaptive protocol. To this end, it suffices to consider pure states (with a unit Bloch vector, $\vec{n}_x$) and non-trivial extremal POVMs. The latter are written $
M_{b|y}=\frac{1}{2}\left(\lvert \vec{v}_{by} \rvert \mathds{1}+\vec{v}_{by}\cdot\vec{\sigma}\right)$
where $\sum_b \lvert \vec{v}_{by} \rvert =2$ by normalisation and $ \sum_b \vec{v}_{by}=0$ by completeness. From Born's rule, we have $p(b|x,y)=\frac{1}{2}\left(\lvert \vec{v}_{by} \rvert+\vec{n}_x\cdot\vec{v}_{by}\right)$.
To simulate $p(b|x,y)$ using a bit and the state $\phi^+$, let Alice measure her qubit in the Bloch sphere direction $\vec{n}'_x$, where the prime means flipping the sign of the $Y$-component in $\vec{n}_x$. This leaves Bob's system in the state $\pm \vec{n}_x$, with the sign determined by Alice's outcome, $m=\pm1$. Alice then sends $m$ to Bob who reads it and selects his measurement as $M_{b|y,m}=\frac{1}{2}\left(\lvert \vec{v}_{by} \rvert \mathds{1}+m\vec{v}_{by}\cdot\vec{\sigma}\right)$. We see that if $m=+1$ then Bob receives the same qubit state as in the original protocol and performs the same measurement. Thus, he gets the right statistics. However, when $m=-1$, he neither receives the same state nor performs the same measurement, but these two offsets cancel to nevertheless return the same statistics $p(b|x,y)$. Bob's choice of measurement thus depends on the message $m$.

This need for adaptivity in the qubit simulation depends on the allowed shared resource. If entanglement is replaced by arbitrary nonsignalling correlations, Ref.~\cite{Nogueira2026} shows that a $d$-valued classical message with non-adaptive nonsignalling assistance can reproduce the statistics of any unassisted $d$-dimensional quantum message for arbitrary preparations and POVMs. Thus, the simulation extends to all dimensions without requiring Bob's input to the shared resource to depend on Alice's message. As above, the statement concerns reproducing PM statistics, not transmitting the quantum state itself.

Another interesting regime in which adaptive protocols are qualitatively important is in PM scenarios where Bob has no independent receiver input, i.e.~his alphabet is $|Y|=1$. In such scenarios, we learned in Sec.~\hyperref[sec:FWtheorem]{II-B-1} that there exists no advantage in sending quantum messages over classical messages. Is it possible to break the constraints of classical models if we substitute our quantum messages with entanglement-assisted classical communication? If such a protocol is non-adaptive, the answer is negative. This follows immediately from the necessity of Bell nonlocality for a quantum advantage: quantum nonlocality without an independent measurement input for Bob is well-known to be impossible. However, in an adaptive protocol, Bob could use Alice's message as his input, thereby raising the question of whether adaptive protocols could produce violations of classicality. The answer is affirmative and this was first shown in Ref.~\cite{Frenkel2022}. Several more examples have followed~\cite{Vieira2023, Mir2023}, including ones that explicitly use the connection to nonlocality~\cite{Rout2025}.  

Given that adaptive advantages are possible without independent receiver inputs, it is natural to ask what is the smallest input/output scenario in which an advantage comes about. This was answered in Ref.~\cite{Svegborn2026}. Since the message cannot be smaller than a bit and Alice needs more than two inputs to avoid trivialising the problem, one might guess that the smallest scenario is $(3,2,1,3)$. However, it turns out that the only non-trivial facet of the classical polytope is the state discrimination inequality given back in Eq.~\eqref{facet3213}. In Box~\ref{Box:EAstatediscrimination}, we show that no state discrimination inequality can be violated by means of entanglement-assisted classical communication. Therefore, we proceed  to give Bob one more output, corresponding to the scenario $(3,2,1,4)$. In this scenario, one new facet inequality appears \cite{Doolittle2021, Vieira2023}. It reads
\begin{equation}\label{facetElna}
\sum_{x=0,1,2} 2p(x|x)+p(3|x)\leq 4,    
\end{equation}
which can be interpreted as Bob trying to guess Alice's input but with the option of rejecting the game. If he guesses correctly (incorrectly) he gets two (zero) points, but if he chooses to reject the game he is guaranteed to score one point. The largest violation of this inequality turns out to be  achieved by distributing a four-dimensional non-maximally entangled state~\cite{Svegborn2026}. 

\begin{examplebox}[label={Box:EAstatediscrimination}]{Entanglement-assisted state discrimination}
In Box~\ref{Box:statediscriminationfacets} we discussed how state discrimination tasks lead to facets of the classical polytope. We now prove that none of these inequalities can be violated using entanglement-assisted classical communication. Following Eq.~\eqref{EACCcorr}, we must place an upper bound on the quantum correlations
\begin{equation}\label{calculateSD}
    S\equiv \sum_{x=1}^n p(b=x|x)=\sum_{x=1}^n\sum_{m=1}^d \tr\left(A_{m|x}\otimes B_{x|m}\ \rho_{AB}\right).
\end{equation}
To this end, consider positive semidefinite operators $X_A$, $Y_B$ and $Z_{AB}$, and let $X_A=\sum_i \lambda_i\ketbra{\lambda_i}$ be the spectral decomposition of $X$. It then follows that $\tr(X_A\otimes Y_B\ Z_{AB})=\sum_i \lambda_i \tr\left(\ketbra{\lambda_i}_A\otimes Y_B\ Z_{AB}\right)$. Using the trivial inequality $\lambda_i\leq \lambda_{\max}(X_A)$, it follows that $\tr(X_A\otimes Y_B\ Z_{AB})\leq \lambda_{\max}(X_A)\tr(Y_BZ_B)$, where $Z_B=\tr_A(Z_{AB})$. We use this inequality for every pair $(m,x)$ in Eq.~\eqref{calculateSD}. It gives
\begin{equation}
    S\leq \sum_{x=1}^n\sum_{m=1}^d  \lambda_{\max}(A_{m|x}) \tr\left(B_{x|m}\ \rho_{B}\right).
\end{equation}
Since $A_{m|x}$ are POVM-elements, their largest eigenvalue cannot exceed unity, i.e.~$\lambda_{\max}(A_{m|x})\leq 1$. Using this, it follows that
\begin{equation}
    S\leq \sum_{m=1}^d  \tr\Big(\sum_{x=1}^n B_{x|m}\ \rho_{B}\Big)=\sum_{m=1}^d \tr(\rho_B)=d.
\end{equation}
This proves that the  inequality $ \sum_{x=1}^n p(b=x|x) \leq d$ is satisfied.

\end{examplebox}

Even though this is the minimal example of an entanglement-assisted advantage, we are going to focus more on what turns out to be the second-simplest case. The reason is that quantum advantages without independent receiver inputs have a natural application. Because only adaptive protocols are compatible with a quantum advantage, any violation of classicality can be interpreted as a black-box proof of adaptive measurement capabilities~\cite{Svegborn2026}. In practice, it is not ideal to use the minimal scenario because the quantum protocol suffers both from noise-sensitivity and the  fact that it is not optimally implemented with maximally entangled states. It turns out that both these obstacles are overcome in the second-simplest scenario, namely $(4,2,1,4)$.  The relevant facet inequality reads~\cite{Doolittle2021}
\begin{equation}\label{eq:witness_cse}
\begin{aligned}
W = \sum_{\substack{x,a=0\\a\neq x}}^{2} p(a|x) + p(3|3) \leq 3.
\end{aligned}
\end{equation}
Its largest violation is $W=3.2247$ which is achieved with a four-dimensional maximally entangled state \cite{Svegborn2026}. However, by restricting to a pair of qubits, its largest violation can be tied to the maximal violation of the CHSH inequality. To this end, let Alice associate $x=0,1$ to the Pauli observables $\sigma_Z$ and $\sigma_X$ respectively and the outcome to the message $m\in\{0,1\}$, with $m=0$ corresponding to the eigenvalue $+1$. In contrast, for $x=2,3$, she does not measure her qubit at all but deterministically outputs $m=0$ and $m=1$ respectively. Take the shared state to be $\ket{\Psi^+}=2^{-\nicefrac12}(\ket{01}+\ket{10})$. When Bob receives $m$, he implements the observable $\frac{1}{\sqrt{2}}\left(\sigma_Z+(-1)^m \sigma_X\right)$ and the outcome is denoted $\pm 1$. He now maps this outcome to his final output, $b$, as follows: if $m=0$ then $(+1,-1)\rightarrow (0,1)$ while if $m=1$ then $(+1,-1)\rightarrow (2,3)$. Evaluating this protocol step by step leads to the violation $W=\frac{5+\sqrt{2}}{2}\approx 3.2071$. Observe that the measurements performed by Alice and Bob when $x=0,1$ are precisely those encountered in the maximal violation of the CHSH inequality, whereas for $x=2,3$ Alice's measurements are non-interactive and in that sense trivial.

Finally, we remarked earlier that if Alice and Bob had access to stronger-than-quantum nonlocality, they could achieve remarkably good performance in communication tasks. Interestingly, adaptivity has a simpler structure when the shared resource is allowed to be an arbitrary nonsignalling correlation. Any advantage of adaptive over non-adaptive nonsignalling assistance can already be witnessed for a fixed value of Bob's input~\cite{Nogueira2026}. Thus, unlike in the entanglement-assisted case, nonsignalling assistance does not exhibit genuinely multi-setting adaptive advantages.

\subsubsection{Network scenarios}\label{Sub:network-scenarios}
\begin{figure}[!tbp]
    \centering
    \includegraphics{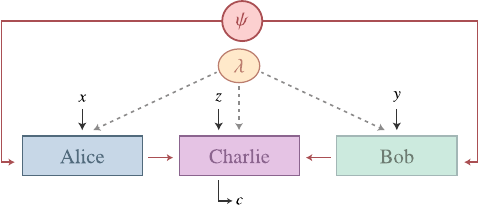}
    \caption{\emph{Two-sender entanglement-assisted PM network}. Alice and Bob receive inputs $x$ and $y$, respectively, and share a bipartite state $\psi$. They encode their inputs into quantum messages sent to Charlie, who receives an input $z$ and produces an outcome $c$. The parties may share classical randomness $\lambda$.}
    \label{F-symmetric-PAM}
\end{figure}

Similarly to the network prepare-and-measure scenarios discussed in Sec.~\hyperref[subsubsec:networkquantummessage]{II-B-7}, entanglement-assisted prepare-and-measure protocols have also been extended to network settings. These generalisations reveal new manifestations of nonclassicality that are absent in the standard bipartite scenario. Here we discuss two representative examples. The first considers a network with two senders and one receiver, where bound entangled states are shown to provide a genuine communication advantage~\cite{Carceller2025,Marton2025,TavakoliCarceller2026} (see Fig.~\ref{F-symmetric-PAM} for an illustration). The second concerns a network with one sender and two receivers, known as the \emph{unrelated confounders} (UC) scenario~\cite{evans2016graphs,lauand2023witnessing}, which provides one of the simplest network generalisations of Bell nonclassicality.
Bound entangled states are entangled states from which no maximally entangled state can be distilled by local operations and classical communication. Although they represent a weak form of entanglement, they nevertheless exhibit genuine quantum features. In Bell experiments, the known violations obtained from bound entangled states are exceedingly small and highly sensitive to noise. Remarkably, the situation changes in network entanglement-assisted prepare-and-measure scenarios, where bound entangled states generate robust violations of communication inequalities.

The network considered in Ref.~\cite{Carceller2025} consists of two senders, Alice and Bob, communicating with a single receiver, Charlie. Alice and Bob receive classical inputs $x$ and $y$, respectively, encode them into $d$-dimensional quantum systems, and send them to Charlie, who receives an input $z$ and produces an output $c$. In addition to communication, Alice and Bob may share an arbitrary bipartite entangled state $\Psi$, giving rise to correlations of the form
\begin{equation}
p_E(c|x,y,z) = \sum_\lambda q_\lambda \, \tr \!\left[ (\Lambda_x^\lambda\otimes\Omega_y^\lambda) [\Psi]\, M_{c|z}^\lambda \right],
\label{eq:boundent-pm}
\end{equation}
where $\Lambda_x^\lambda$ and $\Omega_y^\lambda$ are completely positive trace-preserving encoding maps, $M_{c|z}^\lambda$ are Charlie's measurements, and $\lambda$ denotes shared randomness.

To certify that shared entanglement is responsible for the observed advantage, Eq.~(\ref{eq:boundent-pm}) is compared with the strongest protocol that allows quantum communication but no shared entanglement, that is,
\begin{equation}
p_Q(c|x,y,z) = \sum_\lambda q_\lambda\,\tr\!\left[(\rho_x^\lambda\otimes\sigma_y^\lambda)M_{c|z}^\lambda\right].
\end{equation}

To distinguish these two models, one can employ the communication game introduced in Ref.~\cite{Bakhshinezhad2024}, in which Alice receives $x=(x_0,x_1)\in\{0,\cdots,d-1\}^2$, Bob receives $y=(y_0,y_1)\in\{0,\cdots,d-1\}^2$, Charlie receives $z\in\{0,\ldots,d\}$, and must output
\begin{equation}
w_z(x,y)=
\begin{cases}
x_1+y_1-2z(x_0-y_0)\!\!\!\!\pmod d,
&
z\neq d,
\\[1mm]
x_0-y_0\!\!\!\!\pmod d,
&
z=d.
\end{cases}
\end{equation}

The corresponding average success probability is
\begin{equation}
R_d=\frac{1}{d^4(d+1)}\sum_{x,y,z}p(c=w_z(x,y)|x,y,z).
\end{equation}

For odd prime dimensions (proved explicitly for $d=3,5,7$), every protocol without shared entanglement satisfies
\begin{equation}
R_d\le\frac{2}{d+1},
\label{eq:boundent-ineq}
\end{equation}
making Eq.~(\ref{eq:boundent-ineq}) a witness of entanglement-assisted communication.

Ref.~\cite{Carceller2025} constructs an explicit positive-partial-transpose (PPT) bound-entangled state of two qutrits together with product measurements built from mutually unbiased bases. This strategy achieves $R_3= \tfrac14+\tfrac{1}{2\sqrt3}\approx0.5387>\tfrac12$, thus violating Eq.~(\ref{eq:boundent-ineq}) and certifying bound entanglement as a genuine communication resource. Beyond the violation itself, an important feature is its robustness. Considering the noisy state $\rho_\nu= (1-\nu)\frac{\mathds{1}}{9}+\nu\rho_{\mathrm{BE}}$, the violation survives down to $\nu_c= 1-\frac{9-4\sqrt3}{11} \approx 0.8117$, corresponding to nearly $19\%$ white noise. The noise tolerance for certifying bound entanglement was subsequently improved in \cite{Marton2025} by considering a different task and four-dimensional systems. By generalising those tasks, one can even show that the quantum correlation advantage due to bound entanglement can become unlimited by considering many copies of such states \cite{TavakoliCarceller2026}.

A complementary direction is obtained by increasing the number of receivers instead of senders. One of the simplest examples is the \emph{unrelated confounders} (UC) network \cite{evans2016graphs,lauand2023witnessing}, recently investigated in Ref.~\cite{lauand2025minimal}. The network consists of two independent sources connected through a central node, Bob, who broadcasts a classical message $m$ to two spatially separated parties, Alice and Charlie. Classically, the observed correlations admit the decomposition 
\begin{equation}
p(a,c,m)=\sum_{\lambda_1,\lambda_2}q(\lambda_1)q(\lambda_2)p(a|m,\lambda_1)p(m|\lambda_1,\lambda_2)p(c|m,\lambda_2),
\end{equation}
where the independence of the hidden variables is expressed by $q(\lambda_1,\lambda_2)=q(\lambda_1)q(\lambda_2)$.

Replacing the hidden variables by two independent bipartite quantum states yields the quantum model
\begin{equation}
P(a,c,m)=\tr\!\left[(\psi_1\otimes\psi_2)(E_{a|m}\otimes E_m\otimes E_{c|m})\right],
\label{eq:quan}
\end{equation}
where Bob performs a joint measurement on his subsystems before broadcasting the outcome.

To demonstrate a quantum advantage, Ref.~\cite{lauand2025minimal} considers two maximally entangled pairs distributed by the sources, while Bob performs a two-outcome entangling measurement projecting onto $|\psi_\theta\rangle=\sin\theta\,|01\rangle+\cos\theta\,|10\rangle$.

For Bob's two outcomes, Alice and Charlie perform local projective measurements, giving the unnormalised correlators $\langle A^i C^j\rangle_m=\sum_{a,c} (-1)^{ia+jc} p(a,c,m)$, where $a,c\in\{0,1\}$ and $m\in\{1,-1\}$: $\langle AC\rangle_{1}=\frac{\sin(2\theta)}{4}$, $\langle A\rangle_{1}=\langle C\rangle_{1}=0$, $\langle AC\rangle_{-1}=\frac14$, $\langle C\rangle_{-1}=-\langle A\rangle_{-1}=\frac{\cos(2\theta)}4$, together with $\langle M\rangle=-\tfrac12$. Defining $I_+=1+4\langle AC\rangle_1$ and $I_-=\tfrac18+\tfrac12\langle C\rangle_{-1}$, the nonlinear network inequality
\begin{equation}
\mathcal I=\sqrt{I_+}+3\sqrt{I_-}\le\frac{3}{\sqrt2}+\frac{\sqrt7}{6}
\end{equation}
holds for every classical UC model. While a maximally entangled measurement satisfies the inequality, a partially entangled measurement with $\theta=\tfrac\pi8$ gives $\mathcal I\approx2.692>2.562$, revealing quantum nonclassicality in this network.

\subsubsection{Bounding quantum correlations}\label{sec:NPAEACC}
Let us now turn to the question of giving outer bounds for the set of correlations predicted by quantum theory. This question parallels the one discussed in Sec.~\hyperref[sec:boundingNV]{II-C-2}, but with the following crucial difference. Before, our goal was to bound the set of correlations achievable by Alice sending a $d$-dimensional quantum system to Bob, whereas now our goal is to bound the set of correlations achievable by Alice and Bob sharing an entangled state and Alice sending a $d$-dimensional classical message. These two quantum resources are different and therefore they require distinct methods for bounding the set of associated correlations. 

Let us begin by considering the most general class of protocols, namely those whose correlations are described by Eq.~\eqref{EACCcorr}. As argued below that equation, we can view the correlations as arising from post-selection and marginalisation from an associated Bell scenario. By that argument, adaptive protocols for the PM scenario with alphabet $(|X|,d,|Y|,|B|)$ are associated with quantum nonlocality in the Bell scenario in which Alice has $|X|$ inputs and $d$ outputs, and Bob has $d|Y|$ inputs and $|B|$ outputs. This means we need only a method for bounding the set of quantum nonlocal correlations in a given input/output scenario. This is achieved by the so-called Navascu\'es-Pironio-Ac\'in (NPA) hierarchy~\cite{Navascues2007, Navascues2008}.

Suppose that we are given some arbitrary quantum correlations in the Bell scenario. These can be written 
\begin{equation}
   p(a,b|x,y)=\bracket{\psi}{A_{a|x}\otimes B_{b|y}}{\psi},
\end{equation}
where $(x,a)$ is the input and output of Alice and $(y,b)$ is the input and output of Bob. Here, $\ket{\psi}$ is an arbitrary bipartite quantum state and $\{A_{a|x}\}$ and $\{B_{b|y}\}$ are POVMs. Without loss of generality, we can assume that the POVMs are projective, because had they been non-projective we could always use Naimark's dilation theorem to obtain the same correlations from projective measurements. This is possible because of the unbounded Hilbert space dimension. Let us denote by $\mathcal{Q}$  the set of all such quantum correlations. Our goal now is to give outer approximations of $\mathcal{Q}$ that can, at least in principle, be computed. To this end, we first relax the tensor-product structure between Alice and Bob, and replace it with a commutation constraint. In other words, we allow  $\{A_{a|x}\}$ and $\{B_{b|y}\}$ to be bipartite  projective measurements, but with the constraint that $[A_{a|x},B_{b|y}]=0$. We write $\bar{\mathcal{Q}}$ for the associated set of correlations. It is trivial that $\mathcal{Q}\subset \bar{\mathcal{Q}}$ but whether the inclusion is strict or not was for a long time an open question known as Tsirelson's problem~\cite{Scholz2008}. It was answered in the affirmative in Ref.~\cite{Ji2022}, i.e.~commutation cannot be replaced with tensor-product without loss of generality. Modulo the strict relaxation associated with the commutation structure between Alice and Bob, the NPA hierarchy provides a sequence of SDP relaxations that converges to the set $\bar{\mathcal{Q}}$.

The NPA hierarchy works as follows. Start by listing the measurement operators of Alice and Bob. This list reads $L=\{\mathds{1}\} \cup\{A_{a|x}\}_{a,x}\cup \{B_{b|y}\}_{b,y}$. Here, these operators should be understood as projective POVMs on a single Hilbert space and satisfying commutation between Alice and Bob. Next, define a list of monomials computed over $L$. This set, $S$, therefore contains products of all the operators appearing in $L$. For simplicity, let us include all products of length at most $K$. We then define a moment matrix as follows $\Gamma_{u,v}=\bracket{\psi}{u^\dagger v}{\psi}$, for every $u,v\in S$. We call $\Gamma$ a moment matrix because it is an $|S|\times |S|$ matrix whose rows and columns are indexed by the monomials appearing in $S$. Importantly, the moment matrix is positive semidefinite. This can be verified straightforwardly by writing
\begin{align}\nonumber
    \Gamma&= \sum_{u,v}\bracket{\psi}{u^\dagger v}{\psi}\ketbra{u}{v}\\
    &=\bra{\psi}\Big(\sum_u u^\dagger \ket{u}\Big)\Big(\sum_v v \bra{v}\Big)\ket{\psi}\succeq 0.
\end{align}

\begin{figure}[t]
    \centering
    \includegraphics{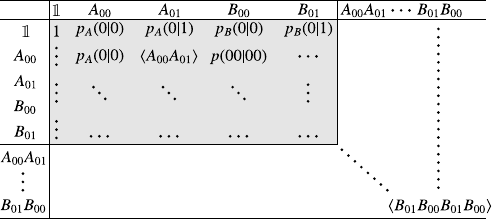}
    \caption{\emph{Moment matrix at an intermediate level of the NPA hierarchy}. Rows and columns are indexed by monomials of Alice's and Bob's measurement operators. The highlighted block contains normalisation, marginal probabilities, and observed joint probabilities. It corresponds to the first level of the hierarchy. The remaining entries represent higher-order moments constrained by projector relations and commutation. Feasibility of a positive semidefinite completion gives an outer approximation to the quantum set.}
    \label{F-matrix-moment-2}
\end{figure}

The next key feature of the moment matrix is that the quantum probabilities appear as specific entries, namely $p(a,b|x,y)=\Gamma_{A_{a|x},B_{b|y}}$. Consequently, these specific elements of $\Gamma$ can be fixed to the probabilities under consideration. Furthermore, the moment matrix has several constraints that are inherited from the structure of quantum theory. For instance, $\Gamma_{\mathds{1},\mathds{1}}=1$ due to normalisation, $\Gamma_{A_{a|x},A_{a'|x}}=\delta_{a,a'} \Gamma_{\mathds{1},A_{a|x}}$ due to the orthogonality of projectors, and $\Gamma_{A_{a|x}B_{b'|y'},B_{b|y}}=\Gamma_{B_{b'|y'}A_{a|x},B_{b|y}}$ due to commutativity. Once all such constraints are taken into account, the  feasibility of $\Gamma$ can be evaluated as an SDP. Labelling the associated sets by $\bar{\mathcal{Q}}_K$, we therefore have the hierarchy $\bar{\mathcal{Q}}_1 \supset \bar{\mathcal{Q}}_2\supset\ldots\supset \bar{\mathcal{Q}}_\infty =\bar{\mathcal{Q}}$.

The last equality means that in the limit of large $K$, the hierarchy converges to $\bar{\mathcal{Q}}$, which was shown in Ref.~\cite{Navascues2008}. Say now that we select some specific $K$ and compute the feasibility SDP. If we find no feasible solution, then we conclude that $p\notin \bar{\mathcal{Q}}_K$ which implies that $p\notin \mathcal{Q}$, i.e.~we have proven that the correlations are not achievable in quantum nonlocality. Instead, if the SDP is feasible, then we can increase $K$ and repeat the procedure in hope of now having a sufficiently strong condition to falsify a quantum realisation. In Fig.~\ref{F-matrix-moment-2}, we illustrate the moment matrix associated with an intermediate level of the NPA hierarchy.

Next, we may ask how we can use these types of SDP relaxation methods to prove that a certain quantum distribution $p(b|x,y)$ can only be achieved by adaptive protocols. This means modifying the SDP relaxation hierarchy so that it applies only to non-adaptive protocols. Thus, infeasibility at some level of the hierarchy proves that no non-adaptive realisation exists, hence certifying the use of adaptive devices. To this end, we must supplement the NPA hierarchy with the constraints associated with non-adaptive protocols. The key insight is that because Bob is non-adaptive, for a given $y$ it is possible to pre-compute all the possible outputs associated with the $d$ possible values of $m$. That is, Bob could output a vector $\vec{b}=(b_1,\ldots,b_d)$, where each $b_k\in\{1,\ldots,|B|\}$. Only when the message arrives does he select to output the element $b_m$. Hence, we may write $B_{b|y,m}=\sum_{\vec{b}} \delta_{b_m,b}B_{\vec{b}|y}$. Consequently, the correlations in non-adaptive protocols become
\begin{equation}
    p(b|x,y)=\sum_{m=1}^d \sum_{\vec{b}} \delta_{b_m,b}\bracket{\psi}{A_{m|x}\otimes B_{\vec{b}|y}}{\psi}.
\end{equation}
The constraint we have introduced on Bob's measurements can equally well be viewed in a common projective dilation as the commutation condition
\begin{equation}\label{commconstraint}
    \big[B_{b|y,m},B_{b'|y,m'} \big]=0,
\end{equation}
for all $y,b,b',m,m'$. The commuting projective representation follows by dilating the parent POVM, and its sufficiency follows from the fact that all commuting projective measurements admit a parent POVM of the form $B_{\vec{b}|y}=B_{b_1|y,1}\ldots B_{b_d|y,d}$. The benefit of phrasing the non-adaptiveness in terms of the commutation constraint in Eq.~\eqref{commconstraint} is that it can easily be introduced in the NPA relaxation. Notice, however, that one typically needs to deal with a heavier SDP than in the adaptive case, because Bob's measurements now include all possible outcomes associated with the different choices of the message. Some of this challenge can be remedied using SDP symmetrisation techniques~\cite{Ioannou2022}.

A drawback of the NPA-based method to bound the set of correlations under entanglement-assisted classical communication is that the associated Bell scenario grows rapidly with the system's dimension. Specifically, both Alice's output  and Bob's auxiliary input are of size $d$. Hence, even at the lowest level, it quickly becomes expensive to evaluate the SDP relaxation beyond the few smallest choices of $d$. There exists an alternative approach to bounding the correlation set that circumvents this problem. This approach is based on the notion of informationally restricted correlations~\cite{Tavakoli2020informationally, Chaturvedi2020quantum, Tavakoli2022informationally} and SDP hierarchies developed to bound the associated set of quantum correlations [we discuss these in Sec.~\hyperref[Sub:information-restriction]{IV-G}]. As we will see in that section, one can relax the set of correlations obtained under entanglement-assisted classical communication to the set of correlations obtained under particular informational limitations. This relaxation is strict, and can therefore negatively affect the in-principle achievable accuracy of the method. However, it can be computationally advantageous since its computational requirements are independent of $d$, and it can therefore easily be computed at low relaxation levels for small scenarios. It is shown in Ref.~\cite{TavakoliPauwels2021} that this can indeed be practically favourable over  the NPA-based methods already in quite small scenarios.

\subsection{Quantum correlations from information principles}

In spite of its remarkable empirical success, quantum theory is formulated in terms of mathematical postulates that are often perceived as abstract and lacking direct physical intuition. This has motivated the search for informational principles capable of singling out quantum correlations among all logically consistent probabilistic theories. Rather than taking the Hilbert space formalism as fundamental, this line of research aims to derive, or at least constrain, the structure of quantum theory from physically well-motivated requirements grounded in information-processing tasks.

A natural framework for this investigation is provided by Bell scenarios. In particular, as discussed in Box~\ref{Box:CHSH}, classical correlations satisfy the CHSH inequality $\text{CHSH} \le 2$, while quantum theory allows for stronger correlations up to the Tsirelson bound $\text{CHSH} \le 2\sqrt{2}$. Remarkably, even stronger correlations are logically possible without contradicting relativistic causality. A paradigmatic example is given by the Popescu--Rohrlich (PR) box~\cite{popescu1994quantum}, which achieves the algebraic maximum $\text{CHSH}=4$ while still obeying the \emph{nonsignalling} principle, i.e.~the impossibility of faster-than-light communication (see Box~\ref{Box:PR}).

\begin{backgroundbox}[label={Box:PR}]{The PR-box}
As discussed in Box~\ref{Box:CHSH}, in a bipartite Bell scenario, two distant parties, Alice and Bob, receive inputs $x,y \in \{0,1\}$ and produce outputs $a,b \in \{0,1\}$ according to some conditional probability distribution $p(a,b|x,y)$. A central physical requirement is the \emph{nonsignalling} condition, which ensures that no information can be transmitted instantaneously between the parties. Formally, this is expressed as
\begin{align}
\sum_b p(a,b|x,y) &= \sum_b p(a,b|x,y') \quad \forall a,x,y,y', \\
\sum_a p(a,b|x,y) &= \sum_a p(a,b|x',y) \quad \forall b,x,x',y.
\end{align}
That is, Alice's (Bob's) marginal distribution does not depend on Bob's (Alice's) choice of input.

\medskip

A particularly important example of a nonsignalling correlation is the Popescu--Rohrlich (PR) box~\cite{popescu1994quantum}. It is defined by the correlations $a \oplus b = xy$, with uniform marginals, i.e.,
\begin{equation}
p(a,b|x,y) =
\begin{cases}
\frac{1}{2}, & \text{if } a \oplus b = xy, \\
0, & \text{otherwise}.
\end{cases}
\end{equation}

The PR-box exhibits correlations that are strictly stronger than those allowed by quantum theory. This can be readily seen by evaluating the CHSH expression introduced in Box~\ref{Box:CHSH}, which yields the algebraic maximum $\text{CHSH} = 4$, thereby surpassing the Tsirelson bound $\text{CHSH} \le 2\sqrt{2}$ imposed by quantum theory.
\end{backgroundbox}

The existence of such superquantum correlations demonstrates that the nonsignalling principle alone is not sufficient to characterise quantum theory. This observation motivates the search for additional constraints that further restrict the set of admissible correlations. Informational principles provide precisely such constraints: they typically bound the efficiency of communication or other information-processing tasks, are satisfied by quantum correlations, yet are violated by PR-box-type resources.

Within this programme, entanglement-assisted classical communication and, in particular, prepare-and-measure scenarios have emerged as central frameworks. They furnish the operational backbone for principles such as information causality (IC)~\cite{pawlowski2009information} and for foundational limitations arising in communication complexity~\cite{van2013implausible,brassard2006limit,linden2007quantum}. As will be discussed below, these principles impose nontrivial restrictions on the strength of correlations and play a crucial role in distinguishing quantum theory from more general nonsignalling models.

\subsubsection{Information causality}\label{sec:infocausality}

Information causality (IC)~\cite{pawlowski2009information} can be viewed as an entropic constraint associated with a prepare-and-measure causal structure~\cite{chaves2015information}. Alice receives a string of independent random bits $\vec{x}=(x_1,\ldots,x_N)$, while Bob receives an index $i\in\{1,\ldots,N\}$. Alice sends a classical message $m$, and Bob produces a guess $g_i$ for $x_i$. In quantum theory, Bob's total information is bounded by the entropy of the communicated message:
\begin{equation}
\label{eq:IC}
\sum_{i=1}^N I(x_i : g_i) \le H(m),
\end{equation}
Here $H(m)=-\sum_m p(m)\log_2 p(m)$ is the Shannon entropy, and $I(x_i:g_i)=H(x_i)+H(g_i)-H(x_i,g_i)$ is the mutual information. The proof uses the von Neumann entropy $S(\rho)=-\tr(\rho\log_2\rho)$, which reduces to Shannon entropy for a classical register: $S(x)=H(x)$.

For clarity, consider $N=2$, with inputs labelled $x_0,x_1$. Alice and Bob initially share a state $\rho_{AB}$, and $B$ denotes Bob's subsystem, with reduced state $\rho_B$. To simplify the notation, let $X=(x_0,x_1)$ and $R=(m,B)$, and take all sums below to run over $i=0,1$. By data processing, $I(x_i:g_i)\le I(x_i:R)$, since Bob's guess is obtained from $R$. Expanding the mutual information gives
\begin{align}
    \sum_i I(x_i:g_i) &\le \sum_i I(x_i:R)\nonumber\\ &=\sum_i S(x_i)+2S(R)-\sum_i S(x_i,R).
\end{align}
Strong subadditivity of the von Neumann entropy~\cite{lieb1973convex,lindblad1975completely} implies $\sum_i S(x_i,R)\ge S(R)+S(X,R)$. Moreover, adding a classical register cannot decrease entropy, so $S(X,R)\ge S(X,B)$. Since the inputs are independent of the shared state, $S(X,B)=S(X)+S(B)$, where $S(B)=S(\rho_B)$. Combining these observations gives
\begin{align}
\sum_i I(x_i:g_i)
&\le \sum_i S(x_i)+S(R)-S(X,R)\nonumber\\
&\le I(x_0:x_1)+S(R)-S(B)\nonumber\\
&\le I(x_0:x_1)+H(m),
\end{align}
where the last step follows from subadditivity, $S(R)\le H(m)+S(B)$. For independent inputs, $I(x_0:x_1)=0$, recovering Eq.~\eqref{eq:IC} for $N=2$. The argument extends to arbitrary $N$.

If Alice instead sends a quantum register $M$, the classical-register inequality is replaced by $S(X,M,B)\ge S(X,B)-S(M)$. The same reasoning then gives $\sum_i I(x_i:g_i)\le 2S(M)$ for independent inputs. The factor of two reflects the enhancement possible through dense coding (Sec.~\hyperref[sec:denseCoding]{III-C-1}).

As shown in Box~\ref{Box:PRboxviolatesIC}, a PR box can maximally violate IC. In the original IC construction, deriving Tsirelson's bound uses a nested random-access-code (RAC) protocol with exponentially many copies of the nonlocal resource. Alice holds a random $2^n$-bit string, and Bob aims to recover the bit at a chosen position (see Sec.~\hyperref[sec:QRAC]{II-B-3}). For a nonsignalling distribution $p(a,b\mid x,y)$, the concatenated RAC protocol implies Uffink's quadratic inequality~\cite{uffink2002quadratic}, $E_1^2+E_2^2\le 1$, where $E_k=2P_k-1$. Writing $q_{xy}=p(a\oplus b=0\mid x,y)$, the relevant success probabilities are $P_1=\tfrac12(q_{00}+q_{10})$ and $P_2=\tfrac12(q_{01}+1-q_{11})$. 

For the depolarised PR-box family $p_v=v\,p_{\mathrm{PR}}+(1-v)p_W$, with $p_W(a,b\mid x,y)=\tfrac14$, one finds $E_1=E_2=v$ and $\mathrm{CHSH}=4v$. Thus Uffink's inequality is violated when $v>\tfrac{1}{\sqrt{2}}$, or equivalently $\mathrm{CHSH}>2\sqrt{2}$. This recovers Tsirelson's bound along this family of correlations.

\begin{examplebox}[label={Box:PRboxviolatesIC}]{A PR box violates IC}
Alice inputs $x=x_0\oplus x_1$ into her side of a PR box, obtains output $a$, and sends the one-bit message $m=x_0\oplus a$ to Bob. Given the index $y\in\{0,1\}$, Bob inputs $y$ into his side, obtains $b$, and guesses $g_y=m\oplus b$. Using the PR-box relation $a\oplus b=xy$, we find
\begin{align}
g_y &=x_0\oplus(a\oplus b) =x_0\oplus(x_0\oplus x_1)y =x_y.
\end{align}
For independent, uniform inputs, perfect guessing gives $I(x_0:g_0)=I(x_1:g_1)=1$. Yet Alice sends only one bit, so
\begin{equation}
I(x_0:g_0)+I(x_1:g_1)=2>1=H(m),
\end{equation}
a maximal violation of IC.
\end{examplebox}

Following its initial formulation, several generalisations of the IC principle have been proposed, for instance to multipartite scenarios \cite{pollyceno2023information,hsu2012multipartite}, as well as approaches interpreting IC as one among potentially infinitely many entropic principles \cite{chaves2015information} emerging from imposing a given causal structure on physical theories. A major difficulty in the original formulation of IC, as well as in many of its extensions, is that its practical application requires the construction of a suitable communication protocol making use of the correlations under investigation. In particular, the strongest known bounds are typically obtained through concatenation protocols, in which several copies of a nonlocal resource are combined in order to amplify the amount of accessible information in stronger-than-quantum theories. Such procedures are often technically involved and highly scenario-dependent.

Recently, however, it was shown that concatenation protocols involving multiple copies of the nonlocal resource can be replaced by protocols employing only a single pair of devices, provided the communication channel between sender and receiver is taken to be noisy rather than ideal \cite{miklin2021information}. In this approach, the role previously played by concatenation is effectively transferred to the communication channel capacity. More precisely, the standard IC inequality $\sum_{i=1}^N I(x_i:g_i)\le H(m)$ is replaced by the noisy-channel version
$\sum_{i=1}^N I(x_i:g_i)\le C$, with $C$ denoting the capacity of the communication channel under consideration. Remarkably, by suitably tuning the channel noise, one can recover the same bounds previously obtained through arbitrarily deep concatenation schemes, including Tsirelson's bound for the CHSH scenario (see Box~\ref{Box:ICnoconcatenation}).

\begin{examplebox}[label={Box:ICnoconcatenation}]{IC principle without concatenation protocols}
The original IC protocol assumes that Alice's one-bit message is transmitted through a perfect communication channel, whose capacity is therefore $C=1$. In this case, the IC condition reads $I(x_0:g_0)+I(x_1:g_1)\le 1$. However, if the box is noisy, then the PR-box relation $a \oplus b =xy$ holds only with probability $p$. Therefore, Bob guesses the desired bit correctly with probability $p_{\mathrm{guess}}=p$. Thus, since each guess succeeds with probability $p$, the mutual information between Alice's bit and Bob's guess is $I(x_i:g_i)=1-h(p)$, where $h(p)=-p\log_2 p -(1-p)\log_2(1-p)$ is the binary entropy. Hence, the IC principle implies directly, without the need for any concatenation protocol, that $2(1-h(p))\leq 1$. This already gives a nontrivial bound on $p$, but it is weaker than Tsirelson's bound.

\medskip
To recover Tsirelson's bound, suppose now that Alice's message is sent through a binary symmetric channel with transmission probability $p_c$, meaning that the transmitted bit is flipped with probability $1-p_c$. In this case, Bob succeeds whenever either: i) the nonlocal box works correctly and the channel does not flip the message, or ii) both the box and the channel fail simultaneously.
Therefore, the total success probability becomes $p_{\mathrm{succ}} = p_c p +(1-p_c)(1-p)$. Using that the capacity of a binary symmetric channel is $C=1-h(p_c)$, the IC condition is then translated to 
\begin{equation}
2\Big[1-h(p_{\mathrm{succ}})\Big]\le1-h(p_c).
\end{equation}
As \(p_c\to 1/2\), the channel capacity approaches zero and the resulting IC bound on $p$ converges to Tsirelson's bound $p \to\frac12\left(1+\frac1{\sqrt2}\right)$ \cite{miklin2021information}.
\end{examplebox}

\subsubsection{No-collapse of communication complexity}

Communication complexity, introduced in Sec.~\hyperref[sec:quantumCCP]{II-B-6}, quantifies the minimal amount of communication required for two or more distant parties to evaluate a distributed function. Concretely, suppose Alice holds an $n$-bit string $\vec{x}$ and Bob an $n$-bit string $\vec{y}$. The task is to compute a Boolean function $f:\{0,1\}^n \times \{0,1\}^n \rightarrow \{0,1\}$, while minimising the number of communicated bits.

A paradigmatic example is the inner product function (modulo $2$), $IP_n(\vec{x},\vec{y}) = \bigoplus_{i=1}^{n} x_i y_i$.  It is known that, even if Alice and Bob share an arbitrary amount of entanglement, computing $IP_n$ requires $\Theta(n)$ bits of communication~\cite{cleve2013quantum}. In particular, no quantum protocol improves the asymptotic scaling of the classical strategy in which (say) Bob sends $\vec{y}$ to Alice.

In contrast, the availability of PR-box correlations trivialises this task. From the defining relation of a PR-box, $a \oplus b = xy$, if Alice and Bob input $x_i$ and $y_i$ into the $i$-th PR-box, their outputs $a_i$ and $b_i$ satisfy $a_i \oplus b_i = x_i y_i$. By sharing $n$ independent PR-boxes and repeating this procedure for all $i=1,\dots,n$, Bob can compute the parity of his outcomes, $m = \bigoplus_{i=1}^{n} b_i$, and send this single bit to Alice. Alice then computes
\begin{equation}
\bigoplus_{i=1}^{n} a_i \oplus m = \bigoplus_{i=1}^{n} (a_i \oplus b_i)=\bigoplus_{i=1}^{n} x_i y_i = IP_n(\vec{x},\vec{y}),
\end{equation}
thus evaluating the function with only one bit of communication.

As shown in Box~\ref{Box:PRboxcollapses}, the same argument can be extended to any Boolean function at the cost of an exponential number of PR-boxes shared between Alice and Bob~\cite{van2013implausible}. The striking consequence is that access to PR-box correlations leads to a collapse of communication complexity: every Boolean function can be computed with only one bit of communication. Since such a collapse is widely regarded as implausible from a physical perspective, the no-collapse of communication complexity has been proposed as an informational principle ruling out these superquantum correlations~\cite{van2013implausible,brassard2006limit,linden2007quantum}.

\begin{examplebox}[label={Box:PRboxcollapses}]{PR-boxes collapse communication complexity}
Any Boolean function can be expressed as a sum (modulo $2$) of products of local functions,
\begin{equation}
f(\vec{x},\vec{y}) = \bigoplus_{i=1}^{2^n} P_i(\vec{x}) Q_i(\vec{y}),
\end{equation}
where $P_i$ and $Q_i$ are Boolean functions (not necessarily monomials) of $\vec{x}$ and $\vec{y}$, respectively. Since $P_i(\vec{x})$ is known to Alice and $Q_i(\vec{y})$ to Bob, they can use one PR-box per term to generate outputs satisfying
\begin{equation}
a_i \oplus b_i = P_i(\vec{x}) Q_i(\vec{y}),
\end{equation}
and proceed as in the inner product case. Although this may require up to $2^n$ PR-boxes, only a \emph{single} bit of communication is needed.

As an illustrative example, consider
\begin{equation}
f(x_1,x_2,y_1,y_2) = x_1 y_2 \oplus x_2 y_1 \oplus y_1.
\end{equation}
This function can be computed using three PR-boxes with inputs $(x_1,y_2)$, $(x_2,y_1)$, and $(1,y_1)$, respectively, followed by the same aggregation procedure.
\end{examplebox}
\begin{figure}[!tbp]
    \centering
    \includegraphics{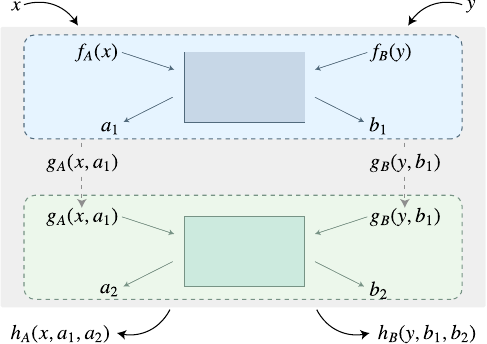}
   \caption{\emph{Schematic representation of a nonlocality distillation protocol.}
  Illustration of a wiring protocol between two probability distributions (represented as boxes).}
    \label{F:NL-distillation}
\end{figure}
Crucially, Brassard \emph{et al.}~\cite{brassard2006limit} demonstrated that this collapse is not restricted to perfect PR-boxes. By considering a probabilistic communication complexity framework, they proved that even imperfect nonlocal boxes would render communication complexity trivial, provided they simulate the PR-box correlations with a success probability greater than $(3+\sqrt{6})/6$. Therefore, this result establishes a tighter boundary on the strength of correlations, beyond which distributed computation would become trivial. Note, however, that there remains an open gap between the maximum success probability allowed by quantum theory (roughly $85.4\%$) and the threshold that triggers this collapse ($(3+\sqrt{6})/6\approx90.8\%$).

Subsequent works~\cite{brunner2009nonlocality,rai2019geometry,brito2019nonlocality,eftaxias2023advantages,botteron2024algebra} have extended this phenomenon to broader classes of nonsignalling correlations employing nonlocality distillation protocols. Representing the correlations as a box with inputs and outputs, a nonlocality distillation protocol can be basically
understood as a wiring among many copies of such boxes. In Fig. \ref{F:NL-distillation}, for instance, we illustrate a particular example with two boxes, the outcomes of the first being used as the input for the second one. In this case, if we wire two distributions as  $W(p(a,b\vert x,y), p'(a,b\vert x,y))$, we obtain a new distribution $q(a,b\vert x,y)$ given by
\begin{eqnarray}
 q(a,b\vert x,y) & & =  W\!\left(p(a,b \vert x,y),\, p'(a,b\vert x,y)\right) \nonumber \\ \nonumber
& &=
\sum_{\substack{
a_1,a_2,b_1,b_2\\
x_1,x_2,y_1,y_2
}}
\chi_x(a,a_1,a_2,x_1,x_2)
\chi_y(b,b_1,b_2,y_1,y_2) \\  & & \quad\quad \times p(a_1, b_1\vert x_1, y_1)\,
p'(a_2, b_2\vert x_2, y_2).   
\end{eqnarray}
where $\chi_x(a,a_1,a_2,x_1,x_2)$ represents the wiring performed
locally by Alice and $\chi_y(b,b_1,b_2,y_1,y_2)$ the wiring performed
by Bob. Here $x_i$ and $y_j$ are respectively the $i$th and $j$th inputs
to Alice's and Bob’s boxes, $a_i$ and $b_j$ are the outcomes from
the respective boxes, and $a$ and $b$ are the final respective outputs in the distilled distribution $q(a,b\vert x,y)$ (see Box~\ref{Box:PRboxdistillation} for an explicit example).

\begin{examplebox}[label={Box:PRboxdistillation}]{Distilling nonlocal correlations}
Consider a mixture of the PR-box, $p_{\mathrm{PR}}(a,b\vert x,y)=\tfrac12\delta_{a\oplus b,xy}$, and a classically correlated box, $p_{\mathrm{C}}(a,b\vert x,y)=\tfrac12\delta_{a\oplus b,0}$, namely
\begin{equation}
    p_{\mathrm{PR-C}}(a,b\vert x,y) = v\, p_{\mathrm{PR}}(a,b\vert x,y) + (1-v)\, p_{\mathrm{C}}(a,b\vert x,y).
\end{equation}
For this family one obtains $\operatorname{CHSH}(v)=2+2v$, and therefore the CHSH inequality [Eq.~\eqref{CHSHcorr}] is violated for every visibility $v > 0$.

\medskip
Now consider the wiring protocol introduced in Ref.~\cite{brunner2009nonlocality}, in which two copies of $p_{\mathrm{PR-C}}$ are used together with the local wirings (specifying, respectively, the input to the first box, the input to the second box and the final output) $f_A(x)=x$, $g_A(x,a_1)=xa_1$, $h_A(x,a_1,a_2)=a_1\oplus a_2$, and $f_B(y)=y$, $g_B(y,b_1)=yb_1$, and $h_B(y,b_1,b_2)=b_1\oplus b_2$. The resulting post-wired box is again of the same form,
\begin{equation}
    p_{\mathrm{Post}}(a,b\vert x,y) = v^{\prime} p_{\mathrm{PR}}(a,b\vert x,y) + (1-v^{\prime}) p_{\mathrm{C}}(a,b\vert x,y),
\end{equation}
but with an enhanced visibility $v^{\prime}=\frac{v(3-v)}{2}$. Consequently, $\operatorname{CHSH}(v^{\prime})=2+2v^{\prime}=2+3v-v^2$, which is strictly larger than the original value for every $0<v<1$. Hence, whenever the initial distribution is nonlocal but not maximally nonlocal, the distilled box obtained from the wiring protocol exhibits a stronger CHSH violation. In this sense, nonlocality has been distilled from the original two copies.

\medskip
Moreover, by employing multiple copies of $p_{\mathrm{PR-C}}(a,b\vert x,y)$ and concatenating this two-copy distillation protocol, one can approach arbitrarily closely a perfect PR-box, thereby once again collapsing communication complexity. Importantly, this remains true for any $v>0$, i.e.~even for distributions arbitrarily close to the classical bound $\operatorname{CHSH}=2$. This demonstrates the post-quantum character of correlations lying well below Tsirelson's bound, $\operatorname{CHSH}=2\sqrt{2}$.

\end{examplebox}

\subsection{Quantum communication}
We now proceed to introduce quantum communication in the entanglement-assisted PM scenario by following Ref.~\cite{TavakoliPauwels2021}. Like before, we consider that Alice and Bob share an entangled state, $\rho_{AB}$, but now the message sent from Alice to Bob is a quantum state of dimension $d$. This means that Alice's input, $x$, is associated with a completely positive trace-preserving map $\Lambda^{A\rightarrow C}_x$ which transforms her part of $\rho_{AB}$ into a quantum message that lives on a $d$-dimensional Hilbert space labelled $C$. Hence, the total state given to Bob is
\begin{equation}\label{tauEAQC}
    \tau_x^{CB}=\left(\Lambda_x^{A\rightarrow C} \otimes\mathds{1}_B\right)[\rho_{AB}].
\end{equation}
Since both Alice's encoding maps and the shared state can be arbitrary, the set of possible states $\{\tau_x\}_x$ is fully characterised by bipartite density matrices defined over a space of dimension $d\times D$, where $D$ is arbitrary, such that
\begin{equation}\label{tauEAQC2}
    \tau_x^B\equiv \tr_C\left(\tau_x^{CB}\right)=\tau^B \qquad \forall x.
\end{equation}
The quantum correlations between Alice and Bob are therefore given by $p(b|x,y)=\tr\left(\tau_x^{CB} M_{b|y}^{CB}\right)$, where $\{M_{b|y}\}_b$ is a POVM that acts jointly on systems $C$ and $B$. Note that shared randomness  has already been accounted for implicitly, because we have placed no bound on the dimension of systems $A$ and $B$. For example, we could write the shared state as $\rho_{AB}=\tilde{\rho}_{A_1B_1}\otimes \sum_\lambda q(\lambda) \ketbra{\lambda,\lambda}_{A_2B_2}$, where $A=(A_1,A_2)$ and $B=(B_1,B_2)$. Thus, the quantum part of the state is confined to system $A_1B_1$ whereas the part $A_2B_2$ holds the shared randomness; Alice and Bob may simply read systems $A_2$ and $B_2$ before acting on systems $A_1$ and $B_1$.  In what follows, we discuss various aspects of these entanglement-assisted quantum communication scenarios.

\subsubsection{Dense coding}\label{sec:denseCoding}
We begin discussing the paradigmatic dense coding protocol. It addresses a state discrimination problem, analogous to that previously identified as a facet of the classical polytope in Box~\ref{Box:statediscriminationfacets}. In this task, Alice receives as input a uniformly random pair $x=(x_1, x_2)\in \{0,\ldots,d-1\}^2$ and her goal is to communicate $x$ to Bob by sending only a $d$-dimensional system. We have seen before that neither by means of quantum communication [see Eq.~\eqref{mesdquantum}] nor by means of entanglement-assisted classical communication [see Box~\ref{Box:EAstatediscrimination}], can Alice and Bob perform this task with a higher average success probability than in a purely classical protocol. The dense coding protocol  circumvents these limitations by combining shared entanglement with quantum communication, and this leads to a deterministic solution to the communication task.

The dense coding protocol provides the following recipe to solve the problem~\cite{bennett1992communication}. Let Alice and Bob share the maximally entangled state of dimension $d$, namely $\ket{\phi^+}=\frac{1}{\sqrt{d}}\sum_{k=0}^{d-1}\ket{k,k}$. Next, let us define the unitaries 
\begin{equation}
    X=\sum_{j=0}^{d-1}\ketbra{j+1}{j}\quad \text{and} \quad Z=\sum_{j=0}^{d-1} e^{\frac{2\pi i}{d}j}\ketbra{j}{j},
\end{equation}
where addition is computed modulo $d$. Alice associates her input, $x$, to the unitary operator $U_x=X^{x_1}Z^{x_2}$, which takes the form 
\begin{equation}\label{Eq:dense-encoding-weyl}
    U_{x} = \sum_{j=0}^{d-1}e^{\frac{2\pi i}{d}j x_2} \ketbra{j+x_1}{j}.
\end{equation}
After applying $U_x$, Alice sends her system to Bob. He then holds the total state 
\begin{equation}\label{densecodingstates}
    \ket{\tau_x}=\left(U_x\otimes\mathds{1}\right)\ket*{\phi^+}=\frac{1}{\sqrt{d}}\sum_{k=0}^{d-1}e^{\frac{2\pi i}{d}k x_2} \ket{k+ x_1,k}.
\end{equation}
By construction, each state $\ket{\tau_x}$ is maximally entangled. Moreover, one can straightforwardly verify that $\braket{\tau_x}{\tau_{x'}}=\delta_{x,x'}$. Consequently, the collection of $d^2$ maximally entangled states, $\{\ket{\tau_x}\}_{x\in\{0,\ldots,d-1\}^2}$, forms an orthonormal basis.  The projective measurement in this basis is usually referred to as a  Bell state measurement. Thus, if Bob measures both his quantum systems using a Bell state measurement, he can perfectly distinguish between the states $\ket{\tau_x}$ and therefore deterministically recover $b=x$. Equivalently, we have achieved a unit success probability, namely $p_\text{succ}=1$, where
\begin{equation}
    p_{\text{succ}} = \frac{1}{|X|}\sum_{x\in X} p\left(b=x|x\right),
\end{equation}
where $|X|=d^2$ is the number of inputs for Alice.

The $d$-dimensional maximally entangled state is crucial for the dense coding protocol. This has been formalised by showing that dense coding schemes and teleportation schemes have a one-to-one correspondence \cite{Werner2001b}. Alternatively, one can show that $p_{\text{succ}}=1$ self-tests a maximally entangled state when Alice's share of the initial state has dimension $d$~\cite{Moreno2021}. A more quantitative approach is to consider whether we can deduce a lower bound on the dimension of the entangled state just from inspecting the success probability of the task \cite{Moreno2021, TavakoliPauwels2021}. Such a semi-device-independent certification is possible and can be based on the Schmidt number of $\rho_{AB}$. For the unfamiliar reader, a minimal overview of the Schmidt number appears in Box~\ref{Box:SchmidtNumber}.

\begin{backgroundbox}[label={Box:SchmidtNumber}]{Schmidt number}
The Schmidt number of a bipartite density matrix is commonly interpreted as the dimensionality of its entanglement. It is defined as follows. First, consider a pure state $\ket{\psi}_{AB}$. In its Schmidt decomposition, it can be written as 
\begin{equation}\label{SchmidtDecomposition}
    \ket{\psi}_{AB}=\sum_{i}\sqrt{\lambda_i} \ket{\alpha_i}\otimes\ket{\beta_i},
\end{equation}
where  $\{\ket{\alpha_i}\}_i$ and $\{\ket{\beta_i}\}_i$ are orthonormal bases and $\{\sqrt{\lambda_i}\}_i$ are called Schmidt coefficients. The number of non-zero Schmidt coefficients (i.e.~the rank of the reduced density matrix) is called the Schmidt rank of $\ket{\psi}_{AB}$ and it represents the entanglement dimension of the state. 

\medskip

This is extended to mixed states, $\rho_{AB}$, through the notion of Schmidt number \cite{Terhal2000}. To this end, we decompose the state into a mixture of pure states, $\rho_{AB}=\sum_i p_i \ketbra{\psi_i}$. Then, we compute the Schmidt rank, $r_i$, of each state $\ket{\psi_i}$. The largest Schmidt rank is labelled $r=\max_i r_i$. Therefore, we say that in order to generate $\rho$ through this decomposition, we would have required access to a pure state with entanglement dimension $r$. However, the pure-state decomposition of $\rho_{AB}$ is not unique. Hence, we must minimise $r$ over all possible such decompositions, and the result is called the Schmidt number of $\rho_{AB}$.
\end{backgroundbox}

Let us derive the corresponding dense-coding bound. Suppose that the shared state has Schmidt number at most $r$. Since the success probability is linear in the realised states, it is enough to prove the bound for pure states of Schmidt rank at most $r$ and then extend it by convexity. Write such a pure state as in Eq.~\eqref{SchmidtDecomposition}, and let $\Pi_B=\sum_{i=0}^{r-1}\ketbra{\beta_i}{\beta_i}$ be the projector onto the support of Bob's reduced state. Alice's encoding acts only on her subsystem and produces a $d$-dimensional message system, which we denote by $C$. Therefore, after the encoding, every state $\tau_x^{CB}$ is supported on $\mathbb{C}^d\otimes\Pi_B$. Consequently,
 \begin{align}\nonumber\label{DCSN}
     p_\text{succ}&=\frac{1}{|X|}\sum_{x=1}^{|X|} \tr\left(\tau^{CB}_x M^{CB}_x\right) \\\nonumber
     &=\frac{1}{|X|}\sum_{x=1}^{|X|} \tr\left(\left(\mathds{1}_C\otimes \Pi_B\right) \tau^{CB}_x\left(\mathds{1}_C\otimes \Pi_B\right)  M^{CB}_x\right)\\\nonumber
     &\leq \frac{1}{|X|}\sum_{x=1}^{|X|} \tr\left(\left(\mathds{1}_C\otimes \Pi_B\right)  M^{CB}_x\right)\\
     &=\frac{1}{|X|}\tr\left(\mathds{1}_C\otimes \Pi_B\right)=\frac{dr}{|X|}.
 \end{align}
In the second line we used the fact that $\tau_x^{CB}$ is supported on $\mathbb{C}^d\otimes\Pi_B$. In the inequality we used $\tau_x^{CB}\preceq\mathds{1}_{CB}$, and in the last line we used $\sum_x M_x^{CB}=\mathds{1}_{CB}$. Thus, if an experiment observes $p_{\mathrm{succ}}>\tfrac{dr}{|X|}$, the shared state must have Schmidt number larger than $r$.

For the standard dense-coding task, $|X|=d^2$, so the bound becomes $p_{\mathrm{succ}}\leq \frac{r}{d}$. In particular, if the shared state is separable, then $r=1$ and $p_{\mathrm{succ}}\leq \tfrac{1}{d}$. This is exactly the benchmark obtained by sending an ordinary $d$-dimensional message without useful entanglement. Hence, any value above $\tfrac{1}{d}$ certifies entanglement, and any value above $\tfrac{r}{d}$ certifies Schmidt number at least $r+1$. The perfect dense-coding value $p_{\mathrm{succ}}=1$ certifies Schmidt number at least $d$ and, when Alice's initial share has dimension $d$, self-tests the maximally entangled state up to the usual equivalences~\cite{Moreno2021}. This makes dense coding an expensive but powerful semi-device-independent entanglement witness. It is expensive because Bob must distinguish $d^2$ maximally entangled states, which generally requires an entangled measurement on the received system and his share of the resource. It is powerful because the conclusion uses only the dimension of Alice's communicated system and the observed success probability, rather than a tomography of Bob's subsystem. Closely related schemes can also be used to certify entanglement properties of measurements~\cite{Abbott2018}. At the same time, one should keep an important caveat in mind: since Bob has only a single measurement input in the dense-coding discrimination task, the certification is tied to the assumed prepare-and-measure causal structure and should not by itself be interpreted as a cryptographic security statement.

Dense coding also leads to a clean self-testing statement. Suppose Alice's initial share has dimension $d$ and that in the $\vert X \vert=d^2$ task one observes the maximal value $p_{\text{succ}}=1$. Then, the $d^2$ preparations must be perfectly distinguishable, and together with the common-marginal constraint $\tr_C(\tau^{CB}_x) = \tr_C(\tau^{CB}_{x'})$ for all $x, x'$, this forces the shared resource to contain a maximally entangled $d$-dimensional factor up to local isometries and auxiliary systems. This is conceptually different from the self-testing result discussed in Sec.~\hyperref[sec:self_test]{II-D-1}: there, the object being certified was typically the set of preparations and measurements used in a PM experiment, whereas here the object being certified is the bipartite state itself. At the same time, one should keep an important caveat in mind. Because Bob has only a single measurement input, this certification depends sensitively on the assumed causal structure and does not by itself guarantee cryptographic security against external adversaries. Dense coding thus serves both as a powerful entanglement witness and as an instructive illustration of the scope and limitations of semi-device-independent communication scenarios.

The success of the dense coding protocol hinges on the fact that local unitaries applied to half of a maximally entangled state can lead to encoding that makes use of the global Hilbert space. In its simplest case, it is based on a  maximally entangled two-qubit state and single-qubit unitaries. What kind of quantum correlations can be generated by protocols that rely only on such entanglement and unitary encoding? Perhaps surprisingly, for PM scenarios with arbitrary alphabets, there exists an isomorphism between the correlations achievable in such entanglement-assisted protocols and the correlations achievable in  PM scenarios without shared entanglement. Ref.~\cite{Pauwels2022many} showed that
\begin{quote}
\textit{Correlations in the PM scenario are achievable with unitary encoding on half of a maximally entangled two-qubit state if and only if they are achievable by sending a four-dimensional quantum system supported only on a real-valued Hilbert space.}
\end{quote}
The meaning of this result is that, to characterise the set of correlations, we can without loss of generality ignore the entanglement and instead consider a PM scenario of the type discussed in Sec.~\ref{S:PM-setting} but restrict the quantum communication to real-valued ququarts. That allows us to recycle the tools known for these simpler PM scenarios to analyse the quantum correlations. In particular, one can apply the heuristic SDP sampling method from Sec.~\hyperref[sec:boundingNV]{II-C-2} with the minor modification of changing the sampling space from complex to real Hilbert spaces. Moreover, while dense coding showed us that two bits can be substituted with one qubit and a maximally entangled state, the above result shows that the latter is actually strictly more powerful than the former. This is because there are many examples of PM scenarios in which real-valued ququarts are sufficient to violate the constraints of the classical polytope; see e.g.~Ref.~\cite{Pauwels2022many}. 

Fundamentally, the proof of the result follows from the group isomorphism $[\text{SU}(2)\times \text{SU}(2)]/\mathbb{Z}_2\simeq \text{SO}(4)$, but we will show it in the language of the PM scenarios. Our proof relies on the fact that any set of pure states is fully determined (up to a global unitary) by its Gram matrix. The Gram matrix is the matrix of all the overlaps, namely $G_{x,x'}=\braket{\tau_x}{\tau_{x'}}$. In our case, we have $\ket{\tau_x}=\left(U_x\otimes\mathds{1}\right)\ket{\phi^+}$, where $U_x$ is a qubit unitary. Those can always be written in the form $U_x=e^{i\varphi_x}\vec{z}_x\cdot \vec{\Sigma}$, where $\vec{\Sigma}=(i\mathds{1},\sigma_X,\sigma_Y,\sigma_Z)$ and $\vec{z}_x$ is a four-dimensional unit vector.  Hence, the Gram matrix becomes 
\begin{align}\nonumber\label{GramMatrix}
G_{x,x'}&=\bracket{\phi^+}{U_x^\dagger U_{x'}\otimes\mathds{1}}{\phi^+}=\frac{1}{2}\tr\left(U_x^\dagger U_{x'}\right)\\
&=e^{i(\varphi_{x'}-\varphi_{x})}\vec{z}_x\cdot\vec{z}_{x'}.    
\end{align}
Thus, if we apply the phase shifts $\ket{\tau_x}\rightarrow e^{-i\varphi_x}\ket{\tau_x}$, the Gram matrix would become real-valued. However, such a global phase shift does not change the correlations and hence we can always apply it without loss of generality. This shows that we can without loss of generality consider only a real-valued Gram matrix. Conversely, consider that we have a real-valued $|X|\times |X|$ Gram matrix. This matrix is by construction positive semidefinite, has rank no more than four (the dimension) and has unit entries on the diagonal. This means that we can write  $G_{x,x'}=\vec{z}_x\cdot\vec{z}_{x'}$ for appropriate four-dimensional real unit vectors $\vec{z}_x$. This gives us the form in Eq.~\eqref{GramMatrix}, apart from the phase factors, which as mentioned can be ignored for the correlations. Importantly, when applying this isomorphism the tensor-product structure of the entanglement-assisted scenario is lost. This means that one cannot directly access information about the entanglement structure when considering  the PM scenario with only real ququarts. 

A curious consequence of this result is that sending a qutrit from Alice to Bob can sometimes be a more powerful resource than sending a qubit assisted by $\phi^+$. The latter resource can clearly transmit two bits of information, and since its state space is isomorphic to real ququarts its state-space is nine-dimensional. The former resource can transmit only $\log_2 3$ bits of information, and its state space is only eight-dimensional. Nevertheless, there exists a PM task which is solved deterministically by the former resource and not by the latter resource~\cite{Pauwels2022many}. This task is introduced in Ref.~\cite{Emeriau2022} where it is named the Torpedo game. We briefly introduce it in Box~\ref{Box:torpedo-game}.

\begin{examplebox}[label={Box:torpedo-game}]{The Torpedo game}
The Torpedo game is an information-retrieval task in a prepare-and-measure scenario. Alice receives two dits, $(x,z)\in\mathbb{Z}_d^2$, and sends a single $d$-dimensional message to Bob. Bob receives a question $q\in\{\infty,0,1,\ldots,d-1\}$ and outputs a dit $c\in\mathbb{Z}_d$.

\medskip

The game has a useful geometric picture. Think of Alice's input $(x,z)$ as the position of a one-cell ship on a $d\times d$ toric grid. Bob is ordered to shoot along one of the $d+1$ possible directions: horizontal, vertical, and the $d-1$ diagonal slopes. The question $q$ tells him the direction, but he is free to choose which parallel line in that direction to use. Alice and Bob win if Bob's torpedo misses Alice's ship.

\begin{center}
    \includegraphics[width=0.9\linewidth]{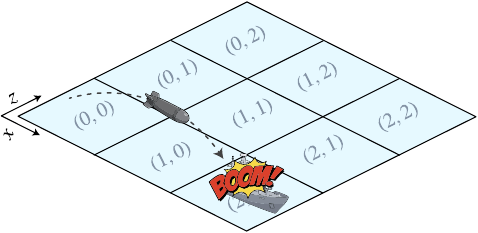}
\end{center}

Algebraically, Bob must avoid exactly one bad output for each question. The winning sets are $ w_\infty(x,z)=\{c\in\mathbb{Z}_d:\,c\neq x\}$ and $ w_q(x,z)=\{c\in\mathbb{Z}_d:\,c\neq qx-z\}$, where $q\in\mathbb{Z}_d$ and all arithmetic is modulo $d$. Thus, for each question, Bob does not need to identify Alice's position exactly; he only needs to avoid the line that passes through it. The average winning probability is
\begin{equation}\label{eq:torpedo_average_score}
   W_{\mathrm T}^{(d)}=\frac{1}{d^2(d+1)}\sum_{x,z,q}p\left(c\in w_q(x,z)\middle|x,z,q\right).
\end{equation}
For $d=3$, a classical trit message can win with probability at most $W_{\mathrm T}^{C}=\tfrac{11}{12}$. In contrast, a qutrit message can win perfectly, $W_{\mathrm T}^{Q}=1$~\cite{Emeriau2022}.

\medskip
Let us describe the perfect qutrit strategy. Write $\omega=e^{2\pi i/3}$ and let $X\ket{k}=\ket{k+1}$ and $Z\ket{k}=\omega^k\ket{k}$ be the qutrit shift and phase operators. Alice starts from the state $\ket{\psi_{0,0}}=(\ket{1}-\ket{2})/\sqrt{2}$ and, on input $(x,z)$, sends the displaced qutrit state $\ket{\psi_{x,z}}=X^xZ^z\ket{\psi_{0,0}}$ up to an irrelevant global phase. Bob measures in one of four mutually unbiased bases, chosen according to $q\in\{\infty,0,1,2\}$. These bases are chosen so that the bad outcome has probability zero: $p(c=x|x,z,\infty)=0$ and $p(c=qx-z|x,z,q)=0$ for $q=0,1,2$. Therefore, Bob never outputs the line that contains Alice's position, and the game is won deterministically. 

\medskip
The relevance for the present discussion is that the Torpedo game separates resources in a slightly counterintuitive way. A qutrit message solves the $d=3$ game perfectly, while a classical trit does not. Combined with the real-ququart equivalence discussed above, this gives a PM task in which sending a qutrit can outperform sending a qubit assisted by a maximally entangled two-qubit state.
\end{examplebox}

\subsubsection{Dimensionality and certification }\label{sec:Dimensionality_and_certification}
As we have seen, protocols in the spirit of dense coding rely on Alice unitarily encoding her classical information into half of a maximally entangled state. However, the general description of entanglement-assisted PM scenarios lets Alice encode her information via any quantum channel, and the shared state could in principle be of a different dimension than the quantum message. At first sight, it may appear meaningless to consider a shared state of local dimension $D>d$, since this would force Alice to apply a compression channel to generate her quantum message. This would mean that the entanglement is decohered in the encoding process, thereby leaving Bob with total states of dimension $d\times D$. Since such asymmetric states have Schmidt number no larger than $d$, one may be led to conclude that it would have been sufficient to restrict $\rho_{AB}$ to local dimension $d$. 

However, this intuition is incorrect because although each individual $\tau_x$ has Schmidt number at most $d$, their collection, $\{\tau_x\}_x$, may span a more complex --- albeit dimension-asymmetric --- Hilbert space. It is indeed the case that in general one cannot restrict entanglement-assisted PM scenarios to be based on shared entanglement of the same  dimension as the quantum message \cite{TavakoliPauwels2021, Pauwels2022many}. To see this, consider the following task. Alice and Bob want to perform the usual random access code, as defined in Eq.~\eqref{RAC4222}, but this time they are each supplemented with one extra input. Thus, Alice has inputs $x\in\{00,01,10,11,5\}$ and Bob has inputs $y\in\{1,2,3\}$. The first four inputs of Alice and the first two inputs of Bob are used for the random access code. However, when Bob chooses $y=3$ his job is to discriminate $x=5$ from $x\neq 5$. We impose this discrimination as a constraint, i.e.~the game is valid only if $ p(b=0|x\neq5,y=3) = p(b=1|x=5,y=3)=1$. Now, if Alice and Bob have dense coding resources, we know from Sec.~\hyperref[sec:denseCoding]{III-C-1} that the correlations are equivalent to those achievable by Alice sending real-valued ququarts. In order to respect the discrimination condition, the state associated with $x=5$ must be orthogonal to the space spanned by the four states associated with $x\neq 5$. Hence, we must constrain the four states relevant to the random access code to a three-dimensional real Hilbert space. If the protocol was classical, we would therefore be tasked with implementing the random access code with a ternary message. It is known that the best classical protocol achieves $W_\text{RAC}=\frac{7}{8}=0.875$, while  a protocol based on a real-valued qutrit can achieve at best $W_\text{RAC}=\frac{5+\sqrt{5}}{8}\approx 0.9045$ \cite{Ahrens2014}. As we now show, it is possible to do better by assisting qubit communication with entanglement of dimension larger than two. 

In order to beat the bound associated with dense coding resources, let the shared state be two copies of the maximally entangled state, i.e.~$\rho_{AB}=\phi^+_{A_1B_1}\otimes \phi^+_{A_2B_2}$. Alice will encode by applying a unitary jointly to systems $A_1$ and $A_2$, and then discard system $A_1$ while relaying $A_2$ to Bob. For $x=00,01,10,11,5$, respectively, we choose her five encoding unitaries as
\begin{align}\nonumber
    &U_1=\mathds{1}\otimes\mathds{1},\\\nonumber
    &U_2=\text{CN}_1 \text{CN}_2,\\ \nonumber
    &U_3=(\mathds{1}\otimes\sigma_X)\text{CN}_1 \text{CN}_2, \\ \nonumber
    & U_4=\mathds{1}\otimes\sigma_X, \\ 
    & U_5=(\mathds{1}\otimes\sigma_Z\sigma_X)\text{CN}_2.
\end{align}
where we have defined the controlled-not gates with control on qubit 1 and qubit 2 as $\text{CN}_1\ket{j,k}=\ket{j,k\oplus j}$ and $\text{CN}_2\ket{j,k}=\ket{k\oplus j,k}$ respectively. The decoherence caused by this non-unitary encoding leads all Bob's states to take the form $\tau_x=\frac{1}{2}\left(\ketbra{\varphi_x}+\ketbra{\phi_x}\right)$, where $\varphi_x$ and $\phi_x$ are orthogonal. In other words, these states are density matrices proportional to rank-two projectors. Since Bob's optimal measurement can be determined by solving an eigenvalue problem [in a way analogous to  Eq.~\eqref{DIDWBrunner}], we may express the average success probability in the random access code only in terms of the states $\tau_x$. Evaluating it leads to $W_\text{RAC}=\frac{6+\sqrt{2}}{8}\approx 0.9268$. Thus, we have outperformed the dense coding resources even though the states given to Bob are noisy and dimension-asymmetric.

The fact that we cannot, without loss of generality, set the dimension of $\rho_{AB}$ equal to $d$ both enriches the scope of entanglement-assisted PM scenarios and poses an additional challenge for the characterisation of the associated set of quantum correlations. Ref.~\cite{TavakoliPauwels2021} presents a converging hierarchy of SDPs to solve the membership problem for this set, but it is reported to be rather impractical at the low levels reachable with reasonable computer resources. To deal with this, the same work proposes a more efficient outer approximation method which is based on the notion of informationally restricted quantum correlations (see Sec.~\hyperref[Sub:information-restriction]{IV-G}) for which SDP relaxation hierarchies have already been proposed \cite{Tavakoli2022informationally}. The main idea here is based on the previously discussed observation in connection with dense coding: that the average success probability of state discrimination cannot exceed $\frac{dr}{|X|}$, where $d$ is the message dimension and $r$ is the largest Schmidt number of the output states $\tau_x$. To justify this output-state version, note that Schmidt number at most $r$ implies $\tau_x^{CB}\preceq r\mathds{1}_C\otimes\tau_x^B$: the inequality follows for pure states by Cauchy--Schwarz in a Schmidt basis and extends by convexity. Since all outputs have the same marginal $\tau_x^B=\tau^B$, we obtain $\sum_x\tr(\tau_xM_x)\le r\tr[(\mathds{1}_C\otimes\tau^B)\sum_xM_x]=rd$. Dividing by $|X|$ gives the bound without assuming a common rank-$r$ support on Bob's side. For quantum communication, this means that the success probability is limited by $\frac{d^2}{|X|}$ and for classical communication [as in Sec.~\hyperref[sec:EACC]{III-A}] it is limited by $\frac{d}{|X|}$, because the Schmidt number of $\tau_x$ is one. Thus, by relaxing the entanglement-assisted PM scenario into an informationally restricted PM scenario (without entanglement) and using its established SDP relaxation methods, we can obtain useful bounds on the original quantum correlation set. In particular, the complexity of these SDPs is independent of $d$, which facilitates their evaluation.

In fact, when we are free to utilise the unrestricted dimension of $\rho_{AB}$, Ref.~\cite{Vieira2023} showed that there exists an elegant one-to-one connection between entanglement-assisted PM scenarios with quantum and classical messages respectively. First, consider that we have an entanglement-assisted scenario where Alice sends a $d$-dimensional quantum message to Bob based on sharing the state $\rho_{AB}$. We could imagine a scenario with classical communication, in which the shared entangled state is $\rho_{A_1B_1}\otimes\phi^+_{A_2B_2}$. Now, Alice applies her encoding channel to system $A_1$, but she cannot send $A_1$ to Bob since only classical communication is permitted. To circumvent this, she uses her half of $\phi^+$ to teleport the $d$-dimensional output of the encoding channel to Bob. The classical communication consists of one of the $d^2$ symbols needed to inform Bob about the correction unitary that he must perform on system $B_2$ in order to recover Alice's state. Thus, the state that Alice would have given to Bob via quantum communication now is available to him in system $B_1B_2$. The cost of this equivalence is that we must double the channel capacity and consume an extra maximally entangled state. For the converse, suppose that Alice and Bob share $\rho_{AB}$ and send a classical message of dimension $d^2$. To reproduce the correlations with quantum communication, again let Alice and Bob share the state $\rho_{A_1B_1}\otimes\phi^+_{A_2B_2}$. Then, Alice first measures $A_1$ to generate her classical message. But instead of sending it to Bob she encodes it into system $A_2$ by applying the unitary encodings in dense coding. She then sends this $d$-dimensional quantum system to Bob who performs a Bell state measurement on it to recover the $d^2$-valued message. He can now measure his system $B_1$ in the same way as he would in the case of classical communication. In conclusion, at the cost of an extra copy of a maximally entangled state, $d$-dimensional quantum messages and $d^2$-dimensional classical messages can be exchanged for each other. 

\subsubsection{Simple measurements}\label{sec:simple_measurements}
Many entanglement-assisted protocols in the PM scenario require Bob to measure both his systems in an entangled basis. When the systems are encoded on distinct photons, this presents a formidable challenge. For instance, dense coding is based on Bob making a Bell state measurement, but complete Bell-state discrimination cannot generally be realised using only passive linear optics and photon counting without additional resources, such as ancillary photons or extra degrees of freedom~\cite{Lutkenhaus1999}. Such resources can improve discrimination, but their introduction alone does not guarantee a deterministic measurement. For qubits, ancilla-free linear optics can distinguish at most two Bell states \cite{Calsamiglia2001} and in higher dimension this becomes even more limited. 

In view of this, it appears reasonable to ask how well we can perform the state discrimination task if Bob is limited to separable measurements, i.e.~$M_b=\sum_\mu M_{b,\mu}^C\otimes M^B_{b,\mu}$. However,  with this restriction it is no longer possible to exceed the classical bound, because
\begin{align}\nonumber
    \sum_{x=1}^{|X|}\tr\left(\tau_x M_x\right) &=\sum_{x=1}^{|X|}\sum_\mu \tr\left(\tau_x^{CB} (M_{x,\mu}^C\otimes M^B_{x,\mu})\right)\\\nonumber
    &\leq \sum_{x=1}^{|X|}\sum_\mu \lambda_{\max}[M^C_{x,\mu}] \tr\left(\tau^B_x M^B_{x,\mu}\right)\\\nonumber
    &\leq \sum_{x=1}^{|X|}\sum_\mu \tr\left(M^C_{x,\mu}\right) \tr\left(\tau^B M^B_{x,\mu}\right)\\\nonumber
    &= \tr\left( (\mathds{1}_C\otimes\tau^B) \sum_{x=1}^{|X|}\sum_\mu M^C_{x,\mu}\otimes M^B_{x,\mu}\right)\\
    &=\tr\left( \mathds{1}_C\right) \tr\left(\tau^B\right)=d.
\end{align}
In the second row, we have used the inequality $\tr(\rho^{CB}(Y_C\otimes Z_B))\leq \lambda_{\max}[Y]\tr(\rho^B Z_B)$ for a state $\rho$ and positive semidefinite operators $Y$ and $Z$. In the third row, we have used that $\lambda_{\max}[X]\leq \tr(X)$ for $X\succeq 0$ and the nonsignalling relation in Eq.~\eqref{tauEAQC2}. Thus, entangled measurements are essential for beating the classical limit in tasks of the dense coding type.

Is it at all possible to achieve advantages in entanglement-assisted PM scenarios by using measurements that have no entanglement features?  Ref.~\cite{Piveteau2022} showed that product measurements can sometimes be sufficient. To this end, consider the random access code in which Alice has three input bits $x=(x_1,x_2,x_3)\in\{0,1\}^3$, namely that labelled by $(n,d)=(3,2)$ in Sec.~\hyperref[sec:QRAC]{II-B-3}. If we permit two bits of communication, the optimal strategy is for Alice to send $x_1$ and $x_2$ to Bob, thereby forcing him to guess only when $y=3$. This yields $W_{3,2}=\frac{5}{6}\approx 0.8333$. Using dense coding resources, namely a qubit message and a shared maximally entangled state, we can do better. Using the isomorphism with $\mathbb{R}^4$, we can use the SDP technique from Sec.~\hyperref[sec:boundingNV]{II-C-2} to obtain the  upper bound  $W_{3,2}\leq \frac{1}{2}+\frac{1}{\sqrt{6}}\approx0.9082$ \cite{Navascues2015b}. Consider now the following specific quantum protocol. Alice and Bob share the maximally entangled state $\phi^+$ and Alice encodes her input via the unitaries
\begin{equation}
    U_x = (-1)^{x_1}
\begin{pmatrix}
-\alpha_{x_1}\,\mu_{x_2 x_3}
&
(-1)^{x_2+x_3}\,\alpha_{\bar{x}_1}\,\mu_{x_2 x_3}
\\[6pt]
(-1)^{x_2+x_3}\sqrt{2}\,\alpha_{\bar{x}_1}
&
\sqrt{2}\,\alpha_{x_1},
\end{pmatrix}
\end{equation}
where $\mu_{x_2x_3}=(-1)^{x_2}+i(-1)^{x_3}$ and $\alpha_s=\frac{1}{2}\sqrt{1+(-1)^s\sqrt{\tfrac23}}$ and the bar-sign denotes bit-flip. Next, Bob associates his input with the product observables
\begin{align}\nonumber
    &E_1=\sigma_Z\otimes\sigma_Z\\ \nonumber
    &E_2=\sigma_Y \otimes\frac{\sqrt{3}\sigma_Y+\sigma_Z}{2}\\
    &E_3=\sigma_X \otimes\frac{\sqrt{3}\sigma_Y-\sigma_Z}{2}.
\end{align}
With this strategy, calculating the success probability in the random access code gives $W_{3,2}= \frac{1}{2}+\frac{1}{\sqrt{6}}$, which saturates the upper bound obtained by means of SDP relaxation. In conclusion, not only are product measurements sufficient to beat the classical limit in the random access code, but they are even optimal when restricting to dense coding resources.

This observation, that product measurements can yield maximal advantages, is even more relevant if it can be leveraged for quantum certification in high-dimensional Hilbert spaces. While the first such protocol was introduced in Ref.~\cite{Bakhshinezhad2024}, perhaps the simplest one was proposed in Ref.~\cite{Zhang2025}. It focuses on a stochastic version of dense coding. In this task, just like in dense coding, Alice has an input $x=(x_1,x_2)\in\{0,\ldots,d-1\}^2$ and communicates a $d$-dimensional system to Bob. However, unlike dense coding, Bob's job is not to recover $x$ but to recover either $x_1$ or $x_2$, as marked by his input $y\in\{1,2\}$. This can equally well be interpreted as a $(n,d)=(2,d)$ random access code, but we choose to think of it as stochastic dense coding because we are now only interested in whether the task can be performed deterministically. The answer  to this question is affirmative. To this end, we let Alice and Bob share the $d$-dimensional maximally entangled state and let Alice apply the same encoding unitaries as in dense coding. This leads to the same total states as in dense coding, as given in Eq.~\eqref{densecodingstates}. Now, we let Bob measure each system independently in either the computational basis ($y=1$) or the Fourier basis ($y=2$). Such a measurement has $d^2$ possible outcomes, and therefore Bob must group them into $d$ distinct bins, which are associated with his final outcome $b$. Specifically, we choose the measurements as $M_{b|1}=\sum_{j=0}^{d-1}\ketbra{j+b,j}$ and $M_{b|2}=\left(F\otimes F^\dagger\right)M_{b|1}\left(F^\dagger\otimes F\right)$ respectively, where $F=\tfrac{1}{\sqrt{d}}\sum_{k,l=0}^{d-1}e^{\frac{2\pi i}{d}kl} \ketbra{k}{l}$ is the Fourier transform. A direct calculation reveals that $b=x_y$ and hence Bob always recovers the information of his interest. 

This task can be used also to benchmark entanglement dimensionality. This is analogous to the discussion of Eq.~\eqref{DCSN}, but now without the need for a high-dimensional Bell state measurement. To this end, we ask for the largest average success probability $W_{2,d}$ achievable in an entanglement-assisted protocol based on a state whose Schmidt number is at most $r$. Due to the linearity of the witness function, we can without loss of generality restrict to pure states whose reductions have rank $r$. Thus, after Alice makes her encoding, the state given to Bob is supported on a space of dimension $d\times r$. In general, these states must satisfy Eq.~\eqref{tauEAQC2}, but in order to derive a general bound we are going to relax this condition and hence consider that Alice may send to Bob any quantum state of dimension $rd$. Ref.~\cite{Zhang2025} analyses such a random access code by adapting the techniques introduced in Ref.~\cite{Farkas2025} to prove that
\begin{equation}\label{RACSN}
    W_{2,d}\leq \frac{1}{2}+\frac{1}{2}\sqrt{\frac{r}{d}}.
\end{equation}
Thus, if this inequality is violated, we can conclude that the shared state has Schmidt number at least $r+1$. In particular, for the special case of $r=d$ the bound permits the perfect stochastic communication associated with the maximally entangled state.  

The test in Eq.~\eqref{RACSN} is more noise-fragile than the one based on dense coding [see Eq.~\eqref{DCSN}], but this is far more practical due to the simple measurements employed by Bob. For instance, if we consider a state of the form $\rho_{AB}=v\phi^++\frac{1-v}{d^2}\mathds{1}$, a violation is achieved when 
\begin{equation}
    v> \frac{d-2+\sqrt{dr}}{2(d-1)}.
\end{equation}
Naturally, restricting Bob to product measurements means reducing some  of the power of a general quantum protocol. The limitations associated with product measurements are made more precise in Ref.~\cite{Bakhshinezhad2024}. There, it is shown that if the state $\rho_{AB}$ is unsteerable then it cannot reveal any quantum advantage under product measurements in the PM scenario. Thus, entangled but unsteerable states are out of reach for protocols based on product measurements. However, as shown in Ref.~\cite{Bakhshinezhad2024}, this limitation can be circumvented by an appropriate extension of the PM scenario (see the discussion in Sec.~\hyperref[Sub:network-scenarios]{III-A-3}).

Inspiration drawn from the use of product measurements in entanglement-assisted PM scenarios also paves the way for an interesting avenue to certifying non-projective measurements. The semi-device-independent certification of non-projective measurements was discussed in Sec.~\hyperref[sec:POVMcertification]{II-D-3} for standard PM scenarios (without entanglement). In those scenarios, we saw that projective measurements are responsible for a large share of the quantum advantage, and non-projective measurements offer an additional sliver of quantum advantage. Distinguishing the latter from the former underpinned the certification. Contrasting this, Ref.~\cite{Svegborn2026} shows that by introducing entanglement, one can identify PM scenarios in which non-projective measurements do not just  sometimes supply a bit of extra advantage, but instead are responsible for all quantum advantages possible in the scenario. To understand this, consider a scenario in which Bob has no input ($|Y|=1$). Alice and Bob share $\rho_{AB}$ and Alice applies a channel $\Lambda^{A\rightarrow C}_x$ to encode her qubit message. In the spirit of the adaptive protocols discussed in Sec.~\hyperref[sec:adaptive]{III-A-2}, we require that system $C$ is measured first, and that the outcome ($m$) is used to inform the choice of measurement applied to system $B$. Now, recall that any projective measurement on a qubit can have at most two non-zero outcomes. However, if $m$ is binary and Bob has no input, we might as well move that measurement to Alice's lab and let her send $m$ directly to Bob. This means that any qubit message that is read out via a projective measurement can equally well be replaced by plain classical communication assisted by entanglement. Therefore, in order to observe a genuine advantage from the quantum message, Bob's measurement of the qubit message must have at least three possible outcomes. Extremal three- (or four-) outcome measurements are only associated with non-projective measurements. As an example of this effect, let us return to the facet of the classical polytope defined in Eq.~\eqref{facetElna}. Using the SDP relaxation method in Sec.~\hyperref[sec:NPAEACC]{III-A-4} one can show that entanglement-assisted classical protocols can achieve at most the value $4.174$. However, Ref.~\cite{Svegborn2026} presents a protocol where Alice and Bob share a two-qubit entangled state and Bob measures the message qubit  using a non-projective read-out measurement whose Bloch vectors form an equilateral triangle in the disc of the Bloch sphere, achieving a value of $4.202$. By using four-dimensional entanglement, thereby supplying to  Bob  a qubit-ququart state, the advantage due to non-projective qubit measurements can be boosted up to $4.3608$.  

\subsection{Experiments}
In this section, we review experimental efforts to demonstrate and optimise quantum communication advantages in entanglement-assisted scenarios. We focus primarily on dense coding implementations, where the theoretical capacity enhancement is limited in practice by the difficulty of realising optimal joint measurements, and on more recent approaches that circumvent these limitations through additional degrees of freedom, higher-dimensional systems, and generalised communication tasks. In Table~\ref{tab:Experiments_sec3} we summarise the experiments discussed in this section, highlighting their main contribution. 

\begin{table*}[t]
\centering
\small

\begin{tabular}{lll}
\hline
\textbf{Experiment}
&
\textbf{Platform/method}
&
\textbf{Main significance}
\\
\hline

Mattle \textit{et al.}~\cite{Mattle1996} 
& 
polarisation-entangled photon pairs
& 
First experimental dense coding. \\
(1996)& & Partial Bell measurement: $1.58$ ideal, $1.13$ measured bits. 
\\

Barreiro \textit{et al.}~\cite{Barreiro2008} 
& 
Spin and OAM hyperentanglement 
& 
Enabled complete Bell-state discrimination by exploiting  \\
(2008)& &hyperentanglement, achieving 1.630 bits. 
\\

Williams \textit{et al.}~\cite{Willianms2017} 
& 
Time and polarisation hyperentanglement 
& 
Demonstrated dense coding over an optical fibre link,\\ 
(2017)& & transmitting a $100\times136$ image with high fidelity. 
\\

Hu \textit{et al.}~\cite{Hu2018} 
& 
Path-polarisation ququarts 
& 
First demonstration of high-dimensional dense coding. Transmitted \\
(2018)& &  $2.09$ bits, exceeding the capacity of a classical ququart (and qubits). 
\\

Guo \textit{et al.}~\cite{Guo2025} 
& 
16 path modes (Single photon) 
& 
Operational advantage using high-dimensional entanglement\\
(2025)& & over a restricted qubit channel.
\\

Piveteau \textit{et al.}~\cite{Piveteau2022} 
& 
Photonic qubits 
& 
First demonstration of a communication advantage using \\
(2022)& & simple product measurements. 
\\

Zhang \textit{et al.}~\cite{Zhang2025} 
& 
High-dimensional entangled states 
& 
Scaled product measurements to high-dimensional regimes.\\ 
(2025)& & Certification of the entanglement Schmidt number. 
\\
\hline
\end{tabular}
\caption{Selected experimental milestones showing advances in quantum communication advantage. The progression shows the shift from proof-of-principle experiments with linear optics to advanced hyperentangled states, high-dimensional entanglement, and product measurement schemes.}
\label{tab:Experiments_sec3}
\end{table*}

As mentioned in Sec.~\hyperref[sec:denseCoding]{III-C-1}, for Bob to perfectly distinguish the states sent by Alice in a dense coding protocol, he must perform a Bell state measurement [see Eq.~\eqref{densecodingstates}].  In photonic implementations, however, complete Bell state measurements cannot generally be realised using only passive linear optics and photon counting, without additional resources such as ancillary photons, nonlinear interactions, or extra degrees of freedom, imposing limitations on dense coding protocols based solely on passive linear optics~\cite{Calsamiglia2002}. Therefore, the ideal dense-coding capacity of $2\log_2d$ bits is hard to achieve in this setting.  In the qubit case, this limitation implies that only two of the four Bell states can be individually identified using passive linear optics. Choosing three Bell-state encodings gives three distinguishable messages and raises the optimal deterministic communication capacity to $\log_2 3\approx 1.58$ bits~\cite{Lutkenhaus1999, Calsamiglia2001}. Several approaches have been proposed to overcome these  limitations in the qubit case, for example, the use of ancillary systems~\cite{Grice2011, Ewert2014}. Although these methods can substantially improve the success probability of Bell-state discrimination, they require considerably greater experimental resources and remain significantly more challenging to implement than passive linear-optical schemes. We refer the interested reader to Ref.~\cite{Bianchi2026} for a recent review of Bell measurements in quantum optics and to Ref.~\cite{Guo_2019} for an experimental overview of dense coding.

\textit{First dense coding experiment.} The first experimental demonstration of three-message dense coding using one qubit message and shared entanglement was performed by Mattle \textit{et al.}~\cite{Mattle1996}. They considered a two-photon polarisation-entangled  state produced by parametric down-conversion. The experiment consisted of three distinct parts: the EPR source generating entangled photons in a well-defined state; Alice’s station for encoding the messages by a unitary transformation of her particle; and, finally, Bob’s Bell state measurement to read the signal sent by Alice. In this experiment, Bob could distinguish three messages encoded in three of the four possible Bell states using linear-optical interference. Consequently, Alice could encode three distinguishable messages, increasing the ideal communication capacity from the classical one-qubit limit of one bit to $\log_23\approx 1.58$ bits.  The measured channel capacity was $1.13$ bits.

\textit{Dense coding with hyperentanglement.} Subsequent works aimed to increase the information capacity to reach the the theoretical capacity of two bits. A relevant solution was demonstrated by Barreiro \textit{et al.}~\cite{Barreiro2008}, where they considered systems with extra degrees of freedom. In particular, they used two-photon states simultaneously entangled in polarisation (spin) and orbital angular momentum. These are generally called hyperentangled states. The extra degree of freedom enabled them to discriminate the four Bell states, leading to a transmission of $1.630$ bits, exceeding the ideal three-message capacity of 1.58 bits. However, preparing and controlling hyperentangled states introduces substantial experimental complexity. Moreover, the additional degrees of freedom used in this approach are not naturally compatible with long-distance fibre-based communication.

\textit{Dense coding with fibre links.} As a consequence of the incompatibility mentioned above, the natural follow-up was to consider states that can actually be transmitted using standard hardware. In Ref.~\cite{Willianms2017}, Williams \textit{et al.} report a different experimental approach  using a complete Bell state measurement based on time-polarisation hyperentanglement. This allows them to show dense coding  over optical fibre links. The experiment requires only linear optics and single-photon detectors once the additional degree of freedom is included. With this approach they achieve the transmission of $1.665$ bits, allowing them to transmit a $100 \times 136$ pixel image with a fidelity of $0.87$.  

\textit{Dense coding with high-dimensional entanglement.}  Instead of increasing the distinguishability of the four qubit Bell states, another strategy is to enlarge the Hilbert space. For a $d$-dimensional system, dense coding theoretically allows transmission of $2\log_2d$ bits. Experimentally, however, the challenge becomes implementing high-dimensional Bell state measurements.  In Ref.~\cite{Hu2018}, they tackle this problem using ququarts with path-polarisation entanglement. By considering five distinguishable high-dimensional Bell states, they obtain a theoretical capacity of $\log_2 5 \approx 2.32$. This value exceeds both the dense-coding capacity achievable with entangled qubits and the capacity of an unassisted ququart channel, $\log_2 4 = 2$. They manage to transmit approximately $2.09$ bits of information by performing partial Bell measurements. This also proves that there exists an advantage in communication using entanglement in systems of higher dimensions.  With this approach they transmit a $53 \times 188$ pixel image, with $0.952$ fidelity.

\textit{Entanglement assistance with compressed messages.} The above experiments have indeed increased the experimentally transmitted information beyond the 1.13 bits measured in the initial proof of principle. However, generating high-dimensional states for transmission through channels relies on expensive experimental setups. A first idea is to reduce this cost by using high-dimensional systems in a qubit channel. As mentioned in Sec.~\ref{sec:Dimensionality_and_certification} for a dense coding protocol, this yields no advantage since the compression of the transmitted state to qubits decoheres the state. However, in Ref.~\cite{Guo2025} they show that there exist other protocols; in particular, they consider a hybrid protocol between quantum state discrimination and quantum  random access codes, where high-dimensional entanglement does constitute an advantage. They experimentally show that qubit messages assisted by four-dimensional entanglement  outperform all protocols based on qubit entanglement to assist the qubit channel.  They do this by exploiting 16 path modes for a heralded single photon.

\textit{First demonstration of product measurement advantage.}
Finally, as one of the main problems in dense coding protocols is the implementation of Bell measurements in a photonic framework, a natural research line studies communication advantage by considering simple measurements (see Sec.~\hyperref[sec:simple_measurements]{III-C-3}). In Ref.~\cite{Piveteau2022}, the authors consider a RAC task with $8$ preparations and $3$ measurements, while sharing a Bell pair. Remarkably, the optimal measurements for the protocol are product measurements and with them they can attain advantage over any $2$-bit protocol. Furthermore, it outperforms any protocol sending a five-valued classical message. These last two results motivate departing from the standard dense coding protocol to broader communication tasks.

These experiments represent a change of strategy: instead of making the measurement more complex to approach the ideal dense-coding capacity, one can design communication tasks whose optimal strategies naturally require only simple measurements.

\textit{High-dimensional product measurements.} In Ref.~\cite{Zhang2025}, the authors follow this approach and demonstrate experimentally, using the $(2,d)$ RAC with $8$-dimensional entangled states, that product measurements can achieve the optimal value. This experiment demonstrates the scalability of the protocol to higher-dimensional entangled states and shows that it can also be used to certify the dimensionality of entanglement through the Schmidt number. Since higher Schmidt numbers allow for higher success probabilities, the observed performance provides a witness of the dimensionality of the shared entanglement.

\section{Alternative semi-device-independent assumptions}\label{sec:non-dimension_constraints}

Dimensionally restricted communication is in the literature the most common avenue to studying PM scenarios. It is conceptually powerful, information-theoretically natural and has meaningful applications in the certification of quantum devices. However, there is also  strong motivation for studying PM scenarios beyond the dimension-assumption, i.e.~to understand  correlations when the communication is subject to other restrictions.  This is partly  motivated by conceptual interest and partly by practical considerations.  From a practical point of view, dimension assumptions have several noteworthy drawbacks. First, an upper bound on the Hilbert space dimension is not an observable. It can therefore not be operationally verified from uncharacterised PM statistics alone but instead justified through inspecting the experimental setup itself. Second, qubits are often only approximations of what is actually a more complicated quantum system. This creates a mismatch between the theoretical assumptions and the experimental reality. Third, in quantum cryptography, which is one of the main applications of PM scenarios, dimension-restricted systems are typically associated with single photons. But single photons usually cannot be generated at  high rates, thereby limiting the efficiency of the protocols based on them.

To address  both the conceptual and practical aspects of going beyond dimension restrictions, the last decade has seen the exploration of PM scenarios studied under a variety of communication assumptions. In this section, we introduce these different avenues for the PM scenario and the SDP methods used to analyse the associated quantum correlations.

\subsection{Overview of cryptographic applications}\label{sec:Crypto_overview}

For information security, randomness is a key prerequisite. Quantum systems are particularly well suited for cryptographic applications, due to the inherent randomness in quantum measurements and the impossibility of perfectly copying unknown quantum states. In general, to establish cryptographic security, one aims to bound the knowledge an adversary may hold about the data or secret key in question, as measured by an entropic quantity. For quantum protocols, such bounds can be established rigorously under the reasonable assumption that the adversary is constrained by the laws of quantum mechanics (or, in some cases, even just nonsignalling). 

Protocols aiming to certify randomness generally comprise three elements: i) a physical process generating raw data (from measurements on quantum states), ii) a method for lower-bounding the entropy of the raw data based on assumptions about the generation process and the capabilities of adversaries, and iii) a classical post-processing step (randomness extraction) compressing the raw data into a shorter output which is uniformly random to within a target tolerance. What is achieved by such protocols is commonly referred to as quantum random-number generation (QRNG), although randomness certification or randomness expansion may be more accurate terms, as some amount of seed randomness is usually required in the data generation and randomness extraction steps. See Refs.~\cite{Ma2016,Herrero-Collantes2017,Mannalath2023,Mahato2026} for reviews on QRNG.

Once high-quality randomness is available, it can be employed in more elaborate cryptographic applications. In particular, quantum key distribution (QKD) enables secure communication by establishing shared, secret keys between distant parties linked by untrusted channels. In addition to secure randomness, QKD protocols themselves also require measurements on quantum states to generate raw keys and classical post-processing (sifting, error correction, privacy amplification) in order to establish secure output keys. For QKD reviews, see e.g.\ \cite{Pirandola2020,Zapatero2023}.

For QRNG, QKD, and other cryptographic protocols, security can be established under different levels of trust in the devices used, from device-dependent protocols, where the implementation is fully characterised, to the device-independent (DI) setting, where devices are treated as black boxes. In general, the fewer assumptions that are made, the smaller is the risk that the real implementation deviates from the theoretical model, and hence the more robust is the security. However, full DI protocols require devices to demonstrate Bell nonlocality without postselection, which is technologically demanding. PM scenarios allow exploring a rich landscape of trade-offs between trust and ease of implementation in semi-DI regimes, in which devices are partially characterised, still enabling strong security but also fast and robust setups that can be realised with existing technology.

Most work on PM semi-DI protocols has focused on QRNG. Here, for a PM experiment producing a string of raw outputs $\bm{B}$ for inputs $\bm{X}$ and $\bm{Y}$, the key quantity of interest is the conditional smooth min-entropy $H_{\min}^\epsilon(\bm{B}|\bm{XYE})$, where $\bm{E}$ denotes the system of an adversary Eve that may hold (classical or quantum) correlations with the devices and $\epsilon$ is a small security parameter \cite{Tomamichel2011,Mahato2026}. Conceptually, this quantity characterises how unpredictable the raw outputs $\bm{B}$ remain to Eve given the inputs $\bm{X}$ and $\bm{Y}$ and her side information $\bm{E}$. After generation, the raw data are compressed by classical post-processing, known as randomness extraction, into a shorter string that is close to uniform and independent of Eve. Up to finite-size corrections, $H_{\min}^\epsilon(\bm{B}|\bm{XYE})$ (when using the base-2 logarithm) quantifies the number of almost-perfectly-uniform bits relative to Eve that can be extracted from $\bm{B}$, according to the quantum leftover hash lemma \cite{Tomamichel2011}.  

A wide range of different PM semi-DI settings have been explored for QRNG, corresponding to different assumptions placed on the preparation or measurements, other than restricting the dimension, as will be discussed in the following subsections. Some other PM semi-DI applications have also been developed, in particular for QKD. The broad array of measurement-device-independent (MDI) QKD protocols~\cite{Lo2012} can be understood as semi-DI with fully characterised states and untrusted measurements, and QKD protocols based on bounded state overlaps (Sec.~\hyperref[Sub:overlaps]{IV-E}) have also been demonstrated \cite{ioannou2022receiver}. Within the dimension-restricted framework, randomness certification has also been analysed for optical schemes using vacuum--single-photon preparations and homodyne or displacement-based measurements~\cite{Alves2026}. In the following sections, we will see how different physical assumptions on the communication and preparations constrain the possible correlations and thereby enable randomness to be certified.

\subsection{Dimensional deviations}\label{Sub:almost-dimension}

One of the drawbacks of the dimension assumption is that physical systems are rarely perfectly $d$-dimensional. Systems that are commonly modelled as qubits often contain small components outside the intended two-dimensional subspace. For example, polarisation qubits generated by spontaneous parametric down-conversion are not strictly confined to the single-photon polarisation space, since multi-photon emissions can occur. Similarly, weak coherent pulses, which are often used as qubit carriers, are fundamentally infinite-dimensional and only approximate qubits in the regime of low average photon number~(see Box~\ref{Box:coherent-almost-qubit} for an explicit example).

One way to account for dimension deviations is to relax the dimension assumption in a controlled way. Specifically, instead of requiring Alice's preparations to be exactly $d$-dimensional, one requires that almost all of their support lies in some common $d$-dimensional subspace. Following the terminology of Ref.~\cite{Pauwels2022}, such systems are called almost qudits. Let $\Pi_d$ denote the projector onto a $d$-dimensional subspace,
\begin{equation}
    \Pi_d = \sum_{j=1}^{d} \ketbra{j}{j}.
\end{equation}
We say that Alice's preparations, $\rho_x$, represent an almost $d$-dimensional system if
\begin{equation}\label{eq:almost_dimension}
    \tr(\rho_x \Pi_d) \geq 1-\epsilon_x,
\end{equation}
where $\epsilon_x\in[0,1]$ is the parameter that accounts for the dimensional deviation (we write $\epsilon$ when it is the same for all $x$). Its meaning is that all but $\epsilon$ of the states' support is on the $d$-dimensional subspace. Notably, the limiting case $\epsilon=0$ returns the usual dimension-restricted model, while setting $\epsilon=1$ renders the states entirely unconstrained. The relevant regime for physical systems that approximate a specific dimension is therefore  $0<\epsilon\ll 1$.

The correlations are still written in the usual PM form $p(b|x,y)=\tr(\rho_x M_{b|y})$, but now the measurements may act also on the part of the Hilbert space complementary to $\Pi_d$. Thus, if a small component of $\rho_x$ leaks outside the intended qudit subspace, Bob's measurement can in principle be designed to exploit that component. Due to the black-box model of Bob, this dimensional deviation cannot be treated as harmless noise but must instead be viewed as a resource available to enlarge the set of achievable correlations. 

\begin{examplebox}[label={Box:coherent-almost-qubit}]{Coherent state as an almost qubit}
A simple way to build intuition for Eq.~\eqref{eq:almost_dimension} is to look at a weak coherent state. Written in the Fock basis, it is
\begin{equation}\label{coherentstate}
    \ket{\alpha} =  e^{-|\alpha|^2/2} \sum_{n=0}^{\infty} \frac{\alpha^n}{\sqrt{n!}}\ket{n}.
\end{equation}
This state is not a qubit, since it has support on all photon-number states $\ket{n}$. However, if $|\alpha|$ is small, most of its weight lies in the vacuum and one-photon subspace. Let us therefore choose the intended qubit subspace as $\Pi_2=\ketbra{0}{0}+\ketbra{1}{1}$. The weight of $\ket{\alpha}$ inside this subspace is
\begin{equation}
    \tr\left(\ketbra{\alpha}{\alpha}\Pi_2\right) = e^{-|\alpha|^2}\left(1+|\alpha|^2\right).
\end{equation}
Hence, the deviation parameter is 
\begin{equation}
    \epsilon =1-e^{-|\alpha|^2}\left(1+|\alpha|^2\right)
\end{equation}
For small $|\alpha|$, this becomes $\epsilon =\frac{|\alpha|^4}{2}+O(|\alpha|^6)$. Thus, a weak coherent state is a good but not perfect approximation of an exact qubit. 
\end{examplebox}

Let us see this through the running example of the quantum random access code of Sec.~\hyperref[sec:QRAC]{II-B-3}. In the qubit version of the simplest  random access code [associated with $(n,d)=(2,2)$], the optimal success probability is $W_{\mathrm{RAC}}^Q=\tfrac{1}{2}(1+\tfrac{1}{\sqrt{2}})$ and the same correlations can be used for semi-DI QRNG. Under an exact qubit assumption, the optimal value certifies about $0.228$ bits of randomness for a suitable output~\cite{Li2011}. However, if Alice's preparations are only almost qubits, the conclusion changes significantly. Ref.~\cite{Pauwels2022} shows that already for a deviation of order $\epsilon=10^{-3}$, there are almost-qubit strategies compatible with the same random access code score but with only about $0.152$ bits of randomness. Thus, a $0.1\%$ deviation from the qubit assumption can cut the certified randomness by roughly a third.

A similar issue appears in the certification of non-projective measurements. Recall from Sec.~\hyperref[sec:self_test]{II-D-1} and Sec.~\hyperref[sec:Qviolations]{II-B-2} that qubit PM correlations can be used to certify measurements that cannot be reduced to projective measurements on a qubit. In the protocol of Ref.~\cite{Smania2020}, the goal is to certify a genuine four-outcome qubit measurement. Ref.~\cite{Pauwels2022} shows that an almost-qubit model with deviation around $5\times 10^{-4}$ can reproduce the experimentally relevant certificate using only a ternary-outcome measurement. With deviation around $3\times 10^{-3}$, ternary-outcome measurements can even exceed the exact-qubit quantum maximum of the witness. The lesson is not that such protocols fail, but that the dimension assumption must be stated with enough physical precision: ``qubit'' and ``almost qubit'' can lead to substantially different certification statements.

The central technical problem is therefore the following. Given a distribution $p(b|x,y)$ and deviation parameters $\{\epsilon_x\}_x$, how can we decide whether the observed distribution admits an almost $d$-dimensional quantum model? As with exact dimension bounds, a complete analytic characterisation is unrealistic in general. Instead, the dimension deviation condition can, with appropriate adjustments, be incorporated into the semidefinite relaxation methods introduced in Sec.~\hyperref[sec:boundingNV]{II-C-2}.

It is worth briefly recalling the basic recipe before adapting it. Starting from an operator list $L$, built from the identity, the states $\{\rho_x\}_x$ and Bob's measurements $\{M_{b|y}\}_{b,y}$, one assembles a monomial list $J$ by taking products of these operators up to some chosen length, which sets the relaxation level. The moment matrix is then defined entry-wise as $\Gamma_{u,v}=\tr(u^\dagger v)$ for $u,v\in J$, and is positive semidefinite by construction. The observed probabilities appear as the entries $p(b|x,y)=\Gamma_{\rho_x,M_{b|y}}$, while normalisation, projectivity and cyclicity of the trace impose linear identities among the remaining entries. The new ingredient is the almost-dimension constraint. Since the relevant $d$-dimensional subspace is not known in a device-independent way, the projector $\Pi_d$ cannot be inserted as a fixed operator. Instead, we introduce an auxiliary operator $\Pi$ in the relaxation. Its role is to represent the unknown intended subspace. We impose the algebraic relations $\Pi^2=\Pi$ and $\tr(\Pi)=d$, and the almost-dimension assumption becomes the linear constraint $\Gamma_{\rho_x,\Pi}\geq1-\epsilon_x$.

At the level of the relaxation, one can use the usual simplifications. Alice's preparations may be taken pure, so that $\rho_x^2=\rho_x$, and Bob's measurements may be taken projective, so that $M_{b|y}M_{b'|y}=\delta_{b,b'}M_{b|y}$. These are not extra assumptions about the laboratory implementation, but rather reduction rules used to simplify the moment matrix. With these ingredients in place, a basic feasibility relaxation is
\begin{align}\label{eq:almost_dimension_sdp}
\text{find } &\quad \Gamma \nonumber\\
\text{s.t. } 
&\quad \Gamma\succeq 0, \nonumber\\
&\quad \Gamma_{\mathds{1},\,\rho_x}=1, \quad \forall x, \nonumber\\
&\quad \Gamma_{\mathds{1},\,\Pi}=d, \nonumber\\
&\quad \Gamma_{\rho_x,\, \Pi}\geq 1-\epsilon,\quad \forall x, 
\end{align}
where the moment matrix is built from an operator list $L=\{\mathds{1},\Pi,\{\rho_x\}_x,\{M_{b|y}\}_{b,y}\}$ and a chosen list of monomials $J$, with $\Gamma_{u,v}=\tr(u^\dagger v)$ for $u,v\in J$. 

It is worth pausing on what this SDP actually does. It does not ask us to identify the physical deviation modes of the experiment, nor to write down an explicit Hilbert space model for the source. It only tests a necessary condition for the existence of some rank-$d$ subspace that captures at least $1-\epsilon_x$ of the support of each preparation. The auxiliary operator $\Pi$ is the mathematical stand-in for that subspace. This is why the method remains platform-independent: the same formulation applies to photons, trapped ions, superconducting systems, or any other implementation in which a dimension bound is only approximate.

\begin{examplebox}[label={Box:almost-dimension-RAC}]{Dimension deviation as a resource}
As an example, we follow Ref.~\cite{Pauwels2022} and consider the random access code but implemented with almost qubits instead of perfect qubits. Let Alice's four almost qubit preparations be
\begin{align}
    \ket{\phi_{00}} &= \sqrt{1-\epsilon}\ket{0}+\sqrt{\epsilon}\ket{2}, \nonumber\\
    \ket{\phi_{11}} &= \sqrt{1-\epsilon}\ket{1}+\sqrt{\epsilon}\ket{3}, \nonumber\\
    \ket{\phi_{10}} &= \sqrt{1-\epsilon}\,s_{01}^{+}-\sqrt{\epsilon}\,s_{23}^{+}, \nonumber\\
    \ket{\phi_{01}} &= \sqrt{1-\epsilon}\,s_{01}^{-}-\sqrt{\epsilon}\,s_{23}^{-},
\end{align}
where we have defined $s_{ij}^{\pm}:=\frac{1}{\sqrt{2}}(\ket{i}\pm\ket{j})$.  The first term in each of these states corresponds to the qubit state used in the ordinary quantum random access code [see Eq.~\eqref{RACvectors}]. The second term in each state is a coherent contribution in a complementary space  spanned by $\ket{2},\ket{3}$. Thus, these two features are coherently superposed in each of the above states. Since the outcomes of Bob are binary the optimal average success probability can be evaluated analytically from these states. It is
\begin{equation}
W_{\mathrm{RAC}}(\epsilon) = \frac{1}{2} + \frac{1}{4} \left( \sqrt{1+h(\epsilon)}+\sqrt{1-h(\epsilon)} \right),
\end{equation}
where $h(\epsilon)=(1-2\epsilon)\sqrt{1+4\epsilon-4\epsilon^2}$. For small $\epsilon$, this gives
\begin{equation}
    W_{\mathrm{RAC}}(\epsilon)= W_{\mathrm{RAC}}^Q+\frac{\epsilon}{\sqrt{2}}+O(\epsilon^2).
\end{equation}
We have obtained the maximal value achievable with perfect qubits and correction terms that depend on $\epsilon$. Since the correction is relevant to first order, we expect a small dimensional deviation to enable significantly stronger quantum correlations.
\end{examplebox}

The dimension deviation SDP also gives a useful new perspective on exact dimension bounds. Setting all $\epsilon_x=0$ recovers a relaxation for standard $d$-dimensional PM correlations. Compared with the sampling-based relaxations discussed in Sec.~\hyperref[sec:boundingNV]{II-C-2}, this approach is deterministic and its matrix size does not directly grow with $d$. This makes it attractive for intermediate and large dimensions, where explicitly sampling all rank combinations of measurements becomes costly. However, one should keep an important limitation in mind: the hierarchy is an outer relaxation and is not known to converge in general to the exact dimension-$d$ quantum set. The reason is that the moment-matrix construction constrains averaged quantities, such as $\sum_\lambda q(\lambda)\tr(\rho_x^{(\lambda)}\Pi)$, and therefore may fail to rule out shared-randomness decompositions in which some components use dimensions larger than $d$, while the average still respects the constraint. This caveat is closely related to the random-access-code example in Box~\ref{Box:almost-dimension-RAC}, although the interpretation is different. In the box, the out-of-subspace component is coherently present in each preparation, whereas here the issue is averaging over different-dimensional components. Hence infeasibility of the SDP is a rigorous certificate of impossibility, but feasibility should not be read as an explicit physical realisation with a fixed Hilbert-space dimension.

\subsection{Energy constraints}\label{Sub:energy-restrictions}

While the almost-dimension restriction can be seen as a correction to remedy the shortcomings of exact dimension-restrictions, an interesting complementary avenue is to depart entirely from dimension-based ideas and instead focus on communication restrictions that are tailored for practical verification. One approach to this is to consider energetic restrictions, which are particularly natural in optical implementations of PM scenarios. This idea was first proposed in Ref.~\cite{VanHimbeeck2017semidevice}, which formulated the restriction more generally as a bound on the mean value of a chosen observable. When this observable is the Hamiltonian with a non-degenerate ground state and a finite gap, the bound constrains the weight outside the ground state. For the optical mode considered below, that ground state is the vacuum. This is relevant because it allows Alice to encode her data into quantum states that have infinite dimension, such as the weak coherent states defined in Eq.~\eqref{coherentstate}, but insist only that their non-vacuum component remains small. This is an observable quantity that can readily be checked to verify the accuracy of the assumption.   

To formalise this, consider a single optical mode with Fock basis $\{\ket{n}\}_{n=0}^{\infty}$, where $\ket{0}$ is the vacuum, and let $\sigma_0=\ketbra{0}$ be the projector onto it. For an optical mode of angular frequency $\Omega$, each photon carries energy $\hbar\Omega$ above the vacuum, so an upper bound on the mean energy also bounds the mean photon number. More generally, for a Hamiltonian with a non-degenerate ground state and gap $\Delta>0$, the probability of occupying an excited state is no larger than the mean energy above the ground state divided by $\Delta$. The probability that $\rho_x$ is not in the vacuum is $1-\bracket{0}{\rho_x}{0}$. The vacuum-component assumption is that this probability is bounded, or equivalently that
\begin{equation}\label{eq:vacuum_component_assumption}
    \bracket{0}{\rho_x}{0}\geq 1-\omega_x,
    \qquad \forall x.
\end{equation}
Thus Alice's $x$th preparation has non-vacuum probability at most $\omega_x$. Formally, Eq.~\eqref{eq:vacuum_component_assumption} resembles the dimension-deviation condition specialised to $d=1$ and $\Pi_1=\sigma_0$, but the physical interpretation is different. In Sec.~\hyperref[Sub:almost-dimension]{IV-B}, the intended low-dimensional subspace was where the useful information was encoded, and the deviation $\epsilon$ was an unwanted leakage. Here it is the opposite as the vacuum carries no information about $x$, so the message is encoded precisely in the small non-vacuum component. The parameter $\omega_x$ controls how distinguishable the preparations can be in the first place. A bound on the vacuum component therefore does not say that the message is a qubit, nor that the Hilbert space has been truncated. The state may have support on arbitrarily high photon-number states; the assumption only says that the total probability of leaving the vacuum is small.

Because Alice and Bob may use shared randomness, the correlations can take the form $p(b|x,y)=\sum_\lambda q(\lambda)\tr(\rho_x^{(\lambda)}M_{b|y}^{(\lambda)})$. Measured averages over many runs cannot verify a separate bound for every unobserved value of $\lambda$ without further assumptions about the source. The variant of the vacuum assumption that is most natural for verification is to consider it on average over the shared randomness, 
\begin{equation}\label{eq:max_average_vacuum}
\sum_\lambda q(\lambda)\tr\left[\rho_x^{(\lambda)}(\mathds{1}-\sigma_0)\right] \leq \omega_x.
\end{equation}
Equivalently, introducing the average preparation $\rho_x=\sum_\lambda q(\lambda)\rho_x^{(\lambda)}$, Eq.~\eqref{eq:max_average_vacuum} is nothing but the single-state assumption~\eqref{eq:vacuum_component_assumption} imposed on $\rho_x$, i.e.~$\bracket{0}{\rho_x}{0}\geq 1-\omega_x$. It is therefore weaker than requiring that every hidden-variable component $\rho_x^{(\lambda)}$ separately remain close to the vacuum, since it allows the source to occasionally emit a brighter pulse provided this is compensated by many near-vacuum emissions. Crucially, the constrained quantity is directly accessible in the lab. The non-vacuum weight is upper bounded by the mean photon number, $\tr[\rho_x(\mathds{1}-\sigma_0)]=\Pr[n\geq 1]\leq \sum_{n\geq 1} n\,p_n=\langle n\rangle_x$, and a calibrated photon-rate measurement can be used to bound this mean photon number. A power meter can instead bound the average optical energy per pulse if the pulse rate is known. If the signal frequency is known or bounded from below, that energy bound gives an upper bound on $\langle n_x\rangle$ and hence certifies Eq.~\eqref{eq:max_average_vacuum}. In the rest of this subsection, we use this average-level formulation. We restrict our attention to the vacuum component for now. Constraints on individual photon-number components, or on the (truncated) mean photon number, are also possible and we return to them later in Sec.~\hyperref[Sub:overlaps]{IV-E}.

Let us now see what the vacuum constraint implies in the simplest PM scenario. Alice has two possible preparations, $x\in\{1,2\}$, and Bob has no input and a binary output $b\in\{\pm1\}$. We write Bob's observable as $M=M_{+1}-M_{-1}$, with $-\iden \preceq M\preceq \mathds{1}$, and define the two expectation values $E_x=\tr(M\rho_x)$ for $x=1,2$. The quantity $E_-:=E_1-E_2$ measures how well Bob can distinguish Alice's two preparations. Indeed, $\frac{1}{2}|E_-|$ is the total variation distance between Bob's two output distributions, and the optimal probability of guessing Alice's input from Bob's output is $p_{\mathrm{guess}} = \frac{1}{2}\left(1+\tfrac{|E_-|}{2}\right)$. Thus, the vacuum constraint should limit how large $|E_-|$ can become. We derive this limit, together with the full set of quantum-achievable pairs $(E_1,E_2)$, in Box~\ref{Box:vacuum-two-state}.

In this two-preparation scenario one may, without loss of generality, take the preparations to be pure and supported on a single qubit, spanned by the vacuum and one excited direction~\cite{VanHimbeeck2017semidevice}.

\begin{examplebox}[label={Box:vacuum-two-state}]{Two states close to the vacuum}
The derivation of the quantum region in this simplest scenario has three steps: \emph{(i)} two states close to the vacuum cannot be too far apart, so their overlap is bounded from below by the vacuum weights; \emph{(ii)} the achievable pairs $(E_1,E_2)$ for two pure states with given overlap form a known region; \emph{(iii)} composing the two facts gives the vacuum-constrained quantum region.
\smallskip

\emph{Step (i).} Consider two pure states $\phi_1=\ketbra{\phi_1}{\phi_1}$ and $\phi_2=\ketbra{\phi_2}{\phi_2}$ satisfying
    \begin{equation}
    \bracket{0}{\phi_x}{0}\geq 1-\omega_x \:\: \text{for} \:\: x =1,2.
\end{equation}
Assume for simplicity that $\omega_1+\omega_2\leq1$, so that the two states cannot be made orthogonal. Writing $\ket{\phi_x}=\sqrt{1-\beta_x}\ket{0}+\sqrt{\beta_x}\ket{\xi_x}$ with $\ket{\xi_x}\perp\ket{0}$ and $\beta_x\leq\omega_x$, we obtain
\begin{equation}
|\braket{\phi_1}{\phi_2}|\geq\sqrt{1-\omega_1}\sqrt{1-\omega_2}-\sqrt{\omega_1\omega_2}=:h(\omega_1,\omega_2).
\end{equation}
\emph{Step (ii).} For two pure states with fixed overlap $\gamma=|\braket{\phi_1}{\phi_2}|$, it is a standard result that the set of expectation pairs $(E_1,E_2)$ that an arbitrary binary POVM can produce is exactly characterised by $g(E_1,E_2)\geq \gamma$, with
\begin{equation}\label{eq:gE1E2}
 g(E_1,E_2)=\frac{1}{2}\left(\sqrt{1+E_1}\sqrt{1+E_2}+\sqrt{1-E_1}\sqrt{1-E_2}\right).
\end{equation}
\emph{Step (iii).} Combining steps \emph{(i)} and \emph{(ii)} gives the vacuum-constrained quantum region
\begin{equation}\label{eq:vacuum_quantum_region}
    g(E_1,E_2)\geq h(\omega_1,\omega_2),
\end{equation}
with $|E_1|,|E_2|\leq1$. This is a complete characterisation in the two-preparation, binary-output scenario~\cite{VanHimbeeck2017semidevice}. Geometrically, for $\omega_1=\omega_2=\omega$, the region shrinks towards the diagonal $E_1=E_2$ as $\omega\to0$---vacuum-dominated preparations are nearly indistinguishable---and grows to fill the full square as $\omega\to1/2$, when the constraint ceases to restrict these correlations.

\smallskip
The maximal distinguishability allowed by the vacuum bounds is
\begin{equation}\label{eq:vacuum_quantum_Eminus}
|E_-|_{\mathrm{Q}} = 2\left( \sqrt{\omega_1(1-\omega_2)}+ \sqrt{\omega_2(1-\omega_1)} \right).
\end{equation}
For symmetric bounds $\omega_1=\omega_2=\omega$, this becomes $|E_-|_{\mathrm{Q}}=4\sqrt{\omega(1-\omega)}$. Thus, even very weak optical pulses can produce a sizeable distinction between the two preparations, scaling as $\sqrt{\omega}$ for small $\omega$.
\end{examplebox}

This example should be compared with a deterministic, or ``classical'', explanation of the same binary-output experiment. Here classical does not mean that Alice sends a classical dit of fixed dimension. Rather, it means that once the shared variable $\lambda$ is fixed, Bob's output is already predetermined for both possible preparations. In other words, each hidden-variable realisation has $E^{(\lambda)}_1,E^{(\lambda)}_2\in\{\pm1\}$. Under the average-level vacuum constraint these deterministic strategies satisfy
\begin{equation}\label{eq:vacuum_classical_bound}
    |E_1-E_2|\leq 2(\omega_1+\omega_2).
\end{equation}
For $\omega_1=\omega_2=\omega$, this is $|E_-|\leq4\omega$. Comparing this with Eq.~\eqref{eq:vacuum_quantum_Eminus}, we see that for $0<\omega<\tfrac12$ the quantum value can be larger than the deterministic one. This gives a vacuum-component analogue of a Bell or dimension-witness inequality, i.e.~observing a value of $|E_-|$ above Eq.~\eqref{eq:vacuum_classical_bound} proves that Bob's outcome could not have been fixed in advance for both inputs.

\begin{figure}[t]
    \centering
    \includegraphics{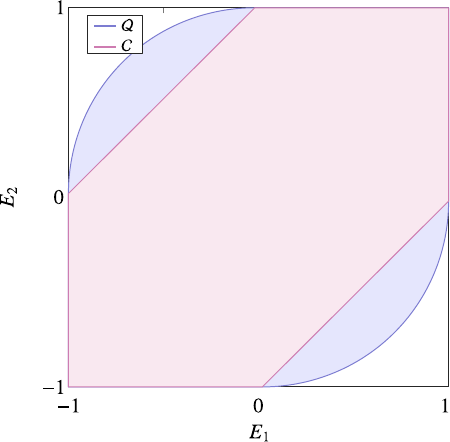}
    \caption{\emph{Correlation space under a vacuum-component constraint.} In the two-preparation, binary-output scenario, each behaviour is represented by a point $(E_1,E_2)$. The quantum region $\mathcal Q$ is larger than the deterministic, or classical, region $\mathcal C$ allowed under the max-average assumption. The transverse direction $E_-=E_1-E_2$ quantifies how well Bob's output distinguishes Alice's two preparations. For the illustrative choice $\omega_1=\omega_2=0.15$, deterministic behaviours obey $|E_-|\leq0.6$, while quantum behaviours can reach $|E_-|\simeq1.43$. Adapted from Ref.~\cite{VanHimbeeck2017semidevice}.}
    \label{F:vacuum-correlation-space}
\end{figure}
Figure~\ref{F:vacuum-correlation-space} makes the separation especially transparent. The classical region is cut off by a straight bound on $E_-$, while the quantum region has a curved boundary inherited from the minimum overlap between two states that are both close to the vacuum. Moving away from the diagonal $E_1=E_2$ therefore corresponds to increasing distinguishability between Alice's preparations. The vacuum constraint limits this distinguishability, but quantum measurements can still exploit the allowed non-vacuum component more efficiently than any deterministic strategy.

To connect these formulas with optical implementations, we recall two schemes analysed in Ref.~\cite{VanHimbeeck2017semidevice}. These examples are particularly useful in the present context for two reasons: they rely on standard components from quantum optics, and in the first scheme, the observed correlations leave the deterministic region over a relevant range of optical amplitudes. Consequently, we can say that they provide simple experimental realisations of the semi-device-independent setting.
\begin{enumerate}
    \item \emph{Binary phase-shift keying (BPSK).} Alice prepares one of the coherent states $\ket{+\alpha}$ or $\ket{-\alpha}$, with $\alpha>0$, and Bob measures the sign of the $X$ quadrature by homodyne detection. The non-vacuum component of both preparations is $\omega=1-e^{-|\alpha|^2}$, and the resulting correlations are $E_1=\operatorname{erf}(\sqrt{2}\alpha)$ and $E_2=-\operatorname{erf}(\sqrt{2}\alpha)$. For suitable $\alpha$, the observed point lies outside the deterministic region of Eq.~\eqref{eq:vacuum_classical_bound} and therefore certifies genuinely random behaviour.
 
    \item \emph{On-off keying.} Alice sends either a weak coherent state $\ket{\alpha}$ or the vacuum $\ket{0}$, and Bob uses a photon detector that outputs $+1$ when it clicks and $-1$ otherwise. Ideally $E_1=1-2e^{-|\alpha|^2}$ and $E_2=-1$; with detector efficiency $\eta$, the first value becomes $E_1=1-2e^{-\eta|\alpha|^2}$, while $E_2=-1$ is unaffected. This implementation is extremely simple and is useful for building intuition, but under the average-level vacuum assumption it lies on the boundary of the deterministic region rather than giving the clean separation obtained by BPSK.
\end{enumerate}

Both implementations illustrate the practical appeal of the vacuum-component framework: simple optical hardware, a single physically meaningful source parameter ($\omega$ or $|\alpha|^2$), and, for BPSK, a robust route to certified randomness. Semi-DI QRNG based on on-off keying  with interference between the signal and a local oscillator before photon detection and an energy bound was demonstrated experimentally in Ref.~\cite{Rusca2019}, achieving an asymptotically extractable secure bit rate of 1.25 Mbits/s.

The discussion so far has focused only on the vacuum probability. More generally, one may have access to additional information about the photon-number statistics of the source, for example bounds on the one- or two-photon components, or on relevant moments of the photon-number distribution. Such constraints provide a more detailed characterisation of the source while following the same underlying principle, namely to constrain experimentally accessible properties of its occupation in the Fock basis without imposing a restriction on the dimension of the communicated system. In particular, restricting the weight outside a finite set of photon-number states connects naturally with the dimension-deviation framework discussed in the previous subsection.

These more general photon-number constraints can be incorporated into semidefinite relaxations of the corresponding correlation set (see Ref.~\cite{Roch2025} for a systematic treatment). Using more detailed information about the source can tighten the resulting bounds and improve the performance of tasks such as state discrimination and semi-device-independent randomness certification. In coherent-state protocols, one can likewise make better use of the measurement data by retaining a finer-grained binning of the detector outcomes rather than reducing them immediately to a binary variable. Importantly, these refinements affect the analysis and classical processing, without changing the basic physical assumption underlying the protocol.

The vacuum-component constraint can also be used in the entanglement-assisted PM scenarios of Sec.~\ref{sec:EAPM}. Alice and Bob then share an entangled state, while the bound in Eq.~\eqref{eq:vacuum_component_assumption} applies to the marginal state of the message Alice sends. Bob may measure this message together with his share of the entangled state. Under this restriction, entanglement gives no advantage for some standard classical communication tasks, but it can improve probabilistic bit transmission with quantum messages when Alice uses non-unitary encodings~\cite{RochiCarceller_2026}. Entanglement can also enlarge the set of correlations compatible with the same energy bound and reduce the randomness certified by an analysis that allows only shared randomness~\cite{davino2026entanglementenergyconstrainedprepareandmeasurescenario}. Thus, the randomness bounds discussed above need to be revised if the devices may share entanglement. 

\subsection{Fidelity constraints}\label{Sub:fidelity-constraints}
A different avenue to semi-device-independence adopts the idea that Alice aims to prepare certain target states. In practice, however, her lab states only approximately correspond to the target states. This degree of approximation can be quantified by, for instance, the fidelity between the target state and the lab state. In particular, while the target state is fixed and therefore has a specific dimension, the lab state can potentially have an unbounded dimension. This perspective is experimentally motivated, as the experimenter only needs to estimate the fidelity of her preparation in order to proceed with the analysis, without having to claim that the source has a known dimension. This leads to a framework in which the degree of distrust, i.e.~the infidelity in the preparation device, can be used to benchmark semi-device-independent protocols~\cite{Tavakoli2021b}.

Let Alice's target states be denoted by $\{\ket{\psi_x}\}_x$, which we take to be $d$-dimensional. In an ideal implementation, input $x$ would make Alice prepare the pure state $\psi_x=\ketbra{\psi_x}{\psi_x}$. In the actual experiment, however, the prepared state is some unknown density matrix $\rho_x$ acting on a Hilbert space of some unknown but countable dimension $D\geq d$. The target states are viewed as embedded in the larger $D$-dimensional space via $\ket{\psi_x}\oplus\mathbf{0}$, so that the fidelity with the target state is
\begin{equation}\label{eq:fidelity_constraint_fidelity}
    F_x \equiv \bracket{\psi_x}{\rho_x}{\psi_x}.
\end{equation}
The fidelity-constrained assumption is that these fidelities obey
\begin{equation}\label{eq:fidelity_constraint}
    F_x \geq 1-\delta_x, \quad \forall x
\end{equation}
where $\delta_x\in[0,1]$ is the distrust associated with the preparation of input $x$. The case $\delta_x=0$ corresponds to perfect trust in that preparation, while larger values of $\delta_x$ allow the lab state to deviate more strongly from the target state.

The operational meaning of Eq.~\eqref{eq:fidelity_constraint} is simple. The fidelity $F_x$ is the probability that the two-outcome test $\{\psi_x,\mathds{1}-\psi_x\}$ accepts the state $\rho_x$. Concretely, an upper bound on $\delta_x$ can be obtained by running, on a sub-sample of preparations, a trusted calibration measurement that includes $\psi_x$ as one outcome and counting the fraction of ``fail'' events. This is the calibration step that turns the fidelity assumption into something an experimentalist can estimate. It is also one of the main reasons fidelity constraints are appealing: the assumption is tied to an experimentally observable quality of the source. It is worth noting that this is different from the other two semi-device-independent parameters introduced earlier in this section. The parameter $\epsilon_x$ in Sec.~\hyperref[Sub:almost-dimension]{IV-B} measures uncontrolled leakage out of the intended subspace, while the parameter $\omega_x$ in Sec.~\hyperref[Sub:energy-restrictions]{IV-C} measures the non-vacuum component of a state. The parameter $\delta_x$ above instead estimates the quality of the source with respect to a calibrated reference. Notice that if we were to choose all target states as the vacuum, namely $\ket{\psi_x}=\ket{0}$, then the fidelity assumption reduces to the vacuum-component assumption of Eq.~\eqref{eq:vacuum_component_assumption}. In this sense, the vacuum-component framework is a special case of the fidelity-constrained framework, though the two are motivated by quite different physical pictures.

When shared randomness is allowed, the correlations take the usual convex form [Eq.~\eqref{quantumModel}]. The fidelity constraint is imposed on the average state $\rho_x=\sum_\lambda q(\lambda)\rho_x^{(\lambda)}$, which matches what an experimenter can estimate from repeated tests of the source. As in other semi-device-independent assumptions, one must also assume that no side-channel is used to communicate $x$ outside the tested degrees of freedom.

How does one bound correlations under this assumption? Since the constraint concerns Alice's source while Bob's measurements remain uncharacterised, the correlations without shared randomness are $p(b|x,y)=\tr(\rho_xM_{b|y})$ subject only to Eq.~\eqref{eq:fidelity_constraint}. As for most quantum correlation sets in the PM scenario, this one is hard to characterise. A useful modern way of providing outer bounds on the set is to use the block-moment-matrix relaxations of Ref.~\cite{dalessandro2026semidefiniteblockmatrixrelaxationscomputing}. While it is also possible to use sampling-based SDP relaxation heuristics~\cite{Tavakoli2021b} similar to those discussed in Sec.~\hyperref[sec:boundingNV]{II-C-2}, the block-moment-matrix approach has several advantages: it does not rely on heuristics, does not need to separately treat different rank-combinations for Bob's measurements, requires no extra assumptions, and its computational cost is independent of the lab dimension $D$.

The idea is to keep the trusted part of the Hilbert space explicit and to compress everything else into a single scalar variable. Let $\mathcal{H}_{\mathrm{tar}}=\operatorname{span}\{\ket{\psi_x}\}_x$ be the trusted target subspace, of dimension $d_\psi\leq |X|$. The completely positive map
\begin{equation}\label{eq:fidelity_bmm_map}
    \Theta=\mathrm{id}_{d_\psi}\oplus \tr_{\perp}
\end{equation}
keeps the part of an operator acting on $\mathcal{H}_{\mathrm{tar}}$ and replaces the part acting on the orthogonal complement by its trace. The operator list is $L=\{\rho_x,M_{b|y}\}$. From products of these operators one builds a monomial list $J$, and then a block moment matrix whose entries are
\begin{equation}
\Gamma_{u,v}=\Theta(u^\dagger v) \quad \text{for } u,v\in J.
\end{equation}
Each entry $\Gamma_{u,v}$ is itself a $(d_\psi+1)\times(d_\psi+1)$ matrix: a $d_\psi\times d_\psi$ block acting on $\mathcal{H}_{\mathrm{tar}}$ direct-summed with a scalar storing the trace of the part orthogonal to it. The whole $\Gamma$ is positive semidefinite by construction.

The probabilities and fidelities appear as simple linear functions of these blocks. At the lowest non-trivial level, the three constraints that carry the physics are
\begin{equation}\label{eq:fidelity_bmm_constraints}
\begin{aligned}
    p(b|x,y) &= \tr\,\Gamma_{\rho_x,M_{b|y}},\\
    \bracket{\psi_x}{\Gamma_{\mathds{1},\rho_x}}{\psi_x} &\geq 1-\delta_x,\\
    \Gamma_{\mathds{1},\rho_x} &\succeq \Gamma_{\rho_x,\rho_x}.
\end{aligned}
\end{equation}
The first line identifies the observed correlations as the trace of a single block. The second line is the fidelity assumption, expressed as a linear functional on the $d_\psi\times d_\psi$ part of $\Gamma_{\mathds{1},\rho_x}$. The third line is the relaxation-level version of the operator inequality $\rho_x\succeq\rho_x^2$, which allows the preparations to be mixed. For the last point, unlike in the dimension-bounded scenario, one should not simply assume that the preparations are pure. Mixed states may be relevant once the source is constrained by fidelities rather than by an exact Hilbert-space dimension~\cite{dalessandro2026semidefiniteblockmatrixrelaxationscomputing}.

The remaining SDP constraints are the familiar ones: $\Gamma\succeq0$, normalisation of states and measurements, projectivity relations for Bob's measurements, and the known relations among the target states inside $\mathcal{H}_{\mathrm{tar}}$. Since the dimension outside the target subspace is uncharacterised, Bob's POVMs may be treated as projective without loss of generality by the usual dilation argument. Shared randomness is also included automatically: if each hidden-variable branch yields a feasible block moment matrix, then their convex mixture is feasible as well. Increasing the monomial list $J$ gives stronger outer approximations. As usual for moment-matrix methods, infeasibility is a rigorous certificate that no quantum realisation satisfies the prescribed fidelity bounds, while feasibility should be read only as an outer test.

For lower bounds on a linear witness [Eq.~\eqref{linearWitness}], one can still use the same alternating convex search methods as in Sec.~\hyperref[sec:seesaw]{II-B-5}. With fixed measurements, optimising the states subject to the fidelity constraints is an SDP. With fixed states, optimising the measurements is again an SDP. Alternating the two steps gives explicit quantum strategies and therefore lower bounds on the achievable value of the witness. The block moment matrix relaxation gives the complementary upper bounds.

There is a useful way to picture what the SDP is doing. The target states define the directions in Hilbert space that Alice claims to be preparing. The lab states are allowed to move away from those directions, but only inside fidelity balls whose radii are controlled by the parameters $\delta_x$. The SDP asks how well uncharacterised measurements can exploit that freedom. This is quite different from a dimension assumption: instead of restricting the size of the message space, we restrict how far the preparations can move away from a calibrated target ensemble.

\begin{examplebox}[label={Box:fidelity-state-discrimination}]{State discrimination with fidelity constraints}
The simplest way to see the effect of fidelity constraints is to revisit two-state discrimination. Let Alice have two target qubit states with Bloch vectors $\vec n_1=(0,0,1)$ and $\vec n_2=(\sin\theta,0,\cos\theta)$, where $\theta\in[0,\pi]$. Bob has no input and wants to guess which state Alice prepared. The average success probability is $W_{\mathrm{sd}}=\frac{1}{2}p(1|1)+\frac{1}{2}p(2|2)$. When $\delta=0$, Alice's states are fixed and this is the usual Helstrom problem.

\medskip

Geometrically, the fidelity constraint confines each lab state to a spherical cap of half-angle $\alpha=\arccos(1-2\delta)$ around its target. Inside the cap, the lab state is free to move---and the worst-case strategy uses that freedom in the most adversarial way: it pushes the two lab states away from each other, opening the angular gap between them from $\theta$ to $\theta+2\alpha$. The Helstrom bound for this enlarged gap then gives the answer. The expression is valid as long as the two caps do not overlap antipodally; once they do, the lab states can be made orthogonal and Bob recovers Alice's input with certainty.

\medskip

Concretely, let the actual pure states have Bloch vectors $\vec m_x=(\sin\nu_x,0,\cos\nu_x)$. The fidelity constraints $\bracket{\psi_x}{\rho_x}{\psi_x}\geq 1-\delta$ become $\cos\nu_1\geq 1-2\delta$ and $\cos(\theta-\nu_2)\geq 1-2\delta$. The optimal choice (when allowed) is $\nu_1=-\alpha$ and $\nu_2=\theta+\alpha$, with $\alpha=\arccos(1-2\delta)$. For two pure states separated by an angle $\Delta$, the Helstrom success probability is
\begin{equation}
    W_{\mathrm{sd}}=\frac{1}{2}\left(1+\sin\frac{\Delta}{2}\right).
\end{equation}
Using $\Delta=\theta+2\alpha$, we obtain
\begin{equation}\label{eq:fidelity_state_discrimination}
   W_{\mathrm{sd}}^Q = \frac{1}{2}\left(1+\sin\frac{\theta}{2}\right) + \sqrt{\delta(1-\delta)}\cos\frac{\theta}{2}-\delta\sin\frac{\theta}{2}.
\end{equation}
This holds whenever $\delta\leq \frac{1}{2}\left(1-\sin\frac{\theta}{2}\right)$; for larger $\delta$, the lab states can be made orthogonal while still satisfying the fidelity constraints, and $W_{\mathrm{sd}}^Q=1$.

\medskip

The takeaway is simple: bounded distrust turns the textbook Helstrom bound into a family of bounds parametrised by the allowed source imperfection. The more freedom Alice's actual preparations have, the easier Bob's discrimination task can become.
\end{examplebox}
Since the source is still described relative to quantum target states, the most natural ``classical'' comparison is not to make Alice classical, but to make Bob's measurement device classical. This means that, in each hidden-variable branch, all of Bob's measurements are diagonal in a common basis. Equivalently, Bob first performs one fixed projective measurement and then classically post-processes its outcome depending on $y$. The resulting correlations take the form
\begin{equation}\label{eq:fidelity_classical_measurement}
    p(b|x,y)=\sum_\lambda q(\lambda)\sum_k p(b|y,k,\lambda)
    \bracket{e_k^{(\lambda)}}{\rho_x^{(\lambda)}}{e_k^{(\lambda)}}.
\end{equation}
If the observed correlations cannot be written in this form, then Bob's measurement device cannot be interpreted as a classical post-processing of a single underlying measurement. In the BMM relaxation, this classical-measurement restriction is implemented by adding commutation relations among Bob's measurement operators.

To see how such a quantum-over-classical separation can arise, consider the minimal scenario that already admits it: Alice has three preparations and Bob has two binary measurements. This is the smallest setting in which Bob's measurement choice is non-trivial and in which neither of his measurements can be viewed as a post-processing of the other. Writing $E_{xy}=p(0|x,y)-p(1|x,y)$, consider the linear witness\footnote{The same witness is written as $-E_{32}$ instead of $-E_{31}$ in Ref.~\cite{Tavakoli2021b}. With the Bloch vectors given below, only the version stated here achieves the quoted optimal quantum value $1+2\sqrt 2$.}
\begin{equation}\label{eq:fidelity_W322}
    W_{322}=E_{11}+E_{12}+E_{21}-E_{22}-E_{31}.
\end{equation}
The target states are three qubit states in the $XZ$-plane, with Bloch vectors
\begin{equation}
    \vec n_1=(0,0,1), \:\:\:
    \vec n_2=(1,0,0), \:\:\:
    \vec n_3=\frac{1}{\sqrt{2}}(-1,0,-1).
\end{equation}
For perfect preparations, the optimal quantum value is $W_{322}^Q=1+2\sqrt{2}$. Under uniform distrust, $\delta_x=\delta$, the classical-measurement bound catches up with this ideal quantum value once $\delta$ is roughly $3.1\%$. The quantum value remains strictly above the classical-measurement bound up to a distrust of about $33\%$. The first threshold tells us how accurate the source must be if the ideal quantum value alone is to certify a nonclassical measurement device. The second threshold tells us when the fidelity assumption has become too weak for this witness to distinguish quantum measurements from classical post-processing. These thresholds are numerical values from Ref.~\cite{Tavakoli2021b}; they should not be read as closed-form constants.

As a final takeaway, fidelity constraints provide a semi-device-independent assumption tailored to the actual states one aims to prepare. They do not replace dimension bounds in every application, but they are often closer to what is experimentally calibrated. Formally, the vacuum-component assumption of the previous subsection can be viewed as the special case in which all target states are the vacuum. Operationally, however, the interpretation is different: for vacuum constraints the useful information is carried by the small non-vacuum component, whereas for fidelity constraints the useful information is carried by staying close to a calibrated target ensemble. In the next subsection, we turn to another way of constraining the source, based not on fidelities with target states but on the overlaps between the preparations themselves.

\subsection{Overlap constraints}\label{Sub:overlaps}

A closely related restriction to the fidelity constraints reviewed above is the constraint on the inner products of the preparations. In this case, instead of aiming for a specific target, Alice's prepared states $\{\ket{\psi}_x\}_x$ are only required to satisfy $|\braket{\psi_x}{\psi_{x'}}| =  |G_{xx'}| \geq g_{xx'}$, where the matrix formed by the entries $G_{xx'}$ is known as the Gram matrix of the prepared states (also introduced in Eq.~\eqref{GramMatrix}), and $g_{xx'}$ is a lower bound on the overlap between the states $\ket{\psi_x}$ and $\ket{\psi_{x'}}$.  This overlap information characterises the non-orthogonality of the encodings, which  limits how well they can be distinguished. Constraining the overlap provides an experimentally robust alternative that bounds the set of quantum-realisable correlations without necessitating a strict assumption on the dimension of the underlying Hilbert space. While overlap constraints are often introduced in standard bipartite prepare-and-measure scenarios involving a single sender and a single receiver~\cite{Brask2017,Shi_2019}, the Gram-matrix formalism can also be generalised to more complex network configurations. In particular, it enables the construction of semidefinite relaxation hierarchies for multipartite prepare-and-measure networks, such as broadcasting scenarios where a single encoding is distributed to several independent receivers~\cite{wang2019characterising}.

The central goal of a PM quantum network is to send some classical message over a quantum network to a group of receivers. To encode information in a quantum system, an initial input $x$ dictates the preparation of a specific pure state from a predefined set of $n$ options: $\{|\psi_x\rangle\}_{x=1}^n$.  The quantum systems go through a broadcasting channel described by a general isometry $U$, resulting in a set of $k$-partite states  $\{|\phi_x\rangle\}_{x=1}^n$. Note that the state $\ket{\phi_x}$ is a state of, in principle, arbitrary dimension dictated by the  application of $U$ to $\ket{\psi_x}$. To decode the sent information, receivers perform a specific selection of measurements on the transmitted quantum signal $|\phi_x\rangle$. In a standard PM network with $k$ different receivers in separate locations, the expression $p(b_1b_2\dots b_k |y_1y_2\dots y_k, x)$ defines the joint probability that each receiver $i$ observes outcome $b_i$ when using measurement setting $y_i$, given Alice's input $x$. 

\begin{figure}[t]
    \centering
    \includegraphics{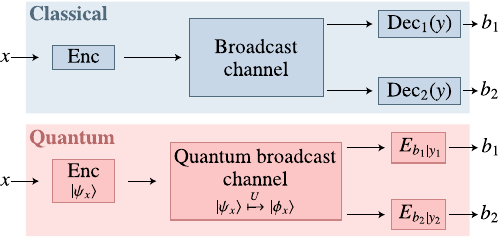}
   \caption{\emph{Two-receiver PM network}. Alice encodes a classical input $x$ into an initial quantum state $\ket{\psi_x}$, which is then distributed through a broadcasting channel (modelled by some isometry $U$) to  $2$ independent receivers, resulting in a joint state $\ket{\phi_x}$. Each receiver applies a local measurement $E_{b_i|y_i}$ dictated by their input setting ($y_1, y_2$) to produce a classical outcome ($b_1, b_2$). Crucially, even if the physical dimension of the broadcast state is unbounded or unknown, the resulting network correlations $p(b_1b_2|y_1y_2,x)$ can be strictly bounded using only the overlap information (Gram matrix) of the initial preparations. Note that in the classical case, one can distribute exact copies of the transformed message $x$.}
    \label{F:broadcasting_2_parties}
\end{figure}

The question now is which probabilities admit a quantum realisation when we restrict the set $\{|\psi_x\rangle\}_{x=1}^n$ to have overlaps defined by a Gram matrix $G$.  Here the full complex Gram matrix $G$ is fixed; if only the lower bounds $g_{xx'}$ are known, one must optimise over all compatible Gram matrices, including their phases. Wang et al. \cite{wang2019characterising} showed that this set features a hierarchy of converging relaxations closely resembling the NPA hierarchy discussed in Sec.~\hyperref[sec:NPAEACC]{III-A-4}. To be concise, let us consider a 2-receiver scenario as depicted in Fig.~\ref{F:broadcasting_2_parties}. 

The observed behaviour is governed by $p(b_1b_2|y_1y_2,x) = \langle\phi_x|E_{b_1|y_1} E_{b_2|y_2} |\phi_x\rangle$ where the measurement operators satisfy standard completeness ($\sum_{b_i} E_{b_i|y_i} = \iden$), orthogonality ($E_{b_i|y_i} E_{b_i'|y_i} = 0$ for $b_i \neq b_i'$), idempotence ($E_{b_i|y_i}^2  = E_{b_i|y_i}$), and commutativity ($[E_{b_1|y_1}, E_{b_2|y_2}] = 0$).   Importantly, note that although the dimension may change, the overlaps are invariant and therefore the overlaps of the transformed states satisfy $\braket{\phi_x}{\phi_{x'}}=\braket{\psi_x}{\psi_{x'}} =G_{xx'}$. 

Furthermore, because the dimension of the purified states $\ket{\phi_x}$ is unbounded, directly optimising over all possible Hilbert spaces and measurement operators to characterise the quantum set is computationally intractable. To bypass this, we can construct a hierarchy of semidefinite relaxations that bounds the set of quantum-realisable correlations from the outside, using only the known overlaps $G_{xx'}$.
We define a finite set of operators $\mathcal{S} = \{S_1, \dots, S_m\}$. Each element in $\mathcal{S}$ is a linear combination of products of the measurement operators $\{E_{b_1|y_1}, E_{b_2|y_2}\}$. From this set, we construct an $nm \times nm$ block moment matrix, which we will call $\Gamma$, to capture the correlation structure across all possible inputs. The full matrix is expressed as
\begin{equation}
    \Gamma = \sum_{x,x'=1}^n \Gamma^{xx'} \otimes \ket{e_x}\bra{e_{x'}},
\end{equation}
where $\{\ket{e_x}\}_{x=1}^n$ forms the standard orthonormal basis of $\mathbb{C}^n$, and the individual matrix entries are defined by the inner products
\begin{equation}
    \Gamma_{(i,j)}^{xx'} = \langle\phi_x|S_i^\dagger S_j|\phi_{x'}\rangle.
\end{equation}
By construction, this matrix must be Hermitian and positive semidefinite.
The  power of this formulation lies in mapping the physical constraints of our prepare-and-measure network directly to linear constraints on the entries of $\Gamma$. Specifically, any valid quantum realisation must satisfy: $(i)$ when $S_i = S_j = \iden$, the matrix entries must exactly reproduce the known state overlaps: $\Gamma_{(I,I)}^{xx'} = \braket{\phi_x}{\phi_{x'}} = G_{xx'}$. $(ii)$ For operators corresponding to the observed measurements, the entries must match the empirical probability distributions: $\Gamma_{(b_1|y_1), (b_2|y_2)}^{xx} = p(b_1 b_2|y_1 y_2, x)$. And finally, $(iii)$ the structural rules of the projectors impose further linear dependencies among the elements of $\Gamma$. For instance, commutativity ensures that $\Gamma_{(b_1|y_1, b_2|y_2)}^{xx'} = \Gamma_{(b_2|y_2, b_1|y_1)}^{xx'}$. 

The hierarchy arises by increasing the operator set $\mathcal S$. For the first level of the hierarchy one defines $\mathcal{S}_1$ as the set containing $\iden$ and the individual measurement operators $\{E_{b_1|y_1}, E_{b_2|y_2}\}$. The second level now considers the set $\mathcal{S}_2$ which now  includes products of two operators, such as $\{E_{b_1|y_1}E_{b_2|y_2}\}$. As the level $k$ increases, the matrix $\Gamma^{(k)}$ grows larger, encapsulating more constraints. Since a valid Gram matrix at step $k+1$ implies validity at step $k$, the approximated sets of probabilities nest within one another, naturally forming a sequence of converging outer bounds to the commuting-operator correlation set. In the supplemental material of Ref.~\cite{wang2019characterising} the convergence of the hierarchy to this set is proved.  

As an application of this formalism, one can study distributed random access codes. In this task, two random bits $x_0x_1$ are encoded into a quantum state $\ket{\psi_{x_0x_1}}$ and are sent to parties $B_1$ and $B_2$ for decoding.  For the decoding part of the protocol, the parties $B_1$ and $B_2$ are each given a random position bit and their goal is to guess the input bit that is associated with the position bit. For example, if $B_1$ receives $y_1=1$, he applies a measurement to his share of $\ket{\psi_{x_0x_1}}$ to guess the input $x_1$. By restricting the overlaps to be those of the optimal $(2,2)$ strategy [see Eq.~\eqref{RACvectors}], one can ask what is the maximum probability $p_1$ of $B_1$ guessing correctly given that $B_2$'s guessing probability $p_2$ is fixed. Here $p_i=\tfrac18\sum_{x_0,x_1,y_i} p(b_i=x_{y_i}|x_0x_1,y_i)$ is the average over uniform inputs and position bits. Their solution depicts a similar curve to that of the sequential random access code trade-off shown in   Fig.~\ref{fig_sequential_tradeoff},
\begin{equation}
    (2p_1-1)^2 + (2p_2-1)^2 \leq \frac{1}{2}.
\end{equation}

Beyond complex network topologies, bounding overlaps in a standard bipartite PM scenario provides a highly practical, semi-DI approach to quantum measurement certification~\cite{Shi_2019}, QRNG \cite{Brask2017}, or QKD \cite{ioannou2022receiver}.  Let us consider QRNG where the overlap constraint was first introduced. Consider a task where Alice prepares one of two pure states, $\ket{\psi_0}$ and $\ket{\psi_1}$, subject to a known overlap lower bound $|\braket{\psi_0}{\psi_1}| \ge \delta$. Bob subsequently aims to identify the state by performing an unambiguous state discrimination measurement~\cite{Barnett09} [see Box \ref{Box:unambiguous-state-discrimination} for a summary of this task]. The objective is to correctly identify the prepared quantum state as frequently as possible ($p(b=x|x)$) while completely eliminating false positives ($p(b=\neg x|x)=0$). 
\begin{backgroundbox}[label={Box:unambiguous-state-discrimination}]{Unambiguous state discrimination}
    In standard quantum state discrimination, an observer attempts to guess which of two non-orthogonal pure states, $\ket{\psi_0}$ or $\ket{\psi_1}$, was prepared, inevitably making some errors. Unambiguous state discrimination  takes a different approach: it strictly forbids errors (zero false positives) \cite{Barnett09}. To achieve this for both states, the measurement must have three outcomes: $0$, $1$, and $\emptyset$ (inconclusive). The conditions for unambiguous state discrimination are:
    \begin{enumerate}
        \item $p(0|1) = 0$ (If the state was $\ket{\psi_1}$, outcome $0$ never clicks),
        \item and $p(1|0) = 0$ (If the state was $\ket{\psi_0}$, outcome $1$ never clicks)
    \end{enumerate} 
Because the states are non-orthogonal, quantum mechanics implies that the observer cannot be certain every time. The penalty for demanding zero errors is that the measurement must yield the inconclusive outcome $\emptyset$ with a probability bounded by the state overlap.    
\end{backgroundbox}

Although quantum mechanics permits error-free measurements of this kind, it dictates a fundamental trade-off: a certain fraction of the measurements must yield an inconclusive result ($p(b=\emptyset|x)$), bounded for equal input probabilities by $\tfrac12[p(b=\emptyset|0)+p(b=\emptyset|1)]\geq\delta$. Note that it is not possible to predict in advance whether a particular round of the experiment will be conclusive or inconclusive, resulting in a source of genuine randomness. Therefore, define now a bit $c$ that encodes whether the result was conclusive, $c=0$ if $b=0,1$ or inconclusive, $c=1$ if $b=\emptyset$. Our goal is to bound the amount of randomness in $c$ given the state overlap $\delta$ and the observed distribution $p(b|x)$. In the ideal case, in which Alice sends two states $\ket{\psi_0},\ket{\psi_1}$ with overlap $|\braket{\psi_0}{\psi_1}| =\delta$, and Bob implements an optimal symmetric unambiguous state discrimination measurement, it can be shown that the probability $p(c=1|x)$ of obtaining an inconclusive result is $p(c=1|x) = \delta$, which for $|\braket{\psi_0}{\psi_1}| = \frac{1}{2}$ gives a completely random guess. We show explicitly this ideal case in the example Box~\ref{Box:ideal_case_randomness_USD}.

\begin{examplebox}[label={Box:ideal_case_randomness_USD}]{QRNG with bounded overlap}
To see how bounding overlaps generates randomness, consider an ideal scenario where Alice's preparation device emits one of two states, $\ket{\psi_0}$ or $\ket{\psi_1}$ (with equal prior probabilities $p(x=0)=p(x=1)=1/2$). We assume the states have a known overlap of exactly $|\braket{\psi_0}{\psi_1}| = \delta$. 

\medskip
Bob's measurement device implements a perfect symmetric unambiguous state discrimination measurement. By definition, a measurement of this type produces no false positives, meaning the error probabilities are strictly zero
\begin{equation}
    p(0|1) = p(1|0) = 0,
\end{equation}
but the probability of obtaining the inconclusive outcome $\emptyset$ is exactly equal to the state overlap ~\cite{Barnett09}
\begin{equation}
    p(\emptyset|0) = p(\emptyset|1) = \delta.
\end{equation}

By normalisation, the conclusive probabilities for Bob are therefore $p(0|0) = p(1|1) = 1 - \delta$. 

\medskip

Now suppose that an eavesdropper Eve is trying to predict the bit $c$ indicating whether Bob's outcome  is inconclusive. Without additional side information, her maximum guessing probability, $p_g$, is simply the highest probability among the possible values of $c$. Assuming a completely random input $x$, the total probabilities for Bob's outcomes are:
\begin{align*}
    p(b = \emptyset) &= p(\emptyset|0)p(x=0) + p(\emptyset|1)p(x=1) = \delta,  \\
    p(b = 0) &= p(0|0)p(x=0) + p(0|1)p(x=1) = \frac{1 - \delta}{2}, \\
    p(b = 1) &= p(1|0)p(x=0) + p(1|1)p(x=1) = \frac{1 - \delta}{2}.
\end{align*}
For the bit $c$, the conclusive and inconclusive probabilities are $1-\delta$ and $\delta$, respectively. Thus, $p_g(c)=\max\{1-\delta,\delta\}$ in this ideal case. In the specific case where Alice tunes her states such that the overlap is $\delta = \tfrac12$, Eve's maximum guessing probability for $c$ in the ideal case is $p_g = \tfrac12$. By calculating the min-entropy, $H_{\min} = -\log_2 p_g = -\log_2\tfrac12 = 1$, we can obtain one bit of min-entropy per round under ideal conditions.
\end{examplebox}

While the ideal case provides a clear intuition, practical implementations of prepare-and-measure protocols inevitably suffer from technical imperfections, noise, and losses. Consequently, the experimentally observed measurement statistics $p(b|x)$ will always deviate from a perfect unambiguous state discrimination. A powerful feature of this overlap-restricted framework is its semi-device-independent nature: one can bound the genuinely quantum randomness directly from the noisy empirical data without needing to model the imperfections of the measurement device~\cite{Brask2017}. To certify randomness in the noisy case considered here, we assume that the source emits fixed states $\rho_0$ and $\rho_1$, independently of the measurement device and Eve, and that Eve holds classical side information $\lambda$ specifying the measurement strategy used in each run. Alice chooses $x$ uniformly and independently of $\lambda$; Eve may learn $x$ when guessing the output. The trusted source constraint is $F(\rho_0,\rho_1)\ge\delta^2$. Under these assumptions, a conservative reduction uses fixed reference pure qubit states $\sigma_x=\ketbra{\chi_x}$, where $\ket{\chi_0}=\ket{0}$ and $\ket{\chi_1}=\delta\ket{0}+\sqrt{1-\delta^2}\ket{1}$. A channel can map these reference states to the allowed preparations, so its action can be absorbed into the uncharacterised measurements. The resulting optimisation over measurements compatible with $p(b|x)$ is an SDP. This argument does not cover arbitrary correlations between source decompositions and Eve: a fidelity bound on averaged states alone would not suffice in that case. By solving the dual of this SDP, one finds a set of real coefficients $\nu_{bx}$ that define a rigorous linear upper bound on the guessing probability
\begin{equation}
p_g \le \sum_{b,x} \nu_{bx} p(b|x).
\end{equation}
To guarantee the validity of this upper bound, the coefficients $\nu_{bx}$ must be chosen such that there exist four $2 \times 2$ Hermitian matrices $H^{\lambda_0, \lambda_1}$ (where $\lambda_0, \lambda_1 \in \{0, 1\}$) satisfying the following operator inequality for every outcome $b$ and every pair $(\lambda_0,\lambda_1)\in\{0,1\}^2$. Each pair labels a deterministic guessing strategy: $\lambda_x=0$ denotes a guess that the outcome is inconclusive, and $\lambda_x=1$ denotes a guess that it is conclusive. The factors $\tfrac12$ assume uniform input probabilities:
\begin{align}
\sum_x \sigma_x &\left(\frac{1}{2}\delta_{\lambda_x, 0}\delta_{b, \emptyset} + \frac{1}{2}\delta_{\lambda_x, 1}(1-\delta_{b, \emptyset}) - \nu_{bx}\right) \nonumber\\&\hspace{1.55cm}+ H^{\lambda_0, \lambda_1} - \frac{1}{2}\tr[H^{\lambda_0, \lambda_1}]\iden \le 0.\end{align}

The practical advantage of this dual formulation is that one does not need to run a computationally heavy SDP for every new batch of data. For a fixed state overlap $\delta$, the optimal coefficients $\nu_{bx}$ that satisfy the above constraints can be pre-computed. The bound on the guessing probability $p_g$ then becomes only a linear function of the experimental data $p(b|x)$. This allows for online entropy estimation, enabling semi-DI QRNG at megahertz rates using common optical components~\cite{Brask2017}.

Despite their usefulness, overlap assumptions also present some limitations. First, the overlaps $|\braket{\psi_x}{\psi_{x'}}|$ are formulated for pure-state preparations, while real implementations always have some mixedness. Second, compared to overlap constraints, assumptions based on physically measurable observables such as energy or mean photon number are often closer to experimental practice and can be verified directly from the emitted states~\cite{VanHimbeeck2017semidevice, Roch2025} while state overlaps typically require an independent characterisation of the source through swap tests or randomised measurements~\cite{Ekert_2002,  Hendrych2003, Fanizza_2020}. As a consequence, overlap constraints can be viewed as theoretically appealing but experimentally less direct than assumptions based on measurable physical observables.

\subsection{Spacetime symmetries and speed limits}\label{Sub:spacetime-symmetries-speed-limits}
Thus far, the semi-device-independent protocols we have discussed restrict communication by bounding internal, information-theoretic properties of the physical system, such as its Hilbert space dimension or state overlaps. In those frameworks, the inputs act merely as abstract labels to select different, uncharacterised preparations or measurements.

A fundamentally different approach is to use macroscopic spacetime operations as our inputs, constraining the system through its physical response to these external operations. In this framework, we still treat the preparation and measurement devices as uncharacterised black boxes, but the choice of input corresponds to a known, controlled physical operation applied to the device itself, such as a physical spatial rotation or a temporal delay~\cite{Jones2026theoryindependent,Aloy2024,jones2025certifiedrandomnessquantumspeed}. Consequently, the semi-device-independent assumption shifts from an abstract information capacity to fundamental physical invariants, such as an upper bound on the system's generalised spin or its energy uncertainty. These constraints, rooted in spacetime symmetries and dynamical speed limits, enable new certification tasks, in particular, secure randomness generation.

Let us begin by considering restrictions in spatial degrees of freedom~\cite{Jones2026theoryindependent}. Assume a prepare-and-measure scenario with a preparation device producing a single physical system $\rho_1=\ketbra{\phi_1}{\phi_1}$. The device receives an input $x \in \{1, 2\}$ which triggers a spatial rotation by an angle $\alpha_x \in \{0, \alpha\}$, with $0\leq \alpha \leq \pi$, around a fixed axis relative to the measurement device. The generated physical system is then sent to the measurement device which performs a single two-outcome POVM described by $\{M_b\}_b$ with outcomes $b \in \{\pm 1\}$. 

Because the preparation device is physically rotated, the covariance of physical laws dictates that this macroscopic rotation must act as a reversible transformation on the state space of the transmitted system. In quantum theory, this corresponds to a unitary transformation $U_{\alpha}$ that furnishes a projective representation of the $SO(2)$ symmetry group. As shown in Ref.~\cite{Jones2026theoryindependent}, any finite-dimensional projective representation of this rotation can be decomposed into a direct sum 
\begin{equation}
    U_{\alpha} = \bigoplus_{j=-J}^{J} e^{ij\alpha} \iden_{n_j}, \label{eq:Unitary_representation}
\end{equation}
where $j$ runs over either integers or half-integers, $n_j$ denotes the multiplicity of the $j$-th irreducible representation, and $J$ represents the maximum representation label. Here lies the core assumption: rather than bounding the number of dimensions the system occupies, we assume an upper bound on $J$, which physically corresponds to the maximum generalised spin or angular momentum quantum number of the transmitted system.  Deriving Eq.~\eqref{eq:Unitary_representation} assumes that each irreducible sector is finite dimensional. However, the total Hilbert space dimension need not be bounded, since the multiplicities $n_j$ may be arbitrarily large.

With this constraint in place, we can now characterise the set of achievable quantum correlations. Let us define the expectation value of the measurement $M = M_{+1} - M_{-1}$ for a given input $x$ as $E_x = p(+1|x) - p(-1|x)$. The preparation for the second input is geometrically related to the first by $\rho_2 = U_\alpha \rho_1 U_\alpha^\dagger$.  As mentioned in Box~\ref{Box:vacuum-two-state}, for two pure states with fixed overlap $\gamma=|\braket{\phi_1}{\phi_2}|$,  the largest set of correlations $(E_1,E_2)$ arising from arbitrary measurements satisfies the inequality $g(E_1,E_2)\geq\gamma_{\min}$, where $g(E_1,E_2)$ was defined in Eq.~\eqref{eq:gE1E2} and $\gamma_{\min}$ is the minimum overlap possible. In Box~\ref{Box:Overlap_SpinBounded} we show that the minimum overlap $\gamma_{\min}$ is given by 
\begin{equation}\label{eq:gammamin_rotations}
\gamma_{\min} = \min |\langle \phi_1 | \phi_2 \rangle| 
= \begin{cases} \cos(J\alpha) & \text{if } |J\alpha| < \frac{\pi}{2} \\ 
0 & \text{if } |J\alpha| \ge \frac{\pi}{2} \end{cases} 
\end{equation}
Together, the last two equations define the set of all pure-state quantum correlations $\mathcal{Q}_{J,\alpha}$ restricted to a maximum spin value $J$.

\begin{examplebox}[label={Box:Overlap_SpinBounded}]{Minimum overlap for a spin-bounded system}

To find the minimum overlap $\gamma := \min |\langle\phi|U_\alpha|\phi\rangle|$ over all pure states in a spin-$J$ bounded system, we decompose any normalised state into the irreducible representation basis: $|\phi\rangle = \sum_{j=-J}^{J} \phi_j |\psi_j\rangle$, where $U_\alpha|\psi_j\rangle = e^{ij\alpha}|\psi_j\rangle$.  Using the property $|z| \ge \text{Re}(z)$, the overlap with the rotated state is  bounded by:
 \begin{align}
     |\langle\phi| U_\alpha |\phi\rangle| &= \left| \sum_j |\phi_j|^2 e^{ij\alpha} \right| \nonumber\\&\ge |\phi_0|^2 + \sum_{j>0} \cos(j\alpha) \left(|\phi_{-j}|^2 + |\phi_j|^2\right).
 \end{align}
For the regime where $|\alpha| \le \frac{\pi}{2J}$, the factor $\cos(j\alpha)$ strictly decreases as $j$ increases.  The expression is therefore minimised when the state's weight is concentrated in the extreme spin projections, meaning $|\phi_{-J}|^2 + |\phi_J|^2 = 1$. If the $j=0$ sector is absent, $|\phi_0|^2$ in the preceding equation is understood to be zero. Geometrically, the overlap is a convex combination of points on the unit circle. Once the angular spread $2J\alpha$ reaches $\pi$, the origin lies inside the convex hull, making a vanishing overlap possible.

\medskip
Combining these two regimes yields the tight bound~\eqref{eq:gammamin_rotations}.
\end{examplebox}
Remarkably, one does not even need to assume the validity of quantum mechanics to derive this exact boundary. To show this, the transmitted system can be treated within the broad mathematical framework of generalised probabilistic theories (GPTs, see Box~\ref{Box:GPT}), which formalises the physical setup without relying on Hilbert spaces or operators.

A physical spatial rotation by an angle $\alpha$ acts on this system as a reversible linear transformation $T_\alpha : V \to V$. Because successive rotations compose, the mapping $\alpha \mapsto T_\alpha$ must furnish a continuous real linear representation of the symmetry group $SO(2)$. By the standard representation theory of compact groups over real vector spaces, $T_\alpha$ can always be block-diagonalised into a direct sum of invariant subspaces. Specifically, it acts as a trivial block and $2 \times 2$ planar rotation matrices $R(j\alpha)$ with frequencies $j$:
\begin{equation}
T_\alpha = \iden_{n_0}\oplus \bigoplus_{j=1}^{2J} R(j\alpha)^{\oplus n_j}, \quad R(j\alpha) = \begin{pmatrix} 
\cos(j\alpha) & -\sin(j\alpha) 
\\ \sin(j\alpha) & \cos(j\alpha) 
\end{pmatrix},
\end{equation}
where again, $n_j$ represents the multiplicity of the $j$-th irreducible representation. Note the nontrivial indices go from $j=1,\dots,2J$, while in the quantum case they take the values $j=-J,\dots,J$. This is because probabilities involve differences of quantum spin labels, whose absolute values can reach $2J$. In this framework, the semi-device-independent assumption of a maximum generalised spin $J$ mathematically corresponds to truncating this representation such that $\max(j) \le 2J$. Because the measurement effects $e$ are linear functionals of the state, this  truncation places an observable limit on the laboratory statistics. Regardless of the underlying theory or the system's dimension, the probability of obtaining any outcome $b$ given a rotation $\alpha$ is forced to be a trigonometric polynomial of maximum degree $2J$~\cite{Aloy2024}:
\begin{equation}
P(b|\alpha) = \sum_{k=0}^{2J} \left[c_k^{(b)} \cos(k\alpha) + s_k^{(b)} \sin(k\alpha) \right],\label{eq:GPT_prob_J}
\end{equation}
where $c_k^{(b)}$ and $s_k^{(b)}$ are real coefficients. 
This defines the set of rotation boxes $\mathcal{R}_{J,\alpha}$ as the set of all correlations $(E_1,E_2)$ with $E_x =p(1|\alpha_x)-p(-1|\alpha_x)$ for $x\in\{1,2\}$ such that Eq.~\eqref{eq:GPT_prob_J} is true and $0\leq p(b|\theta)\leq 1$ for every outcome $b$ and angle $\theta$.

\begin{backgroundbox}[label={Box:GPT}]{Generalised probabilistic theories}
    Generalised probabilistic theories provide a mathematical framework to describe operational physics without assuming the mathematical structure of quantum mechanics (such as Hilbert spaces). Within this framework, physical systems are characterised by three basic elements: states, measurements, and transformations:
    \begin{enumerate}
        \item The state of a physical system is represented by a vector $\omega$ belonging to a convex state space $\Omega$, embedded in a real vector space $V$.
        \item Measurements are described by effects $e$, which are linear functionals $e:V\rightarrow\mathbb{R}$ satisfying $0\le e(\omega)\le1$ for every normalised physical state $\omega\in\Omega$, with probabilities $p=e(\omega)$.
        \item Reversible physical operations, such as spatial rotations, act as linear transformations $T: V \to V$ that map the state space onto itself.
    \end{enumerate}
\end{backgroundbox}
Since quantum theory is a specific instance of a GPT, it is trivially true that $\mathcal{Q}_{J,\alpha} \subseteq \mathcal{R}_{J,\alpha}$. Remarkably, however, the reverse inclusion also holds. To prove this, we observe that since the expectation values are bounded trigonometric polynomials, their rate of variation with the rotation angle cannot become arbitrarily large. Bernstein-type inequalities make this statement precise. In particular we have that if $T(x)$ is a trigonometric polynomial of degree $n$ bounded by $-1 \le T(x) \le 1$, its derivative satisfies:
\begin{equation}
T'(x)^2 + n^2 T(x)^2 \le n^2.
\end{equation}
Applying this to the expectation value $E(\theta)=p(+1|\theta)-p(-1|\theta)$, which is a trigonometric polynomial of degree $n=2J$ in the rotation angle $\theta$, with $E(0)=E_1$ and $E(\alpha)=E_2$, we can rearrange the inequality as:
\begin{equation}
\frac{|E'(\theta)|}{\sqrt{1 - E(\theta)^2}} \le 2J,
\end{equation}
Note that the left-hand side is the absolute value of the derivative of $\arcsin[E(\theta)]$. Integrating this inequality with respect to $\theta$ from our first input ($0$) to our second input ($\alpha$) yields:
\begin{align}
&\Biggl|\int_0^\alpha \drm\theta\,\frac{E'(\theta)}{\sqrt{1 - E(\theta)^2}}\Biggl| \le \int_0^\alpha \drm \theta\,2J  \nonumber \\
&\implies \frac{1}{2} \left| \arcsin E_2 - \arcsin E_1 \right| \le J\alpha. \label{eq:arcsin_bound}
\end{align}

If the rotation is sufficiently large ($J\alpha \ge \pi/2$), this inequality places no restriction on the expectation values, and $\mathcal{R}_{J,\alpha}$ contains all possible correlations $[-1, 1]^2$, matching the quantum case. For smaller angles ($J\alpha < \pi/2$), taking the cosine of both sides of Eq.~\eqref{eq:arcsin_bound} and applying elementary trigonometric identities [see Appendix I in Ref.~\cite{Jones2026theoryindependent}] exactly reproduces the correlation boundary $g(E_1, E_2) \ge \cos(J\alpha)$ derived in the quantum scenario [see Eq.~ \eqref{eq:gammamin_rotations}]. Therefore, $(E_1, E_2) \in \mathcal{Q}_{J,\alpha}$, proving that $\mathcal{Q}_{J,\alpha} = \mathcal{R}_{J,\alpha}$. Therefore, the exact curved boundary of the quantum correlation set is not an exclusive signature of quantum mechanics, but a mathematical consequence of the interplay between generalised probabilities and spacetime symmetries. 
\\

This spatial framework finds a direct and elegant counterpart in the temporal domain~\cite{jones2025certifiedrandomnessquantumspeed}. Instead of spatial rotations, consider a prepare-and-measure scenario where the input $x \in \{1, 2\}$ dictates the precise time at which the untrusted preparation device is triggered, resulting in states prepared at $t_1 = 0$ and $t_2 = \Delta t$. Consider that the time translation is generated by some Hamiltonian $\hat{H}$. Then, the macroscopic temporal delay acts on the state space as the unitary $U_{\Delta t} = e^{-i\hat{H}\Delta t/\hbar}$. Consequently, the two preparations are geometrically related by $\rho_2 = U_{\Delta t} \rho_1 U_{\Delta t}^\dagger$. While the spatial framework bounded the maximum eigenvalue of the symmetry generator (the spin $J$), the temporal framework restricts the fluctuations of the generator. Specifically, the semi-device-independent assumption is an upper bound on the energy uncertainty of the prepared state: $\Delta E \le \mathfrak{E}$. This replaces the need to characterise the full Hamiltonian or assume a finite Hilbert space dimension. This bound on the energy fluctuations limits the speed of the system's dynamical evolution via the Mandelstam-Tamm quantum speed limit (see Box~\ref{Box:QSL}). Because the cosine function bounding the state overlap is strictly decreasing in the interval $[0, \tfrac{\pi}{2}]$, evaluating it at our trusted upper bound $\mathfrak{E}$ provides a rigorous lower bound for the unknown dynamics, since $\cos(\Delta E \Delta t/\hbar) \ge \cos(\mathfrak{E}\Delta t/\hbar)$. Thus, replacing the exact uncertainty $\Delta E$ with the upper bound $\mathfrak{E}$ yields the minimum overlap allowed by the quantum speed limit:
\begin{equation}
\gamma_{\min} = \begin{cases} \cos(\mathfrak{E}\Delta t/\hbar) & \text{if } \frac{\mathfrak{E}\Delta t}{\hbar} < \frac{\pi}{2} \\ 0 & \text{if } \frac{\mathfrak{E}\Delta t}{\hbar} \ge \frac{\pi}{2} \end{cases}.
\end{equation}

By comparing this to the spatial overlap in Eq.~\eqref{eq:gammamin_rotations}, a direct correspondence emerges: one simply substitutes the generalised spin $J$ with $\mathfrak{E}/\hbar$, and the rotation angle $\alpha$ with the temporal delay $\Delta t$. Thus, an upper bound on energy uncertainty combined with a known temporal delay produces the exact same boundary for quantum correlations $\mathcal{Q}_{\mathfrak{E}, \Delta t}$. 

\begin{backgroundbox}[label={Box:QSL}]{Quantum speed limits}
Quantum speed limits bound how fast a physical system can evolve in time. For a quantum state evolving under a Hamiltonian $\hat H$, the Mandelstam-Tamm bound relates the distinguishability of the evolved state to the energy uncertainty:
\begin{equation}
|\langle \phi | e^{-i\hat H t/\hbar} | \phi \rangle|
\ge
\cos\left(\frac{\Delta E\, t}{\hbar}\right),
\end{equation}
valid for $\Delta E\, t/\hbar \le \pi/2$, where $\Delta E = \sqrt{\langle \hat H^2 \rangle - \langle \hat H \rangle^2}$. 

\medskip
Operationally, this means that states with small energy uncertainty cannot become rapidly distinguishable under time evolution. Equivalently, the energy fluctuations constrain the speed at which information encoded in the state can evolve.
\end{backgroundbox}

Furthermore, this geometric boundary provides an operational route to randomness certification. For the following argument, assume that the strictly positive overlap bound $\gamma_{\min}>0$ holds for every hidden-variable realisation. The deterministic pairs $(E_1,E_2)=(1,-1)$ and $(-1,1)$ are then excluded, as they would perfectly distinguish the preparations. Hence, every allowed deterministic assignment satisfies $E_1=E_2$, and mixtures of such assignments remain on the diagonal. Observing $E_1\ne E_2$ thus rules out a fully deterministic explanation. If the physical restriction holds only on average, the adversarial optimisation must account for this weaker constraint, and the argument no longer applies directly.

By optimising the adversary's guessing probability over all strategies compatible with the observed correlations and the spacetime bound, one can  quantify the secure random bits generated per round~\cite{Jones2026theoryindependent, jones2025certifiedrandomnessquantumspeed}. This approach anchors the security of quantum cryptography not in the detailed characterisation of internal device components, but in the symmetries and limits of spacetime itself.

\subsection{Informational constraints}\label{Sub:information-restriction}
Correlations arising in the PM scenario must come at a cost associated with the informational content carried by the states sent from Alice to Bob. In Sec.~\ref{S:PM-setting} this cost was measured in the classical and quantum dimension. However, even in dimensions greater than two, a particular ensemble of classical or quantum states may carry no more than one bit of information about the input. Consequently, a natural avenue to capture the informational cost of creating correlations is to depart from dimension-based quantifiers and instead adopt entropic notions of information. 

Analysing PM scenarios under such informational restrictions was first proposed in Refs.~\cite{chaves2015device,Tavakoli2020informationally, Chaturvedi2020quantum} (see also the more complete analysis in Ref.~\cite{Tavakoli2022informationally}). A natural starting point is to answer the following question: ``information about what?'' Because in PM scenarios Bob's job is to correlate himself with Alice's input, a natural answer is to consider how much information Alice's communication, $\{\rho_x\}_x$, carries about her input $x$. This can be quantified in terms of min-entropy information
\begin{equation}\label{accessibleinformation}
    I_{\min(X:B)}=H_{\min}(X)-H_{\min}(X|B),
\end{equation}
where $B$ denotes the register containing Alice's states and $X$ denotes the classical register containing  Alice's input. This equation can be interpreted as follows: the information made accessible to Bob due to Alice's communication is the difference in the uncertainty about $X$ that Bob had before and after receiving Alice's states. This differs from conventional accessible information, which maximises Shannon mutual information over Bob's measurement. Importantly, and in contrast to what we encountered in the analysis of dimension-restricted PM scenarios, the quantity $I_{\min}(X:B)$ depends on Alice's prior distribution which we denote  $p_X(X=x)=p(x)$. 

When making the definition in Eq.~\eqref{accessibleinformation}, one could in principle replace $H_{\min}$ with another entropy function, with a consistently defined conditional entropy. However, following Ref.~\cite{Tavakoli2020informationally}, we focus on using the min-entropy, denoted $H_{\min}$. This is because the min-entropy admits a natural interpretation in terms of guessing games in PM scenarios. The min-entropy of a random variable is one-to-one with the highest probability of correctly guessing its value, specifically  $H_{\min}(X)=-\log_2\left(\max_x p(x)\right)$. Here, $\max_x p(x)$ is the probability associated with the most likely value of $X$. This means that the first term in Eq.~\eqref{accessibleinformation} is only a function of Alice's prior. The second term, however, addresses how the system $B$ updates the min-entropy. This conditional min-entropy admits an elegant interpretation in terms of the optimal success probability in state discrimination, namely, $H_{\min}(X|B)=-\log_2 P_g$, where~\cite{Konig2009}
\begin{align}
 P_g=\max_{\{N_x\}} \sum_{x} p(x) \tr(\rho_x N_x). \label{eq:guessprob_iformation_restricted}
\end{align}
Here the maximisation is over all POVMs. For simplicity, we call $P_g$ the guessing probability. We see that for a given prior, $P_g$ is one-to-one with the information content. Moreover, while it is in general hard to analytically compute the guessing probability, it can notably be cast as an SDP over the measurement $\{N_x\}_x$. In Box~\ref{Box:infobound} we also derive a simple analytical upper bound on $P_g$. 

\begin{examplebox}[label={Box:infobound}]{Information bound}
Let $\{\rho_x\}_{x=1}^n$ be any set of $d$-dimensional quantum states and let them be sampled with prior $p(x)$.  The success probability for any POVM $\{N_x\}$ satisfies

\begin{align*}
&\sum_{x} p(x)\tr\left(\rho_x N_x\right) \leq \sum_{x} p(x)\lambda_{\max}(\rho_x)\tr(N_x)\\
&\leq  \big[\max_x p(x)\lambda_{\max}(\rho_x)\big] \sum_{x} \tr(N_x)\\
& =  d \max_x p(x)\lambda_{\max}(\rho_x).
\end{align*}

It then follows from Eq.~\eqref{accessibleinformation} that
\begin{align*}
    I_{\min}(X:B)\leq \log_2 d+\log_2\left( \frac{\max_x p(x)\lambda_{\max}(\rho_x)}{\max_x p(x)}\right).
\end{align*}
This inequality is tight if there is a POVM $\{N_x\}$ such that every non-zero $N_x$ is supported on a maximal-eigenvalue eigenspace of $\rho_x$, and $p(x)\lambda_{\max}(\rho_x)$ equals $\max_j p(j)\lambda_{\max}(\rho_j)$ for every such $x$. One way of systematically generating such states is to measure half of a maximally entangled state of dimension $RK$ with a $K$-outcome projective measurement whose elements have rank $R$. 
\end{examplebox}

We now use this notion of informational content to understand how dimension restrictions fit into this broader picture. Systems of dimension $d$ should reasonably not carry more than $\log_2 d$ bits of information, independently of the number of states and Alice's prior $p(x)$. To verify this, we  extend the argument in Eq.~\eqref{mesdquantum} so that it applies to our employed quantity $I_{\min}(X:B)$. Let $\{\rho_x\}_{x=1}^n$ be any set of $d$-dimensional quantum states. The probability of state discrimination is bounded as 
\begin{align}\nonumber\label{Pguessdim}
&\sum_{x} p(x)p(b=x|x)=\sum_{x}p(x)\tr\left(\rho_x N_x\right) \\
& \leq \sum_{x} p(x)\lambda_{\max}(N_x)\leq \sum_{x} p(x)\tr(N_x) \nonumber\\
& \leq \big[\max_x p(x)\big]\sum_x\tr(N_x)=d \max_x p(x).
\end{align}
Maximising over $\{N_x\}$, we conclude that $H_{\min}(X|B)\geq -\log d- \log(\max_x p(x))$. The second term is precisely $H_{\min}(X)$. Hence, when inserted in Eq.~\eqref{accessibleinformation} we obtain the desired result, namely
\begin{align}\label{dimbound}
    I_{\min}(X:B)\leq \log_2 d.
\end{align}

Equipped with the informational content of Alice's communication, we can now deploy it to study correlations in the PM scenario. Specifically, for a given theory of states and measurements, we are interested in the set of correlations $p(b|x,y)$ that can be generated in the PM scenario if Alice's communication is informationally restricted by allowing only some values for $I_{\min}(X:B)$. Equivalently, once the prior is fixed, we can express this as a bound $P_g\leq G$ on the guessing probability. In quantum and classical models, there are some key differences in the properties of informationally restricted correlations as compared to most other semi-device-independent frameworks. We now discuss two of these properties. 

When dealing with informationally restricted correlations, shared randomness is always included for free. The reason is that in an informationally restricted picture, there is no difference between an independent source emitting the shared randomness to Alice and Bob, and Alice generating it in her lab and sending it to Bob as a part of her communication. In the first case, Alice's states are $\rho_x^{(\lambda)}$, where $\lambda$ is distributed as $q(\lambda)$. The total guessing probability is therefore the convex combination of the guessing probabilities associated to each $\lambda$.    In the second case, Alice's states are 
    \begin{equation}\label{infostate}
        \rho_x=\sum_\lambda q(\lambda) \rho_x^{(\lambda)}\otimes \ketbra{\lambda}{\lambda}.
    \end{equation}
The guessing probability for these states becomes 
    \begin{align}\nonumber
    P_g&=\max_{\{N_x\}} \sum_\lambda q(\lambda)\sum_{x} p(x)\tr\left(\rho_x^{(\lambda)}\otimes \ketbra{\lambda}{\lambda} N_x\right)\\ \nonumber
    &=\max_{\{N_{x|\lambda}\}} \sum_\lambda q(\lambda)\sum_{x} p(x)\tr\left(\rho_x^{(\lambda)} N_{x|\lambda}\right)\\ \nonumber
    &=\sum_\lambda q(\lambda) \max_{\{N_{x|\lambda}\}} \sum_{x} p(x)\tr\left(\rho_x^{(\lambda)} N_{x|\lambda}\right)\\
    &=\sum_\lambda q(\lambda) P_g(\mathcal{E}_{\lambda}),
    \end{align}
where we have defined the set of POVMs $\{N_{x|\lambda}\}$ and  written $\mathcal{E}_{\lambda}=\{p(x),\rho_x^{(\lambda)}\}_x$ for Alice's set of states conditioned on $\lambda$. We see that $P_g$ is again the convex combination of the guessing probabilities conditioned on $\lambda$. In other words, Alice can include the shared randomness in her communication without increasing the informational content.

Another interesting property is that unlike  most other semi-device-independent frameworks, we cannot without loss of generality assume that Alice's states are pure when conditioning on $\lambda$. For example, if in a dimension-restricted setting Alice were to send a mixed state $\rho_x^{(\lambda)}$, then she could decompose each state into a mixture of pure states and absorb the randomness of that decomposition into $q(\lambda)$. This works because each pure state in the decomposition has the same dimension as $\rho_x^{(\lambda)}$ and hence the assumption remains respected. However, under informational restrictions this argument no longer goes through. Specifically, if we were to assume that $\rho_x^{(\lambda)}$ in Eq.~\eqref{infostate} is pure, we would impose an effective extra restriction on Alice's communication \cite{Tavakoli2022informationally}. Once decomposing the mixed states into pure states, there is no guarantee that the information content does not increase.

Moreover, these properties are not unique to quantum theory but apply also to classical systems. In a classical approach to informationally restricted correlations, we need only to impose that the states $\rho_x^{(\lambda)}$ in Eq.~\eqref{infostate} commute. In analogy with Eq.~\eqref{PMdimclassical2},  classical correlations admit the form $p(b|x,y)=\sum_m p(m|x)p(b|m,y)$ --- where we have excluded the shared randomness because we now know it can be absorbed into the communication without loss of generality. The informational restriction can then be represented as a bound of the form
\begin{equation}
    P_g = \sum_m \max_x p(x) p(m|x)\leq G.
\end{equation}

\begin{figure}[t]
    \centering
    \includegraphics{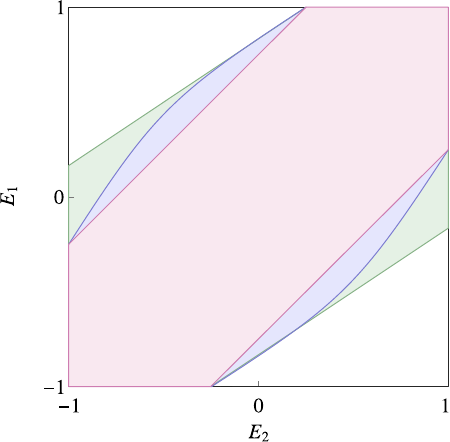}
   \caption{\emph{Informationally restricted correlations}. Sets of correlations under informational restrictions in the scenario $|X|=|B|=2$ and $|Y|=1$. Alice's prior has been selected as $p=0.6$ and $G=3/4$. The green polytope is the space of mixed correlations in classical, quantum and post-quantum models alike. When classical models are restricted to pure states, one obtains the pink polytope. The same restriction for quantum theory leads to the blue set.}
    \label{F:Sets_informational_restricted}
\end{figure}

To exemplify the importance of allowing $\rho_{x}^{(\lambda)}$ to be mixed in both the classical and quantum setting, consider the input/output scenario corresponding to $(|X|,|Y|,|B|)=(2,1,2)$ and let us define $E_x=p(1|x)-p(2|x)$. It is shown in Ref.~\cite{Tavakoli2022informationally} that the space of $(E_1,E_2)$ in both classical and quantum theory is given by 
\begin{equation}
    |pE_1 - (1-p)E_2| \le 2G - 1,
\end{equation}
where $p=p(x=1)$. In fact, this represents the full space of correlations possible in any theory subject to the informational restriction. In contrast, if we were to assume that $\rho_{x}^{(\lambda)}$ is pure, then the classical correlation space is given by 
\begin{equation}\label{pureC}
    |E_1 - E_2| \le 2 \frac{G -p_{\max}}{p_{\min}},
\end{equation}
where $p_{\max} = \max\{p,1-p\}$ and $p_{\min} = \min\{p,1-p\}$, whereas the quantum one is given by an infinite family of linear inequalities:
\begin{equation}\label{pureQ}
    |c_1E_1 - c_2E_2| \le \sqrt{1 - \frac{4c_1c_2}{p(1-p)}G(1 - G)}.
\end{equation}
where $c_1 + c_2 = 1$ in the range $p_{\min} \leq c_1, c_2 \leq p_{{\max}}$. For suitable asymmetric priors and nontrivial information bounds, the pure quantum set defined by Eq.~\eqref{pureQ} is strictly larger than the pure classical set defined by Eq.~\eqref{pureC}, while the identical mixed classical and quantum sets are strictly larger still. These strict inclusions are not universal: for example, when $p=\tfrac12$, all three sets coincide and obey $|E_1-E_2|\le4G-2$. This is illustrated in  Fig.~\ref{F:Sets_informational_restricted}. In view of this, one must in  the analysis of informationally restricted correlations in general allow mixed states even when conditioning on $\lambda$.

\begin{backgroundbox}[label={Box:FWinfo}]{Frenkel-Weiner theorem for informational restrictions}
It is not an accident that the classical and quantum sets of correlations in the $(2,1,2)$ scenario are identical. In fact, they are identical whenever  Bob only has a single input ($|Y|=1$) \cite{Manna2026}. It can be thought of as the theorem of Frenkel and Weiner, discussed in Sec.~\hyperref[sec:FWtheorem]{II-B-1}, but now for informationally restricted scenarios. Conveniently, the proof is more straightforward than in the original dimension-restricted setting.

\medskip

Let us fix the prior of Alice. In a quantum model, she  generates $\rho_x$ and sends it to Bob who applies the measurement $\{M_{b}\}_b$ yielding $p(b|x)=\tr(\rho_x M_b)$. The associated guessing probability is denoted $P_g$. Now, to classically reproduce the correlations, we need only to let Alice perform Bob's measurement in her lab. This is possible because Bob has no input. Once Alice has obtained the outcome $b$, she sends it to Bob as a classical message and he just outputs it as his outcome. The question is whether the guessing probability for the classical message is still no larger than $P_g$. The answer is affirmative because the states sent by Alice in the classical protocol are $\tilde{\rho}_x=\sum_b \tr(\rho_x M_b)\ketbra{b}{b}$. For the guessing probability, we can adopt the view that the quantum states, $\rho_x$, first pass through the measurement $\{M_b\}_b$ and then through the optimal state discrimination measurement for the post-measurement classical states. Compared to $P_g$, this means imposing an extra constraint on the optimal measurement for discriminating $\{\rho_x\}_x$. Hence, the guessing probability in the classical protocol can be no larger than in the quantum one. The classical simulation therefore succeeds. 
\end{backgroundbox}

We have seen that $d$-dimensional quantum systems have at most $\log_2 d$ bits of information. If the prior is uniform and $d<|X|$, this corresponds to $P_g\leq \frac{d}{|X|}$. A central conceptual question becomes whether there exist correlations that cost no more information than that corresponding to  $P_g=\frac{d}{|X|}$ but which cannot be realised with $d$-dimensional systems. The answer is affirmative: it was first shown for quantum theory \cite{Tavakoli2020informationally} and later also for classical models \cite{Svegborn2026}. Focusing on the quantum case, we do not discuss explicit examples here but instead aim to qualitatively explain this phenomenon. To this end, let us make a more general connection between informational restrictions and entanglement-assisted PM scenarios. Back in Sec.~\hyperref[sec:denseCoding]{III-C-2} we proved in Eq.~\eqref{DCSN} that if Alice's input alphabet is $|X|$ and she sends $d$-dimensional quantum messages to Bob, with all total states having  Schmidt number no larger than $r$, then the success probability of Bob guessing her input is no larger than $P_g\leq \frac{dr}{|X|}$. On the one hand, for a classical protocol in the entanglement-assisted PM (EAPM) scenario, the Schmidt number is $r=1$ because half the total state is classical. On the other hand, a quantum protocol for the EAPM scenario would allow any Schmidt number, which due to the communication channel is limited as $r=d$. Consequently, a classical (quantum) communication in the EAPM scenario effectively corresponds to a communication in the PM scenario that costs at most a guessing probability of $\frac{d}{|X|}$ ($\frac{d^2}{|X|}$). This means that every proof of entanglement-assisted communication with $d$-dimensional classical systems  outperforming  communication in the PM scenario with $d$-dimensional quantum systems implies that informationally restricted quantum correlations can outperform their dimension-restricted counterparts. Such examples are known, as mentioned for instance in Sec.~\hyperref[sec:adaptive]{III-A-2}. Moreover, as mentioned in Sec.~\hyperref[sec:denseCoding]{III-C-1} and~\ref{sec:Dimensionality_and_certification}, this relaxation of entanglement-assisted classical or quantum protocols into informationally restricted protocols paves the way for bounding correlations associated with the former through techniques developed for the latter. This can be viewed as an independent application of informationally restricted quantum correlations.

Let us now discuss tools for bounding the set of informationally restricted correlations. In the classical case, the space of correlations is a polytope which can be characterised through an LP  (see Ref.~\cite{Tavakoli2022informationally} for details). We focus on the quantum case and start with a search for inner approximations (lower bounds) in the spirit of the seesaw heuristic from Sec.~\hyperref[sec:seesaw]{II-B-5}. While in the dimension-restricted scenario, for example, the seesaw included iterative optimisation over states and measurements respectively, it may  not be immediately clear how to perform the analogous optimisation of a witness $W = \sum_{x,y,b} c_{bxy} \tr(\rho_x M_{b|y})$ over states when they are subjected to an informational restriction. Indeed, the guessing probability, which now acts as a constraint, is itself  an SDP quantity \eqref{eq:guessprob_iformation_restricted}. The key idea \cite{Tavakoli2022informationally} for dealing with this  is to represent the guessing probability in its dual formulation. The dual SDP is 
\begin{align}\nonumber
P_g=\min_\sigma & \quad \tr(\sigma)\\
\text{s.t} &\quad \sigma\succeq p(x)\rho_x\quad \forall x.
\end{align}
We will include the auxiliary variable $\sigma$ in the SDP. Its job is to impose the information restriction through the semidefinite constraint $\sigma\succeq p(x)\rho_x$. For our choice of prior, the assumed information bound can now be introduced in the SDP as the linear constraint $\tr(\sigma)\leq G$.  Note that this type of approach can be used also for other problems where linear quantities are optimised over constraints on a quantity that itself is SDP-computable.

The explicit use of  $\sigma$ is also crucial for constructing a hierarchy of SDP relaxations for the set of quantum correlations. The optimisation of an arbitrary witness can  be expressed as 
\begin{align}\label{SDP_primal}
\max_{\rho_x, \sigma, M_{b|y}} \quad & \sum_{x,y,b} c_{bxy} \tr(\rho_x M_{b|y}) \nonumber \\ \nonumber
\text{s.t.} \quad & \rho_x - \rho_x^2 \succeq 0, \quad \quad \quad\quad\quad \tr(\rho_x) = 1, \\ \nonumber
& \sigma - p(x) \rho_x \succeq 0,  \:\: \quad\quad\quad \tr(\sigma) \leq G, \\ 
& M_{b|y} M_{b'|y} = \delta_{bb'} M_{b|y}, \quad\:\: \sum_b M_{b|y} = \iden,
\end{align}
where notably the Hilbert space dimension is unrestricted. Upper bounds on this problem can be obtained by means of SDP relaxation. To this end, let $J_{k}$ denote the set of all monomials up to length $k$ generated by the list of operators $L=\{\mathds{1},\sigma, \{\rho_x\}_x, \{M_{b|y}\}_{b,y}\}$. At a chosen relaxation step $k$, we construct a moment matrix $\Gamma^k$ indexed by the monomials $u, v \in J_{k}$ with entries $(\Gamma^k)_{u,v} = \tr(u^\dagger v)$. Additionally, as in Sec.~\hyperref[Sub:energy-restrictions]{IV-C}, we construct localising matrices to enforce our specific constraints; for example, the constraint $\sigma - p(x)\rho_x \succeq 0$ is enforced via a localising matrix $\Upsilon_k^{(\sigma - p(x)\rho_x)}$ with entries $(\Upsilon_k^{(\sigma - p(x)\rho_x)})_{u,v} = \tr\left(u^\dagger (\sigma - p(x)\rho_x) v\right)$ which is required to satisfy $\Upsilon_k^{(\sigma - p(x)\rho_x)} \succeq 0$. The outer relaxation of the quantum set is then computed via the SDP: 
\begin{align}
\max_{\Upsilon_k, \Gamma^k} \quad & \sum_{x,y,b} c_{bxy} \Gamma^k_{\rho_x, M _{b|y}} \nonumber \\ \nonumber
\text{s.t.} \quad & \Gamma^k\succeq 0,\quad \Gamma^k_{\mathds{1}, \rho_x} = 1, \quad \Gamma^k_{\mathds{1}, \sigma} \le G, \\ \nonumber
& \Upsilon_k^{( \rho_x - \rho_x^2)} \succeq 0, \quad \Upsilon_k^{(\sigma - p(x)\rho_x)} \succeq 0,\nonumber\\ \nonumber &\Upsilon_k^{ (G\iden - \sigma)} \succeq 0, \\ 
& \Gamma^k_{u,v} = \Gamma^k_{u',v'} \quad \text{if } \tr(u^\dagger v) = \tr(u'^\dagger v').
\end{align}
The last condition ensures that operational substitution rules are respected: for example, the idempotency of projectors and the cyclicity of the trace impose linear relations among the elements of $\Gamma^k$. As mentioned already, as $k$ increases, this constructs a sequence of monotonically tightening semidefinite relaxations that bound the informationally restricted quantum set from the outside \cite{Tavakoli2022informationally}; however, it is not known to converge.

Note that by replacing the positivity constraints $\rho_x-\rho_x^2\succeq0$ by the strict polynomial constraint $\rho_x-\rho_x^2=0$, the hierarchy now approximates the pure quantum set. Furthermore, allowing for commutativity of all operators, one can also approximate the classical set. 

So far, our focus has been restrictions on the amount of information Alice can send about her input $x$. We now discuss the interest in considering information about other quantities, which we can represent as functions of her input, namely $f(x)$. As discussed in Ref.~\cite{InfoContextuality}, appropriate choices of $f$ boil the informationally restricted framework down to scenarios for testing preparation contextuality. A brief introduction to this notion is given in Box~\ref{Box:Operational_Otological}. Tests of preparation contextuality can be cast in the PM scenario and they rely on Alice's states obeying certain equivalence relations. Specifically,  certain combinations of her states must be perfectly indistinguishable. An example is based on the alphabet $(|X|,|Y|,|B|)=(4,2,2)$ and the usual random access code task of Sec.~\hyperref[sec:QRAC]{II-B-3}, but with the modification that the only constraint on Alice's states is that the parity $x_1 \oplus x_2$ must be hidden. For uniformly distributed inputs, this means $\rho_{00}+\rho_{11}=\rho_{01}+\rho_{10}$ in a quantum model. This condition is equivalent to permitting only a minimal guessing probability for Alice's states with respect to the quantity $f(x)=x_1\oplus x_2$. In other words, it imposes $I_{\min}\left(f(X):B\right)=0$. Tests of preparation contextuality can include many more equivalence relations between Alice's states (see e.g.~Ref.~\cite{Spekkens2009}) and each of these can be associated with a specific $f$ and a corresponding minimal guessing probability. It is then possible to use the SDP hierarchy for informationally restricted correlations to bound the set of preparation contextual quantum correlations from the exterior, as well as to quantify how much the indistinguishability must be violated in order to explain correlations in one theory with those in a more restricted  theory \cite{InfoContextuality}. Notably, different SDP hierarchies for addressing this problem have been proposed  without adopting the informationally restricted view of preparation contextuality \cite{Chaturvedi2021}.

\begin{backgroundbox}[label={Box:Operational_Otological}]{Preparation contextuality}

An operational theory describes preparations $P$, measurements $M$, and outcome probabilities $p(k|P,M)$. Two preparations are said to be operationally equivalent if they give identical statistics for every measurement
\begin{equation}
p(k|P,M)=p(k|P',M) \qquad \forall\; M,k.
\end{equation}
 A preparation $P$ is represented by a distribution $\mu(\lambda|P)$ over the underlying ontic states $\lambda$. A model is said to be \emph{preparation noncontextual} if operationally equivalent preparations are represented identically at the ontological level:
\begin{equation}
    P \simeq P' \quad \Longrightarrow \quad \mu(\lambda|P)=\mu(\lambda|P').
\end{equation}
If no such representation exists, the model is said to exhibit \emph{preparation contextuality}. See~\cite{Spekkens2005} for an introduction to operational theories and generalised contextuality.
\end{backgroundbox}

Lastly, in a broader perspective one can view informational restrictions as an assumption to which several other semi-device-independent assumptions can be related; see the discussion in Ref.~\cite{Pauwels2025informationcapacity}. We have already seen that a dimension restriction implies an information restriction [see Eq.~\eqref{dimbound}]. We have also seen that such communication when assisted by entanglement (for both  classical and quantum channels) can be related to an informational restriction. In the same spirit, one can consider the vacuum overlap restriction discussed in Sec.~\hyperref[Sub:energy-restrictions]{IV-C}. If one assumes pure states, the optimal information bound associated with $n$ equally likely states with non-vacuum component at most $\omega$ is
\begin{equation}
    P_g(\omega)\leq \frac{1}{n}\left(\sqrt{\omega(n-1)}+\sqrt{1-\omega}\right)^2,
\end{equation}
for $0\leq \omega\leq \frac{n-1}{n}$. Similarly, for real overlaps satisfying $\braket{\psi_x}{\psi_{x'}}\geq t$ for all $x\ne x'$, with $0\le t\le1$ (as in Sec.~\hyperref[Sub:overlaps]{IV-E}) one has
\begin{equation}
    P_g(t)\leq \frac{1}{n^2}\left[ (n-1)\sqrt{1-t}+\sqrt{1+t(n-1)}\right]^2.
\end{equation}
To find this, the best choice is to consider equiangular states and the so-called pretty-good measurement \cite{Montanaro2007}. Furthermore, one can consider the case of the fidelity assumption and almost-dimension assumption (see sections~\hyperref[Sub:fidelity-constraints]{IV-D} and \hyperref[Sub:almost-dimension]{IV-B} respectively) and arrive at the bound 
\begin{equation}
    P_g(\epsilon)\leq P_g+(1-2P_g)\epsilon +2\sqrt{P_g(1-P_g)\epsilon(1-\epsilon)},
\end{equation}
This expression applies for $0\leq \epsilon\leq1-P_g$. For larger $\epsilon$, the bound is $P_g(\epsilon)\leq 1$. Here $\epsilon$ is either the distrust parameter or the dimensional deviation parameter, and $P_g$ is the guessing probability, in either the fidelity or almost-dimension case, associated with $\epsilon=0$. This bound is not tight for the fidelity assumption (since the target states can be arbitrary) but is conjectured to be tight for almost-dimension, in which case one has simply $P_g=\frac{d}{n}$. 

The reason that informational restrictions can be related to all these other communication constraints is that the assumption is qualitatively different. Dimension, almost dimension, distrust and vacuum overlap are all about bounding the weight of the states on various subspaces of Hilbert space. Contrasting this, informational restrictions do not favour any particular subspace but instead concern the capacity of the states to perform a task (specifically state discrimination). Such a capacity can always exhibit a meaningful relation to other classes of states established by alternative assumptions.

\section{Outlook, Challenges and Open Problems}\label{S:conclusions} 

A central challenge of quantum information science is to understand the conditions under which classical descriptions might fail and quantum advantages might emerge. Within this context, the aim of this tutorial is to provide a comprehensive and self-contained overview of quantum prepare-and-measure (PM) scenarios, establishing their central role not only in  understanding the boundary between classical and quantum predictions but also in providing quantum advantages across a variety of communication protocols and serving as a practical tool to certify various properties of quantum devices. While fully device-independent (DI) protocols based on Bell nonlocality represent the most stringent benchmark, by providing guarantees without relying on trusted hardware, their dependence on distributed and high-fidelity entanglement and  highly efficient detectors remains a formidable experimental challenge. PM scenarios offer a natural and practical middle ground, bridging the gap between device-dependent protocols and fully DI setups by enforcing realistic physical constraints on transmitted messages and in some cases without requiring any entanglement at all.

A key strength of PM frameworks is indeed their flexibility to accommodate a variety of communication constraints that accurately reflect physical and experimental limitations.  While the standard and most employed formulation bounds the Hilbert space dimension of the message to establish dimension witnesses and certify non-classicality, practical implementations frequently require alternative physical assumptions. For instance, energy constraints bounding the average photon number or mean energy naturally extend protocols to continuous-variable photonic systems with infinite-dimensional Hilbert spaces. Remarkably, this yields quantum advantages even within fully Gaussian regimes using coherent states and homodyne detection, a capability strictly impossible in Bell tests. Similarly, overlap constraints limit the mutual fidelity or distinguishability between prepared states, ensuring robust certification despite environmental noise or mode mismatch. Overall, this highlights a fundamental conceptual shift: whereas Bell nonlocality forces experimental setups to be meticulously adapted to rigid theoretical conditions, prepare-and-measure scenarios do the exact opposite by tailoring the theoretical framework directly to the existing physical platform.

This operational versatility translates into the rich landscape of practical applications detailed throughout this tutorial. In quantum random access codes, dimension-restricted quantum states allow for classical information to be encoded and retrieved with a strictly higher success probability than is possible classically. When applied to distributed tasks, quantum nonclassicality can yield significant improvements in communication complexity, most notably the exponential efficiency gains exemplified by the hidden-matching problem. In cryptographic settings, the PM scenario naturally describes semi-device-independent quantum key distribution and randomness generation protocols.  Beyond quantum communication, the PM framework has also found fertile ground in certification tasks, advancing robust self-testing protocols and offering new ways to verify quantum resources, ranging from the non-projective and entangled nature of measurement devices to the presence of entanglement or nonstabiliserness in the prepared states.

When shared entanglement is allowed between the preparation and measurement devices, the PM framework expands even further, allowing conceptual and practical links to Bell nonlocality while also unlocking new features like adaptivity, where measurements adapt in real time to incoming signals across multi-party network architectures. On the foundational side, entanglement-assisted PM scenarios serve as the natural operational setting for paradigmatic informational principles designed to explain the limits of quantum correlations, such as information causality and the non-triviality of communication complexity, both of which are natively cast as prepare-and-measure tasks. In terms of applications, the entanglement-assisted extension underpins paradigmatic quantum communication protocols like superdense coding, while offering certification tools to verify measurement devices and shared quantum resources.

Complementing these theoretical advances, experimental realisations across diverse physical platforms have demonstrated the practical viability of the prepare-and-measure framework. As reviewed in this tutorial, experimental implementations --spanning discrete-variable photonic systems, continuous-variable optical setups, and integrated quantum photonic chips-- have successfully realised high-dimensional quantum random access codes, semi-device-independent key distribution and randomness generation, and robust dimension witnessing. By achieving certified quantum advantages without the stringent loophole-free requirements of Bell tests, these experimental milestones demonstrate that prepare-and-measure protocols provide a mature and feasible approach for present-day quantum technologies.

Bridging foundational principles and real-world applications ultimately relies on a powerful computational toolkit to analyse and bound potential quantum advantages. While membership in classical message polytopes can be formulated as a linear program, whose size may grow exponentially with the scenario alphabets, characterising optimal quantum strategies requires more sophisticated computational approaches. To construct inner approximations of the set of quantum correlations, the tutorial details heuristic algorithms like the seesaw method, which alternates between optimising state preparations through eigenvalue problems and measurement operators via semidefinite programming to uncover achievable quantum strategies. To establish outer approximations, the tutorial introduces semidefinite programming hierarchy relaxations based on moment matrices. To overcome the exponentially scaling bottlenecks inherent to these hierarchies, we also highlight the role of symmetry reduction, for instance, employing representation theory and specialised computational packages to render previously intractable high-dimensional scenarios computationally feasible.

Despite substantial progress across conceptual, practical, and experimental fronts, quantum prepare-and-measure scenarios remain a dynamic and rapidly evolving field with key questions left to address. To map out the road ahead and inspire future developments spanning quantum foundations, quantum communication, and device certification, in the following we highlight the challenges and open directions currently facing the field, organised into three broad themes:
\begin{enumerate}
    \item \emph{Fundamental concepts}: the structure of classical, quantum and entanglement-assisted correlation sets, and the resources needed to simulate them.
    \item \emph{Computational methods}: convex programming techniques for characterising quantum sets and tight witnesses beyond small scenarios.
    \item \emph{Applications}: robust self-testing, certification and cryptographic protocols.  
\end{enumerate}

\begin{center}
\small{\emph{\textbf{Fundamental concepts}}}
\end{center}

\begin{enumerate}
    \item Consider dimension-restricted PM scenarios and find families of facets of the classical polytope (Sec.~\hyperref[sec:ClassicalPolytope]{II-A-1}) that are valid in any dimension and admit a corresponding quantum violation.

    \item Consider some monotonically increasing and unbounded sequence of dimensions, $\{d_k\}_k$, associated with a dimension-restricted PM scenario (with shared randomness, see Sec.~\hyperref[sec:ClassicalModels]{II-A}). In the limit of large $k$, is it possible to find quantum correlations that do not belong to the classical polytope when the associated quantum states have any non-zero visibility? Specifically, if the states undergo the transformation $\rho_x\rightarrow v\rho_x +\frac{1-v}{d_k}\mathds{1}$ associated with a depolarisation channel, can we for any given $v>0$ find a sufficiently large $k$ to generate nonclassical correlations?

     \item Consider that Alice sends some set of states $\{\rho_x\}_x$ to Bob in a dimension-restricted PM scenario with shared randomness. Find criteria for proving that for any set of measurements performed by Bob, the resulting correlations will always admit a classical model. In other words, the states cannot generate quantum correlations in the PM scenario.

     \item Simulating all ququart PM statistics without error requires an unbounded classical message alphabet, even with unlimited shared randomness~\cite{deGois2026infinite}. Can one construct explicit sequences of finite PM tasks for which the classical simulation cost grows without bound while the quantum message dimension remains fixed at four, and derive quantitative lower bounds on this cost in terms of the number of settings and the allowed simulation error? In particular, how rapidly must the error decrease as the number of settings increases for the classical cost to diverge?

     \item Consider PM scenarios with classically uncorrelated devices (Sec.~\hyperref[sec:NoSR]{II-E}) but without assuming that the rounds are i.i.d. How can we best model the influence of reasonable (non-adversarial) memory effects?

    \item In Sec.~\hyperref[subsubsec:networkquantummessage]{II-B-7}, the activation of nonclassicality in the prepare-and-broadcast (PB) scenario was established relative to a restricted family of PM tests. Now, we would like to find a set of preparations $\{\rho_x\}_x$ whose PM correlations admit a classical $d$-value model (with shared randomness) for all POVMs, yet whose PB correlations violate the $\mathcal{CQ}$ bound. More generally, which ensembles can be activated by some network? Any ensemble that admits a decomposition $\rho_x=\sum_\lambda q(\lambda)\sum_{m=1}^{d}p(m|x,\lambda) \sigma_{m,\lambda}$, which can be decided by an SDP, remains classical in every network in which the parties receiving Alice's message share randomness with her, since they can re-prepare $\sigma_{m,\lambda}$ themselves. Is the converse true, in analogy with the network certification of all entangled states~\cite{Buscemi2012,Bowles2018entanglement}?

    \item In the PM scenario, two classical bits assisted by shared randomness reproduce every qubit protocol. How much classical communication must Alice send to a quantum broadcaster to reproduce the PB correlations of a $d$-dimensional message? 
    
    \item Characterise the correlation sets for the prepare-and-broadcast scenarios defined in Sec.~\hyperref[subsubsec:networkquantummessage]{II-B-7} extended to $N>2$ receivers. In particular, derive full-correlator inequalities that tightly bound the classical set in these scenarios.
    
    \item Consider a PB scenario with $N$ receivers. Can two or more receivers simultaneously observe nonclassical PM correlations in their marginals? 
     
    \item We have seen in Sec.~\hyperref[sec:adaptive]{III-A-2} that shared entanglement can reduce communication complexity and that adaptive protocols can outperform those based on Bell inequality violations followed by classical communication. Is it possible to design a communication complexity problem such that in the limit of many inputs adaptive protocols have an unbounded performance advantage over non-adaptive quantum protocols?

    \item We have seen in Sec.~\hyperref[sec:denseCoding]{III-C-1} that correlations obtained from  unitary encodings on a maximally entangled two-qubit state are equivalent to those obtained from sending real-valued ququarts. Are there other such one-to-one mappings between entanglement-assisted and standard PM scenarios?

    \item Consider the entanglement-assisted PM scenario with quantum messages of dimension $d$ (Sec.~\hyperref[sec:denseCoding]{III-C-1}). In order to generate the full set of quantum correlations, is it possible to restrict the shared entanglement to some finite dimension? 

    \item As seen in Sec.~\hyperref[sec:infocausality]{III-B-1}, information causality is an information-theoretic principle designed to bound physical correlations, natively rooted in a bipartite prepare-and-measure setup. Could network generalisations of the prepare-and-measure framework unlock new foundational principles? Specifically, in prepare-and-broadcast scenarios --where a single sender distributes quantum or classical messages across multiple receivers-- can one derive generalised informational principles that constrain multi-party quantum correlations beyond standard bipartite limits, or rule out post-quantum operational theories in complex network topologies?

    \item While pure-state overlap constraints provide a powerful tool for semi-device-independent certification (Sec.~\hyperref[Sub:overlaps]{IV-E}), experimental preparations inevitably suffer from mixedness. An open challenge is to generalise the overlap-based PM scenario to mixed states,  closing the gap between theory and noisy experimental implementations. 

    \item In Sec.~\hyperref[Sub:information-restriction]{IV-G} we discussed informationally restricted correlations and used the min-entropy as the information quantifier. Are there other  choices of the information quantifier that are well-motivated to consider? If so, what new features do these reveal?

    \item We have seen that correlations achievable in the PM scenario with two-dimensional communication are more restricted than those achievable with sending states containing only one bit of information (in the sense of Sec.~\hyperref[Sub:information-restriction]{IV-G}). Is it possible that states carrying only a finite amount of information (with uniform prior) can generate correlations in the PM scenarios that cannot be reproduced by sending states of any finite dimension?

    \item Develop a systematic method to compare the sets of quantum correlations in PM scenarios under different communication assumptions (Sec.~\ref{sec:non-dimension_constraints}). When does one assumption lead to more powerful correlations than another? Can the various assumptions, under appropriate limitations, be organised into a hierarchy according to the correlation sets they generate? 

    \item In section~\ref{sec:non-dimension_constraints} we discussed several avenues for semi-device-independent quantum information. Are there also other interesting communication constraints that ought to be studied? 

\begin{center}
\small{\emph{\textbf{Computational methods}}}
\end{center}

    \item In Appendix~\ref{AppFastLP} we discuss an LP for determining whether a given $p(b|x,y)$ is a member of the classical polytope in dimension-restricted PM scenarios. This LP becomes expensive to compute when both the number of states and the number of measurements are large. Find more efficient ways to either prove membership of or separation from the polytope.

    \item  We have discussed SDP relaxations for outer bounds on the set of dimension-restricted quantum correlations (Sec.~\hyperref[sec:boundingNV]{II-C-2}). Find such relaxations that are tailored for obtaining useful results at reasonable computational cost, without relying on heuristics. 

    \item To bound the set of correlations obtained from $d$-dimensional quantum states assisted by entanglement, there are currently two methods \cite{TavakoliPauwels2021}. One relaxes the problem to informational restrictions and uses SDP hierarchies tailored for the latter problem. The other uses a field-theoretic variant of the NPA hierarchy, which provably converges but appears to be of little practical use. Find a practically more useful and converging hierarchy of SDP relaxations for this problem.

    \item In Sec.~\hyperref[Sub:energy-restrictions]{IV-C}, we discussed PM scenarios where states have a restricted vacuum component. If Alice has $|X|$ inputs, is it sufficient to restrict her states $\rho_x$ to be pure in order to describe the full set of quantum correlations? In the special case of $|X|=2$ the answer is affirmative \cite{VanHimbeeck2017semidevice}. 

    \item Find efficient and practically accurate methods for bounding both the conditional von Neumann entropy and the conditional Shannon entropy in PM scenarios. This is important for cryptographic applications discussed in Sec.~\hyperref[sec:Crypto_overview]{IV-A}.

    \item Is it possible to develop a unified SDP relaxation method that can bound many, if not all, quantum correlation sets arising in the PM scenario under various semi-device-independent assumptions (Sec.~\ref{sec:non-dimension_constraints})? 

    \item We have discussed many different SDP relaxations (e.g., in Sec.~\hyperref[sec:boundingNV]{II-C-2} and Sec.~\hyperref[sec:NPAEACC]{III-A-4}). Increasing the level is the standard way to make a given relaxation more precise. However, this usually means a significant increase in computational cost. Is it possible to achieve a similar tightening of the relaxation by only including a few well-selected monomials? This would require an algorithmic method to select monomials, perhaps most conveniently when given a specific witness. Such an approach could make the evaluation of the SDP relaxations more practical.

\begin{center}
\small{\emph{\textbf{Applications}}}
\end{center}
    
    \item The known protocols for self-testing of non-projective measurements (Sec.~\hyperref[sec:POVMcertification]{II-D-3}) in the PM scenario have poor robustness to noise, and it typically gets worse as the dimension increases. Is it possible to find self-testing protocols with substantially more  noise-robustness? 

    \item Find a scheme for robustly self-testing (Sec.~\hyperref[sec:robust_Self_test]{II-D-2}) any entangled measurement in an appropriate variation of the PM scenario.

    \item How do self-tests in the dimension-restricted PM scenario change when we relax the strict dimension assumption and replace it with the almost-qudit notion from Sec.~\hyperref[Sub:almost-dimension]{IV-B}? How accurately can one certify the quantum devices once the $\epsilon$-leakage is taken into account?     
    
    \item As discussed in Sec.~\hyperref[subsubsec:nonstab]{II-D-4}, certain random access codes, such as the $(3,2)$ qubit code, require nonstabiliser preparations to attain their optimal quantum value. Can we identify other communication tasks that are similarly powered by specific quantum resources? Investigating this in the limit of large input alphabets is of particular interest, as it would enable the semi-device-independent certification of complex multi-qubit resources and many-body properties.

    \item Finite-energy Gottesman--Kitaev--Preskill (GKP) encodings provide multipartite Bell can lead to multipartite Bell violations using homodyne detection with fixed periodic binning~\cite{yang2026}. Can this approach be developed into robust semi-device-independent PM protocols? For preparations within a common two-dimensional GKP code subspace and the same type of homodyne readout, determine the trade-offs between squeezing, optical loss, thermal noise and violations of classical communication bounds. Can the resulting correlations certify useful amounts of randomness, and how do the guarantees change when the preparations have bounded leakage outside the code subspace, as in Sec.~\hyperref[Sub:almost-dimension]{IV-B}?
    
    \item The loss of particles, due to lossy hardware, detector inefficiency or long-distance communication, is a major obstacle in quantum cryptography (see e.g., the discussion of detection inefficiency in Sec.~\ref{sec:limitations-qubit-advantage}). Investigate the robustness to loss under the various assumptions relevant for the PM scenario. Can one design witness tests that permit quantum violations at high loss? Are there fundamental limitations to loss-tolerance in the PM scenario?

    \item Most practical semi-device-independent quantum random number generators (Sec.~\hyperref[sec:Crypto_overview]{IV-A}) currently assume that Alice and Bob share no entanglement. However, even if true at the beginning of the protocol, entanglement can be built up over the rounds. Find practically well-motivated protocols that are robust to shared entanglement.

    \item Quantum key distribution in semi-device-independent settings (Sec.~\ref{sec:non-dimension_constraints}) is currently little explored. Is there a practically appealing pathway for such applications? 

    \item As established in Sec.~\hyperref[subsubsec:nonstab]{II-D-4}, the dimension-restricted PM scenario can be employed as an SDI witness for the nonstabiliserness of prepared states. Can this framework be extended to certify nonstabiliserness in measurement or transformation devices?

\end{enumerate}

\begin{acknowledgments}
We thank Elna Svegborn for providing us with the argument in Box~\ref{Box:statediscriminationfacets} and  the calculations in Appendix~\ref{App:ElnasConstruction}. A.T.~is supported by the Knut and Alice Wallenberg Foundation through the Wallenberg Center for Quantum Technology (WACQT), the Swedish Research Council under Contract No.~2023-03498 and the Swedish Foundation for Strategic Research. A. O. J. and J. B. B. acknowledge financial support from the  EU Horizon Europe (QSNP, grant no. 101114043) and from the Danish National Research Foundation grant bigQ (DNRF 142), VILLUM FONDEN through a research grant (40864). S.Z. acknowledges the Coordenação de Aperfeiçoamento de Pessoal de Nível Superior - Brasil (CAPES) – Finance Code 001 and the support of the Natural Sciences and Engineering Research Council of Canada (NSERC), (CIRTA-P-614385-2026). R. C. acknowledges the Simons Foundation (Grant No. 1023171, R.C.), the Brazilian National Council for Scientific and Technological Development (CNPq, Grants No. 403181/2024-0, and 301687/2025-0), the National Institute of Science and Technology for Applied Quantum Computing through CNPq Process No. 408884/2024-0, the Financiadora de Estudos e Projetos (Grant No. 1699/24 IIF-FINEP) and a guest professorship from the Otto M\o nsted Foundation.
\end{acknowledgments}
\newpage

\bibliography{2-references}

@book{Scarani2019,
  author = {Valerio Scarani},
  title = {{Bell} Nonlocality},
  series = {Oxford Graduate Texts},
  year = {2019},
  publisher = {Oxford University Press},
  isbn = {9780198788416},
  doi = {10.1093/oso/9780198788416.001.0001},
  url = {https://doi.org/10.1093/oso/9780198788416.001.0001},
}

@article{Buscemi2012,
  author = {Buscemi, Francesco},
  title = {All Entangled Quantum States Are Nonlocal},
  journal = {Phys. Rev. Lett.},
  volume = {108},
  number = {20},
  pages = {200401},
  numpages = {5},
  year = {2012},
  month = {May},
  publisher = {American Physical Society},
  doi = {10.1103/PhysRevLett.108.200401},
  url = {https://link.aps.org/doi/10.1103/PhysRevLett.108.200401},
}

@article{Bowles2018entanglement,
  author = {Bowles, Joseph and {\v{S}}upi{\'{c}}, Ivan and Cavalcanti, Daniel and Ac{\'{\i}}n, Antonio},
  title = {Device-Independent Entanglement Certification of All Entangled States},
  journal = {Phys. Rev. Lett.},
  volume = {121},
  number = {18},
  pages = {180503},
  numpages = {6},
  year = {2018},
  month = {Oct},
  publisher = {American Physical Society},
  doi = {10.1103/PhysRevLett.121.180503},
  url = {https://link.aps.org/doi/10.1103/PhysRevLett.121.180503},
}

@misc{Renner2026private,
  author = {Martin J. Renner},
  title = {Private communication},
  year = {2026},
}

@article{Grothendieck1952,
  author = {Grothendieck, Alexander},
  title = {Résumé des résultats essentiels dans la théorie des produits tensoriels topologiques et des espaces nucléaires},
  journal = {Ann. Inst. Fourier},
  volume = {4},
  pages = {73--112},
  year = {1952},
  publisher = {MathDoc/Centre Mersenne},
  doi = {10.5802/aif.46},
  url = {https://doi.org/10.5802/aif.46},
}

@misc{leone2026,
  author = {Lorenzo Leone and Jens Eisert and Salvatore F. E. Oliviero},
  title = {The unbearable hardness of deciding about magic},
  year = {2026},
  eprint = {2602.22330},
  archivePrefix = {arXiv},
  primaryClass = {quant-ph},
  url = {https://arxiv.org/abs/2602.22330},
}

@article{bell1964einstein,
  author = {Bell, John S},
  title = {On the {Einstein} {Podolsky} {Rosen} paradox},
  journal = {Physics Physique Fizika},
  volume = {1},
  number = {3},
  pages = {195},
  year = {1964},
  publisher = {APS},
  doi = {10.1103/PhysicsPhysiqueFizika.1.195},
  url = {https://doi.org/10.1103/PhysicsPhysiqueFizika.1.195},
}

@misc{sarubi2026randomness,
  author = {Tailan S. Sarubi and Mois{\'{e}}s Alves and Vin{\'{i}}cius F. Alves and Santiago Zamora and Diogo H. G. Duarte and A. de Oliveira Junior and Rafael Chaves},
  title = {Randomness Certification and Trade-offs in the Prepare-and-Broadcast Scenario},
  year = {2026},
  eprint = {2608.08329},
  archivePrefix = {arXiv},
  primaryClass = {quant-ph},
  url = {https://arxiv.org/abs/2608.08329},
}

@misc{varela2026predicting,
  author = {Varela, J. M. and Keller, L. L. and de Oliveira Junior, A. and Moreira, D. A. and Chaves, R. and Mac{\^e}do, R. A.},
  title = {Predicting Magic from Very Few Measurements},
  year = {2026},
  eprint = {2602.18939},
  archivePrefix = {arXiv},
  primaryClass = {quant-ph},
  url = {https://arxiv.org/abs/2602.18939},
}

@article{Frenkel2015,
  author = {Frenkel, P{\'e}ter E. and Weiner, Mih{\'a}ly},
  title = {Classical Information Storage in an n-Level Quantum System},
  journal = {Commun. Math. Phys.},
  volume = {340},
  number = {2},
  pages = {563--574},
  year = {2015},
  month = sep,
  publisher = {Springer},
  doi = {10.1007/s00220-015-2463-0},
  url = {https://doi.org/10.1007/s00220-015-2463-0},
}

@article{Li2011,
  author = {Li, Hong-Wei and Yin, Zhen-Qiang and Wu, Yu-Chun and Zou, Xu-Bo and Wang, Shuang and Chen, Wei and Guo, Guang-Can and Han, Zheng-Fu},
  title = {Semi-device-independent random-number expansion without entanglement},
  journal = {Phys. Rev. A},
  volume = {84},
  number = {3},
  pages = {034301},
  numpages = {4},
  year = {2011},
  month = {Sep},
  publisher = {American Physical Society},
  doi = {10.1103/PhysRevA.84.034301},
  url = {https://link.aps.org/doi/10.1103/PhysRevA.84.034301},
}

@article{Hendrych2012,
  author = {Hendrych, Martin and Gallego, Rodrigo and Mi{\v{c}}uda, Michal and Brunner, Nicolas and Ac{\'i}n, Antonio and Torres, Juan P.},
  title = {Experimental estimation of the dimension of classical and quantum systems},
  journal = {Nat. Phys.},
  volume = {8},
  number = {8},
  pages = {588--591},
  year = {2012},
  month = aug,
  doi = {10.1038/nphys2334},
  url = {https://doi.org/10.1038/nphys2334},
}

@article{eftaxias2023advantages,
  author = {Eftaxias, Giorgos and Weilenmann, Mirjam and Colbeck, Roger},
  title = {Advantages of multicopy nonlocality distillation and its application to minimizing communication complexity},
  journal = {Phys. Rev. Lett.},
  volume = {130},
  number = {10},
  pages = {100201},
  year = {2023},
  publisher = {APS},
  doi = {10.1103/PhysRevLett.130.100201},
  url = {https://doi.org/10.1103/PhysRevLett.130.100201},
}

@book{boyd2004convex,
  author = {Boyd, Stephen and Vandenberghe, Lieven},
  title = {Convex Optimization},
  year = {2004},
  publisher = {Cambridge University Press},
  isbn = {9780521833783},
  doi = {10.1017/CBO9780511804441},
  url = {https://doi.org/10.1017/CBO9780511804441},
}

@article{pironio2005lifting,
  author = {Pironio, Stefano},
  title = {Lifting {Bell} inequalities},
  journal = {J. Math. Phys.},
  volume = {46},
  number = {6},
  pages = {062112},
  year = {2005},
  month = jun,
  publisher = {AIP Publishing},
  doi = {10.1063/1.1928727},
  url = {https://doi.org/10.1063/1.1928727},
}

@article{Muhammad2014,
  author = {Muhammad, Sadiq and Tavakoli, Armin and Kurant, Maciej and Paw\l{}owski, Marcin and {\.{Z}}ukowski, Marek and Bourennane, Mohamed},
  title = {Quantum Bidding in Bridge},
  journal = {Phys. Rev. X},
  volume = {4},
  number = {2},
  pages = {021047},
  numpages = {10},
  year = {2014},
  month = {Jun},
  publisher = {American Physical Society},
  doi = {10.1103/PhysRevX.4.021047},
  url = {https://link.aps.org/doi/10.1103/PhysRevX.4.021047},
}

@article{grossi2023one,
  author = {Grossi, R and Brugger, Lucas L and Rizzuti, B. F. and Duarte, C},
  title = {One Hundred Years Later: {Stern-Gerlach} Experiment and Dimension Witnesses},
  journal = {Rev. Bras. Ensino Fis.},
  volume = {45},
  pages = {e20220227},
  year = {2023},
  publisher = {SciELO},
  doi = {10.1590/1806-9126-RBEF-2022-0227},
  url = {https://doi.org/10.1590/1806-9126-RBEF-2022-0227},
}

@article{Young1804,
  author = {Young, Thomas},
  title = {I. {The} {Bakerian} Lecture. {Experiments} and calculations relative to physical optics},
  journal = {Phil. Trans. R. Soc. Lond.},
  volume = {94},
  pages = {1--16},
  year = {1804},
  month = {12},
  issn = {0261-0523},
  doi = {10.1098/rstl.1804.0001},
  url = {https://doi.org/10.1098/rstl.1804.0001},
}

@incollection{Gerlach1989,
  author = {Gerlach, Walther and Stern, Otto},
  editor = {Bachmann, Hans-Reinhard and Rechenberg, Helmut},
  title = {{Der experimentelle Nachweis der Richtungsquantelung im Magnetfeld}},
  booktitle = {Walther Gerlach (1889--1979)},
  pages = {26--29},
  year = {1989},
  publisher = {Springer Berlin Heidelberg},
  doi = {10.1007/978-3-642-74813-4_4},
  url = {https://doi.org/10.1007/978-3-642-74813-4_4},
}

@incollection{Wheeler1978,
  author = {Wheeler, John Archibald},
  title = {The ``Past'' and the ``Delayed-Choice'' Double-Slit Experiment},
  booktitle = {Mathematical Foundations of Quantum Theory},
  pages = {9--48},
  year = {1978},
  publisher = {Elsevier},
  doi = {10.1016/B978-0-12-473250-6.50006-6},
  url = {https://www.sciencedirect.com/science/chapter/edited-volume/pii/B9780124732506500066},
}

@article{popescu1994quantum,
  author = {Popescu, Sandu and Rohrlich, Daniel},
  title = {Quantum nonlocality as an axiom},
  journal = {Found. Phys.},
  volume = {24},
  number = {3},
  pages = {379--385},
  year = {1994},
  month = mar,
  publisher = {Springer},
  doi = {10.1007/BF02058098},
  url = {https://doi.org/10.1007/BF02058098},
}

@article{chaves2018causal,
  author = {Chaves, Rafael and Lemos, Gabriela Barreto and Pienaar, Jacques},
  title = {Causal modeling the delayed-choice experiment},
  journal = {Phys. Rev. Lett.},
  volume = {120},
  number = {19},
  pages = {190401},
  year = {2018},
  month = may,
  publisher = {American Physical Society},
  doi = {10.1103/PhysRevLett.120.190401},
  url = {https://link.aps.org/doi/10.1103/PhysRevLett.120.190401},
}

@article{rossi2017restrictions,
  author = {Rossi Jr, R},
  title = {Restrictions for the causal inferences in an interferometric system},
  journal = {Phys. Rev. A},
  volume = {96},
  number = {1},
  pages = {012106},
  year = {2017},
  publisher = {American Physical Society},
  doi = {10.1103/PhysRevA.96.012106},
  url = {https://link.aps.org/doi/10.1103/PhysRevA.96.012106},
}

@article{Ambainis2002,
  author = {Ambainis, Andris and Nayak, Ashwin and Ta-Shma, Amnon and Vazirani, Umesh},
  title = {Dense quantum coding and quantum finite automata},
  journal = {J. ACM},
  volume = {49},
  number = {4},
  pages = {496--511},
  numpages = {16},
  year = {2002},
  month = jul,
  publisher = {Association for Computing Machinery},
  doi = {10.1145/581771.581773},
  url = {https://doi.org/10.1145/581771.581773},
}

@inproceedings{Ambainis1999,
  author = {Ambainis, Andris and Nayak, Ashwin and Ta-Shma, Amnon and Vazirani, Umesh},
  title = {Dense quantum coding and a lower bound for 1-way quantum automata},
  booktitle = {Proceedings of the Thirty-First Annual ACM Symposium on Theory of Computing},
  series = {STOC '99},
  pages = {376--383},
  numpages = {8},
  year = {1999},
  publisher = {Association for Computing Machinery},
  isbn = {1581130678},
  doi = {10.1145/301250.301347},
  url = {https://doi.org/10.1145/301250.301347},
}

@misc{Ambainis2008,
  author = {Andris Ambainis and Debbie Leung and Laura Mancinska and Maris Ozols},
  title = {Quantum random access codes with shared randomness},
  year = {2008},
  eprint = {0810.2937},
  archivePrefix = {arXiv},
  primaryClass = {quant-ph},
  url = {https://arxiv.org/abs/0810.2937},
}

@article{Tavakoli2015,
  author = {Tavakoli, Armin and Hameedi, Alley and Marques, Breno and Bourennane, Mohamed},
  title = {Quantum Random Access Codes Using Single $d$-Level Systems},
  journal = {Phys. Rev. Lett.},
  volume = {114},
  number = {17},
  pages = {170502},
  numpages = {5},
  year = {2015},
  month = apr,
  publisher = {American Physical Society},
  doi = {10.1103/PhysRevLett.114.170502},
  url = {https://link.aps.org/doi/10.1103/PhysRevLett.114.170502},
}

@article{Anwer2020,
  author = {Anwer, Hammad and Muhammad, Sadiq and Cherifi, Walid and Miklin, Nikolai and Tavakoli, Armin and Bourennane, Mohamed},
  title = {Experimental Characterization of Unsharp Qubit Observables and Sequential Measurement Incompatibility via Quantum Random Access Codes},
  journal = {Phys. Rev. Lett.},
  volume = {125},
  number = {8},
  pages = {080403},
  numpages = {7},
  year = {2020},
  month = aug,
  publisher = {American Physical Society},
  doi = {10.1103/PhysRevLett.125.080403},
  url = {https://link.aps.org/doi/10.1103/PhysRevLett.125.080403},
}

@article{Sun2026,
  author = {Sun, Yong-Nan and Ma, Meng-Yun and Su, Qi-Ping and Sun, Zhe and Yang, Chui-Ping and Nori, Franco},
  title = {Experimental Certification of Ensembles of High-Dimensional Quantum States with Independent Quantum Devices},
  journal = {Phys. Rev. Lett.},
  volume = {136},
  number = {6},
  pages = {060804},
  year = {2026},
  month = feb,
  publisher = {American Physical Society},
  doi = {10.1103/q85z-wl38},
  url = {https://link.aps.org/doi/10.1103/q85z-wl38},
}

@article{Carmeli2020,
  author = {Carmeli, Claudio and Heinosaari, Teiko and Toigo, Alessandro},
  title = {Quantum random access codes and incompatibility of measurements},
  journal = {EPL},
  volume = {130},
  number = {5},
  pages = {50001},
  year = {2020},
  month = jun,
  publisher = {IOP Publishing},
  doi = {10.1209/0295-5075/130/50001},
  url = {https://doi.org/10.1209/0295-5075/130/50001},
}

@article{Kaniewski2016,
  author = {Kaniewski, J{\k e}drzej},
  title = {Analytic and Nearly Optimal Self-Testing Bounds for the {Clauser-Horne-Shimony-Holt} and {Mermin} Inequalities},
  journal = {Phys. Rev. Lett.},
  volume = {117},
  number = {7},
  pages = {070402},
  year = {2016},
  month = aug,
  publisher = {American Physical Society},
  doi = {10.1103/PhysRevLett.117.070402},
  url = {https://link.aps.org/doi/10.1103/PhysRevLett.117.070402},
}

@article{Cleve1997,
  author = {Cleve, Richard and Buhrman, Harry},
  title = {Substituting quantum entanglement for communication},
  journal = {Phys. Rev. A},
  volume = {56},
  number = {2},
  pages = {1201--1204},
  year = {1997},
  month = aug,
  publisher = {American Physical Society},
  doi = {10.1103/PhysRevA.56.1201},
  url = {https://link.aps.org/doi/10.1103/PhysRevA.56.1201},
}

@article{brask2017bell,
  author = {Brask, Jonatan Bohr and Chaves, Rafael},
  title = {{Bell} scenarios with communication},
  journal = {J. Phys. A: Math. Theor.},
  volume = {50},
  number = {9},
  pages = {094001},
  year = {2017},
  publisher = {IOP Publishing},
  doi = {10.1088/1751-8121/aa5840},
  url = {https://doi.org/10.1088/1751-8121/aa5840},
}

@article{chaves2015unifying,
  author = {Chaves, Rafael and Kueng, Richard and Brask, Jonatan Bohr and Gross, David},
  title = {Unifying framework for relaxations of the causal assumptions in {Bell}’s theorem},
  journal = {Phys. Rev. Lett.},
  volume = {114},
  number = {14},
  pages = {140403},
  year = {2015},
  publisher = {APS},
  doi = {10.1103/PhysRevLett.114.140403},
  url = {https://doi.org/10.1103/PhysRevLett.114.140403},
}

@article{yu2019realization,
  author = {Yu, Shang and Sun, Yong-Nan and Liu, Wei and Liu, Zhao-Di and Ke, Zhi-Jin and Wang, Yi-Tao and Tang, Jian-Shun and Li, Chuan-Feng and Guo, Guang-Can},
  title = {Realization of a causal-modeled delayed-choice experiment using single photons},
  journal = {Phys. Rev. A},
  volume = {100},
  number = {1},
  pages = {012115},
  year = {2019},
  publisher = {APS},
  doi = {10.1103/PhysRevA.100.012115},
  url = {https://doi.org/10.1103/PhysRevA.100.012115},
}

@article{polino2019device,
  author = {Polino, Emanuele and Agresti, Iris and Poderini, Davide and Carvacho, Gonzalo and Milani, Giorgio and Lemos, Gabriela Barreto and Chaves, Rafael and Sciarrino, Fabio},
  title = {Device-independent test of a delayed choice experiment},
  journal = {Phys. Rev. A},
  volume = {100},
  number = {2},
  pages = {022111},
  year = {2019},
  publisher = {APS},
  doi = {10.1103/PhysRevA.100.022111},
  url = {https://doi.org/10.1103/PhysRevA.100.022111},
}

@article{huang2019compatibility,
  author = {Huang, He-Liang and Luo, Yi-Han and Bai, B and Deng, Y-H and Wang, H and Zhao, Q and Zhong, H-S and Nie, Y-Q and Jiang, W-H and Wang, X-L and others},
  title = {Compatibility of causal hidden-variable theories with a delayed-choice experiment},
  journal = {Phys. Rev. A},
  volume = {100},
  number = {1},
  pages = {012114},
  year = {2019},
  publisher = {APS},
  doi = {10.1103/PhysRevA.100.012114},
  url = {https://doi.org/10.1103/PhysRevA.100.012114},
}

@article{pironio2003violations,
  author = {Pironio, Stefano},
  title = {Violations of {Bell} inequalities as lower bounds on the communication cost of nonlocal correlations},
  journal = {Phys. Rev. A},
  volume = {68},
  number = {6},
  pages = {062102},
  year = {2003},
  publisher = {APS},
  doi = {10.1103/PhysRevA.68.062102},
  url = {https://doi.org/10.1103/PhysRevA.68.062102},
}

@article{ioannou2022receiver,
  author = {Ioannou, Marie and Sekatski, Pavel and Abbott, Alastair A and Rosset, Denis and Bancal, Jean-Daniel and Brunner, Nicolas},
  title = {Receiver-device-independent quantum key distribution protocols},
  journal = {New J. Phys.},
  volume = {24},
  number = {6},
  pages = {063006},
  year = {2022},
  month = {jun},
  publisher = {IOP Publishing},
  doi = {10.1088/1367-2630/ac71bc},
  url = {https://doi.org/10.1088/1367-2630/ac71bc},
}

@article{wang2019characterising,
  author = {Wang, Yukun and Primaatmaja, Ignatius William and Lavie, Emilien and Varvitsiotis, Antonios and Lim, Charles Ci Wen},
  title = {Characterising the correlations of prepare-and-measure quantum networks},
  journal = {npj Quantum Inf.},
  volume = {5},
  number = {1},
  pages = {17},
  year = {2019},
  month = {Feb},
  publisher = {Nature Publishing Group},
  doi = {10.1038/s41534-019-0133-3},
  url = {https://doi.org/10.1038/s41534-019-0133-3},
}

@article{jia2025characterizing,
  author = {Jia, Yanning and Guo, Fenzhuo and Wang, Yukun and Dong, Haifeng and Gao, Fei},
  title = {Characterizing the set of quantum correlations in prepare-and-measure quantum multichain-shaped networks},
  journal = {Phys. Rev. A},
  volume = {111},
  number = {2},
  pages = {022439},
  numpages = {11},
  year = {2025},
  month = {Feb},
  publisher = {American Physical Society},
  doi = {10.1103/PhysRevA.111.022439},
  url = {https://link.aps.org/doi/10.1103/PhysRevA.111.022439},
}

@article{lauand2025minimal,
  author = {Lauand, Pedro and Poderini, Davide and Rabelo, Rafael and Chaves, Rafael},
  title = {Minimal Example of Quantum Nonclassicality without Freedom of Choice},
  journal = {Phys. Rev. Lett.},
  volume = {135},
  number = {9},
  pages = {090201},
  year = {2025},
  publisher = {APS},
  doi = {10.1103/3854-r395},
  url = {https://doi.org/10.1103/3854-r395},
}

@article{evans2016graphs,
  author = {Evans, Robin J},
  title = {Graphs for margins of {Bayesian} networks},
  journal = {Scand. J. Stat.},
  volume = {43},
  number = {3},
  pages = {625--648},
  year = {2016},
  publisher = {Wiley Online Library},
  doi = {10.1111/sjos.12194},
  url = {https://www.jstor.org/stable/24886954},
}

@article{lauand2023witnessing,
  author = {Lauand, Pedro and Poderini, Davide and Nery, Ranieri and Moreno, George and Pollyceno, Lucas and Rabelo, Rafael and Chaves, Rafael},
  title = {Witnessing Nonclassicality in a Causal Structure with Three Observable Variables},
  journal = {PRX Quantum},
  volume = {4},
  number = {2},
  pages = {020311},
  numpages = {18},
  year = {2023},
  month = {Apr},
  publisher = {American Physical Society},
  doi = {10.1103/PRXQuantum.4.020311},
  url = {https://link.aps.org/doi/10.1103/PRXQuantum.4.020311},
}

@article{Brukner2004,
  author = {Brukner, {\v C}aslav and {\.Z}ukowski, Marek and Pan, Jian-Wei and Zeilinger, Anton},
  title = {{Bell}'s Inequalities and Quantum Communication Complexity},
  journal = {Phys. Rev. Lett.},
  volume = {92},
  number = {12},
  pages = {127901},
  numpages = {4},
  year = {2004},
  month = mar,
  publisher = {American Physical Society},
  doi = {10.1103/PhysRevLett.92.127901},
  url = {https://link.aps.org/doi/10.1103/PhysRevLett.92.127901},
}

@article{Brukner2002,
  author = {Brukner, {\v C}aslav and {\.Z}ukowski, Marek and Zeilinger, Anton},
  title = {Quantum Communication Complexity Protocol with Two Entangled Qutrits},
  journal = {Phys. Rev. Lett.},
  volume = {89},
  number = {19},
  pages = {197901},
  numpages = {4},
  year = {2002},
  month = oct,
  publisher = {American Physical Society},
  doi = {10.1103/PhysRevLett.89.197901},
  url = {https://link.aps.org/doi/10.1103/PhysRevLett.89.197901},
}

@article{Mermin1990,
  author = {Mermin, N. David},
  title = {Extreme quantum entanglement in a superposition of macroscopically distinct states},
  journal = {Phys. Rev. Lett.},
  volume = {65},
  number = {15},
  pages = {1838--1840},
  year = {1990},
  month = oct,
  publisher = {American Physical Society},
  doi = {10.1103/PhysRevLett.65.1838},
  url = {https://link.aps.org/doi/10.1103/PhysRevLett.65.1838},
}

@inproceedings{Buhrman1998,
  author = {Buhrman, Harry and Cleve, Richard and Wigderson, Avi},
  title = {Quantum vs. classical communication and computation},
  booktitle = {Proceedings of the Thirtieth Annual ACM Symposium on Theory of Computing},
  series = {STOC '98},
  pages = {63--68},
  numpages = {6},
  year = {1998},
  publisher = {Association for Computing Machinery},
  isbn = {0897919629},
  doi = {10.1145/276698.276713},
  url = {https://doi.org/10.1145/276698.276713},
}

@inproceedings{Raz1999,
  author = {Raz, Ran},
  title = {Exponential separation of quantum and classical communication complexity},
  booktitle = {Proceedings of the Thirty-First Annual ACM Symposium on Theory of Computing},
  series = {STOC '99},
  pages = {358--367},
  numpages = {10},
  year = {1999},
  publisher = {Association for Computing Machinery},
  isbn = {1581130678},
  doi = {10.1145/301250.301343},
  url = {https://doi.org/10.1145/301250.301343},
}

@article{Baroni2025,
  author = {Baroni, Matilde and Diamanti, Eleni and Markham, Damian and {\v S}upi{\'c}, Ivan},
  title = {Translating {Bell} nonlocality to prepare-and-measure scenarios under dimensional constraints},
  journal = {Phys. Rev. A},
  volume = {112},
  number = {6},
  pages = {062220},
  year = {2025},
  month = dec,
  publisher = {American Physical Society},
  doi = {10.1103/6k4f-p3bb},
  url = {https://link.aps.org/doi/10.1103/6k4f-p3bb},
}

@article{Brukner2003,
  author = {Brukner, {\v{C}}aslav and Paterek, Tomasz and {\.{Z}}ukowski, Marek},
  title = {Quantum communication complexity protocols based on higher-dimensional entangled systems},
  journal = {Int. J. Quantum Inf.},
  volume = {01},
  number = {04},
  pages = {519--525},
  year = {2003},
  doi = {10.1142/S0219749903000395},
  url = {https://doi.org/10.1142/S0219749903000395},
}

@article{Frenkel2022,
  author = {Frenkel, P{\'e}ter E. and Weiner, Mih{\'a}ly},
  title = {On entanglement assistance to a noiseless classical channel},
  journal = {Quantum},
  volume = {6},
  pages = {662},
  year = {2022},
  month = mar,
  publisher = {Verein zur F{\"o}rderung des Open Access Publizierens in den Quantenwissenschaften},
  doi = {10.22331/q-2022-03-01-662},
  url = {https://doi.org/10.22331/q-2022-03-01-662},
}

@article{Vieira2023,
  author = {Vieira, Carlos and de Gois, Carlos and Pollyceno, Lucas and Rabelo, Rafael},
  title = {Interplays between classical and quantum entanglement-assisted communication scenarios},
  journal = {New J. Phys.},
  volume = {25},
  number = {11},
  pages = {113004},
  year = {2023},
  month = nov,
  publisher = {IOP Publishing},
  doi = {10.1088/1367-2630/ad0526},
  url = {https://doi.org/10.1088/1367-2630/ad0526},
}

@article{Navascues2007,
  author = {Navascu\'es, Miguel and Pironio, Stefano and Ac\'{\i}n, Antonio},
  title = {Bounding the Set of Quantum Correlations},
  journal = {Phys. Rev. Lett.},
  volume = {98},
  number = {1},
  pages = {010401},
  numpages = {4},
  year = {2007},
  month = jan,
  publisher = {American Physical Society},
  doi = {10.1103/PhysRevLett.98.010401},
  url = {https://link.aps.org/doi/10.1103/PhysRevLett.98.010401},
}

@article{Barreiro2008,
  author = {Barreiro, Julio T. and Wei, Tzu-Chieh and Kwiat, Paul G.},
  title = {Beating the channel capacity limit for linear photonic superdense coding},
  journal = {Nat. Phys.},
  volume = {4},
  number = {4},
  pages = {282--286},
  year = {2008},
  month = mar,
  publisher = {Springer},
  doi = {10.1038/nphys919},
  url = {https://doi.org/10.1038/nphys919},
}

@article{Willianms2017,
  author = {Williams, Brian P. and Sadlier, Ronald J. and Humble, Travis S.},
  title = {Superdense Coding over Optical Fiber Links with Complete {Bell}-State Measurements},
  journal = {Phys. Rev. Lett.},
  volume = {118},
  number = {5},
  pages = {050501},
  numpages = {5},
  year = {2017},
  month = feb,
  publisher = {American Physical Society},
  doi = {10.1103/PhysRevLett.118.050501},
  url = {https://link.aps.org/doi/10.1103/PhysRevLett.118.050501},
}

@article{Hu2018,
  author = {Xiao-Min Hu and Yu Guo and Bi-Heng Liu and Yun-Feng Huang and Chuan-Feng Li and Guang-Can Guo},
  title = {Beating the channel capacity limit for superdense coding with entangled ququarts},
  journal = {Sci. Adv.},
  volume = {4},
  number = {7},
  pages = {eaat9304},
  year = {2018},
  doi = {10.1126/sciadv.aat9304},
  url = {https://www.science.org/doi/abs/10.1126/sciadv.aat9304},
}

@article{Guo2025,
  author = {Guo, Yu and Tang, Hao and Pauwels, Jef and Cruzeiro, Emmanuel Zambrini and Hu, Xiao-Min and Liu, Bi-Heng and Huang, Yun-Feng and Li, Chuan-Feng and Guo, Guang-Can and Tavakoli, Armin},
  title = {Compression of Entanglement Improves Quantum Communication},
  journal = {Laser Photonics Rev.},
  volume = {19},
  number = {10},
  pages = {2401110},
  year = {2025},
  month = feb,
  publisher = {Wiley},
  doi = {10.1002/LPOR.202401110},
  url = {https://doi.org/10.1002/LPOR.202401110},
}

@article{Zhang2025,
  author = {Zhang, Chao and Miao, Jia-Le and Hu, Xiao-Min and Pauwels, Jef and Guo, Yu and Li, Chuan-Feng and Guo, Guang-Can and Tavakoli, Armin and Liu, Bi-Heng},
  title = {Quantum Stochastic Communication via High-Dimensional Entanglement},
  journal = {Phys. Rev. Lett.},
  volume = {135},
  number = {12},
  pages = {120802},
  numpages = {8},
  year = {2025},
  month = sep,
  publisher = {American Physical Society},
  doi = {10.1103/rq78-1qbh},
  url = {https://link.aps.org/doi/10.1103/rq78-1qbh},
}

@article{Zukowski2017,
  author = {Tavakoli, Armin and {\.Z}ukowski, Marek},
  title = {Higher-dimensional communication complexity problems: Classical protocols versus quantum ones based on {Bell}'s theorem or prepare-transmit-measure schemes},
  journal = {Phys. Rev. A},
  volume = {95},
  number = {4},
  pages = {042305},
  numpages = {6},
  year = {2017},
  month = apr,
  publisher = {American Physical Society},
  doi = {10.1103/PhysRevA.95.042305},
  url = {https://link.aps.org/doi/10.1103/PhysRevA.95.042305},
}

@misc{Ioannou2022,
  author = {Marie Ioannou and Denis Rosset},
  title = {Noncommutative polynomial optimization under symmetry},
  year = {2022},
  eprint = {2112.10803},
  archivePrefix = {arXiv},
  primaryClass = {quant-ph},
  url = {https://arxiv.org/abs/2112.10803},
}

@article{Piveteau2022,
  author = {Piveteau, Am{\'e}lie and Pauwels, Jef and H{\aa}kansson, Emil and Muhammad, Sadiq and Bourennane, Mohamed and Tavakoli, Armin},
  title = {Entanglement-assisted quantum communication with simple measurements},
  journal = {Nat. Commun.},
  volume = {13},
  number = {1},
  pages = {7878},
  year = {2022},
  month = dec,
  doi = {10.1038/s41467-022-33922-5},
  url = {https://doi.org/10.1038/s41467-022-33922-5},
}

@article{Mattle1996,
  author = {Mattle, Klaus and Weinfurter, Harald and Kwiat, Paul G. and Zeilinger, Anton},
  title = {Dense Coding in Experimental Quantum Communication},
  journal = {Phys. Rev. Lett.},
  volume = {76},
  number = {25},
  pages = {4656--4659},
  year = {1996},
  month = jun,
  publisher = {American Physical Society},
  doi = {10.1103/PhysRevLett.76.4656},
  url = {https://link.aps.org/doi/10.1103/PhysRevLett.76.4656},
}

@misc{Ji2022,
  author = {Zhengfeng Ji and Anand Natarajan and Thomas Vidick and John Wright and Henry Yuen},
  title = {{$\mathrm{MIP}^{*}=\mathrm{RE}$}},
  year = {2022},
  eprint = {2001.04383},
  archivePrefix = {arXiv},
  primaryClass = {quant-ph},
  url = {https://arxiv.org/abs/2001.04383},
}

@misc{Scholz2008,
  author = {V. B. Scholz and R. F. Werner},
  title = {{Tsirelson}'s Problem},
  year = {2008},
  eprint = {0812.4305},
  archivePrefix = {arXiv},
  primaryClass = {math-ph},
  url = {https://arxiv.org/abs/0812.4305},
}

@article{Navascues2008,
  author = {Navascu{\'e}s, Miguel and Pironio, Stefano and Ac{\'i}n, Antonio},
  title = {A convergent hierarchy of semidefinite programs characterizing the set of quantum correlations},
  journal = {New J. Phys.},
  volume = {10},
  number = {7},
  pages = {073013},
  year = {2008},
  month = jul,
  doi = {10.1088/1367-2630/10/7/073013},
  url = {https://doi.org/10.1088/1367-2630/10/7/073013},
}

@article{Mir2023,
  author = {Alimuddin, Mir and Chakraborty, Ananya and Sidhardh, Govind Lal and Patra, Ram Krishna and Sen, Samrat and Chowdhury, Snehasish Roy and Naik, Sahil Gopalkrishna and Banik, Manik},
  title = {Advantage of {Hardy}'s nonlocal correlation in reverse zero-error channel coding},
  journal = {Phys. Rev. A},
  volume = {108},
  number = {5},
  pages = {052430},
  numpages = {9},
  year = {2023},
  month = nov,
  publisher = {American Physical Society},
  doi = {10.1103/PhysRevA.108.052430},
  url = {https://link.aps.org/doi/10.1103/PhysRevA.108.052430},
}

@misc{Rout2025,
  author = {Sumit Rout and Anubhav Chaturvedi and Some Sankar Bhattacharya and Pawe{\l} Horodecki},
  title = {Facets of Non-locality and Advantage in Entanglement-Assisted Classical Communication Tasks},
  year = {2025},
  eprint = {2507.10830},
  archivePrefix = {arXiv},
  primaryClass = {quant-ph},
  url = {https://arxiv.org/abs/2507.10830},
}

@article{sarubi2026prepare,
  author = {Sarubi, Tailan S. and Alves, Mois{\'e}s and Zamora, Santiago and Alves, Vin{\'{\i}}cius F. and de Oliveira Junior, A. and Chaves, Rafael},
  title = {Prepare and broadcast scenario},
  journal = {Phys. Rev. A},
  volume = {114},
  number = {4},
  pages = {042404},
  numpages = {20},
  year = {2026},
  month = {Oct},
  publisher = {American Physical Society},
  doi = {10.1103/6yzk-2q6j},
  url = {https://link.aps.org/doi/10.1103/6yzk-2q6j},
}

@article{Bennet2014,
  author = {Bennet, Adam and V\'ertesi, Tam\'as and Saunders, Dylan J. and Brunner, Nicolas and Pryde, G. J.},
  title = {Experimental Semi-Device-Independent Certification of Entangled Measurements},
  journal = {Phys. Rev. Lett.},
  volume = {113},
  number = {8},
  pages = {080405},
  numpages = {5},
  year = {2014},
  month = {Aug},
  publisher = {American Physical Society},
  doi = {10.1103/PhysRevLett.113.080405},
  url = {https://link.aps.org/doi/10.1103/PhysRevLett.113.080405},
}

@article{Chaturvedi2020quantum,
  author = {Chaturvedi, Anubhav and Saha, Debashis},
  title = {Quantum prescriptions are more ontologically distinct than they are operationally distinguishable},
  journal = {Quantum},
  volume = {4},
  pages = {345},
  year = {2020},
  month = oct,
  publisher = {Verein zur F{\"o}rderung des Open Access Publizierens in den Quantenwissenschaften},
  doi = {10.22331/q-2020-10-21-345},
  url = {https://doi.org/10.22331/q-2020-10-21-345},
}

@article{Konig2009,
  author = {Konig, Robert and Renner, Renato and Schaffner, Christian},
  title = {The Operational Meaning of Min- and Max-Entropy},
  journal = {IEEE Trans. Inf. Theory},
  volume = {55},
  number = {9},
  pages = {4337–4347},
  year = {2009},
  month = 9,
  publisher = {Institute of Electrical and Electronics Engineers (IEEE)},
  issn = {0018-9448},
  doi = {10.1109/tit.2009.2025545},
  url = {https://doi.org/10.1109/TIT.2009.2025545},
}

@article{Martinez2018,
  author = {Mart{\'i}nez, Daniel and Tavakoli, Armin and Casanova, Mauricio and Ca{\~n}as, Gustavo and Marques, Breno and Lima, Gustavo},
  title = {High-Dimensional Quantum Communication Complexity beyond Strategies Based on {Bell}'s Theorem},
  journal = {Phys. Rev. Lett.},
  volume = {121},
  number = {15},
  pages = {150504},
  year = {2018},
  month = oct,
  publisher = {American Physical Society},
  doi = {10.1103/PhysRevLett.121.150504},
  url = {https://link.aps.org/doi/10.1103/PhysRevLett.121.150504},
}

@misc{davino2026entanglementenergyconstrainedprepareandmeasurescenario,
  author = {Raffaele D'Avino and Gabriel Senno and Mir Alimuddin and Antonio Ac{\'{i}}n},
  title = {Entanglement in the energy-constrained prepare-and-measure scenario: applications to randomness certification and channel discrimination},
  year = {2026},
  eprint = {2510.27559},
  archivePrefix = {arXiv},
  primaryClass = {quant-ph},
  url = {https://arxiv.org/abs/2510.27559},
}

@article{RochiCarceller_2026,
  author = {Roch i Carceller, Carles and Tavakoli, Armin},
  title = {The role of entanglement in energy-restricted communication and randomness generation},
  journal = {Quantum Sci. Technol.},
  volume = {11},
  number = {2},
  pages = {025020},
  year = {2026},
  month = {mar},
  publisher = {IOP Publishing},
  doi = {10.1088/2058-9565/ae4d80},
  url = {https://doi.org/10.1088/2058-9565/ae4d80},
}

@inproceedings{Yao1993,
  author = {Chi-Chih Yao, A.},
  title = {Quantum circuit complexity},
  booktitle = {Proceedings of 1993 IEEE 34th Annual Foundations of Computer Science},
  pages = {352--361},
  year = {1993},
  doi = {10.1109/SFCS.1993.366852},
  url = {https://doi.org/10.1109/SFCS.1993.366852},
}

@article{Emeriau2022,
  author = {Emeriau, Pierre-Emmanuel and Howard, Mark and Mansfield, Shane},
  title = {Quantum Advantage in Information Retrieval},
  journal = {PRX Quantum},
  volume = {3},
  number = {2},
  pages = {020307},
  year = {2022},
  month = apr,
  publisher = {American Physical Society},
  doi = {10.1103/PRXQuantum.3.020307},
  url = {https://link.aps.org/doi/10.1103/PRXQuantum.3.020307},
}

@article{Doolittle2021,
  author = {Doolittle, Brian and Chitambar, Eric},
  title = {Certifying the classical simulation cost of a quantum channel},
  journal = {Phys. Rev. Res.},
  volume = {3},
  number = {4},
  pages = {043073},
  numpages = {25},
  year = {2021},
  month = {Oct},
  publisher = {American Physical Society},
  doi = {10.1103/PhysRevResearch.3.043073},
  url = {https://link.aps.org/doi/10.1103/PhysRevResearch.3.043073},
}

@misc{Nogueira2026,
  author = {Jos{\'{e}} Nogueira and Carlos Vieira and Lucas E. A. Porto and Lucas Pollyceno and Rafael Rabelo and Otfried G{\"{u}}hne},
  title = {Non-signaling assistance in prepare-and-measure scenarios with classical communication},
  year = {2026},
  eprint = {2607.01193},
  archivePrefix = {arXiv},
  primaryClass = {quant-ph},
  url = {https://arxiv.org/abs/2607.01193},
}

@misc{Cruzeiro2026,
  author = {Emmanuel Zambrini Cruzeiro},
  title = {Perfect Games in Dimension-Bounded Communication},
  year = {2026},
  eprint = {2608.05092},
  archivePrefix = {arXiv},
  primaryClass = {quant-ph},
  url = {https://arxiv.org/abs/2608.05092},
}

@inproceedings{Yossef2004,
  author = {Bar-Yossef, Ziv and Jayram, T. S. and Kerenidis, Iordanis},
  title = {Exponential separation of quantum and classical one-way communication complexity},
  booktitle = {Proceedings of the Thirty-Sixth Annual ACM Symposium on Theory of Computing},
  series = {STOC '04},
  pages = {128--137},
  numpages = {10},
  year = {2004},
  publisher = {Association for Computing Machinery},
  isbn = {1581138520},
  doi = {10.1145/1007352.1007379},
  url = {https://doi.org/10.1145/1007352.1007379},
}

@article{TavakoliBrukner2020,
  author = {Tavakoli, Armin and {\.Z}ukowski, Marek and Brukner, {\v C}aslav},
  title = {Does violation of a {Bell} inequality always imply quantum advantage in a communication complexity problem?},
  journal = {Quantum},
  volume = {4},
  pages = {316},
  year = {2020},
  month = sep,
  publisher = {Verein zur F{\"o}rderung des Open Access Publizierens in den Quantenwissenschaften},
  doi = {10.22331/q-2020-09-07-316},
  url = {https://doi.org/10.22331/q-2020-09-07-316},
}

@article{Buhrman2001,
  author = {Buhrman, Harry and Cleve, Richard and van Dam, Wim},
  title = {Quantum Entanglement and Communication Complexity},
  journal = {SIAM J. Comput.},
  volume = {30},
  number = {6},
  pages = {1829--1841},
  year = {2001},
  month = jan,
  publisher = {Society for Industrial and Applied Mathematics},
  doi = {10.1137/S0097539797324886},
  url = {https://doi.org/10.1137/S0097539797324886},
}

@article{Mironowicz2019,
  author = {Mironowicz, Piotr and Paw\l{}owski, Marcin},
  title = {Experimentally feasible semi-device-independent certification of four-outcome positive-operator-valued measurements},
  journal = {Phys. Rev. A},
  volume = {100},
  number = {3},
  pages = {030301},
  numpages = {6},
  year = {2019},
  month = sep,
  publisher = {American Physical Society},
  doi = {10.1103/PhysRevA.100.030301},
  url = {https://link.aps.org/doi/10.1103/PhysRevA.100.030301},
}

@article{Sun2020,
  author = {Yong-Nan Sun and Zhao-Di Liu and Joseph Bowles and Geng Chen and Xiao-Ye Xu and Jian-Shun Tang and Chuan-Feng Li and Guang-Can Guo},
  title = {Experimental certification of quantum dimensions and irreducible high-dimensional quantum systems with independent devices},
  journal = {Optica},
  volume = {7},
  number = {9},
  pages = {1073--1078},
  year = {2020},
  month = sep,
  publisher = {Optica Publishing Group},
  doi = {10.1364/OPTICA.396932},
  url = {https://opg.optica.org/optica/abstract.cfm?URI=optica-7-9-1073},
}

@article{Martinez2023,
  author = {Mart{\'i}nez, Daniel and G{\'o}mez, Esteban S. and Cari{\~{n}}e, Jaime and Pereira, Luciano and Delgado, Aldo and Walborn, Stephen P. and Tavakoli, Armin and Lima, Gustavo},
  title = {Certification of a non-projective qudit measurement using multiport beamsplitters},
  journal = {Nat. Phys.},
  volume = {19},
  number = {2},
  pages = {190--195},
  year = {2023},
  month = feb,
  doi = {10.1038/s41567-022-01845-z},
  url = {https://doi.org/10.1038/s41567-022-01845-z},
}

@article{Smania2020,
  author = {Tavakoli, Armin and Smania, Massimiliano and V{\'e}rtesi, Tam{\'a}s and Brunner, Nicolas and Bourennane, Mohamed},
  title = {Self-testing nonprojective quantum measurements in prepare-and-measure experiments},
  journal = {Sci. Adv.},
  volume = {6},
  number = {16},
  pages = {eaaw6664},
  year = {2020},
  month = apr,
  publisher = {American Association for the Advancement of Science},
  doi = {10.1126/sciadv.aaw6664},
  url = {https://doi.org/10.1126/sciadv.aaw6664},
}

@article{Miklin2020,
  author = {Miklin, Nikolai and Borka\l{}a, Jakub J. and Paw\l{}owski, Marcin},
  title = {Semi-device-independent self-testing of unsharp measurements},
  journal = {Phys. Rev. Res.},
  volume = {2},
  number = {3},
  pages = {033014},
  numpages = {15},
  year = {2020},
  month = jul,
  publisher = {American Physical Society},
  doi = {10.1103/PhysRevResearch.2.033014},
  url = {https://link.aps.org/doi/10.1103/PhysRevResearch.2.033014},
}

@article{Foletto2020,
  author = {Foletto, Giulio and Calderaro, Luca and Vallone, Giuseppe and Villoresi, Paolo},
  title = {Experimental demonstration of sequential quantum random access codes},
  journal = {Phys. Rev. Res.},
  volume = {2},
  number = {3},
  pages = {033205},
  numpages = {8},
  year = {2020},
  month = aug,
  publisher = {American Physical Society},
  doi = {10.1103/PhysRevResearch.2.033205},
  url = {https://link.aps.org/doi/10.1103/PhysRevResearch.2.033205},
}

@article{Mohan2019,
  author = {Mohan, Karthik and Tavakoli, Armin and Brunner, Nicolas},
  title = {Sequential random access codes and self-testing of quantum measurement instruments},
  journal = {New J. Phys.},
  volume = {21},
  number = {8},
  pages = {083034},
  year = {2019},
  month = aug,
  publisher = {IOP Publishing},
  doi = {10.1088/1367-2630/ab3773},
  url = {https://doi.org/10.1088/1367-2630/ab3773},
}

@book{SkrzypczykCavalcanti2023,
  author = {Paul Skrzypczyk and Daniel Cavalcanti},
  title = {Semidefinite Programming in Quantum Information Science},
  year = {2023},
  publisher = {IOP Publishing},
  isbn = {9780750333412},
  doi = {10.1088/978-0-7503-3343-6},
  url = {https://doi.org/10.1088/978-0-7503-3343-6},
}

@article{chaves2018quantum,
  author = {Chaves, Rafael and Carvacho, Gonzalo and Agresti, Iris and Di Giulio, Valerio and Aolita, Leandro and Giacomini, Sandro and Sciarrino, Fabio},
  title = {Quantum violation of an instrumental test},
  journal = {Nat. Phys.},
  volume = {14},
  number = {3},
  pages = {291--296},
  year = {2018},
  publisher = {Nature Publishing Group UK London},
  doi = {10.1038/s41567-017-0008-5},
  url = {https://doi.org/10.1038/s41567-017-0008-5},
}

@article{SDPreview,
  author = {Tavakoli, Armin and Pozas-Kerstjens, Alejandro and Brown, Peter and Ara\'ujo, Mateus},
  title = {Semidefinite programming relaxations for quantum correlations},
  journal = {Rev. Mod. Phys.},
  volume = {96},
  number = {4},
  pages = {045006},
  numpages = {68},
  year = {2024},
  month = dec,
  publisher = {American Physical Society},
  doi = {10.1103/RevModPhys.96.045006},
  url = {https://link.aps.org/doi/10.1103/RevModPhys.96.045006},
}

@article{Werner2001,
  author = {Werner, Reinhard F. and Wolf, Michael M.},
  title = {{Bell} inequalities and entanglement},
  journal = {Quantum Inf. Comput.},
  volume = {1},
  number = {3},
  pages = {1--25},
  year = {2001},
  doi = {10.26421/QIC1.3-1},
  url = {https://doi.org/10.26421/QIC1.3-1},
}

@article{Aguilar2018b,
  author = {Aguilar, Edgar A. and Farkas, M\'at\'e and Mart\'{\i}nez, Daniel and Alvarado, Mat\'{\i}as and Cari\~ne, Jaime and Xavier, Guilherme B. and Barra, Johanna F. and Ca\~nas, Gustavo and Paw\l{}owski, Marcin and Lima, Gustavo},
  title = {Certifying an Irreducible 1024-Dimensional Photonic State Using Refined Dimension Witnesses},
  journal = {Phys. Rev. Lett.},
  volume = {120},
  number = {23},
  pages = {230503},
  numpages = {6},
  year = {2018},
  month = jun,
  publisher = {American Physical Society},
  doi = {10.1103/PhysRevLett.120.230503},
  url = {https://link.aps.org/doi/10.1103/PhysRevLett.120.230503},
}

@article{Karthik2025,
  author = {Karthik, H. S. and G\'omez, S. and Quinteros, F. M. and H., Akshata Shenoy and Paw\l{}owski, M. and Walborn, S. P. and Lima, G. and G\'omez, E. S.},
  title = {Noise-resilient quantum random access codes},
  journal = {Phys. Rev. A},
  volume = {111},
  number = {3},
  pages = {032613},
  numpages = {8},
  year = {2025},
  month = mar,
  publisher = {American Physical Society},
  doi = {10.1103/PhysRevA.111.032613},
  url = {https://link.aps.org/doi/10.1103/PhysRevA.111.032613},
}

@article{Ribezzo2024,
  author = {Ribezzo, Domenico and Salazar, Roberto and Czartowski, Jakub and Segur, Flora and Lemmi, Gianmarco and Petitjean, Antoine and Farrugia, Noel and Xuereb, Andr{\'e} and Bacco, Davide and Zavatta, Alessandro},
  title = {Quantum Random Access Codes Implementation for Resource Allocation and Coexistence with Classical Telecommunication},
  journal = {Adv. Quantum Technol.},
  volume = {7},
  number = {4},
  pages = {2300162},
  year = {2024},
  doi = {10.1002/qute.202300162},
  url = {https://advanced.onlinelibrary.wiley.com/doi/abs/10.1002/qute.202300162},
}

@article{Spekkens2009,
  author = {Spekkens, Robert W. and Buzacott, D. H. and Keehn, A. J. and Toner, Ben and Pryde, G. J.},
  title = {Preparation Contextuality Powers Parity-Oblivious Multiplexing},
  journal = {Phys. Rev. Lett.},
  volume = {102},
  number = {1},
  pages = {010401},
  numpages = {4},
  year = {2009},
  month = jan,
  publisher = {American Physical Society},
  doi = {10.1103/PhysRevLett.102.010401},
  url = {https://link.aps.org/doi/10.1103/PhysRevLett.102.010401},
}

@article{Brierley2009,
  author = {Stephen Brierley and Stefan Weigert and Ingemar Bengtsson},
  title = {All Mutually Unbiased Bases in Dimensions Two to Five},
  journal = {Quantum Inf. Comput.},
  volume = {10},
  number = {9-10},
  pages = {803--820},
  year = {2010},
  doi = {10.26421/QIC10.9-10-6},
  eprint = {0907.4097},
  archivePrefix = {arXiv},
  primaryClass = {math-ph},
  url = {https://doi.org/10.26421/QIC10.9-10-6},
}

@article{Acin2016,
  author = {Ac\'{\i}n, Antonio and Pironio, Stefano and V\'ertesi, Tam\'as and Wittek, Peter},
  title = {Optimal randomness certification from one entangled bit},
  journal = {Phys. Rev. A},
  volume = {93},
  number = {4},
  pages = {040102},
  numpages = {5},
  year = {2016},
  month = apr,
  publisher = {American Physical Society},
  doi = {10.1103/PhysRevA.93.040102},
  url = {https://link.aps.org/doi/10.1103/PhysRevA.93.040102},
}

@article{InfoContextuality,
  author = {Tavakoli, Armin and Cruzeiro, Emmanuel Zambrini and Uola, Roope and Abbott, Alastair A.},
  title = {Bounding and Simulating Contextual Correlations in Quantum Theory},
  journal = {PRX Quantum},
  volume = {2},
  number = {2},
  pages = {020334},
  numpages = {19},
  year = {2021},
  month = {Jun},
  publisher = {American Physical Society},
  doi = {10.1103/PRXQuantum.2.020334},
  url = {https://link.aps.org/doi/10.1103/PRXQuantum.2.020334},
}

@misc{Manna2026,
  author = {Satyaki Manna and Ankush Pandit and Anubhav Chaturvedi and Debashis Saha},
  title = {Entanglement assisted communication complexity measured by distinguishability},
  year = {2026},
  eprint = {2603.19105},
  archivePrefix = {arXiv},
  primaryClass = {quant-ph},
  url = {https://arxiv.org/abs/2603.19105},
}

@article{Chaturvedi2021,
  author = {Chaturvedi, Anubhav and Farkas, M{\'{a}}t{\'{e}} and Wright, Victoria J},
  title = {Characterising and bounding the set of quantum behaviours in contextuality scenarios},
  journal = {Quantum},
  volume = {5},
  pages = {484},
  year = {2021},
  month = 6,
  publisher = {Verein zur Forderung des Open Access Publizierens in den Quantenwissenschaften},
  issn = {2521-327X},
  doi = {10.22331/q-2021-06-29-484},
  url = {https://doi.org/10.22331/q-2021-06-29-484},
}

@article{Montanaro2007,
  author = {Montanaro, Ashley},
  title = {On the Distinguishability of Random Quantum States},
  journal = {Commun. Math. Phys.},
  volume = {273},
  number = {3},
  pages = {619--636},
  year = {2007},
  month = {Aug},
  issn = {1432-0916},
  doi = {10.1007/s00220-007-0221-7},
  url = {https://doi.org/10.1007/s00220-007-0221-7},
  day = {01},
}

@article{Tavakoli2020informationally,
  author = {Tavakoli, Armin and Zambrini Cruzeiro, Emmanuel and Bohr Brask, Jonatan and Gisin, Nicolas and Brunner, Nicolas},
  title = {Informationally restricted quantum correlations},
  journal = {Quantum},
  volume = {4},
  pages = {332},
  year = {2020},
  month = sep,
  publisher = {Verein zur F{\"o}rderung des Open Access Publizierens in den Quantenwissenschaften},
  doi = {10.22331/q-2020-09-24-332},
  url = {https://doi.org/10.22331/q-2020-09-24-332},
}

@article{Tavakoli2022informationally,
  author = {Tavakoli, Armin and Zambrini Cruzeiro, Emmanuel and Woodhead, Erik and Pironio, Stefano},
  title = {Informationally restricted correlations: a general framework for classical and quantum systems},
  journal = {Quantum},
  volume = {6},
  pages = {620},
  year = {2022},
  month = jan,
  publisher = {Verein zur F{\"o}rderung des Open Access Publizierens in den Quantenwissenschaften},
  doi = {10.22331/q-2022-01-05-620},
  url = {https://doi.org/10.22331/q-2022-01-05-620},
}

@article{Werner2001b,
  author = {Werner, R F},
  title = {All teleportation and dense coding schemes},
  journal = {J. Phys. A: Math. Gen.},
  volume = {34},
  number = {35},
  pages = {7081–7094},
  year = {2001},
  month = Aug,
  publisher = {IOP Publishing},
  issn = {1361-6447},
  doi = {10.1088/0305-4470/34/35/332},
  url = {https://doi.org/10.1088/0305-4470/34/35/332},
}

@article{Terhal2000,
  author = {Terhal, Barbara M. and Horodecki, Pawe\l{}},
  title = {{Schmidt} number for density matrices},
  journal = {Phys. Rev. A},
  volume = {61},
  number = {4},
  pages = {040301(R)},
  numpages = {4},
  year = {2000},
  month = {Mar},
  publisher = {American Physical Society},
  doi = {10.1103/PhysRevA.61.040301},
  url = {https://link.aps.org/doi/10.1103/PhysRevA.61.040301},
}

@article{Abbott2018,
  author = {Tavakoli, Armin and Abbott, Alastair A. and Renou, Marc-Olivier and Gisin, Nicolas and Brunner, Nicolas},
  title = {Semi-device-independent characterization of multipartite entanglement of states and measurements},
  journal = {Phys. Rev. A},
  volume = {98},
  number = {5},
  pages = {052333},
  year = {2018},
  month = Nov,
  publisher = {American Physical Society (APS)},
  issn = {2469-9934},
  doi = {10.1103/PhysRevA.98.052333},
  url = {https://doi.org/10.1103/PhysRevA.98.052333},
}

@article{Calsamiglia2001,
  author = {Calsamiglia, J. and L{\"{u}}tkenhaus, N.},
  title = {Maximum efficiency of a linear-optical {Bell}-state analyzer},
  journal = {Appl. Phys. B},
  volume = {72},
  number = {1},
  pages = {67–71},
  year = {2001},
  month = Jan,
  publisher = {Springer Science and Business Media LLC},
  issn = {1432-0649},
  doi = {10.1007/s003400000484},
  url = {https://doi.org/10.1007/s003400000484},
}

@article{Lutkenhaus1999,
  author = {L\"utkenhaus, N. and Calsamiglia, J. and Suominen, K.-A.},
  title = {{Bell} measurements for teleportation},
  journal = {Phys. Rev. A},
  volume = {59},
  number = {5},
  pages = {3295--3300},
  numpages = {0},
  year = {1999},
  month = {May},
  publisher = {American Physical Society},
  doi = {10.1103/PhysRevA.59.3295},
  url = {https://link.aps.org/doi/10.1103/PhysRevA.59.3295},
}

@article{Drotos2024,
  author = {Dr\'otos, G\'abor and P\'al, K\'aroly F. and V\'ertesi, Tam\'as},
  title = {Self-testing of semisymmetric informationally complete measurements in a qubit prepare-and-measure scenario},
  journal = {Phys. Rev. A},
  volume = {110},
  number = {3},
  pages = {032427},
  numpages = {11},
  year = {2024},
  month = sep,
  publisher = {American Physical Society},
  doi = {10.1103/PhysRevA.110.032427},
  url = {https://link.aps.org/doi/10.1103/PhysRevA.110.032427},
}

@article{Busch1986,
  author = {Busch, Paul},
  title = {Unsharp reality and joint measurements for spin observables},
  journal = {Phys. Rev. D},
  volume = {33},
  number = {8},
  pages = {2253--2261},
  year = {1986},
  month = apr,
  publisher = {American Physical Society},
  doi = {10.1103/PhysRevD.33.2253},
  url = {https://link.aps.org/doi/10.1103/PhysRevD.33.2253},
}

@article{Miklin2021,
  author = {Miklin, Nikolai and Oszmaniec, Micha{\l{}}},
  title = {A universal scheme for robust self-testing in the prepare-and-measure scenario},
  journal = {Quantum},
  volume = {5},
  pages = {424},
  year = {2021},
  month = apr,
  publisher = {Verein zur F{\"o}rderung des Open Access Publizierens in den Quantenwissenschaften},
  doi = {10.22331/q-2021-04-06-424},
  url = {https://doi.org/10.22331/q-2021-04-06-424},
}

@article{Navascues2023,
  author = {Navascu\'es, Miguel and P\'al, K\'aroly F. and V\'ertesi, Tam\'as and Ara\'ujo, Mateus},
  title = {Self-Testing in Prepare-and-Measure Scenarios and a Robust Version of {Wigner}'s Theorem},
  journal = {Phys. Rev. Lett.},
  volume = {131},
  number = {25},
  pages = {250802},
  numpages = {6},
  year = {2023},
  month = dec,
  publisher = {American Physical Society},
  doi = {10.1103/PhysRevLett.131.250802},
  url = {https://link.aps.org/doi/10.1103/PhysRevLett.131.250802},
}

@article{Farkas2019,
  author = {Farkas, M\'at\'e and Kaniewski, J{\k e}drzej},
  title = {Self-testing mutually unbiased bases in the prepare-and-measure scenario},
  journal = {Phys. Rev. A},
  volume = {99},
  number = {3},
  pages = {032316},
  numpages = {11},
  year = {2019},
  month = mar,
  publisher = {American Physical Society},
  doi = {10.1103/PhysRevA.99.032316},
  url = {https://link.aps.org/doi/10.1103/PhysRevA.99.032316},
}

@article{Tavakoli2018,
  author = {Tavakoli, Armin and Kaniewski, J{\k e}drzej and V\'ertesi, Tam\'as and Rosset, Denis and Brunner, Nicolas},
  title = {Self-testing quantum states and measurements in the prepare-and-measure scenario},
  journal = {Phys. Rev. A},
  volume = {98},
  number = {6},
  pages = {062307},
  numpages = {13},
  year = {2018},
  month = dec,
  publisher = {American Physical Society},
  doi = {10.1103/PhysRevA.98.062307},
  url = {https://link.aps.org/doi/10.1103/PhysRevA.98.062307},
}

@article{Supic2020,
  author = {{\v{S}}upi{\'{c}}, Ivan and Bowles, Joseph},
  title = {Self-testing of quantum systems: a review},
  journal = {Quantum},
  volume = {4},
  pages = {337},
  year = {2020},
  month = sep,
  publisher = {Verein zur F{\"o}rderung des Open Access Publizierens in den Quantenwissenschaften},
  doi = {10.22331/q-2020-09-30-337},
  url = {https://doi.org/10.22331/q-2020-09-30-337},
}

@article{Vicente2019,
  author = {de Vicente, Julio I},
  title = {A general bound for the dimension of quantum behaviours in the prepare-and-measure scenario},
  journal = {J. Phys. A: Math. Theor.},
  volume = {52},
  number = {9},
  pages = {095304},
  year = {2019},
  month = feb,
  publisher = {IOP Publishing},
  doi = {10.1088/1751-8121/aafde7},
  url = {https://doi.org/10.1088/1751-8121/aafde7},
}

@article{Vicente2017,
  author = {de Vicente, Julio I.},
  title = {Shared randomness and device-independent dimension witnessing},
  journal = {Phys. Rev. A},
  volume = {95},
  number = {1},
  pages = {012340},
  numpages = {11},
  year = {2017},
  month = jan,
  publisher = {American Physical Society},
  doi = {10.1103/PhysRevA.95.012340},
  url = {https://link.aps.org/doi/10.1103/PhysRevA.95.012340},
}

@article{Sikora2016,
  author = {Sikora, Jamie and Varvitsiotis, Antonios and Wei, Zhaohui},
  title = {Device-independent dimension tests in the prepare-and-measure scenario},
  journal = {Phys. Rev. A},
  volume = {94},
  number = {4},
  pages = {042125},
  numpages = {4},
  year = {2016},
  month = oct,
  publisher = {American Physical Society},
  doi = {10.1103/PhysRevA.94.042125},
  url = {https://link.aps.org/doi/10.1103/PhysRevA.94.042125},
}

@article{Brunner2013,
  author = {Brunner, Nicolas and Navascu\'es, Miguel and V\'ertesi, Tam\'as},
  title = {Dimension Witnesses and Quantum State Discrimination},
  journal = {Phys. Rev. Lett.},
  volume = {110},
  number = {15},
  pages = {150501},
  numpages = {4},
  year = {2013},
  month = apr,
  publisher = {American Physical Society},
  doi = {10.1103/PhysRevLett.110.150501},
  url = {https://link.aps.org/doi/10.1103/PhysRevLett.110.150501},
}

@article{Aguilar2018,
  author = {Aguilar, Edgar A. and Borka\l{}a, Jakub J. and Mironowicz, Piotr and Paw\l{}owski, Marcin},
  title = {Connections between Mutually Unbiased Bases and Quantum Random Access Codes},
  journal = {Phys. Rev. Lett.},
  volume = {121},
  number = {5},
  pages = {050501},
  numpages = {6},
  year = {2018},
  month = jul,
  publisher = {American Physical Society},
  doi = {10.1103/PhysRevLett.121.050501},
  url = {https://link.aps.org/doi/10.1103/PhysRevLett.121.050501},
}

@article{Rosset2019,
  author = {Tavakoli, Armin and Rosset, Denis and Renou, Marc-Olivier},
  title = {Enabling Computation of Correlation Bounds for Finite-Dimensional Quantum Systems via Symmetrization},
  journal = {Phys. Rev. Lett.},
  volume = {122},
  number = {7},
  pages = {070501},
  numpages = {7},
  year = {2019},
  month = feb,
  publisher = {American Physical Society},
  doi = {10.1103/PhysRevLett.122.070501},
  url = {https://link.aps.org/doi/10.1103/PhysRevLett.122.070501},
}

@inproceedings{Rosset2021,
  author = {Rosset, Denis and Montealegre-Mora, Felipe and Bancal, Jean-Daniel},
  editor = {Paranjape, M.B. and MacKenzie, R. and Thomova, Z. and Winternitz, P. and Witczak-Krempa, W.},
  title = {{RepLAB}: A Computational/Numerical Approach to Representation Theory},
  booktitle = {Quantum Theory and Symmetries},
  series = {CRM Series in Mathematical Physics},
  pages = {643--653},
  year = {2021},
  month = mar,
  publisher = {Springer},
  doi = {10.1007/978-3-030-55777-5_60},
  url = {https://doi.org/10.1007/978-3-030-55777-5_60},
}

@article{Pauwels2022,
  author = {Pauwels, Jef and Pironio, Stefano and Woodhead, Erik and Tavakoli, Armin},
  title = {Almost Qudits in the Prepare-and-Measure Scenario},
  journal = {Phys. Rev. Lett.},
  volume = {129},
  number = {25},
  pages = {250504},
  numpages = {7},
  year = {2022},
  month = dec,
  publisher = {American Physical Society},
  doi = {10.1103/PhysRevLett.129.250504},
  url = {https://link.aps.org/doi/10.1103/PhysRevLett.129.250504},
}

@article{TavakoliPauwels2021,
  author = {Tavakoli, Armin and Pauwels, Jef and Woodhead, Erik and Pironio, Stefano},
  title = {Correlations in Entanglement-Assisted Prepare-and-Measure Scenarios},
  journal = {PRX Quantum},
  volume = {2},
  number = {4},
  pages = {040357},
  numpages = {17},
  year = {2021},
  month = dec,
  publisher = {American Physical Society},
  doi = {10.1103/PRXQuantum.2.040357},
  url = {https://link.aps.org/doi/10.1103/PRXQuantum.2.040357},
}

@article{Navascues2015b,
  author = {Navascu\'es, Miguel and Feix, Adrien and Ara\'ujo, Mateus and V\'ertesi, Tam\'as},
  title = {Characterizing finite-dimensional quantum behavior},
  journal = {Phys. Rev. A},
  volume = {92},
  number = {4},
  pages = {042117},
  numpages = {15},
  year = {2015},
  month = oct,
  publisher = {American Physical Society},
  doi = {10.1103/PhysRevA.92.042117},
  url = {https://link.aps.org/doi/10.1103/PhysRevA.92.042117},
}

@article{Navascues2015,
  author = {Navascu\'es, Miguel and V\'ertesi, Tam\'as},
  title = {Bounding the Set of Finite Dimensional Quantum Correlations},
  journal = {Phys. Rev. Lett.},
  volume = {115},
  number = {2},
  pages = {020501},
  numpages = {5},
  year = {2015},
  month = jul,
  publisher = {American Physical Society},
  doi = {10.1103/PhysRevLett.115.020501},
  url = {https://link.aps.org/doi/10.1103/PhysRevLett.115.020501},
}

@article{bae2015quantum,
  author = {Bae, Joonwoo and Kwek, Leong-Chuan},
  title = {Quantum state discrimination and its applications},
  journal = {J. Phys. A: Math. Theor.},
  volume = {48},
  number = {8},
  pages = {083001},
  year = {2015},
  publisher = {IOP Publishing},
  doi = {10.1088/1751-8113/48/8/083001},
  url = {https://doi.org/10.1088/1751-8113/48/8/083001},
}

@article{Ambainis2024,
  author = {Ambainis, Andris and Kravchenko, Dmitry and Sazim, Sk and Bae, Joonwoo and Rai, Ashutosh},
  title = {Quantum advantages in random access codes},
  journal = {New J. Phys.},
  volume = {26},
  number = {12},
  pages = {123023},
  year = {2024},
  month = dec,
  publisher = {IOP Publishing},
  doi = {10.1088/1367-2630/ad9bdf},
  url = {https://doi.org/10.1088/1367-2630/ad9bdf},
}

@article{Farkas2025,
  author = {Farkas, M{\'{a}}t{\'{e}} and Miklin, Nikolai and Tavakoli, Armin},
  title = {Simple and general bounds on quantum random access codes},
  journal = {Quantum},
  volume = {9},
  pages = {1643},
  year = {2025},
  month = feb,
  publisher = {Verein zur F{\"o}rderung des Open Access Publizierens in den Quantenwissenschaften},
  doi = {10.22331/q-2025-02-25-1643},
  url = {https://doi.org/10.22331/q-2025-02-25-1643},
}

@article{PauwelsAdaptive2022,
  author = {Pauwels, Jef and Pironio, Stefano and Cruzeiro, Emmanuel Zambrini and Tavakoli, Armin},
  title = {Adaptive Advantage in Entanglement-Assisted Communications},
  journal = {Phys. Rev. Lett.},
  volume = {129},
  number = {12},
  pages = {120504},
  numpages = {7},
  year = {2022},
  month = sep,
  publisher = {American Physical Society},
  doi = {10.1103/PhysRevLett.129.120504},
  url = {https://link.aps.org/doi/10.1103/PhysRevLett.129.120504},
}

@article{Tavakoli2016,
  author = {Tavakoli, Armin and Marques, Breno and Paw\l{}owski, Marcin and Bourennane, Mohamed},
  title = {Spatial versus sequential correlations for random access coding},
  journal = {Phys. Rev. A},
  volume = {93},
  number = {3},
  pages = {032336},
  numpages = {6},
  year = {2016},
  month = mar,
  publisher = {American Physical Society},
  doi = {10.1103/PhysRevA.93.032336},
  url = {https://link.aps.org/doi/10.1103/PhysRevA.93.032336},
}

@article{van2019quantum,
  author = {Van Himbeeck, Thomas and Brask, Jonatan Bohr and Pironio, Stefano and Ramanathan, Ravishankar and Sainz, Ana Bel{\'e}n and Wolfe, Elie},
  title = {Quantum violations in the Instrumental scenario and their relations to the {Bell} scenario},
  journal = {Quantum},
  volume = {3},
  pages = {186},
  year = {2019},
  publisher = {Verein zur F{\"o}rderung des Open Access Publizierens in den Quantenwissenschaften},
  doi = {10.22331/q-2019-09-16-186},
  url = {https://doi.org/10.22331/q-2019-09-16-186},
}

@article{Pati2000,
  author = {Pati, Arun K.},
  title = {Minimum classical bit for remote preparation and measurement of a qubit},
  journal = {Phys. Rev. A},
  volume = {63},
  number = {1},
  pages = {014302},
  numpages = {3},
  year = {2000},
  month = dec,
  publisher = {American Physical Society},
  doi = {10.1103/PhysRevA.63.014302},
  url = {https://link.aps.org/doi/10.1103/PhysRevA.63.014302},
}

@article{PawlowskiZukowski2010,
  author = {Paw\l{}owski, Marcin and {\.Z}ukowski, Marek},
  title = {Entanglement-assisted random access codes},
  journal = {Phys. Rev. A},
  volume = {81},
  number = {4},
  pages = {042326},
  numpages = {4},
  year = {2010},
  month = apr,
  publisher = {American Physical Society},
  doi = {10.1103/PhysRevA.81.042326},
  url = {https://link.aps.org/doi/10.1103/PhysRevA.81.042326},
}

@article{van2013implausible,
  author = {Van Dam, Wim},
  title = {Implausible consequences of superstrong nonlocality},
  journal = {Nat. Comput.},
  volume = {12},
  number = {1},
  pages = {9--12},
  year = {2013},
  publisher = {Springer},
  doi = {10.1007/s11047-012-9353-6},
  url = {https://doi.org/10.1007/s11047-012-9353-6},
}

@article{botteron2024algebra,
  author = {Botteron, Pierre and Broadbent, Anne and Chhaibi, Reda and Nechita, Ion and Pellegrini, Cl{\'e}ment},
  title = {Algebra of Nonlocal Boxes and the Collapse of Communication Complexity},
  journal = {Quantum},
  volume = {8},
  pages = {1402},
  year = {2024},
  publisher = {Verein zur F{\"o}rderung des Open Access Publizierens in den Quantenwissenschaften},
  doi = {10.22331/q-2024-07-10-1402},
  url = {https://doi.org/10.22331/q-2024-07-10-1402},
}

@article{brito2019nonlocality,
  author = {Brito, Samura{\'\i} Gomes de Aguiar and Moreno, M. G. M. and Rai, Ashutosh and Chaves, Rafael},
  title = {Nonlocality distillation and quantum voids},
  journal = {Phys. Rev. A},
  volume = {100},
  number = {1},
  pages = {012102},
  year = {2019},
  doi = {10.1103/PhysRevA.100.012102},
  eprint = {1903.09289},
  archivePrefix = {arXiv},
  primaryClass = {quant-ph},
  url = {https://doi.org/10.1103/PhysRevA.100.012102},
}

@article{uffink2002quadratic,
  author = {Uffink, Jos},
  title = {Quadratic {Bell} inequalities as tests for multipartite entanglement},
  journal = {Phys. Rev. Lett.},
  volume = {88},
  number = {23},
  pages = {230406},
  year = {2002},
  publisher = {American Physical Society},
  doi = {10.1103/PhysRevLett.88.230406},
  url = {https://link.aps.org/doi/10.1103/PhysRevLett.88.230406},
}

@article{cleve2013quantum,
  author = {Cleve, Richard and Van Dam, Wim and Nielsen, Michael and Tapp, Alain},
  title = {Quantum entanglement and the communication complexity of the inner product function},
  journal = {Theor. Comput. Sci.},
  volume = {486},
  pages = {11--19},
  year = {2013},
  publisher = {Elsevier},
  doi = {10.1016/j.tcs.2012.12.012},
  url = {https://doi.org/10.1016/j.tcs.2012.12.012},
}

@article{rai2019geometry,
  author = {Rai, Ashutosh and Duarte, Cristhiano and Brito, Samura{\'\i} and Chaves, Rafael},
  title = {Geometry of the quantum set on no-signaling faces},
  journal = {Phys. Rev. A},
  volume = {99},
  number = {3},
  pages = {032106},
  year = {2019},
  publisher = {American Physical Society},
  doi = {10.1103/PhysRevA.99.032106},
  url = {https://link.aps.org/doi/10.1103/PhysRevA.99.032106},
}

@article{brunner2009nonlocality,
  author = {Brunner, Nicolas and Skrzypczyk, Paul},
  title = {Nonlocality distillation and postquantum theories with trivial communication complexity},
  journal = {Phys. Rev. Lett.},
  volume = {102},
  number = {16},
  pages = {160403},
  year = {2009},
  publisher = {American Physical Society},
  doi = {10.1103/PhysRevLett.102.160403},
  url = {https://link.aps.org/doi/10.1103/PhysRevLett.102.160403},
}

@article{hsu2012multipartite,
  author = {Hsu, Li-Yi},
  title = {Multipartite information causality},
  journal = {Phys. Rev. A},
  volume = {85},
  number = {3},
  pages = {032115},
  year = {2012},
  publisher = {APS},
  doi = {10.1103/PhysRevA.85.032115},
  url = {https://doi.org/10.1103/PhysRevA.85.032115},
}

@article{miklin2021information,
  author = {Miklin, Nikolai and Paw{\l}owski, Marcin},
  title = {Information causality without concatenation},
  journal = {Phys. Rev. Lett.},
  volume = {126},
  number = {22},
  pages = {220403},
  year = {2021},
  publisher = {APS},
  doi = {10.1103/PhysRevLett.126.220403},
  url = {https://doi.org/10.1103/PhysRevLett.126.220403},
}

@article{pollyceno2023information,
  author = {Pollyceno, Lucas and Chaves, Rafael and Rabelo, Rafael},
  title = {Information causality in multipartite scenarios},
  journal = {Phys. Rev. A},
  volume = {107},
  number = {4},
  pages = {042203},
  year = {2023},
  publisher = {APS},
  doi = {10.1103/PhysRevA.107.042203},
  url = {https://doi.org/10.1103/PhysRevA.107.042203},
}

@article{brassard2006limit,
  author = {Brassard, Gilles and Buhrman, Harry and Linden, Noah and M{\'e}thot, Andr{\'e} Allan and Tapp, Alain and Unger, Falk},
  title = {Limit on nonlocality in any world in which communication complexity is not trivial},
  journal = {Phys. Rev. Lett.},
  volume = {96},
  number = {25},
  pages = {250401},
  year = {2006},
  publisher = {American Physical Society},
  doi = {10.1103/PhysRevLett.96.250401},
  url = {https://link.aps.org/doi/10.1103/PhysRevLett.96.250401},
}

@article{linden2007quantum,
  author = {Linden, Noah and Popescu, Sandu and Short, Anthony J and Winter, Andreas},
  title = {Quantum nonlocality and beyond: limits from nonlocal computation},
  journal = {Phys. Rev. Lett.},
  volume = {99},
  number = {18},
  pages = {180502},
  year = {2007},
  publisher = {American Physical Society},
  doi = {10.1103/PhysRevLett.99.180502},
  url = {https://link.aps.org/doi/10.1103/PhysRevLett.99.180502},
}

@article{bennett1992communication,
  author = {Bennett, Charles H and Wiesner, Stephen J},
  title = {Communication via one- and two-particle operators on {Einstein-Podolsky-Rosen} states},
  journal = {Phys. Rev. Lett.},
  volume = {69},
  number = {20},
  pages = {2881},
  year = {1992},
  publisher = {American Physical Society},
  doi = {10.1103/PhysRevLett.69.2881},
  url = {https://link.aps.org/doi/10.1103/PhysRevLett.69.2881},
}

@article{lindblad1975completely,
  author = {Lindblad, G{\"o}ran},
  title = {Completely positive maps and entropy inequalities},
  journal = {Commun. Math. Phys.},
  volume = {40},
  number = {2},
  pages = {147--151},
  year = {1975},
  publisher = {Springer},
  doi = {10.1007/BF01609396},
  url = {https://doi.org/10.1007/BF01609396},
}

@article{lieb1973convex,
  author = {Lieb, Elliott H.},
  title = {Convex trace functions and the {Wigner-Yanase-Dyson} conjecture},
  journal = {Adv. Math.},
  volume = {11},
  number = {3},
  pages = {267--288},
  year = {1973},
  month = dec,
  publisher = {Elsevier},
  doi = {10.1016/0001-8708(73)90011-X},
  url = {https://doi.org/10.1016/0001-8708(73)90011-X},
}

@article{ringbauer2017probing,
  author = {Ringbauer, Martin and Chaves, Rafael},
  title = {Probing the non-classicality of temporal correlations},
  journal = {Quantum},
  volume = {1},
  pages = {35},
  year = {2017},
  publisher = {Verein zur F{\"o}rderung des Open Access Publizierens in den Quantenwissenschaften},
  doi = {10.22331/q-2017-11-25-35},
  url = {https://doi.org/10.22331/q-2017-11-25-35},
}

@article{chaves2015device,
  author = {Chaves, Rafael and Brask, Jonatan Bohr and Brunner, Nicolas},
  title = {Device-independent tests of entropy},
  journal = {Phys. Rev. Lett.},
  volume = {115},
  number = {11},
  pages = {110501},
  year = {2015},
  publisher = {APS},
  doi = {10.1103/PhysRevLett.115.110501},
  url = {https://doi.org/10.1103/PhysRevLett.115.110501},
}

@article{pawlowski2009information,
  author = {Paw{\l}owski, Marcin and Paterek, Tomasz and Kaszlikowski, Dagomir and Scarani, Valerio and Winter, Andreas and {\.Z}ukowski, Marek},
  title = {Information causality as a physical principle},
  journal = {Nature},
  volume = {461},
  number = {7267},
  pages = {1101--1104},
  year = {2009},
  publisher = {Nature Publishing Group},
  doi = {10.1038/nature08400},
  url = {https://doi.org/10.1038/nature08400},
}

@article{chaves2015information,
  author = {Chaves, Rafael and Majenz, Christian and Gross, David},
  title = {Information--theoretic implications of quantum causal structures},
  journal = {Nat. Commun.},
  volume = {6},
  number = {1},
  pages = {5766},
  year = {2015},
  publisher = {Nature Publishing Group},
  doi = {10.1038/ncomms6766},
  url = {https://doi.org/10.1038/ncomms6766},
}

@article{Ahrens2012,
  author = {Ahrens, Johan and Badzi{\k{a}}g, Piotr and Cabello, Ad{\'a}n and Bourennane, Mohamed},
  title = {Experimental device-independent tests of classical and quantum dimensions},
  journal = {Nat. Phys.},
  volume = {8},
  number = {8},
  pages = {592--595},
  year = {2012},
  month = jun,
  publisher = {Springer},
  doi = {10.1038/nphys2333},
  url = {https://doi.org/10.1038/nphys2333},
}

@article{Anwer2021,
  author = {Anwer, Hammad and Wilson, Natalie and Silva, Ralph and Muhammad, Sadiq and Tavakoli, Armin and Bourennane, Mohamed},
  title = {Noise-robust preparation contextuality shared between any number of observers via unsharp measurements},
  journal = {Quantum},
  volume = {5},
  pages = {551},
  year = {2021},
  month = sep,
  publisher = {Verein zur F{\"o}rderung des Open Access Publizierens in den Quantenwissenschaften},
  doi = {10.22331/q-2021-09-28-551},
  url = {https://doi.org/10.22331/q-2021-09-28-551},
}

@article{Gouveia2013,
  author = {Jo{\~a}o Gouveia and Pablo A. Parrilo and Rekha R. Thomas},
  title = {Lifts of convex sets and cone factorizations},
  journal = {Math. Oper. Res.},
  volume = {38},
  number = {2},
  pages = {248--264},
  year = {2013},
  doi = {10.1287/moor.1120.0575},
  url = {https://doi.org/10.1287/moor.1120.0575},
}

@article{Fawzi2015,
  author = {Hamza Fawzi and Jo{\~a}o Gouveia and Pablo A. Parrilo and Richard Z. Robinson and Rekha R. Thomas},
  title = {Positive semidefinite rank},
  journal = {Math. Program.},
  volume = {153},
  number = {1},
  pages = {133--177},
  year = {2015},
  doi = {10.1007/s10107-015-0922-1},
  url = {https://doi.org/10.1007/s10107-015-0922-1},
}

@article{Fiorini2015,
  author = {Samuel Fiorini and Serge Massar and Sebastian Pokutta and Hans Raj Tiwary and Ronald de Wolf},
  title = {Exponential Lower Bounds for Polytopes in Combinatorial Optimization},
  journal = {J. ACM},
  volume = {62},
  number = {2},
  pages = {17},
  year = {2015},
  doi = {10.1145/2716307},
  url = {https://doi.org/10.1145/2716307},
}

@article{Tavakoli2020indep,
  author = {Tavakoli, Armin},
  title = {Semi-Device-Independent Certification of Independent Quantum State and Measurement Devices},
  journal = {Phys. Rev. Lett.},
  volume = {125},
  number = {15},
  pages = {150503},
  numpages = {6},
  year = {2020},
  month = oct,
  publisher = {American Physical Society},
  doi = {10.1103/PhysRevLett.125.150503},
  url = {https://link.aps.org/doi/10.1103/PhysRevLett.125.150503},
}

@article{Li2018,
  author = {Li, Hong-Wei and Zhang, Yong-Sheng and An, Xue-Bi and Han, Zheng-Fu and Guo, Guang-Can},
  title = {Three-observer classical dimension witness violation with weak measurement},
  journal = {Commun. Phys.},
  volume = {1},
  number = {1},
  pages = {10},
  year = {2018},
  month = mar,
  doi = {10.1038/s42005-018-0011-x},
  url = {https://doi.org/10.1038/s42005-018-0011-x},
}

@article{Bowles2014,
  author = {Bowles, Joseph and Quintino, Marco T\'ulio and Brunner, Nicolas},
  title = {Certifying the Dimension of Classical and Quantum Systems in a Prepare-and-Measure Scenario with Independent Devices},
  journal = {Phys. Rev. Lett.},
  volume = {112},
  number = {14},
  pages = {140407},
  numpages = {5},
  year = {2014},
  month = apr,
  publisher = {American Physical Society},
  doi = {10.1103/PhysRevLett.112.140407},
  url = {https://link.aps.org/doi/10.1103/PhysRevLett.112.140407},
}

@inproceedings{Nayak1999,
  author = {Nayak, Ashwin},
  title = {Optimal Lower Bounds for Quantum Automata and Random Access Codes},
  booktitle = {Proceedings of the 40th Annual Symposium on Foundations of Computer Science},
  series = {FOCS '99},
  pages = {369},
  year = {1999},
  publisher = {IEEE Computer Society},
  isbn = {0769504094},
  doi = {10.1109/SFFCS.1999.814608},
  url = {https://doi.org/10.1109/SFFCS.1999.814608},
}

@article{Roch2025,
  author = {Roch i Carceller, Carles and Pauwels, Jef and Pironio, Stefano and Tavakoli, Armin},
  title = {Prepare-and-Measure Scenarios with Photon-Number Constraints},
  journal = {Phys. Rev. Lett.},
  volume = {135},
  number = {14},
  pages = {140802},
  numpages = {8},
  year = {2025},
  month = oct,
  publisher = {American Physical Society},
  doi = {10.1103/glty-kkbp},
  url = {https://link.aps.org/doi/10.1103/glty-kkbp},
}

@article{BrunnerReview2014,
  author = {Brunner, Nicolas and Cavalcanti, Daniel and Pironio, Stefano and Scarani, Valerio and Wehner, Stephanie},
  title = {{Bell} nonlocality},
  journal = {Rev. Mod. Phys.},
  volume = {86},
  number = {2},
  pages = {419--478},
  numpages = {60},
  year = {2014},
  month = apr,
  publisher = {American Physical Society},
  doi = {10.1103/RevModPhys.86.419},
  url = {https://link.aps.org/doi/10.1103/RevModPhys.86.419},
}

@article{Abramson01101970,
  author = {Morton Abramson and W. O. J. Moser},
  title = {More Birthday Surprises},
  journal = {Am. Math. Mon.},
  volume = {77},
  number = {8},
  pages = {856--858},
  year = {1970},
  publisher = {Taylor \& Francis},
  doi = {10.1080/00029890.1970.11992600},
  url = {https://doi.org/10.1080/00029890.1970.11992600},
}

@article{jones2025certifiedrandomnessquantumspeed,
  author = {Caroline L. Jones and Albert Aloy and Gerard Higgins and Markus P. M{\"u}ller},
  title = {Certified randomness from quantum speed limits},
  journal = {PRX Quantum},
  volume = {7},
  number = {1},
  pages = {010349},
  year = {2026},
  doi = {10.1103/bnrq-sp5w},
  eprint = {2506.14526},
  archivePrefix = {arXiv},
  primaryClass = {quant-ph},
  url = {https://doi.org/10.1103/bnrq-sp5w},
}

@article{Aloy2024,
  author = {Aloy, Albert and Galley, Thomas D. and Jones, Caroline L. and Ludescher, Stefan L. and M{\"u}ller, Markus P.},
  title = {Spin-Bounded Correlations: Rotation Boxes Within and Beyond Quantum Theory},
  journal = {Commun. Math. Phys.},
  volume = {405},
  number = {12},
  pages = {292},
  year = {2024},
  month = nov,
  doi = {10.1007/s00220-024-05123-2},
  url = {https://doi.org/10.1007/s00220-024-05123-2},
}

@article{TavakoliUolaPauwels2025,
  author = {Tavakoli, Armin and Uola, Roope and Pauwels, Jef},
  title = {Binarization of multioutcome measurements in high-dimensional quantum correlation experiments},
  journal = {Phys. Rev. A},
  volume = {111},
  number = {4},
  pages = {042433},
  numpages = {7},
  year = {2025},
  month = {Apr},
  publisher = {American Physical Society},
  doi = {10.1103/PhysRevA.111.042433},
  url = {https://link.aps.org/doi/10.1103/PhysRevA.111.042433},
}

@article{Jones2026theoryindependent,
  author = {Jones, Caroline L. and Ludescher, Stefan L. and Aloy, Albert and M{\"{u}}ller, Markus P.},
  title = {Theory-independent randomness generation from spatial symmetries},
  journal = {Quantum},
  volume = {10},
  pages = {1966},
  year = {2026},
  month = jan,
  publisher = {Verein zur F{\"o}rderung des Open Access Publizierens in den Quantenwissenschaften},
  doi = {10.22331/q-2026-01-16-1966},
  url = {https://doi.org/10.22331/q-2026-01-16-1966},
}

@article{Bakhshinezhad2024,
  author = {Bakhshinezhad, Pharnam and Mehboudi, Mohammad and Carceller, Carles Roch i and Tavakoli, Armin},
  title = {Scalable Entanglement Certification via Quantum Communication},
  journal = {PRX Quantum},
  volume = {5},
  number = {2},
  pages = {020319},
  numpages = {16},
  year = {2024},
  month = apr,
  publisher = {American Physical Society},
  doi = {10.1103/PRXQuantum.5.020319},
  url = {https://link.aps.org/doi/10.1103/PRXQuantum.5.020319},
}

@article{Brask2017,
  author = {Brask, Jonatan Bohr and Martin, Anthony and Esposito, William and Houlmann, Raphael and Bowles, Joseph and Zbinden, Hugo and Brunner, Nicolas},
  title = {Megahertz-Rate Semi-Device-Independent Quantum Random Number Generators Based on Unambiguous State Discrimination},
  journal = {Phys. Rev. Appl.},
  volume = {7},
  number = {5},
  pages = {054018},
  numpages = {10},
  year = {2017},
  month = may,
  publisher = {American Physical Society},
  doi = {10.1103/PhysRevApplied.7.054018},
  url = {https://link.aps.org/doi/10.1103/PhysRevApplied.7.054018},
}

@article{Pauwels2025informationcapacity,
  author = {Pauwels, Jef and Pironio, Stefano and Tavakoli, Armin},
  title = {Information capacity of quantum communication under natural physical assumptions},
  journal = {Quantum},
  volume = {9},
  pages = {1637},
  year = {2025},
  month = feb,
  publisher = {Verein zur F{\"o}rderung des Open Access Publizierens in den Quantenwissenschaften},
  doi = {10.22331/q-2025-02-18-1637},
  url = {https://doi.org/10.22331/q-2025-02-18-1637},
}

@article{Goeisletter,
  author = {de Gois, Carlos and Moreno, George and Nery, Ranieri and Brito, Samura\'{\i} and Chaves, Rafael and Rabelo, Rafael},
  title = {General Method for Classicality Certification in the Prepare and Measure Scenario},
  journal = {PRX Quantum},
  volume = {2},
  number = {3},
  pages = {030311},
  numpages = {12},
  year = {2021},
  month = jul,
  publisher = {American Physical Society},
  doi = {10.1103/PRXQuantum.2.030311},
  url = {https://link.aps.org/doi/10.1103/PRXQuantum.2.030311},
}

@article{Moreno2021,
  author = {Moreno, George and Nery, Ranieri and de Gois, Carlos and Rabelo, Rafael and Chaves, Rafael},
  title = {Semi-device-independent certification of entanglement in superdense coding},
  journal = {Phys. Rev. A},
  volume = {103},
  number = {2},
  pages = {022426},
  numpages = {11},
  year = {2021},
  month = feb,
  publisher = {American Physical Society},
  doi = {10.1103/PhysRevA.103.022426},
  url = {https://link.aps.org/doi/10.1103/PhysRevA.103.022426},
}

@article{bowles2015testing,
  author = {Bowles, Joseph and Brunner, Nicolas and Paw{\l}owski, Marcin},
  title = {Testing dimension and nonclassicality in communication networks},
  journal = {Phys. Rev. A},
  volume = {92},
  number = {2},
  pages = {022351},
  year = {2015},
  month = aug,
  publisher = {American Physical Society},
  doi = {10.1103/PhysRevA.92.022351},
  url = {https://link.aps.org/doi/10.1103/PhysRevA.92.022351},
}

@article{Gisin1999,
  author = {Gisin, N. and Gisin, B.},
  title = {A local hidden variable model of quantum correlation exploiting the detection loophole},
  journal = {Phys. Lett. A},
  volume = {260},
  number = {5},
  pages = {323--327},
  year = {1999},
  month = sep,
  publisher = {Elsevier},
  doi = {10.1016/S0375-9601(99)00519-8},
  url = {https://doi.org/10.1016/S0375-9601(99)00519-8},
}

@article{Sarubi2026,
  author = {Sarubi, Tailan S. and Zamora, Santiago and Alves, Mois{\'e}s and Alves, Vin{\'i}cius F. and Viswanathan, Gandhimohan M. and Chaves, Rafael},
  title = {Detection Efficiency Bounds in (Semi-)Device-Independent Scenarios},
  journal = {Braz. J. Phys.},
  volume = {56},
  number = {3},
  pages = {136},
  year = {2026},
  doi = {10.1007/s13538-026-02008-8},
  url = {https://doi.org/10.1007/s13538-026-02008-8},
}

@article{Designolle2023,
  author = {Designolle, S\'ebastien and Iommazzo, Gabriele and Besan{\c{c}}on, Mathieu and Knebel, Sebastian and Gel\ss{}, Patrick and Pokutta, Sebastian},
  title = {Improved local models and new {Bell} inequalities via {Frank-Wolfe} algorithms},
  journal = {Phys. Rev. Res.},
  volume = {5},
  number = {4},
  pages = {043059},
  numpages = {6},
  year = {2023},
  month = {Oct},
  publisher = {American Physical Society},
  doi = {10.1103/PhysRevResearch.5.043059},
  url = {https://link.aps.org/doi/10.1103/PhysRevResearch.5.043059},
}

@article{Acin2006G,
  author = {Ac\'{\i}n, Antonio and Gisin, Nicolas and Toner, Benjamin},
  title = {{Grothendieck}'s constant and local models for noisy entangled quantum states},
  journal = {Phys. Rev. A},
  volume = {73},
  number = {6},
  pages = {062105},
  numpages = {5},
  year = {2006},
  month = {Jun},
  publisher = {American Physical Society},
  doi = {10.1103/PhysRevA.73.062105},
  url = {https://link.aps.org/doi/10.1103/PhysRevA.73.062105},
}

@article{Divianszky2023,
  author = {Divi{\'{a}}nszky, P{\'{e}}ter and M{\'{a}}rton, Istv{\'{a}}n and Bene, Erika and V{\'{e}}rtesi, Tam{\'{a}}s},
  title = {Certification of qubits in the prepare-and-measure scenario with large input alphabet and connections with the {Grothendieck} constant},
  journal = {Sci. Rep.},
  volume = {13},
  number = {1},
  pages = {13200},
  year = {2023},
  month = aug,
  publisher = {Springer Sci. Bus. Media LLC},
  doi = {10.1038/s41598-023-39529-0},
  url = {https://doi.org/10.1038/s41598-023-39529-0},
}

@article{pusey2012reality,
  author = {Pusey, Matthew F. and Barrett, Jonathan and Rudolph, Terry},
  title = {On the reality of the quantum state},
  journal = {Nat. Phys.},
  volume = {8},
  number = {6},
  pages = {475--478},
  year = {2012},
  month = may,
  publisher = {Nature Publishing Group},
  doi = {10.1038/nphys2309},
  url = {https://doi.org/10.1038/nphys2309},
}

@article{Trojek2005,
  author = {Trojek, Pavel and Schmid, Christian and Bourennane, Mohamed and Brukner, {\v{C}}aslav and {\.{Z}}ukowski, Marek and Weinfurter, Harald},
  title = {Experimental quantum communication complexity},
  journal = {Phys. Rev. A},
  volume = {72},
  number = {5},
  pages = {050305(R)},
  numpages = {4},
  year = {2005},
  month = {Nov},
  publisher = {American Physical Society},
  doi = {10.1103/PhysRevA.72.050305},
  url = {https://link.aps.org/doi/10.1103/PhysRevA.72.050305},
}

@article{Marton2025,
  author = {M{\'{a}}rton, Istv{\'{a}}n and Bene, Erika and V{\'{e}}rtesi, Tam{\'{a}}s},
  title = {Bound entanglement-assisted prepare-and-measure scenarios based on four-dimensional quantum messages},
  journal = {Quantum Sci. Technol.},
  volume = {10},
  number = {4},
  pages = {04LT02},
  year = {2025},
  month = Oct,
  publisher = {IOP Publishing},
  issn = {2058-9565},
  doi = {10.1088/2058-9565/ae095f},
  url = {https://doi.org/10.1088/2058-9565/ae095f},
}

@article{TavakoliCarceller2026,
  author = {Tavakoli, Armin and Roch i Carceller, Carles and Tendick, Lucas and V{\'{e}}rtesi, Tam{\'{a}}s},
  title = {Unlimited quantum correlation advantage from bound entanglement},
  journal = {New J. Phys.},
  volume = {28},
  number = {3},
  pages = {034503},
  year = {2026},
  month = {mar},
  publisher = {IOP Publishing},
  doi = {10.1088/1367-2630/ae45c5},
  url = {https://doi.org/10.1088/1367-2630/ae45c5},
}

@article{Pauwels2022many,
  author = {Pauwels, Jef and Tavakoli, Armin and Woodhead, Erik and Pironio, Stefano},
  title = {Entanglement in prepare-and-measure scenarios: many questions, a few answers},
  journal = {New J. Phys.},
  volume = {24},
  number = {6},
  pages = {063015},
  year = {2022},
  month = jun,
  publisher = {IOP Publishing},
  doi = {10.1088/1367-2630/ac724a},
  url = {https://doi.org/10.1088/1367-2630/ac724a},
}

@article{Carceller2025,
  author = {Roch i Carceller, Carles and Tavakoli, Armin},
  title = {Bound Entangled States Are Useful in Prepare-and-Measure Scenarios},
  journal = {Phys. Rev. Lett.},
  volume = {134},
  number = {12},
  pages = {120203},
  numpages = {6},
  year = {2025},
  month = mar,
  publisher = {American Physical Society},
  doi = {10.1103/PhysRevLett.134.120203},
  url = {https://link.aps.org/doi/10.1103/PhysRevLett.134.120203},
}

@article{VanHimbeeck2017semidevice,
  author = {Van Himbeeck, Thomas and Woodhead, Erik and Cerf, Nicolas J. and Garc{\'{i}}a-Patr{\'{o}}n, Ra{\'{u}}l and Pironio, Stefano},
  title = {Semi-device-independent framework based on natural physical assumptions},
  journal = {Quantum},
  volume = {1},
  pages = {33},
  year = {2017},
  month = nov,
  publisher = {Verein zur F{\"o}rderung des Open Access Publizierens in den Quantenwissenschaften},
  doi = {10.22331/q-2017-11-18-33},
  url = {https://doi.org/10.22331/q-2017-11-18-33},
}

@article{dalessandro2026semidefiniteblockmatrixrelaxationscomputing,
  author = {D'Alessandro, Nicola and Roch i Carceller, Carles and Tavakoli, Armin},
  title = {Semidefinite Block-Matrix Relaxations for Computing Quantum Correlations},
  journal = {Phys. Rev. X},
  volume = {16},
  number = {3},
  pages = {031050},
  numpages = {21},
  year = {2026},
  month = {Aug},
  publisher = {American Physical Society},
  doi = {10.1103/6rh2-s3y2},
  url = {https://link.aps.org/doi/10.1103/6rh2-s3y2},
}

@article{Tavakoli2021b,
  author = {Tavakoli, Armin},
  title = {Semi-Device-Independent Framework Based on Restricted Distrust in Prepare-and-Measure Experiments},
  journal = {Phys. Rev. Lett.},
  volume = {126},
  number = {21},
  pages = {210503},
  numpages = {6},
  year = {2021},
  month = may,
  publisher = {American Physical Society},
  doi = {10.1103/PhysRevLett.126.210503},
  url = {https://link.aps.org/doi/10.1103/PhysRevLett.126.210503},
}

@article{Hayashi2006,
  author = {Hayashi, M and Iwama, K and Nishimura, H and Raymond, R and Yamashita, S},
  title = {(4,1)-Quantum random access coding does not exist---one qubit is not enough to recover one of four bits},
  journal = {New J. Phys.},
  volume = {8},
  number = {8},
  pages = {129--129},
  year = {2006},
  month = aug,
  publisher = {IOP Publishing},
  doi = {10.1088/1367-2630/8/8/129},
  url = {https://doi.org/10.1088/1367-2630/8/8/129},
}

@article{Ahrens2014,
  author = {Ahrens, Johan and Badzi{\k{a}}g, Piotr and Paw\l{}owski, Marcin and {\.Z}ukowski, Marek and Bourennane, Mohamed},
  title = {Experimental Tests of Classical and Quantum Dimensionality},
  journal = {Phys. Rev. Lett.},
  volume = {112},
  number = {14},
  pages = {140401},
  numpages = {5},
  year = {2014},
  month = apr,
  publisher = {American Physical Society},
  doi = {10.1103/PhysRevLett.112.140401},
  url = {https://link.aps.org/doi/10.1103/PhysRevLett.112.140401},
}

@article{Saha2023,
  author = {Saha, Debashis and Das, Debarshi and Das, Arun Kumar and Bhattacharya, Bihalan and Majumdar, A. S.},
  title = {Measurement incompatibility and quantum advantage in communication},
  journal = {Phys. Rev. A},
  volume = {107},
  number = {6},
  pages = {062210},
  numpages = {14},
  year = {2023},
  month = jun,
  publisher = {American Physical Society},
  doi = {10.1103/PhysRevA.107.062210},
  url = {https://link.aps.org/doi/10.1103/PhysRevA.107.062210},
}

@misc{Brask2026,
  author = {Jonatan Bohr Brask and Nicolas Brunner and Jef Pauwels and Davide Rusca and Armin Tavakoli},
  title = {Quantum correlations in prepare-and-measure scenarios and their semi-device-independent applications},
  year = {2026},
  eprint = {2603.23604},
  archivePrefix = {arXiv},
  primaryClass = {quant-ph},
  url = {https://arxiv.org/abs/2603.23604},
}

@article{CVXPY,
  author = {Steven Diamond and Stephen Boyd},
  title = {{CVXPY}: A {Python}-embedded modeling language for convex optimization},
  journal = {J. Mach. Learn. Res.},
  volume = {17},
  number = {83},
  pages = {1--5},
  year = {2016},
  url = {https://www.jmlr.org/papers/v17/15-408.html},
}

@misc{CVX,
  author = {{CVX Research, Inc.}},
  title = {{CVX}: {MATLAB} Software for Disciplined Convex Programming, version 2.0},
  year = {2012},
  month = aug,
  url = {https://cvxr.com/cvx},
}

@article{JuMP,
  author = {Iain Dunning and Joey Huchette and Miles Lubin},
  title = {{JuMP}: A modeling language for mathematical optimization},
  journal = {SIAM Rev.},
  volume = {59},
  number = {2},
  pages = {295--320},
  year = {2017},
  doi = {10.1137/15M1020575},
  url = {https://doi.org/10.1137/15M1020575},
}

@inproceedings{YALMIP,
  author = {Johan L{\"o}fberg},
  title = {{YALMIP}: A Toolbox for Modeling and Optimization in {MATLAB}},
  booktitle = {Proceedings of the IEEE International Symposium on Computer Aided Control Systems Design},
  year = {2004},
  doi = {10.1109/CACSD.2004.1393890},
  url = {https://doi.org/10.1109/CACSD.2004.1393890},
}

@misc{Svegborn2026,
  author = {Elna Svegborn and Armin Tavakoli},
  title = {Entanglement in prepare-and-measure scenarios without receiver inputs},
  year = {2026},
  eprint = {2603.29625},
  archivePrefix = {arXiv},
  primaryClass = {quant-ph},
  url = {https://arxiv.org/abs/2603.29625},
}

@article{Gallego2010,
  author = {Gallego, Rodrigo and Brunner, Nicolas and Hadley, Christopher and Ac\'{\i}n, Antonio},
  title = {Device-Independent Tests of Classical and Quantum Dimensions},
  journal = {Phys. Rev. Lett.},
  volume = {105},
  number = {23},
  pages = {230501},
  numpages = {4},
  year = {2010},
  month = nov,
  publisher = {American Physical Society},
  doi = {10.1103/PhysRevLett.105.230501},
  url = {https://link.aps.org/doi/10.1103/PhysRevLett.105.230501},
}

@article{PANDA2015,
  author = {Stefan L{\"{o}}rwald and Gerhard Reinelt},
  title = {{PANDA}: a software for polyhedral transformations},
  journal = {EURO J. Comput. Optim.},
  volume = {3},
  number = {4},
  pages = {297--308},
  year = {2015},
  issn = {2192-4406},
  doi = {https://doi.org/10.1007/s13675-015-0040-0},
  url = {https://www.sciencedirect.com/science/article/pii/S2192440621000484},
}

@article{Renner2023,
  author = {Renner, Martin J. and Tavakoli, Armin and Quintino, Marco T\'ulio},
  title = {Classical Cost of Transmitting a Qubit},
  journal = {Phys. Rev. Lett.},
  volume = {130},
  number = {12},
  pages = {120801},
  numpages = {7},
  year = {2023},
  month = mar,
  publisher = {American Physical Society},
  doi = {10.1103/PhysRevLett.130.120801},
  url = {https://link.aps.org/doi/10.1103/PhysRevLett.130.120801},
}

@article{aaronson2004improved,
  author = {Aaronson, Scott and Gottesman, Daniel},
  title = {Improved simulation of stabilizer circuits},
  journal = {Phys. Rev. A},
  volume = {70},
  number = {5},
  pages = {052328},
  numpages = {14},
  year = {2004},
  month = nov,
  publisher = {American Physical Society},
  doi = {10.1103/PhysRevA.70.052328},
  url = {https://link.aps.org/doi/10.1103/PhysRevA.70.052328},
}

@article{Veitch_2014,
  author = {Victor Veitch and S A Hamed Mousavian and Daniel Gottesman and Joseph Emerson},
  title = {The resource theory of stabilizer quantum computation},
  journal = {New J. Phys.},
  volume = {16},
  number = {1},
  pages = {013009},
  year = {2014},
  month = jan,
  publisher = {IOP Publishing},
  doi = {10.1088/1367-2630/16/1/013009},
  url = {https://doi.org/10.1088/1367-2630/16/1/013009},
}

@inproceedings{gottesman1998heisenberg,
  author = {Gottesman, Daniel},
  editor = {Corney, S. P. and Delbourgo, R. and Jarvis, P. D.},
  title = {The {Heisenberg} representation of quantum computers},
  booktitle = {Group22: Proceedings of the XXII International Colloquium on Group Theoretical Methods in Physics},
  pages = {32--43},
  year = {1999},
  publisher = {International Press},
  address = {Cambridge, MA},
  eprint = {quant-ph/9807006},
  archivePrefix = {arXiv},
  url = {https://arxiv.org/abs/quant-ph/9807006},
}

@article{Macedo_2025,
  author = {Mac\^edo, R. A. and Andriolo, P. and Zamora, S. and Poderini, D. and Chaves, R.},
  title = {Witnessing nonstabilizerness with {Bell} inequalities},
  journal = {Phys. Rev. A},
  volume = {112},
  number = {5},
  pages = {L050401},
  numpages = {6},
  year = {2025},
  month = nov,
  publisher = {American Physical Society},
  doi = {10.1103/6srg-723m},
  url = {https://link.aps.org/doi/10.1103/6srg-723m},
}

@misc{zamora2026logical,
  author = {Zamora, Santiago and Chatterjee, Kaustav and Andersen, Ulrik Lund and Guedes, Thiago Lucena de Macedo and Brask, Jonatan Bohr and de Oliveira Junior, A. and Chaves, Rafael},
  title = {The logical set of {Gaussian} states under the stabilizer subsystem decomposition},
  year = {2026},
  eprint = {2609.26898},
  archivePrefix = {arXiv},
  primaryClass = {quant-ph},
  url = {https://arxiv.org/abs/2609.26898},
}

@article{Zamora_2025,
  author = {Zamora, Santiago and Mac\^edo, Rafael A. and Sarubi, Tailan S. and Alves, Mois\'es and Poderini, Davide and Chaves, Rafael},
  title = {Semi-device-independent nonstabilizerness certification in the prepare-and-measure scenario},
  journal = {Phys. Rev. A},
  volume = {112},
  number = {4},
  pages = {042410},
  numpages = {14},
  year = {2025},
  month = oct,
  publisher = {American Physical Society},
  doi = {10.1103/wm7m-xnfq},
  url = {https://link.aps.org/doi/10.1103/wm7m-xnfq},
}

@misc{patra2026certification,
  author = {Patra, Ram Krishna and Taoutioui, Abdelmalek and V{\'e}rtesi, Tam{\'a}s},
  title = {Certification of Absolutely Entangled Sets in the Prepare-and-Measure Scenario},
  year = {2026},
  eprint = {2609.13075},
  archivePrefix = {arXiv},
  primaryClass = {quant-ph},
  url = {https://arxiv.org/abs/2609.13075},
}

@article{Chowdhury_2026,
  author = {Chowdhury, Snehasish Roy and Naik, Sahil Gopalkrishna and Chakraborty, Ananya and Patra, Ram Krishna and Ghosh, Subhendu B. and Ghosal, Pratik and Banik, Manik and Maity, Ananda G.},
  title = {{Gottesman-Knill} limit on one-way communication complexity: Tracing the quantum advantage down to magic resources},
  journal = {Phys. Rev. A},
  volume = {113},
  number = {4},
  pages = {042607},
  numpages = {11},
  year = {2026},
  month = apr,
  publisher = {American Physical Society},
  doi = {10.1103/x741-4kd1},
  url = {https://link.aps.org/doi/10.1103/x741-4kd1},
}

@article{Poderini_2020,
  author = {Poderini, Davide and Brito, Samura\'{\i} and Nery, Ranieri and Sciarrino, Fabio and Chaves, Rafael},
  title = {Criteria for nonclassicality in the prepare-and-measure scenario},
  journal = {Phys. Rev. Res.},
  volume = {2},
  number = {4},
  pages = {043106},
  numpages = {13},
  year = {2020},
  month = oct,
  publisher = {American Physical Society},
  doi = {10.1103/PhysRevResearch.2.043106},
  url = {https://link.aps.org/doi/10.1103/PhysRevResearch.2.043106},
}

@article{Howard_2014,
  author = {Howard, Mark and Wallman, Joel and Veitch, Victor and Emerson, Joseph},
  title = {Contextuality supplies the `magic' for quantum computation},
  journal = {Nature},
  volume = {510},
  number = {7505},
  pages = {351--355},
  year = {2014},
  publisher = {Springer Nature},
  doi = {10.1038/nature13460},
  url = {https://doi.org/10.1038/nature13460},
}

@article{Barnett09,
  author = {Stephen M. Barnett and Sarah Croke},
  title = {Quantum state discrimination},
  journal = {Adv. Opt. Photon.},
  volume = {1},
  number = {2},
  pages = {238--278},
  year = {2009},
  month = {Apr},
  publisher = {Optica Publishing Group},
  doi = {10.1364/AOP.1.000238},
  url = {https://opg.optica.org/aop/abstract.cfm?URI=aop-1-2-238},
}

@article{Shi_2019,
  author = {Shi, Weixu and Cai, Yu and Brask, Jonatan Bohr and Zbinden, Hugo and Brunner, Nicolas},
  title = {Semi-device-independent characterization of quantum measurements under a minimum overlap assumption},
  journal = {Phys. Rev. A},
  volume = {100},
  number = {4},
  pages = {042108},
  year = {2019},
  month = Oct,
  publisher = {American Physical Society (APS)},
  issn = {2469-9934},
  doi = {10.1103/physreva.100.042108},
  url = {https://doi.org/10.1103/PhysRevA.100.042108},
}

@article{Spekkens2005,
  author = {Spekkens, R. W.},
  title = {Contextuality for preparations, transformations, and unsharp measurements},
  journal = {Phys. Rev. A},
  volume = {71},
  number = {5},
  pages = {052108},
  numpages = {17},
  year = {2005},
  month = {May},
  publisher = {American Physical Society},
  doi = {10.1103/PhysRevA.71.052108},
  url = {https://link.aps.org/doi/10.1103/PhysRevA.71.052108},
}

@article{Hendrych2003,
  author = {Martin Hendrych and Miloslav Du{\v{s}}ek and Radim Filip and Jarom{\'{\i}}r Fiur{\'{a}}{\v{s}}ek},
  title = {Simple optical measurement of the overlap and fidelity of quantum states},
  journal = {Phys. Lett. A},
  volume = {310},
  number = {2},
  pages = {95--100},
  year = {2003},
  issn = {0375-9601},
  doi = {https://doi.org/10.1016/S0375-9601(03)00262-7},
  url = {https://www.sciencedirect.com/science/article/pii/S0375960103002627},
}

@article{Ekert_2002,
  author = {Ekert, Artur K. and Alves, Carolina Moura and Oi, Daniel K. L. and Horodecki, Micha\l{} and Horodecki, Pawe\l{} and Kwek, L. C.},
  title = {Direct Estimations of Linear and Nonlinear Functionals of a Quantum State},
  journal = {Phys. Rev. Lett.},
  volume = {88},
  number = {21},
  pages = {217901},
  numpages = {4},
  year = {2002},
  month = {May},
  publisher = {American Physical Society},
  doi = {10.1103/PhysRevLett.88.217901},
  url = {https://link.aps.org/doi/10.1103/PhysRevLett.88.217901},
}

@article{Fanizza_2020,
  author = {Fanizza, M. and Rosati, M. and Skotiniotis, M. and Calsamiglia, J. and Giovannetti, V.},
  title = {Beyond the Swap Test: Optimal Estimation of Quantum State Overlap},
  journal = {Phys. Rev. Lett.},
  volume = {124},
  number = {6},
  pages = {060503},
  numpages = {6},
  year = {2020},
  month = {Feb},
  publisher = {American Physical Society},
  doi = {10.1103/PhysRevLett.124.060503},
  url = {https://link.aps.org/doi/10.1103/PhysRevLett.124.060503},
}

@article{Calsamiglia2002,
  author = {Calsamiglia, John},
  title = {Generalized measurements by linear elements},
  journal = {Phys. Rev. A},
  volume = {65},
  number = {3},
  pages = {030301(R)},
  numpages = {4},
  year = {2002},
  month = {Feb},
  publisher = {American Physical Society},
  doi = {10.1103/PhysRevA.65.030301},
  url = {https://link.aps.org/doi/10.1103/PhysRevA.65.030301},
}

@article{Bianchi2026,
  author = {Bianchi, Luca and Marconi, Carlo and Bacco, Davide},
  title = {{Bell} state measurements in quantum optics: a review of recent progress and open challenges},
  journal = {Quantum Sci. Technol.},
  volume = {11},
  number = {2},
  pages = {023001},
  year = {2026},
  month = {apr},
  publisher = {IOP Publishing},
  doi = {10.1088/2058-9565/ae609e},
  url = {https://doi.org/10.1088/2058-9565/ae609e},
}

@article{Grice2011,
  author = {Grice, W. P.},
  title = {Arbitrarily complete {Bell}-state measurement using only linear optical elements},
  journal = {Phys. Rev. A},
  volume = {84},
  number = {4},
  pages = {042331},
  numpages = {6},
  year = {2011},
  month = {Oct},
  publisher = {American Physical Society},
  doi = {10.1103/PhysRevA.84.042331},
  url = {https://link.aps.org/doi/10.1103/PhysRevA.84.042331},
}

@article{Ewert2014,
  author = {Ewert, Fabian and van Loock, Peter},
  title = {$3/4$-Efficient {Bell} Measurement with Passive Linear Optics and Unentangled Ancillae},
  journal = {Phys. Rev. Lett.},
  volume = {113},
  number = {14},
  pages = {140403},
  numpages = {5},
  year = {2014},
  month = {Sep},
  publisher = {American Physical Society},
  doi = {10.1103/PhysRevLett.113.140403},
  url = {https://link.aps.org/doi/10.1103/PhysRevLett.113.140403},
}

@article{Herrero-Collantes2017,
  author = {Herrero-Collantes, Miguel and Garcia-Escartin, Juan Carlos},
  title = {Quantum random number generators},
  journal = {Rev. Mod. Phys.},
  volume = {89},
  number = {1},
  pages = {015004},
  numpages = {48},
  year = {2017},
  month = {Feb},
  publisher = {American Physical Society},
  doi = {10.1103/RevModPhys.89.015004},
  url = {https://link.aps.org/doi/10.1103/RevModPhys.89.015004},
}

@article{Ma2016,
  author = {Ma, Xiongfeng and Yuan, Xiao and Cao, Zhu and Qi, Bing and Zhang, Zhen},
  title = {Quantum random number generation},
  journal = {npj Quantum Inf.},
  volume = {2},
  pages = {16021},
  year = {2016},
  doi = {10.1038/npjqi.2016.21},
  url = {https://doi.org/10.1038/npjqi.2016.21},
}

@article{Mannalath2023,
  author = {Mannalath, Vaisakh and Mishra, Sandeep and Pathak, Anirban},
  title = {A comprehensive review of quantum random number generators: concepts, classification and the origin of randomness},
  journal = {Quantum Inf. Process.},
  volume = {22},
  pages = {439},
  year = {2023},
  doi = {10.1007/s11128-023-04175-y},
  url = {https://doi.org/10.1007/s11128-023-04175-y},
}

@article{Mahato2026,
  author = {Mahato, Shyam Sundar},
  title = {Certified Quantum Randomness: Device-Independent and Semi-Device-Independent Approaches, Experimental Realizations, and Applications},
  journal = {Int. J. Theor. Phys.},
  volume = {65},
  pages = {71},
  year = {2026},
  doi = {10.1007/s10773-026-06288-3},
  url = {https://doi.org/10.1007/s10773-026-06288-3},
}

@article{Pirandola2020,
  author = {S. Pirandola and U. L. Andersen and L. Banchi and M. Berta and D. Bunandar and R. Colbeck and D. Englund and T. Gehring and C. Lupo and C. Ottaviani and J. L. Pereira and M. Razavi and J. Shamsul Shaari and M. Tomamichel and V. C. Usenko and G. Vallone and P. Villoresi and P. Wallden},
  title = {Advances in quantum cryptography},
  journal = {Adv. Opt. Photon.},
  volume = {12},
  number = {4},
  pages = {1012--1236},
  year = {2020},
  doi = {10.1364/AOP.361502},
  url = {https://doi.org/10.1364/AOP.361502},
}

@article{Zapatero2023,
  author = {Zapatero, V{\'{i}}ctor and van Leent, Tim and Arnon-Friedman, Rotem and Liu, Wen-Zhao and Zhang, Qiang and Weinfurter, Harald and Curty, Marcos},
  title = {Advances in device-independent quantum key distribution},
  journal = {npj Quantum Inf.},
  volume = {9},
  pages = {10},
  year = {2023},
  doi = {10.1038/s41534-023-00684-x},
  url = {https://doi.org/10.1038/s41534-023-00684-x},
}

@article{Tomamichel2011,
  author = {Tomamichel, Marco and Schaffner, Christian and Smith, Adam and Renner, Renato},
  title = {Leftover Hashing Against Quantum Side Information},
  journal = {IEEE Trans. Inf. Theory},
  volume = {57},
  number = {8},
  pages = {5524--5535},
  year = {2011},
  doi = {10.1109/TIT.2011.2158473},
  url = {https://doi.org/10.1109/TIT.2011.2158473},
}

@article{Lo2012,
  author = {Lo, Hoi-Kwong and Curty, Marcos and Qi, Bing},
  title = {Measurement-Device-Independent Quantum Key Distribution},
  journal = {Phys. Rev. Lett.},
  volume = {108},
  pages = {130503},
  year = {2012},
  doi = {10.1103/PhysRevLett.108.130503},
  url = {https://doi.org/10.1103/PhysRevLett.108.130503},
}

@article{Rusca2019,
  author = {Rusca, Davide and van Himbeeck, Thomas and Martin, Anthony and Brask, Jonatan Bohr and Shi, Weixu and Pironio, Stefano and Brunner, Nicolas and Zbinden, Hugo},
  title = {Self-testing quantum random-number generator based on an energy bound},
  journal = {Phys. Rev. A},
  volume = {100},
  pages = {062338},
  year = {2019},
  doi = {10.1103/PhysRevA.100.062338},
  url = {https://doi.org/10.1103/PhysRevA.100.062338},
}

@misc{deGois2026infinite,
  author        = {de Gois, Carlos and Santos, Thyago S. R. and Vieira, Carlos},
  title         = {Quantum communication and {Bell} nonlocality require infinite classical communication to simulate},
  year          = {2026},
  eprint        = {2609.04182},
  archivePrefix = {arXiv},
  primaryClass  = {quant-ph},
  url = {https://arxiv.org/abs/2609.04182}
}

@misc{yang2026,
      title={Robust Bell Nonlocality from Gottesman-Kitaev-Preskill States}, 
      author={Xiaotian Yang and Santiago Zamora and Rafael Chaves and Ulrik L. Andersen and Jonatan Bohr Brask and A. de Oliveira Junior},
      year={2026},
      eprint={2601.16189},
      archivePrefix={arXiv},
      primaryClass={quant-ph},
      url={https://arxiv.org/abs/2601.16189}, 
}

@article{Alves2026,
  title = {Semi-device-independent randomness certification on discretized continuous-variable platforms},
  author = {Alves, Mois{\'e}s and Sena, Vitor L. and Zamora, Santiago and Sarubi, Tailan S. and de Oliveira Junior, A. and Tacla, Alexandre B. and Chaves, Rafael},
  journal = {Phys. Rev. A},
  volume = {113},
  issue = {4},
  pages = {042418},
  numpages = {17},
  year = {2026},
  month = {Apr},
  publisher = {American Physical Society},
  doi = {10.1103/z3s5-bhb7},
  url = {https://link.aps.org/doi/10.1103/z3s5-bhb7}
}

@misc{stratton_2026,
  author = {Benjamin Stratton and Santiago Zamora and Rafael Chaves},
  title = {Quantum State Discrimination With Stabilizer Circuits},
  year = {2026},
  eprint = {2607.27339},
  archivePrefix = {arXiv},
  primaryClass = {quant-ph},
  url = {https://arxiv.org/abs/2607.27339},
}

@misc{Vieira2025outcome,
  author        = {Vieira, Carlos and de Gois, Carlos and Lauand, Pedro and Porto, Lucas E. A. and Designolle, S{\'e}bastien and Quintino, Marco T{\'u}lio},
  title         = {Can outcome communication explain {Bell} nonlocality?},
  year          = {2025},
  eprint        = {2510.12886},
  archivePrefix = {arXiv},
  primaryClass  = {quant-ph},
  url = {https://arxiv.org/abs/2510.12886}
}

@article{Guo_2019,
  author = {Guo, Yu and Liu, Bi-Heng and Li, Chuan-Feng and Guo, Guang-Can},
  title = {Advances in Quantum Dense Coding},
  journal = {Adv. Quantum Technol.},
  volume = {2},
  number = {5-6},
  pages = {1900011},
  year = {2019},
  doi = {https://doi.org/10.1002/qute.201900011},
  url = {https://doi.org/10.1002/qute.201900011},
}

\newpage
\appendix
\onecolumngrid

\section{Weak and strong duality}\label{App:weak-and-strong-duality}

The optimisation methods in this tutorial address two complementary questions: what value can a strategy achieve, and how can we prove that no strategy does better? Duality connects these questions. The original problem, called the primal, searches for feasible choices of its variables, whereas the dual searches for bounds on their performance. Here we explain this connection. All optimisation problems in this appendix are finite-dimensional.

Consider the primal linear program $\beta_{\textrm{P}} = \max_{\v x} \{\v c^{\textsf{T}} \v x : A \v x \leq \v b\,, \, \v x\ge0\}$. The inequalities are understood componentwise, and a vector satisfying all constraints is called \emph{feasible}. To construct an upper bound, we multiply each inequality in $A \v x \leq \v b$ by a nonnegative number $y_i$. Adding these inequalities together gives $\v y^\top A\v x \leq \v y^\top \v b$. If we also choose $A^\top \v y \geq \v c$, then $\v x \geq 0$ implies $\v c^\top \v x\leq(A^\top \v y)^\top \v x = \v y^\top A \v x \leq \v b^\top \v y$. Thus every such $\v y$ certifies an upper bound, whether or not it is optimal. Searching for the smallest bound gives the dual LP from Eq.~\eqref{LPdualdefinition}, $\beta_{\textrm{D}} = \min_{\v y} \{\v b^\top \v y: A^\top \v y \geq \v c\,,\, \v y\geq0\}$. For any primal-feasible $\v x$ and dual-feasible $\v y$, we have $\v c^\top \v x \leq \beta_{\textrm{P}} \leq \beta_{\textrm{D}} \leq \v b^\top \v y$. The inequality $\beta_{\textrm{P}} \leq \beta_{\textrm{D}}$ is called weak duality. The difference between the two feasible objective values bounds how much either candidate could still improve. In particular, matching values prove that both candidates are optimal. Strong duality means that $\beta_{\textrm{P}} = \beta_{\textrm{D}}$, i.e.~the best achievable value equals the best dual bound. For an LP, if the primal is feasible and its objective is bounded above, both programs attain their optima and strong duality holds. Equality concerns the objective values, not the variables, which may even have different numbers of components.

The same idea applies to the SDP in Box~\ref{Box:SDP}. Let $C$, $A_i$ and $X$ be Hermitian matrices, with $\langle C,X\rangle = \tr(CX)$, and write
\begin{eqnarray}
    \beta_{\textrm{P}} = \sup_X \{\langle C,X\rangle : \langle A_i, X\rangle = b_i \: \text{with} \: i=1, \dots, m \: \text{and} \: X\succeq0\}.
\end{eqnarray}
Choose real coefficients $\v y = (y_1, \dots y_m)$ such that $S=\sum_i y_i A_i - C \succeq0$. For every feasible $X$, 
\begin{eqnarray}
    \sum_i b_i y_i -\langle C,X\rangle = \langle S, X\rangle = \tr(X^{\nicefrac12}S X^{\nicefrac12}) \geq0.
\end{eqnarray}
The last step holds because $X^{\nicefrac12}S X^{\nicefrac12}$ is positive semidefinite. Consequently, the dual searches for the best such bound: 
\begin{equation}\label{eq:duality-sdp-dual}
    \beta_{\textrm{D}} = \inf_{\v y\in\mathbbm{R}^m}\left\{\sum_i b_i y_i : \sum_i y_i A_i-C\succeq0\right\}.
\end{equation}
Unlike the LP above, the primal constraints are equalities, so their weights $y_i$ need not be nonnegative. Equation~\ref{eq:duality-sdp-dual} proves weak duality and again makes a matching feasible pair a certificate of optimality. We use $\sup$ and $\inf$ because an SDP need not attain its optimal values. For SDPs, feasibility of both programs alone does not guarantee strong duality. A useful sufficient condition is Slater's condition~\cite{boyd2004convex}. Assuming both programs are feasible, strong duality holds if either the primal has a feasible $X_0\succ0$ or the dual has coefficients satisfying $\sum_i y_i A_i-C\succ 0$. Here $\succ$ means positive definite (all eigenvalues are strictly positive). Strict primal feasibility also guarantees that the dual optimum is attained. Strict dual feasibility  guarantees that the primal optimum is attained. Thus, if both are strictly feasible, both optima are attained. These conditions are sufficient, not necessary and their failure does not by itself imply a nonzero duality gap. 

\section{Efficient linear program for membership in the classical polytope}\label{AppFastLP}
In Sec.~\hyperref[sec:ClassicalPolytope]{II-A-1} we derived the LP for deciding whether a given correlation $p(b|x,y)$ is a member of the classical polytope associated with dimension-restricted communication in the PM scenario. The LP was given by 
\begin{equation}\label{dimLP2}
\begin{aligned}
\text{find } & \quad \{q(\lambda)\}_{\lambda} \\[1ex]
\text{s.t.} &\quad q(\lambda) \ge 0, \quad \forall\, \lambda,\\
& \quad p(b|x,y) = \sum_{\lambda} q(\lambda)\, D_{\lambda}(b|x,y), \quad \forall\, b,x,y.
\end{aligned}
\end{equation}
where $D_\lambda$ are the distributions associated with all possible deterministic input/output assignments for Alice and Bob. There are $d^{|X|}|B|^{d|Y|}$ such assignments, which means that we have equally many variables $q(\lambda)$ in our LP. It is clear that beyond the few smallest alphabets, this number will be too large to make the evaluation of Eq.~\eqref{dimLP2} viable in practice.

Fortunately, as shown in Ref.~\cite{Renner2023}, there is a simple way to significantly speed up the computation of this LP. To this end, as we did in Sec.~\hyperref[sec:ClassicalModels]{II-A}, let us write $\lambda=(\mu,\nu)$ where $\mu$ indexes Alice's deterministic strategies and $\nu$ indexes Bob's deterministic strategies. Let us then write
\begin{align}\label{fastLPderivation}
p(b|x,y) &= \sum_{\lambda} q(\lambda)\, D_{\lambda}(b|x,y) =  \sum_{m=1}^{d}\sum_{\mu=1}^{d^{|X|}}\sum_{\nu=1}^{|B|^{d|Y|}} q(\mu,\nu)\, D_{\mu}(m|x) D_\nu(b|m,y) = \sum_{m=1}^d \sum_{\mu=1}^{d^{|X|}}  D_{\mu}(m|x) \pi(b|m,y,\mu)
\end{align}
where we have defined $\pi(b|m,y,\mu)=\sum_{\nu=1}^{|B|^{d|Y|}} q(\mu,\nu) D_\nu(b|m,y)$. Notice that $\pi(b|m,y,\mu)\geq 0$ and that $\sum_b \pi(b|m,y,\mu)=q(\mu) \quad \forall m,y,\mu$. This represents a full characterisation of $\pi$. We can therefore re-cast our original LP \eqref{dimLP2} as
\begin{equation}\label{dimLP3}
\begin{aligned}
\text{find } & \quad \{\pi(b|m,y,\mu)\}_{m,y,\mu} \quad \text{and} \quad \{q(\mu)\}_\mu \\[1ex]
\text{s.t.} &  \quad \sum_b \pi(b|m,y,\mu)=q(\mu) \quad \forall\ m,y,\mu\\
&\quad  p(b|x,y)=\sum_{m=1}^d \sum_{\mu=1}^{d^{|X|}}  D_{\mu}(m|x) \pi(b|m,y,\mu), \quad \forall\ b,x,y\\
&\quad \pi(b|m,y,\mu) \ge 0, \quad \forall\ b,m,y,\mu.
\end{aligned}
\end{equation}
We can now see why this is advantageous as compared to the LP in Eq.~\eqref{dimLP2}. In total, $q(\mu)$ gives us  $d^{|X|}$ variables and $\pi(b|m,y,\mu)$ gives us another $|B||Y|d^{|X|+1}$ variables. Thus, in total we have $d^{|X|}(1+d|B||Y|)$ variables. This is far fewer than in the LP~\eqref{dimLP2} because the exponential factor $|B|^{d|Y|}$ has been reduced to the much more modest $(1+d|B||Y|)$. Naturally, this comes at a cost: the LP~\eqref{dimLP3} has many more linear constraints than does the LP~\eqref{dimLP2}, but this is not nearly enough to outweigh the computational advantages gained from the variable reduction.

Note that in Eq.~\eqref{fastLPderivation} we chose to eliminate  $\nu$, but we could equally well have re-arranged the expression to eliminate  $\mu$ instead. Doing the latter would have led to an LP with $|B|^{d|Y|}(1+d|X|)$ variables. The choice of whether to eliminate $\nu$ or $\mu$ should be based on the alphabets of the given PM scenario: one should make the choice associated with the largest reduction of the number of variables in the LP.

\section{Proof of the facet inequality introduced in Box~\ref{Box:statediscriminationfacets} }\label{App:ElnasConstruction}

Consider that Alice has $n$ inputs that she encodes into a classical $d$-valued message, with $n>d\geq 2$. From the message, Bob must guess Alice's input. In this appendix we show that the sum-total of the success probabilities satisfies
\begin{equation}
    \sum_{x=1}^n p(b=x|x) \leq d,
\end{equation}
and that this is a facet of the classical polytope in the scenario $(n,d,1,n)$.
First we prove that the upper bound of the sum is indeed $d$. Consider all the possible deterministic strategies $b = h(x) = g \big(f(x)\big)$, where $f(x)=m$ is defined as Alice's deterministic response function $f:\{1,\dots,n\}\to\{1,\dots,d\}$ corresponding to her encoding, while the response function $g:\{1,\dots,d\}\to\{1,\dots,n\}$ corresponds to Bob's decoding $g(m)=b$. Because the intermediate message can take at most $d$ values, the image of $h$ has a size of at most $d$. The success probability for a given input $x$ is $1$ if $h(x)=x$, and $0$ otherwise. Therefore, the total number of successes across all $n$ inputs is exactly the number of fixed points of $h$. Since the image of $h$ is bounded by $d$, $h$ can have at most $d$ fixed points. Consequently, for any deterministic strategy, and thus for any probabilistic classical strategy (which is a convex combination of deterministic ones), we have:
\begin{equation}
    \sum_{x=1}^n p(b=x|x) \leq d,
\end{equation}
which proves the bound.
Now we show that it is indeed a facet of the classical polytope. To show this, we need to prove that there exists a set of $D$ affinely independent probability distributions (where $D$ is the dimension of the polytope) such that all of these points in $\mathbb{R}^D$ satisfy the equality
\begin{equation}
    \sum_{x=1}^n p(b=x|x) = d.
\end{equation}
We prove this by characterising the linear functionals that are constant on the saturating strategies, and then give an explicit construction for the simplest example. First note that any possible probability distribution can be represented by an $n\times n$ matrix $P_{xb}$ where each row corresponds to a particular choice of $x$ and each column to an outcome $b$. As such, any successful guess will imply a $1$ in the diagonal entry $P_{xx}$. Note that the normalisation condition for each $x$ on the probabilities implies that each row sums to 1, $\sum_bP_{xb} = 1$, leaving in total $n(n-1)$ free variables, corresponding to the dimension of the polytope $D=n(n-1)$.

Now, note that if we are looking for strategies that saturate the bound, all of them should have exactly $d$ diagonal elements with $1$. Denote by $E_{ij}$ the matrix with a $1$ in the entry $(i,j)$ and $0$ in all other entries. Therefore, a general deterministic strategy $P$ can be expressed as
\begin{equation}
    P = \sum_{x}E_{xh(x)}.
\end{equation}
Now consider the set $S_{0} = \{1,\dots,d\}$. Define the $P_{(0)}$ strategy as the strategy that acts as the identity on this set,  and maps all other inputs (rows) to column $1$. 
\begin{equation}
    P_{(0)} = \sum_{x\in S_0}E_{xx}  + \sum_{x\notin S_0}E_{x1} = \sum_{x = 1}^dE_{xx}  + \sum_{x=d+1}^{n}E_{x1}
\end{equation}
That is, $S_0$ is the set that defines Bob's $d$ successful guesses. From this first strategy we can build $(n-d)(d-1)$ other strategies $P^{(x,b)}$ by reordering the ones of the rows $x\in\{d+1,\dots,n\}$ in all the available $b\in\{2,\dots,d\}$ columns (note that only these columns are available since Bob has already chosen to guess the outputs in $S_0$). Formally, these strategies can be written as
\begin{equation}
    P^{(x,b)}_{(0)} = P_{(0)} - E_{x1} + E_{xb},
\end{equation}
for $x\in\{d+1,\dots,n\}$ and $b\in\{2,\dots,d\}$. Note that since this moves the $1$ in row $x$ from column $1$ to column $b$, it preserves the  normalisation. As a result, they are all valid strategies. 

To construct the other missing strategies we now change Bob's perfect guessing choices. Define the set $S_{k,l} = (S_0 \setminus\{k\})\cup\{l\}$ for $k\in\{1,\dots,d\}$ and $l\in\{d+1,\dots n\}$. For each of these sets, we generate a first matrix that acts as the identity over $S_{k,l}$, while mapping all the other inputs to column $b=l$:
\begin{equation}
P_{(k,l)} = \sum_{x\in S_{k,l}}E_{xx}  + \sum_{x\notin S_{k,l}}E_{xl}.
\end{equation}
From these matrices again we can build a family of strategies defined by:
\begin{equation}
P^{(x,b)}_{(k,l)} = P_{(k,l)} - E_{xl} + E_{xb},
\end{equation}
for $x\notin S_{k,l}$ and $b\in S_{k,l}\setminus\{l\}$.

Now, to prove affine independence, we take the matrix differences $\Delta P (\lambda) = P(\lambda)-P_{(0)}$, where $P(\lambda)$ is a deterministic strategy as defined above. The normalisation of each row implies that the first column can be safely removed since it can be written in terms of all the other $n-1$ columns.  The first family above provides independent differences; to establish the dimension of the full equality face, we characterise the linear functionals that are constant on all saturating strategies.

For the first family of strategies, we have 
\begin{equation}
    \Delta P^{(x,b)}_{(0)}=-E_{x1}+E_{xb},
\end{equation}
and after removing the first column the only non-zero entry is the coordinate $(x,b)$ where $x\in\{d+1,\dots,n\}$ and $b\in\{2,\dots,d\}$. As such, they form a linearly independent set. Now we continue with the matrices
\begin{equation}
\Delta P^{(x,b)}_{(k,l)} = P_{(k,l)} - P_{(0)} - E_{xl} + E_{xb}.
\end{equation}

Notice that for columns strictly greater than $d$, the newly introduced non-zero elements (after removing the first column) are at the coordinate $(x,b)$ where $x\notin S_{k,l}$ and $b\in S_{k,l}\setminus\{l\}$, as well as a $+1$ in the entry $(l,l)$ corresponding to $l\in S_{k,l}\setminus S_0$.   From these matrices, we only choose those that have a $1$ for $b>d$ where the previous matrices $\Delta P^{(x,b)}_{(0)}$  had a $0$ and that had not appeared previously. Proceeding recursively, we obtain exactly $n(n-1)-1$ linearly independent difference matrices. By simply adding  $P_{(0)}$, this gives $n(n-1)$ affinely independent saturating strategies.

To illustrate this construction, we explicitly write out the basis for the simplest non-trivial scenario. For $n=3$ inputs and $d=2$ messages, the classical probability distributions are $3 \times 3$  matrices. The affine space has $D = n(n-1) = 6$ dimensions. Thus, we require exactly 6 affinely independent matrices that saturate the bound (i.e.~have a trace equal to 2). Following the notation of our general construction, we select the base perfectly guessed set $S_0 = \{1, 2\}$.

$P_{(0)}$ strategy: Row 3 ($x \notin S_0$) is mapped to column 1.
\begin{equation}P_{(0)} = 
\begin{pmatrix}1 & 0 & 0 \\
0 & 1 & 0 \\
1 & 0 & 0
\end{pmatrix}
\end{equation}

$P^{(x,b)}_{(0)}$ strategies: We reorder the mapping of row 3 to the other available column in $S_0$ (column 2).
\begin{equation}P^{(3,2)}_{(0)} = 
\begin{pmatrix}
1 & 0 & 0 \\
0 & 1 & 0 \\
0 & 1 & 0\end{pmatrix}
\end{equation}

$P^{(x,b)}_{(k,l)} $ strategies: We shift the perfectly guessed set by swapping an element out of $S_0$.
First, we shift the set to $S_{1,3} = \{2, 3\}$, meaning rows 2 and 3 are perfectly guessed. The remaining row ($x=1$) is mapped to the available outputs in the new set ($b \in \{2, 3\}$):
\begin{equation}P^{(1,2)}_{(1,3)} = \begin{pmatrix}
0 & 1 & 0 \\
0 & 1 & 0 \\
0 & 0 & 1
\end{pmatrix}, \quad
P^{(1,3)}_{(1,3)} = \begin{pmatrix}
0 & 0 & 1 \\
0 & 1 & 0 \\
0 & 0 & 1\end{pmatrix}
\end{equation}
Finally, we shift the set to $S_{2,3} = \{1, 3\}$, meaning rows 1 and 3 are perfectly guessed. The remaining row ($x=2$) is mapped to the available outputs in this set ($b \in \{1, 3\}$):
\begin{equation}
P^{(2,1)}_{(2,3)} = \begin{pmatrix}
1 & 0 & 0 \\
1 & 0 & 0 \\
0 & 0 & 1
\end{pmatrix}, \quad
P^{(2,3)}_{(2,3)} = 
\begin{pmatrix}
1 & 0 & 0 
\\0 & 0 & 1 
\\0 & 0 & 1\end{pmatrix}.
\end{equation}

To explicitly prove that these strategies span the optimal hyperplane, we evaluate the 5 difference matrices $\Delta P^{(\lambda)} = P^{(\lambda)} - P_{(0)}$:
\begin{align}
\Delta P^{(3,2)}_{(0)} = \begin{pmatrix} 
0 & 0 & 0 \\ 
0 & 0 & 0 \\ 
-1 & 1 & 0 
\end{pmatrix}, \:\:\:
\Delta P^{(1,2)}_{(1,3)} = \begin{pmatrix} 
-1 & 1 & 0 \\ 
0 & 0 & 0 \\ 
-1 & 0 & 1 
\end{pmatrix}, \:\:\:
\Delta P^{(1,3)}_{(1,3)} = \begin{pmatrix} 
-1 & 0 & 1 \\ 
0 & 0 & 0 \\ 
-1 & 0 & 1 
\end{pmatrix}, \:\:\:
\Delta P^{(2,1)}_{(2,3)} = 
\begin{pmatrix} 
0 & 0 & 0 \\ 
1 & -1 & 0 \\ 
-1 & 0 & 1 \end{pmatrix}, \:\:\:
\Delta P^{(2,3)}_{(2,3)} = 
\begin{pmatrix} 
0 & 0 & 0 \\ 
0 & -1 & 1 \\ 
-1 & 0 & 1 \end{pmatrix}.
\end{align}

Note now that, since each difference matrix has rows summing to $0$ and trace $0$, there are effectively only five parameters: we can discard the first column and a diagonal element. We end up with $5$  relevant free coordinates $(x,b)$ of our difference matrices and we can map them  into flat row vectors. Let us select the coordinates in the specific order: $(3,2), (1,2), (1,3), (2,2), (2,3)$. Extracting these specific entries from our $5$ difference matrices yields the following matrix:
\begin{equation}
M = \begin{pmatrix}
1 & 0 & 0 & 0 & 0 \\
0 & 1 & 0 & 0 & 0 \\
0& 0 & 1 & 0 & 0 \\
0 & 0 & 0 & -1 & 0 \\
0 & 0 & 0 & -1 & 1
\end{pmatrix}
\end{equation}
The resulting coordinate matrix $M$ is lower-triangular with non-zero entries on the main diagonal. Its determinant is non-vanishing ($\det M = -1$), which proves that these 5 difference matrices are linearly independent. By adding the base strategy $P_{(0)}$ to this set, we obtain 6 affinely independent matrices that form a complete basis for the hyperplane, concluding the proof that the bound is a facet of the classical polytope.

\end{document}